\documentclass[%
 reprint,
 amsmath,amssymb,
 aps,
floatfix,
]{revtex4-1}

\usepackage{graphicx}
\usepackage{dcolumn}
\usepackage{bm}
\usepackage{slashed}
\usepackage{hyperref}

\begin{document}

\preprint{APS/123-QED}

\title{Calculation of a complete set of spin observables for proton elastic scattering \\from stable and unstable nuclei}

\author{W. A. Yahya}
\email{wasiu.yahya@gmail.com}
\affiliation{%
	Physics Department, Stellenbosch University, South Africa
}%
 \affiliation{%
 	Department of Physics and Materials Science, Kwara State University, Nigeria
 }%
 
\author{B. I. S. van der Ventel}%
 \email{bventel@sun.ac.za}
\affiliation{%
 Physics Department, Stellenbosch University, South Africa
}%

\author{R. A. Bark}%
\email{bark@tlabs.ac.za}
\affiliation{%
	iThemba LABS, Old Faure Road, Cape Town, South Africa
}%

\author{B. C. Kimene Kaya}
\email{christel@sun.ac.za}

\affiliation{%
	Physics Department, Stellenbosch University, South Africa
}%

\date{\today}

\begin{abstract}
A microscopic study of proton elastic scattering from unstable nuclei at intermediate energies using a relativistic formalism is presented. We have employed both the original relativistic impulse approximation (IA1) and the generalised impulse approximation (IA2) formalisms to calculate the relativistic optical potentials, with target densities derived from relativistic mean field (RMF) theory using the NL3 and FSUGold parameter sets. Comparisons between the optical potentials computed using both IA1 and IA2 formalisms, and the different RMF Lagrangians are presented for both stable and unstable targets. The comparisons are required to study the effect of using IA1 versus IA2 optical potentials, with different RMF parameter sets, on elastic scattering observables for unstable targets at intermediate energies. We also study the effect of full-folding versus the factorized form of the optical potentials on elastic scattering observables. As with the case for stable nuclei, we found that the use of the full-folding optical potential improves the scattering observables (especially spin observables) at low intermediate energy (e.g. 200MeV). No discernible difference is found at a projectile incident energy of 500 MeV. To check the validity of using localized optical potential, we calculate the scattering observables using non-local potentials by solving the momentum space Dirac equation. The Dirac equation is transformed to two coupled Lippmann-Schwinger equations, which are then numerically solved to obtain elastic scattering observables. The results are discussed and compared to calculations involving local coordinate-space optical potentials.
\begin{description}
\item[PACS numbers]
21.10.Gv, 24.10.Jv., 24.70.+s, 25.40.Cm

\end{description}
\end{abstract}

\maketitle


\section{\label{sec:level1}Introduction}

The availability of high-intensity radioactive ion beams (RIB) has made elastic and inelastic proton scattering from unstable nuclei available to study and the old theories of nuclear physics are now being tested in new settings, the limits of nuclear stability are being probed, and surprising results have been obtained thus far. Major surprises in low-energy nuclear structure include the disappearance of the normal shell closures observed near the stability valley, appearance of new magic numbers, exotic features of nuclear structure such as nuclear halos and skins, and new regions of deformation \cite{sakaguchi2017,nakamura2013}. Structure and reaction studies of unstable nuclei will have great impact on astrophysics because they are known to play an important role in nucleosynthesis. Radioactive ion beam facilities will make available large amount of unstable nuclei data, and will enhance the study of unstable nuclei via electron and proton scattering.  

One of the reaction processes to study both stable and unstable nuclei is elastic scattering. Employing electron and proton scattering, one can obtain information on the neutron ground state density and transition density distributions \cite{sakaguchi2017,alam04}.
At intermediate energies ($100 - 1000$ MeV), a good tool to probe nucleon density distributions is proton elastic scattering, due to its larger mean free path in the nuclear medium. The mean free path of intermediate energy protons in nuclear matter is large enough to penetrate into the nucleus, thus providing some sensitivity to the nuclear interior. The nuclear reaction mechanism becomes simpler at intermediate energies since the velocity of the projectile is much faster than the Fermi motion of the bound nucleons \cite{suo99, blum01, khan01, rash03}. A considerable number of works have therefore been devoted to proton elastic scattering to determine interactions and nuclear structures even in the nuclear interior. It has been stated that the best energy region to deduce the density distribution in nuclei is between 200 MeV and 400 MeV per nucleon, where the mean free path of the nucleon in nuclei is expected to be large and the scattering does not suffer much from meson production. The new facility at RIKEN (RIBF) will be able to supply the sufficient unstable nuclear beam in this energy region.

Elastic proton scattering yields information on the nuclear matter distributions and the effective nucleon-nucleon potentials. Inelastic scattering towards low lying collective states gives access to transition probabilities and nuclear deformations, and is a well suited tool to scan new regions of deformation. Proton scattering experiments on unstable nuclei are performed in inverse kinematics, where the radioactive beam strikes a target containing the protons. This is because the lifetime of unstable nuclei are too short to prepare as targets in most cases. In direct kinematics the light particle (in our case, proton) is accelerated onto the stationary heavy target, while in inverse kinematics the heavy particle is accelerated, and the light particle (proton) serves as the target. Very good sensitivity and high resolution are required for experiments in inverse kinematics in order to detect rare events with high efficiency and to have the maximum information possible with low statistics \cite{sakaguchi2017,nakamura2013}. It is sometimes experimentally difficult to detect the heavy fragment in inverse kinematics because of the short lifetime of unstable nuclei. Hence, the energy and angle of the recoiling protons are therefore measured for this type of reaction, from which the scattering angle and excitation energy can be deduced.  

Proton elastic and inelastic scattering studies of proton-rich $^{30}$S and $^{34}$Ar isotopes at 53 MeV/A and 47 MeV/A have been performed \cite{khan01}. Secondary beams from the MUST silicon detector array and GANIL facility were used in the experiment. It was found from the study that there was no indication of a proton skin in the two nuclei. Angular distributions of proton elastic scattering at 277--300 MeV per nucleon on $^{9} \mathrm{C}$ was studied in Ref. \cite{mat13}. The experiment was performed in inverse kinematics at GSI Darmstadt, and the relativistic impulse approximation was used to analyse the angular distribution. The recoil angle and recoil energy of the proton were measured using the recoil proton spectrometer they developed. At the same facility, $^{6} \mathrm{He}$, $^{8} \mathrm{He}$, $^{8} \mathrm{B}$, $^{6} \mathrm{Li}$, $^{8} \mathrm{Li}$, $^{9} \mathrm{Li}$, $^{11} \mathrm{Li}$, and $^{12,14} \mathrm{Be}$, have been studied at intermediate energies \cite{dobrovolsky2006,Kiselev2011,Korolev2018,Ilieva2012}. At RIKEN, the proton scattering of $^{16} \mathrm{C}$ at 300 MeV/A has been carried out in inverse kinematics \cite{terashima2014}. 

In this paper, proton elastic scattering from unstable nuclei at intermediate projectile laboratory energy is studied using the relativistic impulse approximation (IA1) and generalised relativistic impulse approximation (IA2) formalisms. To calculate the elastic scattering spin observables needed to study these nuclei, one requires the Lorentz invariant nucleon--nucleon (NN) amplitudes and the bound state wave functions of the target nuclei. The bound state wave functions are calculated using relativistic mean field theory with the NL3 and FSUGold parameter sets. The nucleon--nucleon amplitudes to be employed are those used in the IA1 and IA2 formalisms. It is an open question as to what effect the use of IA1 versus IA2 will have in the study of scattering experiments from unstable nuclei. We also present the calculation of the complete set of spin observables, namely the unpolarized cross section, the analysing power and the spin rotation function.

The outline of the paper is given as follows. In Section \ref{sec2}, the relativistic impulse approximation formalisms employed in this research are presented. This section also contains the descriptions of how the scattering observables are calculated in both position space (using localised optical potentials) and momentum space (using non-local optical potentials). Comparisons of the optical potentials calculated using both IA1 and IA2 formalisms are also presented. Section \ref{secresult} contains results of the elastic scattering observables namely the differential cross section, analysing power, and spin rotation parameters. These scattering observables are first calculated by solving the coordinate space Dirac equation with the localized IA1 and IA2 optical potentials. In this same section, the scattering observables calculated using the different RMF models are compared. The scattering observables obtained using the factorized optical potentials are also compared with the results obtained using the full-folding optical potentials. Finally, the scattering observables calculated using localised optical potentials are compared with the momentum space calculations employing non-local optical potentials.

\section{\label{sec2} FORMALISM}
\subsection{\label{subsec2} Relativistic impulse approximations}

In this section, the generalised relativistic impulse approximation (called IA2) for elastic proton scattering, introduced by Tjon and Wallace \cite{tjon87b} is presented. In this formalism, the relativistic optical potential is constructed by making use of the symmetric Lorentz-invariant nucleon-nucleon amplitudes of Ref. \cite{tjon87}.  Following Ref. \cite{tjon87b}, the first order relativistic optical potential is given in momentum space by
\begin{equation}
\hat{U} (\mathbf{k}', \mathbf{k}) = \frac{-4 \pi i k_{\mathrm{lab}}}{m}  \sum_{a} \int \frac{d^{3} P}{(2 \pi)^{3}} \overline{\psi}_{a} (\mathbf{P}+\frac{1}{2} \mathbf{q}) \hat{\mathcal{F}} \psi_{a} (\mathbf{P} - \frac{1}{2} \mathbf{q}),
\label{gia4}
\end{equation}
where all occupied (proton or neutron single particle) states are included over $a$ and $q=k-k'$ is the momentum transfer, $\hat{\mathcal{F}}$ is the invariant NN amplitude, and $\psi_{a}$ denotes the bound state wave function obtained from relativistic mean field theory. Medium effects are incorporated using the prescription of Ref. \cite{kak09}.

Applying optimal factorization (i.e. evaluating the NN amplitude at $\mathbf{P} = 0$ as it is often assumed that the NN amplitude generally varies slowly compared to the nuclear wave function.), the optical potential becomes (the so-called $t\rho$ form):
\begin{eqnarray}
\hat{U} (\mathbf{k}', \mathbf{k}) = &- \frac{1}{4} \mathrm{Tr}_{2} \left[ \hat{\mathcal{M}}_{pp} \left( k, \frac{1}{2} q ; k', \frac{1}{2} q \right) \hat{\rho}_{p} (\mathbf{q})  \right] \nonumber \\
&- \frac{1}{4} \mathrm{Tr}_{2} \left[ \hat{\mathcal{M}}_{pn} \left( k, \frac{1}{2} q ; k', \frac{1}{2} q \right) \hat{\rho}_{n} (\mathbf{q})  \right] ,
\label{gia8}
\end{eqnarray}
where the nuclear density form factor is written as
\begin{equation}
\hat{\rho} (\mathbf{q}) =  \rho_{S} (q) + \gamma_{2}^{0} \rho_{V} (q) - \frac{ \bm{\alpha}_{2} \cdot \mathbf{q}}{2 m} \rho_{T} (q)
\label{gia9}  	
\end{equation}
and the scalar, vector, and tensor form factors are given, respectively, by \cite{tjon87b}
\begin{eqnarray}
\rho_{S} (q) &=& 4 \pi \int_{0}^{\infty} dr \: r^{2} \rho_{S}(r) j_{0}, \nonumber \\
\rho_{V} (q) &=& 4 \pi \int_{0}^{\infty} dr \: r^{2} \rho_{V}(r) j_{0}, \nonumber \\
\rho_{T} (q) &=& 4 \pi m \int_{0}^{\infty} dr \: r^{2} \rho_{T}(r) \frac{j_{1}}{q} ,
\label{formia2}
\end{eqnarray}
where $j_{0}$ and $j_{1}$ are spherical Bessel functions. The scalar density $\rho_{S} (r)$, vector density $\rho_{V} (r)$, and tensor density, are given as
\begin{eqnarray}
\rho_{S}(r) &=& \sum _{\alpha}^{occ} \left(\frac{2j_{\alpha}+1}{4 \pi r^{2}} \right) \left[ g_{\alpha}^{2}(r)-f_{\alpha}^{2}(r) \right],  \\
\rho_{V}(r) &=& \sum _{\alpha}^{occ} \left(\frac{2j_{\alpha}+1}{4 \pi r^{2}} \right) \left[ g_{\alpha}^{2}(r)+f_{\alpha}^{2}(r) \right],  \\
\rho_{T} (r) &=& \sum_{\alpha}^{\mathrm{occ}} \left(\frac{2 j_{\alpha} + 1}{4 \pi r^{2}}  \right) \left[4 g_{\alpha} (r) f_{\alpha} (r) \right] .
\label{gia23}
\end{eqnarray}
Here, $f_{\alpha}$ and $g_{\alpha}$ are the bound state wave functions calculated using relativistic mean field theory. We have employed both NL3 and FSUGold parametrisations \cite{lal97,tod03,tod05}. The root-mean square radii computed using NL3 and FSUGold parameter sets, are shown and compared with experimental data, where available, in  Table \ref{tab_rms}. There is satisfactory agreement with experiment at the $1 \% $ level. The experimental data for $^{40,48}$Ca, $^{206}$Hg, and $^{132}$Sn are taken from Ref. \cite{angeli13} while the theoretical result for $^{54}$Ca is taken from Ref. \cite{garcia16}. 

\begin{figure}
	\centering
	\includegraphics[width=0.49\linewidth]{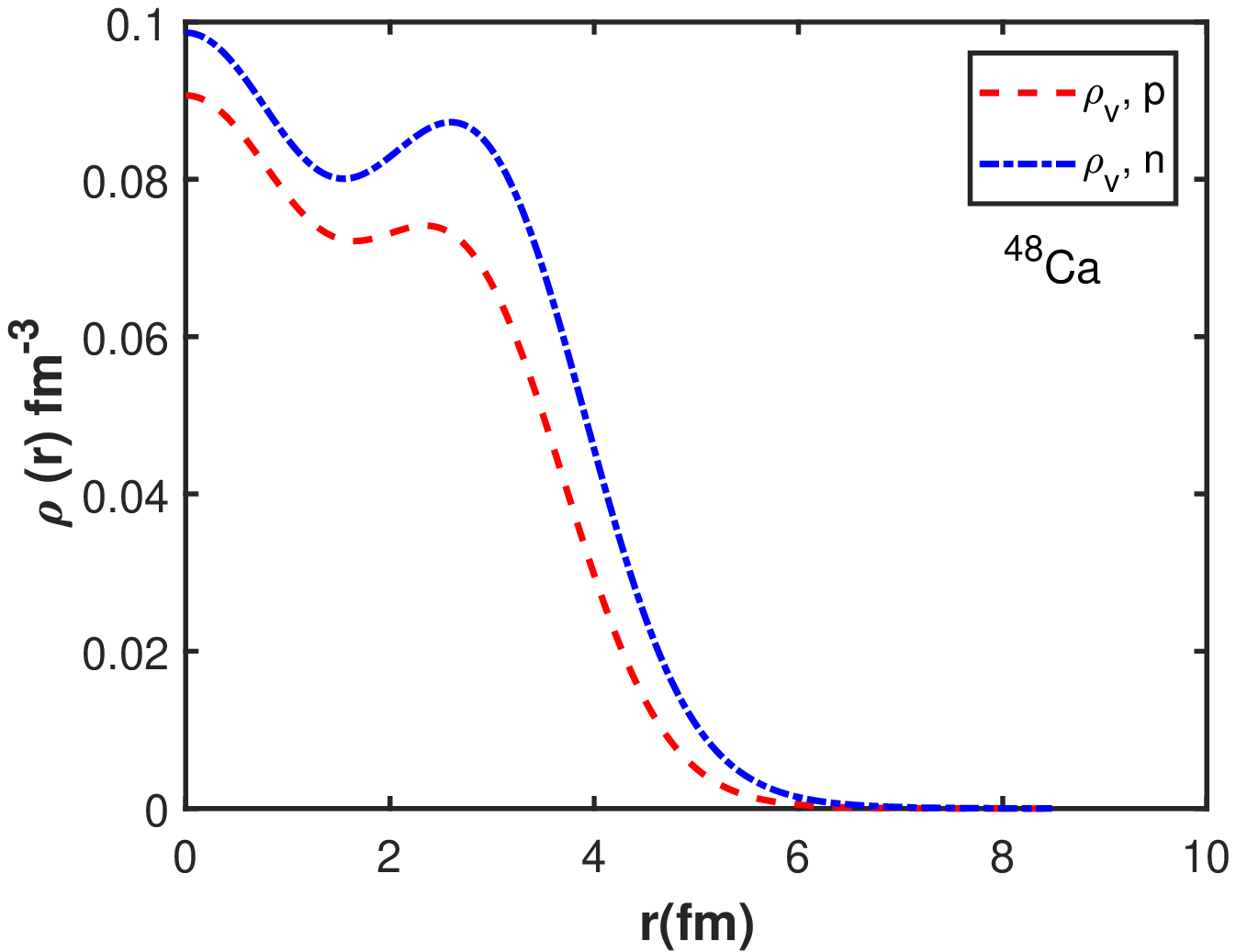}
	\includegraphics[width=0.49\linewidth]{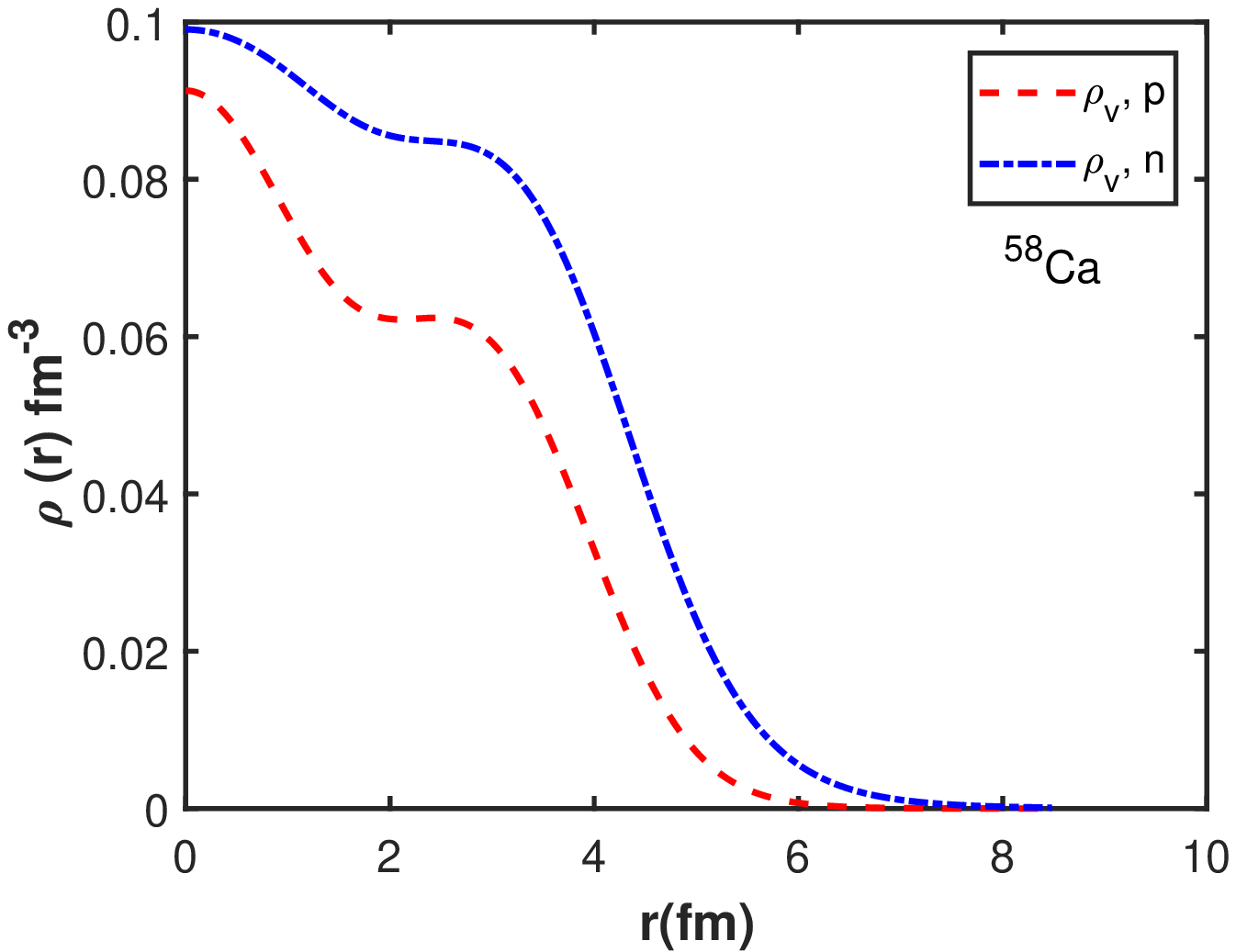}
	\includegraphics[width=0.49\linewidth]{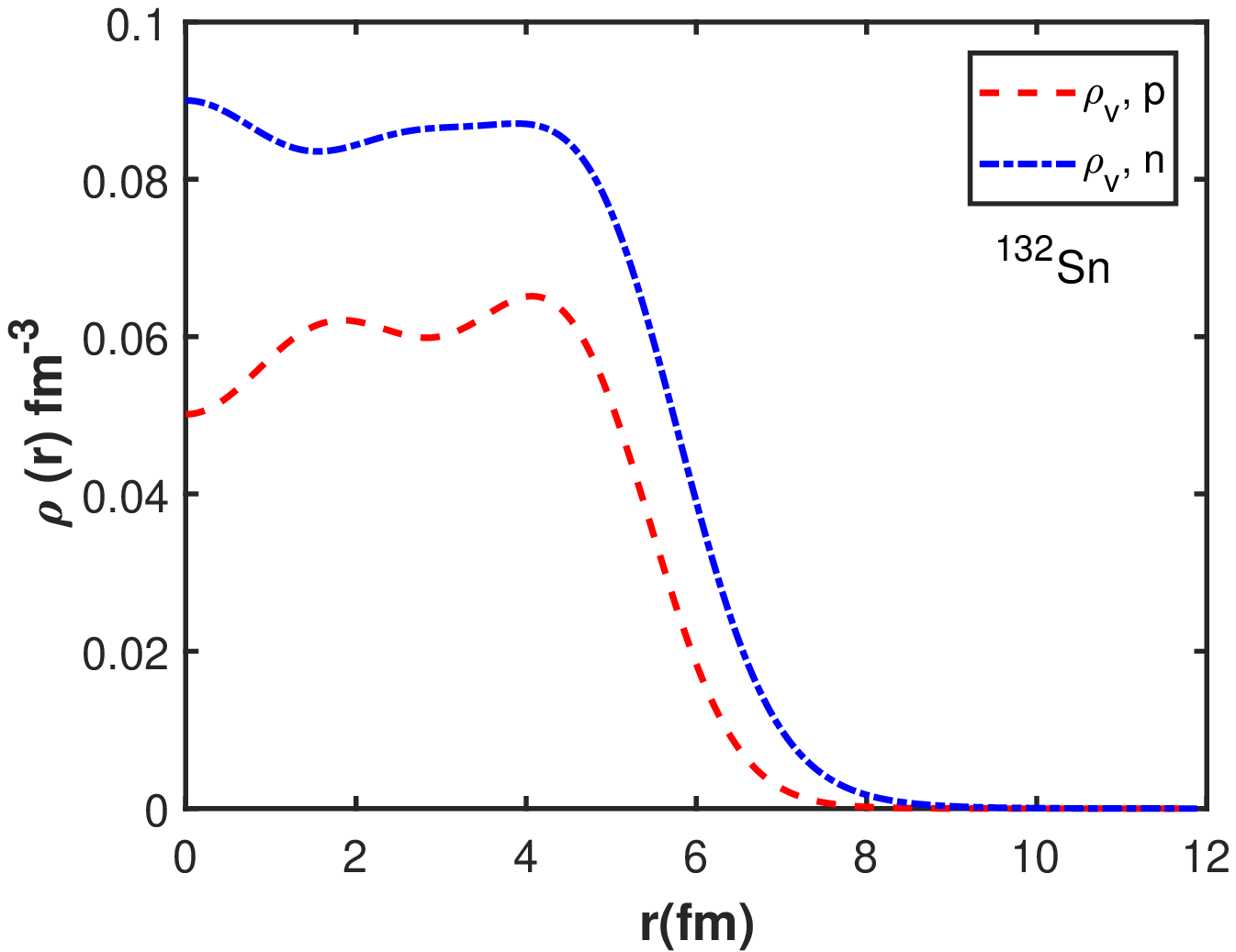}
	\includegraphics[width=0.49\linewidth]{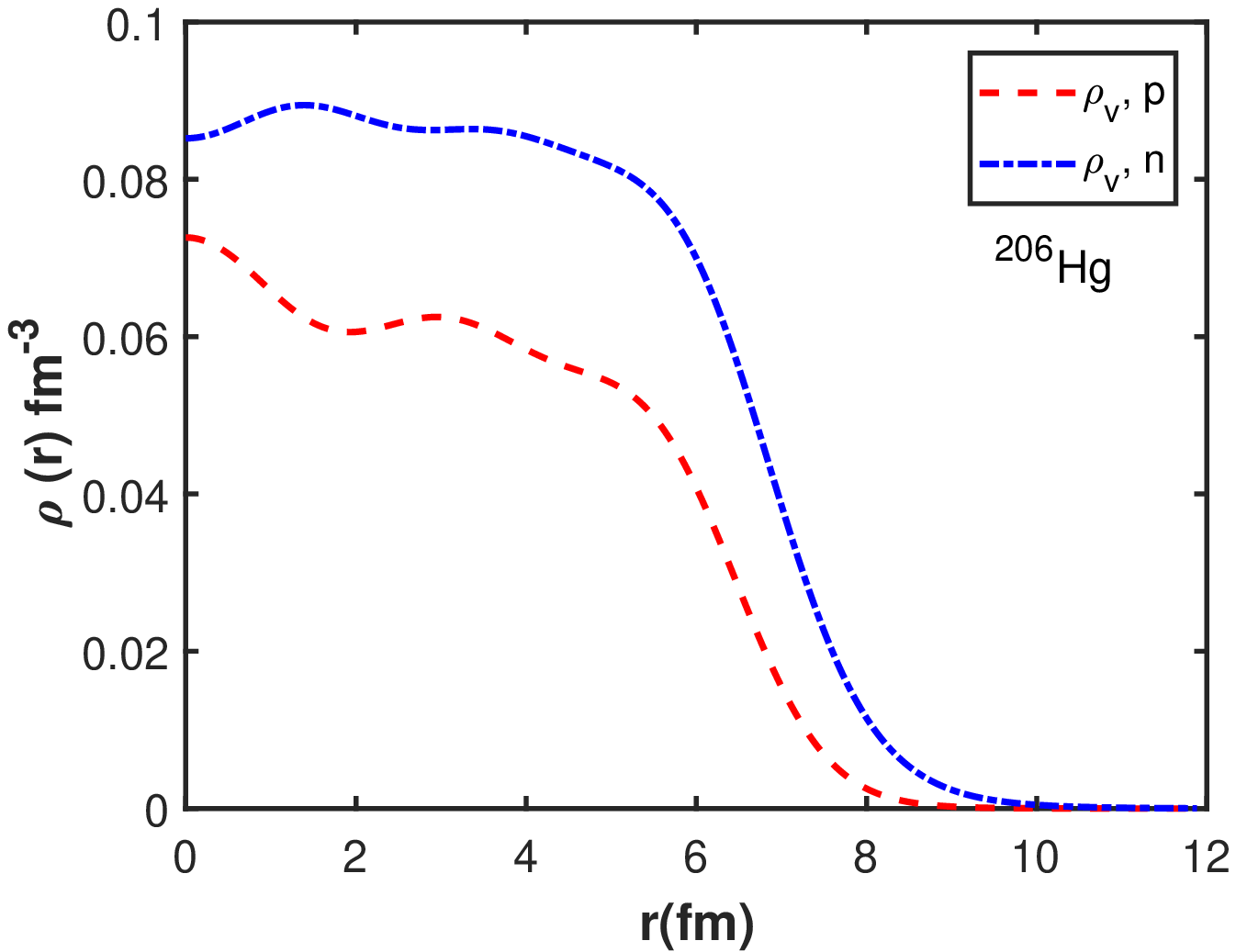}
	\caption{Proton and neutron vector density plots for $^{48,58}$Ca, $^{132}$Sn, and $^{206}$Hg nuclei calculated using the NL3 parametrisation. Proton vector density plots are shown in dashed lines while neutron vector density plots are shown in dot-dashed lines.}
	\label{fig:vectdenspn}
\end{figure}

Figure \ref{fig:vectdenspn} shows plots of the proton and neutron vector densities for $^{48,58}$Ca, $^{132}$Sn, and $^{206}$Hg nuclei calculated using the NL3 parametrisation. Proton vector density plots are shown in dashed lines while neutron vector density plots are shown in dot-dashed lines. One can observe that $^{132}$Sn and $^{206}$Hg are very neutron rich.

In the IA2 formalism, the full NN amplitude is expanded in terms of covariant projection operators $\Lambda_{\rho_{i}}$ to separate positive and negative--energy sectors of the Dirac space and it can be written in terms of the kinematic covariants $K_{n}$ ($n=1 \cdots 13$) as
\begin{eqnarray}
\hat{F} =&& \sum_{\rho'_{1} \rho'_{2} \rho_{1} \rho_{2}} \sum_{n = 1}^{13} F_{n}^{\rho'_{1} \rho'_{2} \rho_{1} \rho_{2}} \left[\Lambda_{\rho'_{1}} (\mathbf{k}'_{1}) \otimes \Lambda_{\rho'_{2}} (\mathbf{k}'_{2}) \right] \: K_{n} \nonumber \\ 
&& \times \left[\Lambda_{\rho_{1}} (\mathbf{k}_{1}) \otimes \Lambda_{\rho_{2}} (\mathbf{k}_{2}) \right],
\label{gia29}
\end{eqnarray}
where $\rho (\rho') = +$ for positive energy initial (final) state or $-$ for negative energy initial (final) state, $\rho_{1}$ is for projectile particle and $\rho_{2}$ for target struck nucleon. The kinematic covariants $K_{n}$ are given in Table II of Ref. \cite{tjon87b}. It should be noted that $\hat{F}^{11}_{\mathrm{IA2}} \neq \hat{F}_{\mathrm{IA1}}$ due to the presence of projection operators in the IA2 $\hat{F}$. The covariant energy projection operators $\Lambda_{\pm} (\mathbf{k})$ allow the separation of the positive and negative energy sectors of the Dirac space, and $Q_{ij, \mu}$ denote the four momenta, where $i=1$ (for nucleon 1, which is the projectile), $i=2$ (for nucleon 2, which is the target struck nucleon). See Refs. \cite{tjon87,tjon87b,kak09,kak14} for details.

Local forms of the optical potentials have been found to be accurate at high energy for nucleon--nucleus elastic scattering due to the diffractive nature of the scattering. In the Dirac optical potential derived above, nonlocalities are present due to projection operators and covariants, as they depend on $\mathbf{k}$ and $\mathbf{k}'$. These are localised by assuming that the momentum operator $\mathbf{k}$ stays near the asymptotic value $\hat{\mathbf{k}}$, i.e. $\mathbf{k} \approx \hat{\mathbf{k}}$. This enables $E (\mathbf{k})$ and $E(\mathbf{k}')$ to be replaced by $E = E(\hat{\mathbf{k}})$. Also
\begin{equation}
\frac{\mathbf{k}}{m} \approx \frac{\hat{\mathbf{k}}}{m} , \; \; \; \frac{\mathbf{k}_{a} \cdot \mathbf{q}}{m} =  \frac{\mathbf{k}^{2} - \mathbf{k}^{' 2}}{2 m} \approx 0.
\label{gia75b}
\end{equation}

The localised coordinate space Dirac equation to be solved is given by
\begin{equation}
\left[E \gamma^{0}  + i \bm{\gamma} \cdot \bm{\nabla} -m - \tilde{U}(\mathbf{r}) \right] \tilde{\Psi} (\mathbf{r}) = 0,
\label{gia83}
\end{equation}
in which case the optical potential $\tilde{U}(\mathbf{r})$ is given by \cite{tjon87b}:
\begin{eqnarray}
\tilde{U}(\mathbf{r}) =&& \tilde{S}(r) + \gamma^{0} \tilde{V}(r) -i \bm{\alpha} \cdot \hat{\mathbf{r}} \: \tilde{T} (r)- \left[\tilde{S}_{LS} (r) + \gamma^{0} \tilde{V}_{LS} (r) \right] \nonumber \\
&& \times \left[ \bm{\alpha} \cdot (-i \mathbf{r} \times \bm{\nabla})\right],
\label{gia84} 
\end{eqnarray}
where the scalar $\tilde{S}(r)$, vector $\tilde{V}(r)$, tensor $\tilde{T}(r)$, scalar spin-orbit $\tilde{S}_{LS}(r)$, and vector spin-orbit $\tilde{V}_{LS}(r)$ potentials are as given in equations (3.4.47) to (3.4.51) of Ref. \cite{yahyaphd2018}. 

%

\begin{small}
	\begin{table}
		\centering
		\caption{Root-mean-square charge radius, proton and neutron root-mean-square radii of some closed shell Calcium isotopes.}
		\begin{tabular}{cccccc}
			\hline \hline \\
			
			Nucleus & Observable & NL3 & FSUGold & Experiment  \\ [0.5ex]
			
			\hline \hline \\
			
			$^{48} Ca$ & $r_{p}$ & 3.3789 & 3.3659 &  \\ [0.5ex]
			
			& $r_{n}$ &  3.6046 & 3.5632 &  \\ [0.5ex]
			
			& $\Delta r = r_{n} - r_{p}$ &  0.22572 & 0.1973 &  \\ [0.5ex]
			
			& $r_{ch}$  & 3.4723 & 3.4597 & 3.4771 \cite{angeli13} \\ [0.5ex]
			
			\hline \\
			$^{54} Ca$ & $r_{p}$ & 3.5037 & 3.4834 &  \\ [0.5ex]
			
			& $r_{n}$ & 3.9008 & 3.8249 &  \\ [0.5ex]
			
			& $\Delta r = r_{n} - r_{p}$ & 0.39704 & 0.3414 &  \\ [0.5ex]
			
			& $r_{ch}$ & 3.5939 & 3.5741 & 3.5640 \cite{garcia16}  \\ [0.5ex]
			
			\hline \\
			
			$^{58} Ca$ & $r_{p}$ & 3.5317 & 3.5191  &  \\ [0.5ex]
			
			& $r_{n}$ & 4.0668 & 3.9950 &  \\ [0.5ex]
			
			& $\Delta r = r_{n} - r_{p}$ &  0.53514 & 0.47589 &  \\ [0.5ex]
			
			& $r_{ch}$ & 3.6212 & 3.6089 &  \\ [0.5ex]
			
			\hline \\
			
			$^{60} Ca $ & $r_{p}$ & 3.5513 & 3.5407 &  \\ [0.5ex]
			
			& $r_{n}$ &  4.1591 & 4.0841 &  \\ [0.5ex]
			
			& $\Delta r = r_{n} - r_{p}$ &  0.60779 & 0.54339 &  \\ [0.5ex]
			
			& $r_{ch}$ & 3.6403 & 3.6300 &  \\ [0.5ex]
			
			\hline \\
			
			$^{132} Sn $ & $r_{p}$ & 4.6435 & 4.6542 &  \\ [0.5ex]
			
			& $r_{n}$ &  4.9891 & 4.9251 &  \\ [0.5ex]
			
			& $\Delta r = r_{n} - r_{p}$ &  0.34558 & 0.27090 &  \\ [0.5ex]
			
			& $r_{ch}$  & 4.7119 & 4.7225 & 4.7093 \cite{angeli13} \\ [0.5ex]
			
			\hline \\
			
			$^{206} Hg $ & $r_{p}$ & 5.3127 & 5.3109 &  \\ [0.5ex]
			
			& $r_{n}$ &  5.7739 & 5.6797 &  \\ [0.5ex]
			
			& $\Delta r = r_{n} - r_{p}$ &  0.46115 & 0.36882 &  \\ [0.5ex]
			
			& $r_{ch}$ &  5.3633 & 5.3615 & 5.4837 \cite{angeli13} \\ [0.5ex]
			
			\hline \hline
		\end{tabular} 
		\label{tab_rms}
	\end{table}
\end{small}

\begin{figure}
	\centering
	\includegraphics[width=0.49\linewidth]{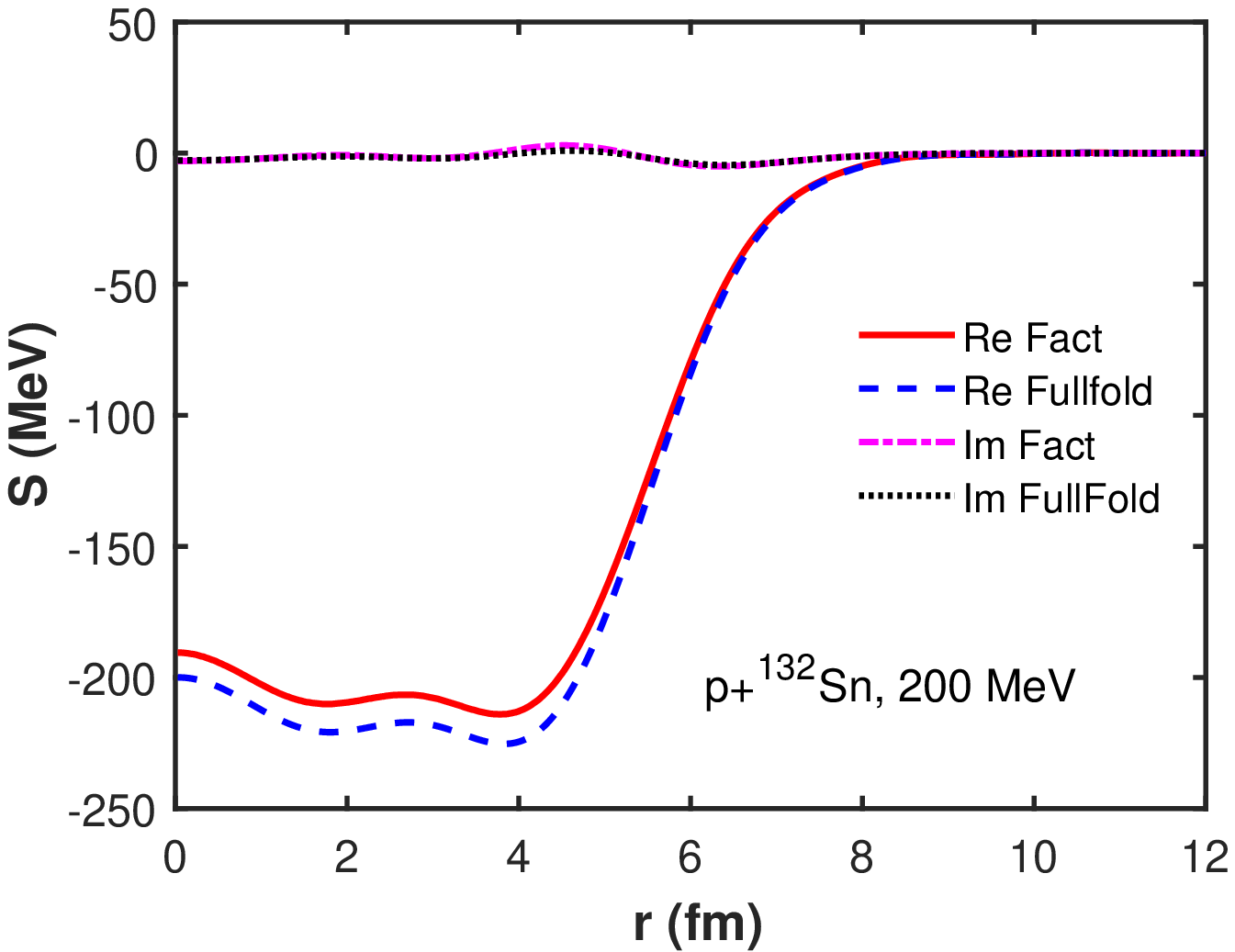}
	\includegraphics[width=0.49\linewidth]{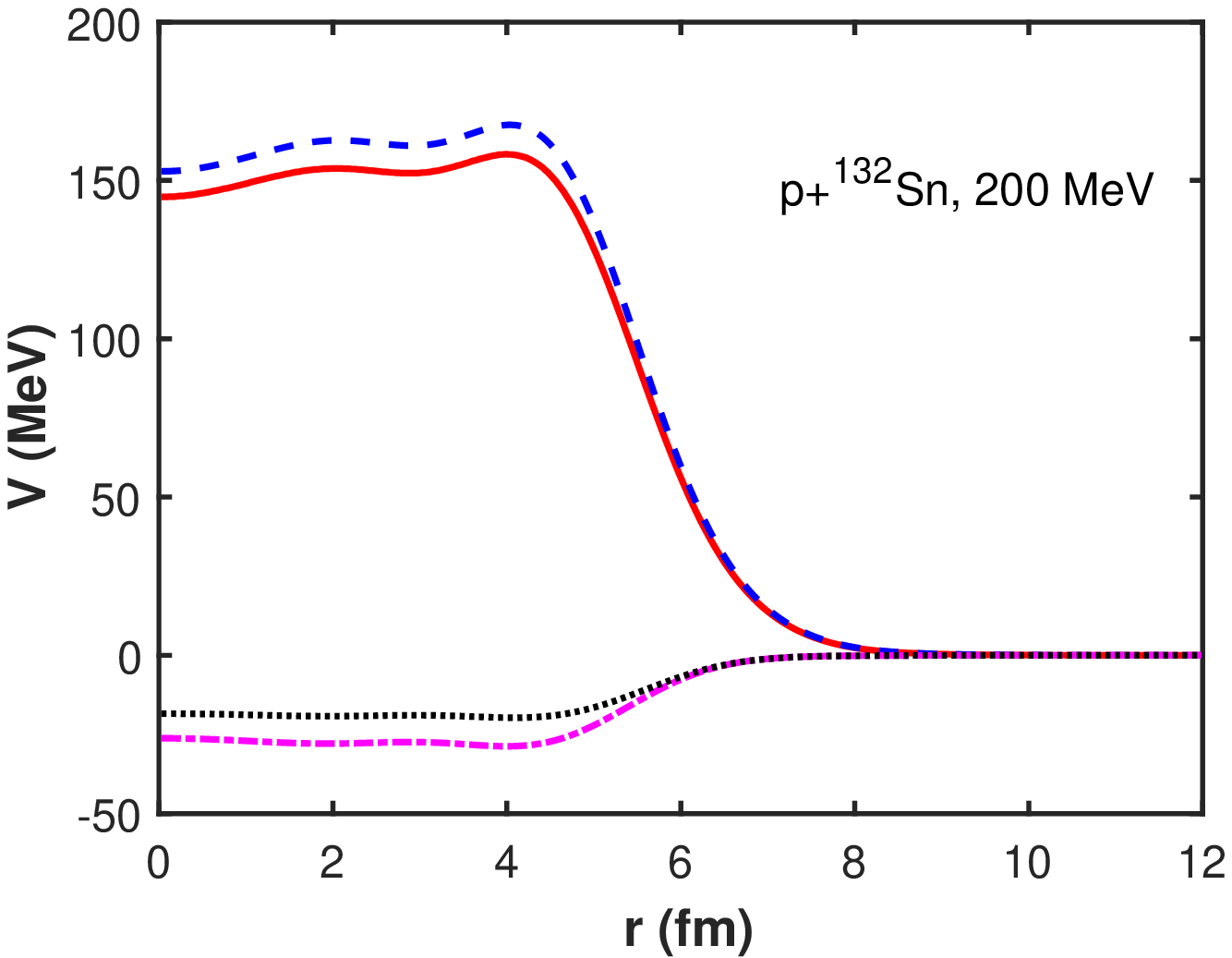}
	\includegraphics[width=0.49\linewidth]{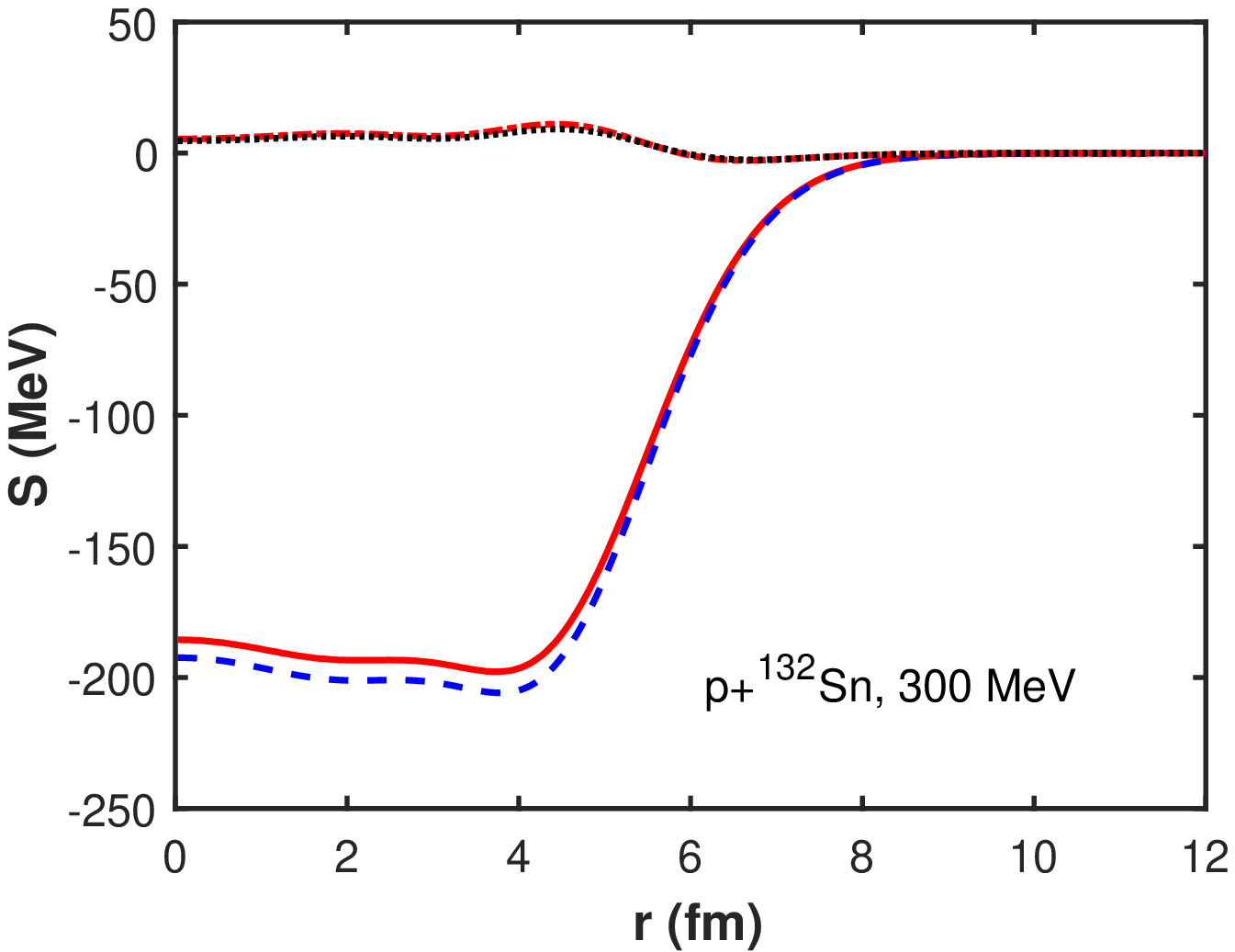}
	\includegraphics[width=0.49\linewidth]{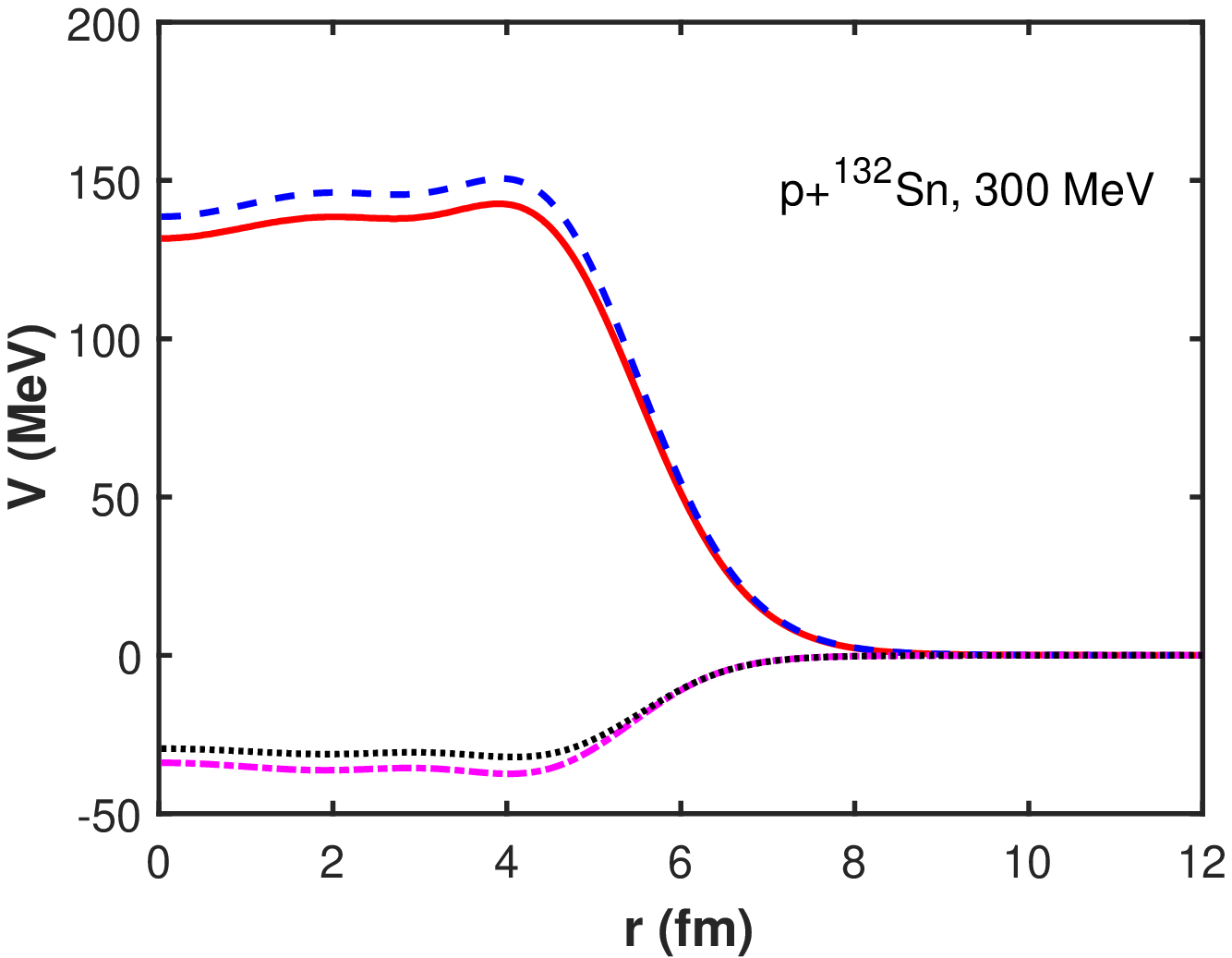}
	\includegraphics[width=0.49\linewidth]{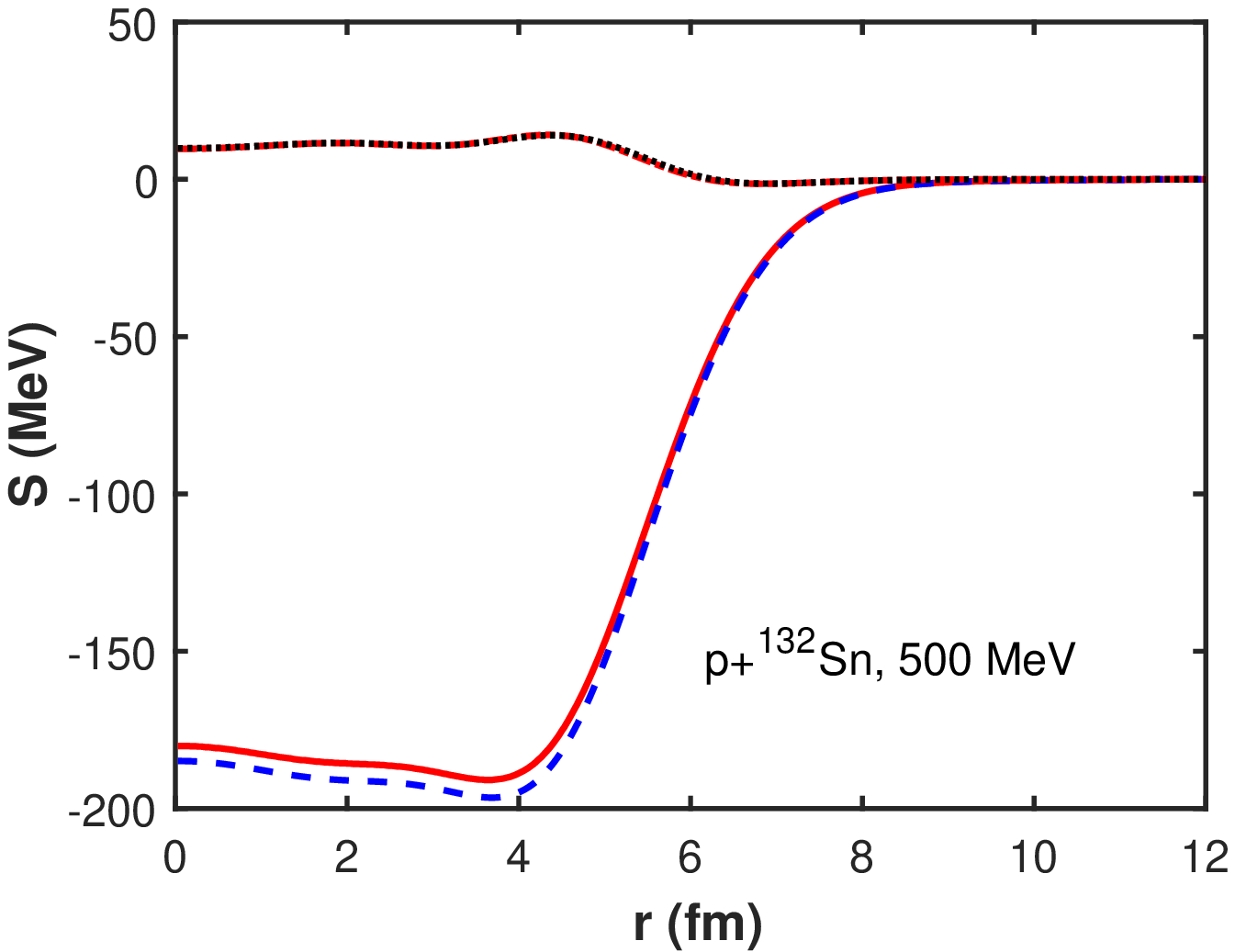}
	\includegraphics[width=0.49\linewidth]{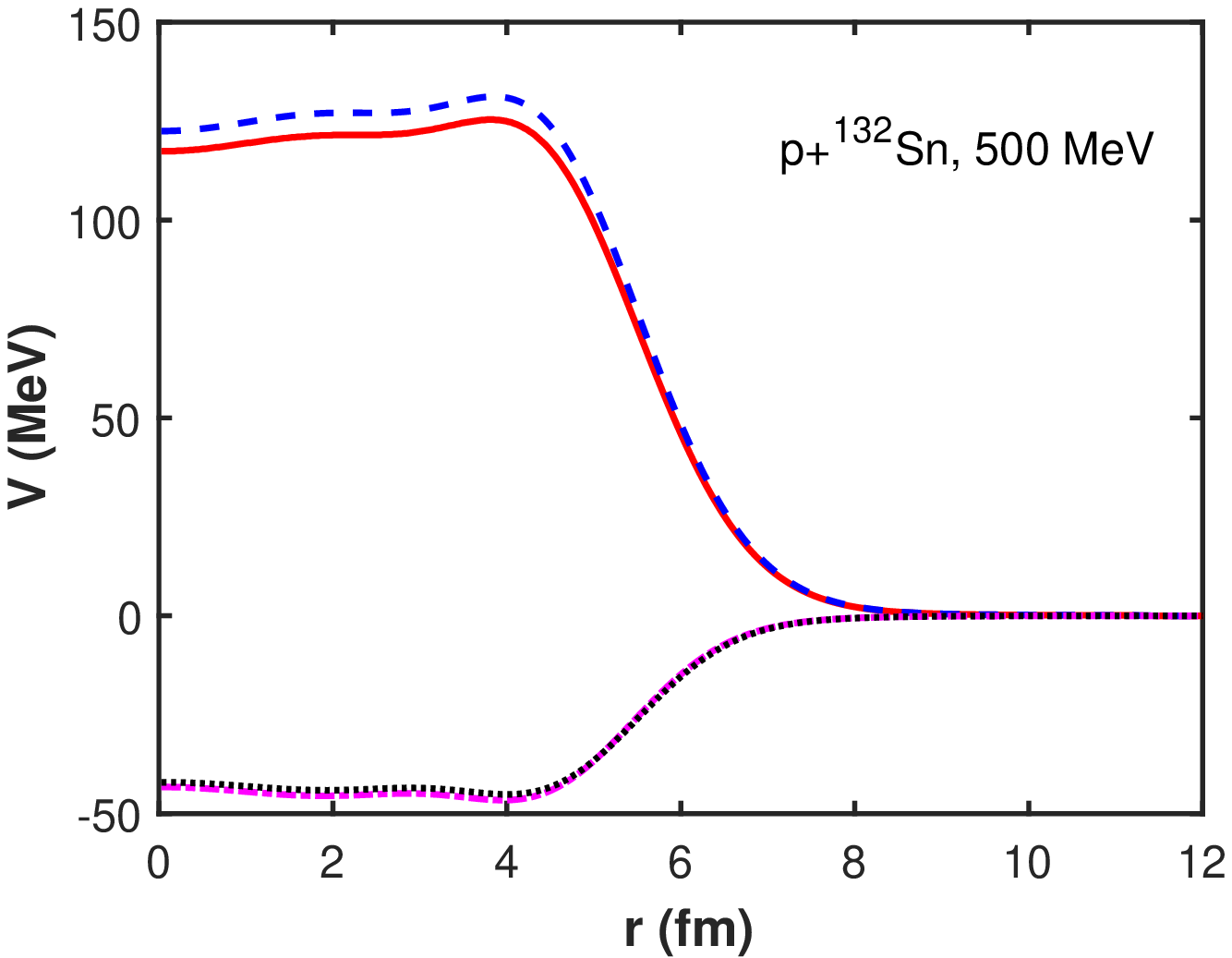}
	\caption{IA2 scalar and vector optical potentials}
	\label{fig:SVpot_Sn132}
\end{figure}

Figure \ref{fig:SVpot_Sn132} shows plots of the IA2 scalar and vector optical potentials calculated for proton scattering on $^{132}$Sn at $T_{\mathrm{lab}} = 200, 300, 500$ MeV using optimally factorised potentials and full-folding potentials.

\begin{figure}
	\centering
	\includegraphics[width=0.49\linewidth]{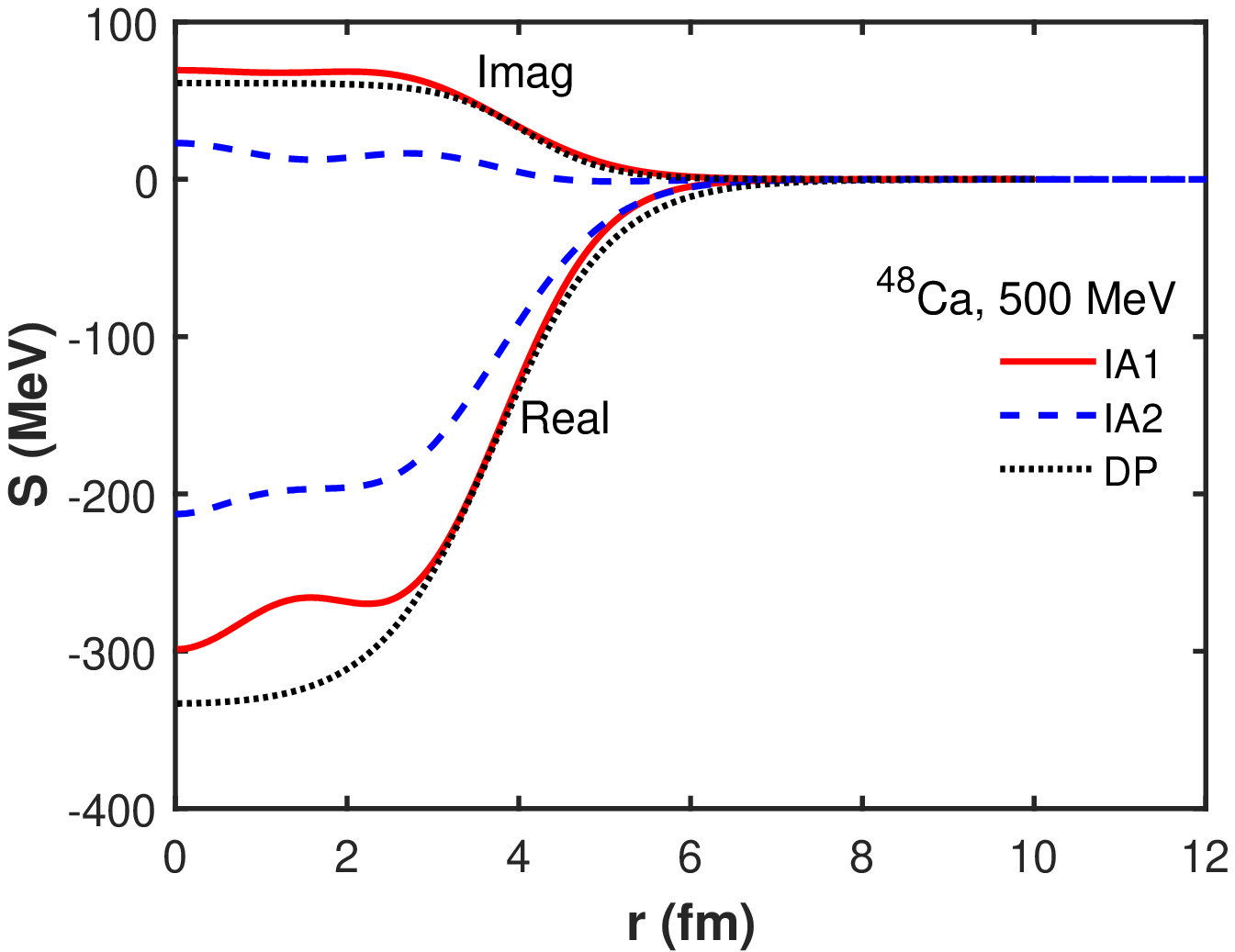}
	\includegraphics[width=0.49\linewidth]{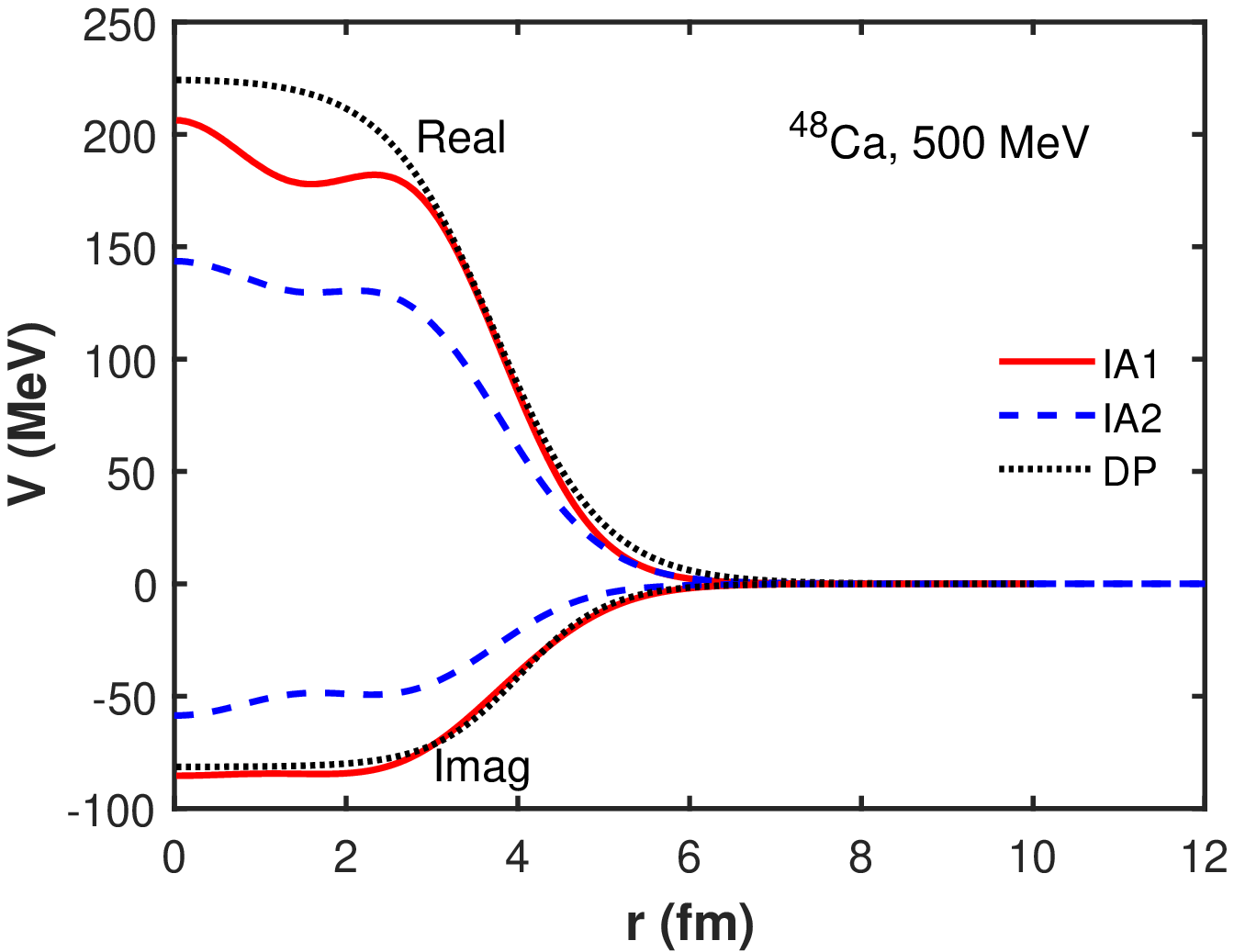}
	\includegraphics[width=0.49\linewidth]{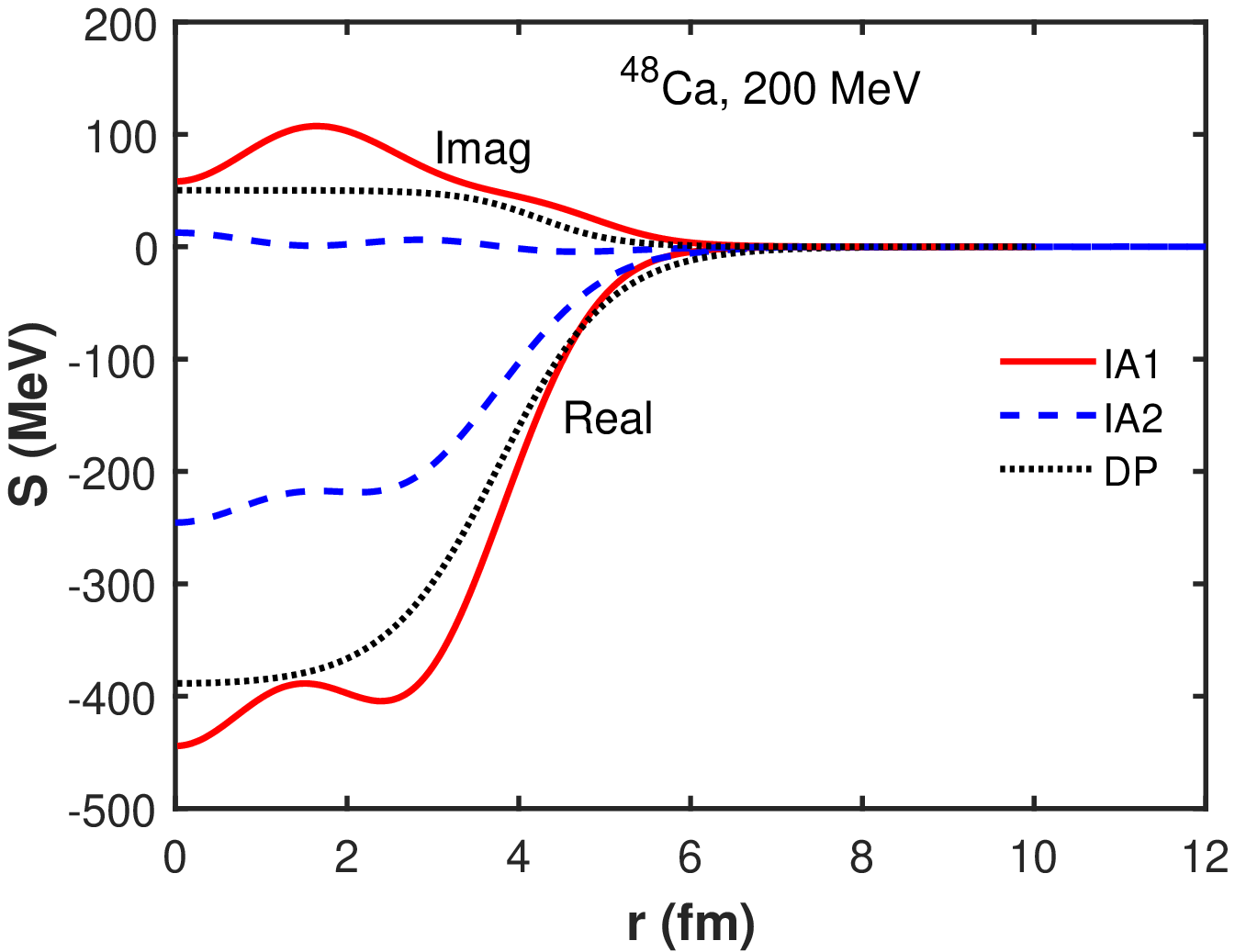}
	\includegraphics[width=0.49\linewidth]{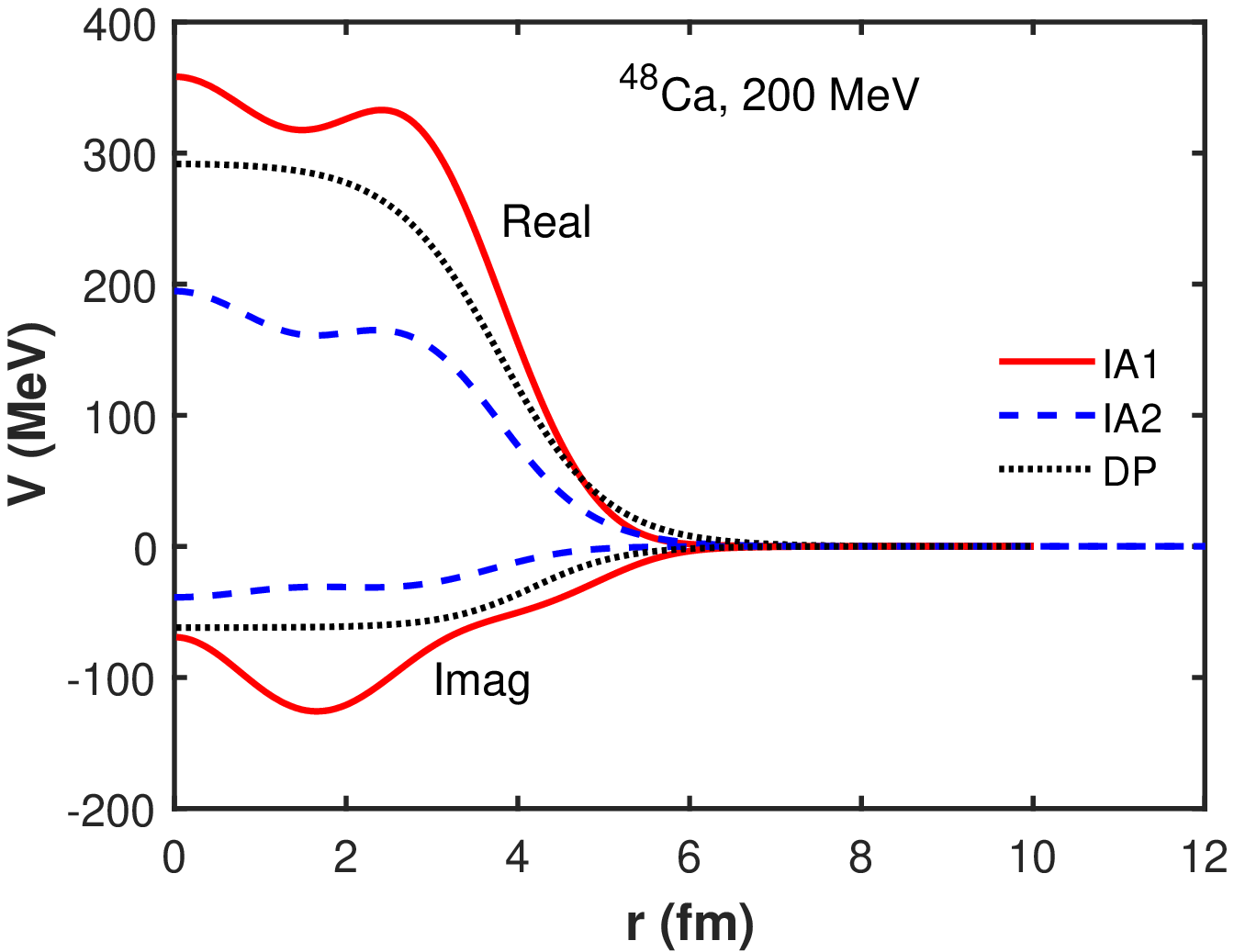}
	\caption{$^{48}$Ca Scalar and vector optical potentials calculated with NL3 parametrisation for IA1, IA2, and Dirac phenomenology at $T_{\mathrm{lab}} = 500$MeV and $200$MeV.}
	\label{fig:SVpot_Ca48}
\end{figure}

\subsection{\label{sec2:level2} Momentum space calculations}
Here, we present the solution of the momentum space Dirac equation using non-local optical potentials. This enables us to check the validity of using localised optical potentials to calculate elastic scattering observables at intermediate energies. We only give a brief description of the procedure. Detailed presentation can be found in Refs. \cite{otten91} and \cite{yahyaphd2018}. The momentum space Dirac equation is transformed to two coupled Lippmann-Schwinger-like equations in momentum space.  The momentum--space integral equation approach to solving scattering problem deals directly with the scattering amplitudes, whose values can be measured experimentally \cite{kadyrov05}. This method also incorporates the required boundary conditions in scattering problems. The two integral equations are numerically solved to calculate the elastic scattering observables. The results obtained are then compared with those calculated using localised optical potentials.

The stationary--state Dirac equation for the scattering of a particle of mass $m$ from an external central field $U$ can be written as
\begin{align}
(\slashed{p} - m) \left. | \Psi \right\rangle &= U \left. | \Psi \right\rangle . \nonumber \\
\left[E \gamma^{0} - \mathbf{k}' \cdot \bm{\gamma} -m \right] \Psi (\mathbf{k}') &= \int \frac{d^{3} k}{(2 \pi)^{3}} \hat{U} (\mathbf{k}', \mathbf{k}) \Psi(\mathbf{k}) = 0,
\label{dir1}
\end{align}
where $E$ is the on--shell energy calculated in the proton--nucleus centre of mass frame, $m$ is mass of the projectile, and $\hat{U} (\mathbf{k}', \mathbf{k})$ is the optical potential. In the static approximation, only three--momentum can be transferred; energy is fixed. If we denote helicity amplitude by $\left\langle \lambda ' | \phi | \lambda \right\rangle $, with incident helicity $\lambda$ and final helicity $\lambda '$, then for elastic proton scattering from a spin--zero nucleus, the two required helicity amplitudes are expanded as follows \cite{kubis72, otten91}:
\begin{align}
\phi_{1} (\theta) & \equiv \left\langle + | \phi | + \right\rangle  \equiv \phi_{1/2 , 1/2} (\mathbf{k}, \mathbf{k}') \nonumber \\
 &= \sum_{j} \frac{2 j +1}{2 \hat{k}} \phi_{1}^{j} d_{1/2 , 1/2}^{j} (\theta), \label{part24} \\
\phi_{2} (\theta) & \equiv \left\langle + | \phi | - \right\rangle  \equiv \phi_{1/2 , -1/2} (\mathbf{k}, \mathbf{k}') \nonumber \\
&= \sum_{j} \frac{2 j +1}{2 \hat{k}} \phi_{2}^{j} d_{-1/2 , 1/2}^{j} (\theta). \label{part25}
\end{align}

The three scattering observables to be calculated are differential cross section ($\sigma$), analysing power ($A_{y}$), and spin--rotation function ($Q$). For elastic proton--nucleus scattering these observables are obtained from the helicity amplitudes using the following relations:
\begin{align}
\sigma &= |\phi_{1}|^{2} + |\phi_{2}|^{2}, \label{part45} \\
A_{y} &= \frac{2 \, \mathrm{Im} (\phi_{1} \phi_{2}^{*})}{|\phi_{1}|^{2} + |\phi_{2}|^{2}} , \label{part46} \\
Q &= \frac{\cos (\theta) \mathrm{Re}(\phi_{1} \phi_{2}^{*}) + \tfrac{1}{2} \sin (\theta) \left[|\phi_{1}|^{2} - |\phi_{2}|^{2} \right]}{|\phi_{1}|^{2} + |\phi_{2}|^{2}} .
\label{part47}
\end{align}

\section{RESULTS}
\label{secresult}

\subsection{Comparison of the RMF densities}
We show here plots of the scattering observables calculated using both NL3 and FSUGold parametrisations.

Here we study the effect of the different forms of the Lagrangian densities on the proton-nucleus scattering observables. Calculations for elastic proton scattering from the stable $^{40,48}$Ca nuclei are included in order to compare our results with existing experimental data; this will also check the validity and reliability of our calculations. 

\begin{figure}
	\centering
	\includegraphics[width=0.49\linewidth]{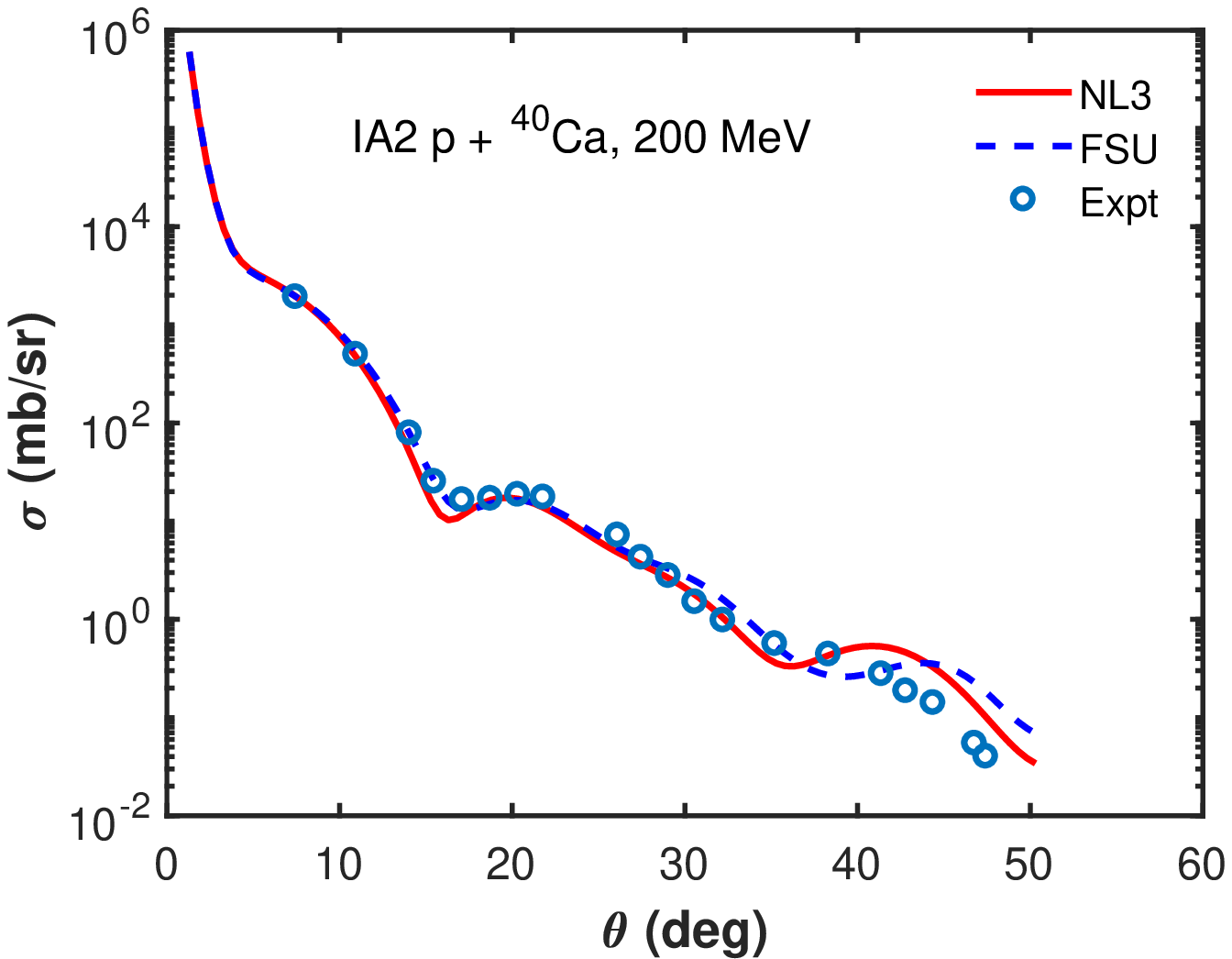}
	\includegraphics[width=0.49\linewidth]{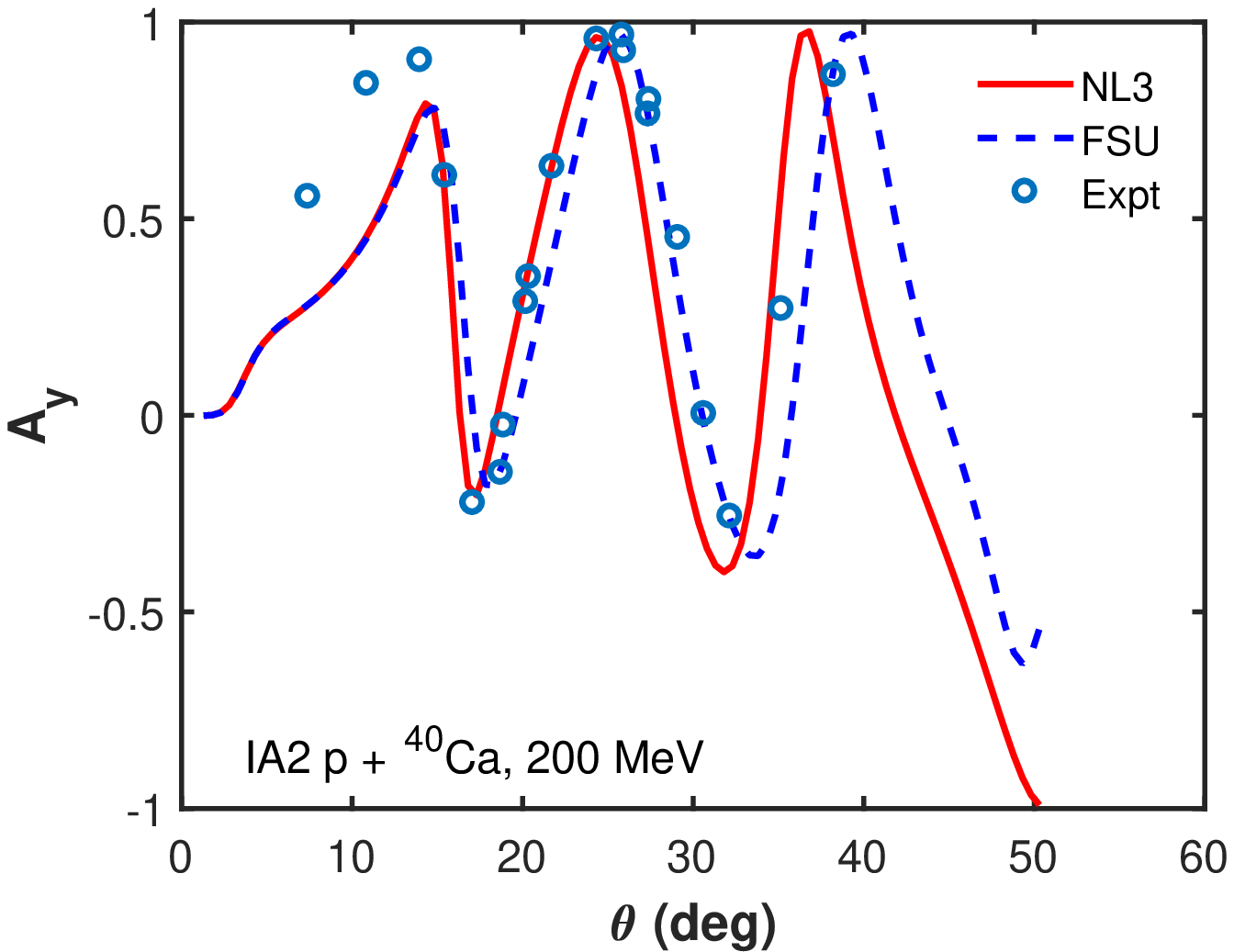}
	\includegraphics[width=0.49\linewidth]{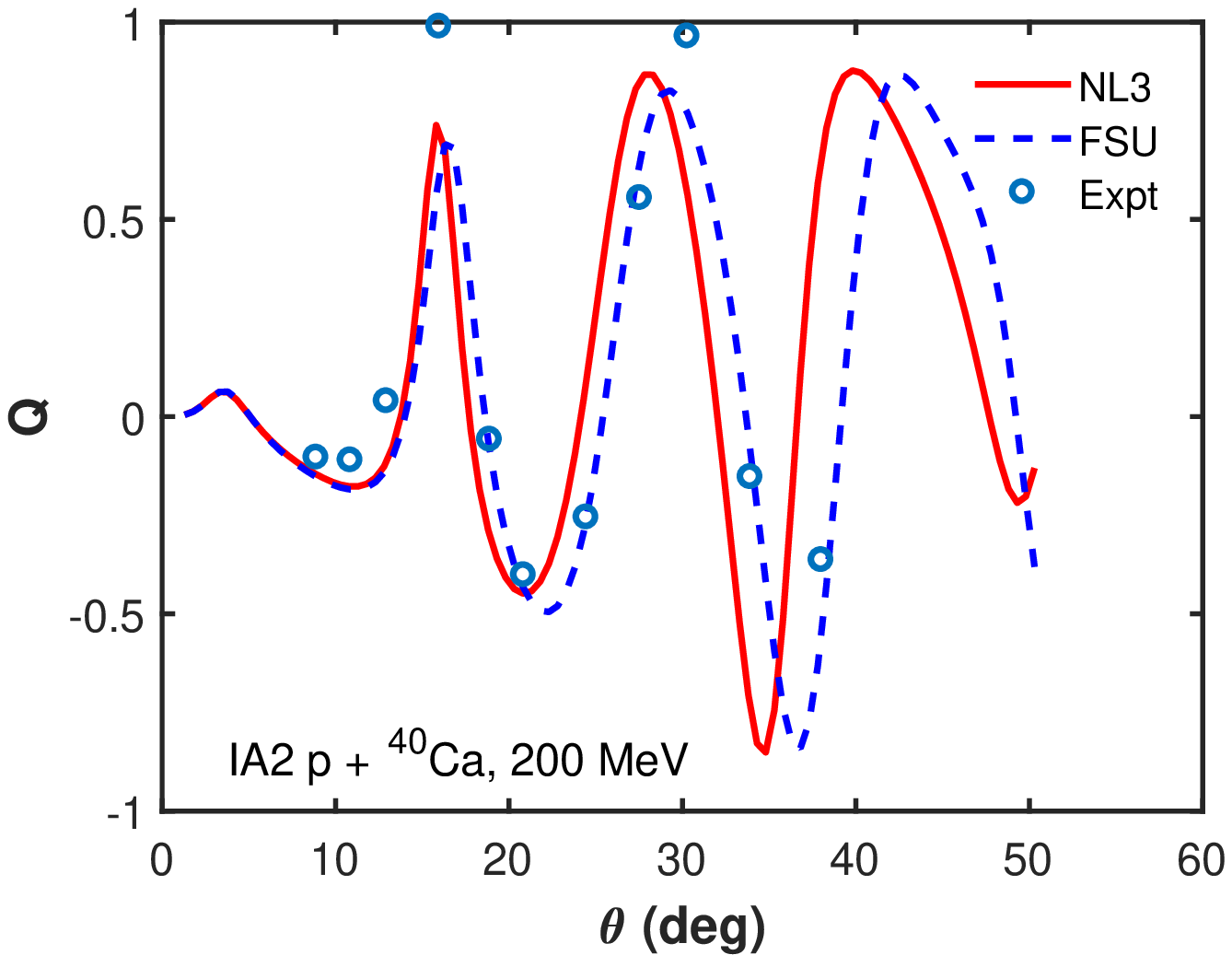}
	\caption{$^{40}$Ca scattering observables calculated with the NL3 and FSUGold parametrisations using IA2 formalism at $T_{\mathrm{lab}} = 200$MeV. The solid lines show the results using NL3 parametrisation, the dashed lines show the results using FSUGold parametrisation, while the experimental data are shown in circles.}
	\label{RMFCa40_200}
\end{figure}

\begin{figure}
	\centering
	\includegraphics[width=0.49\linewidth]{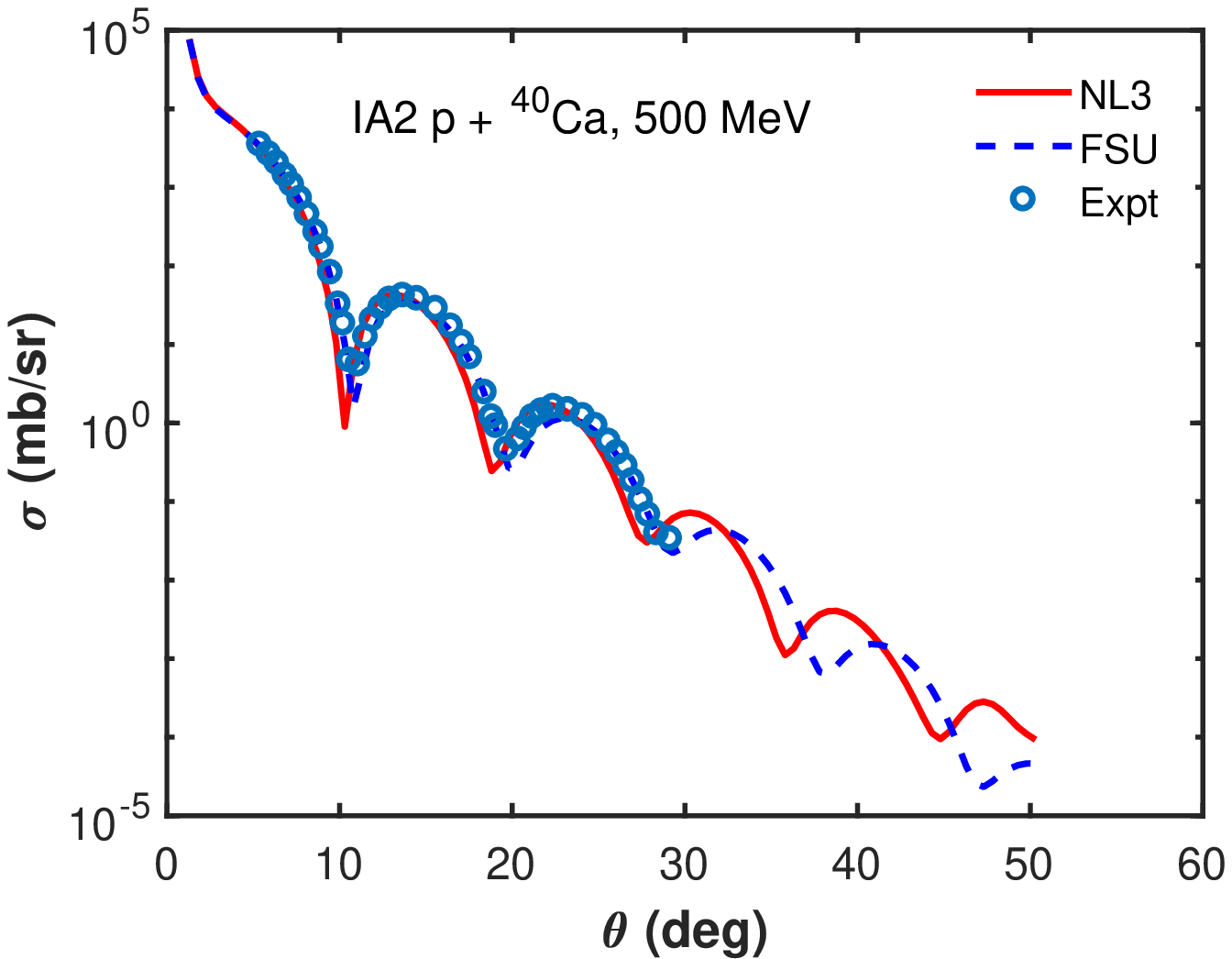}
	\includegraphics[width=0.49\linewidth]{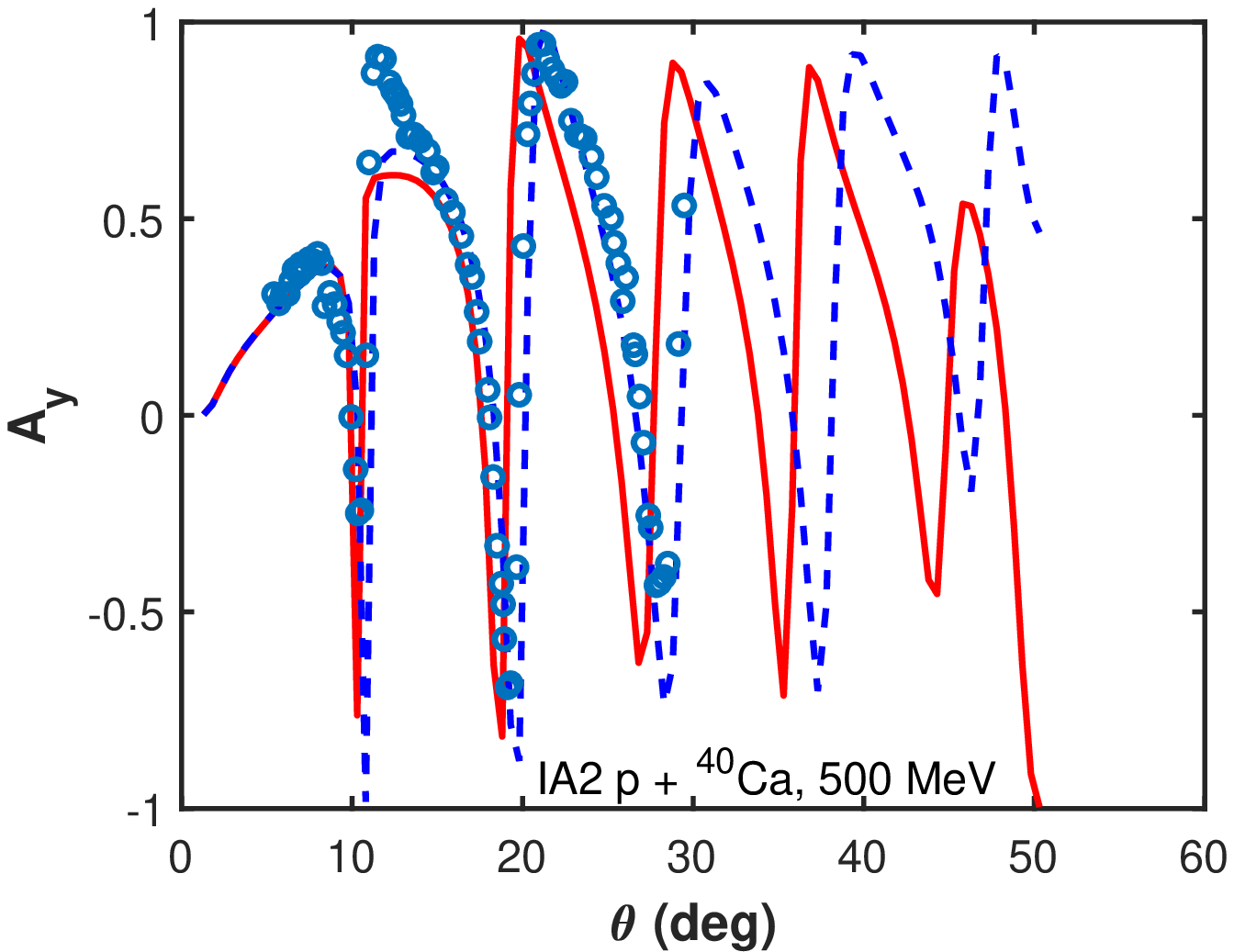}
	\includegraphics[width=0.49\linewidth]{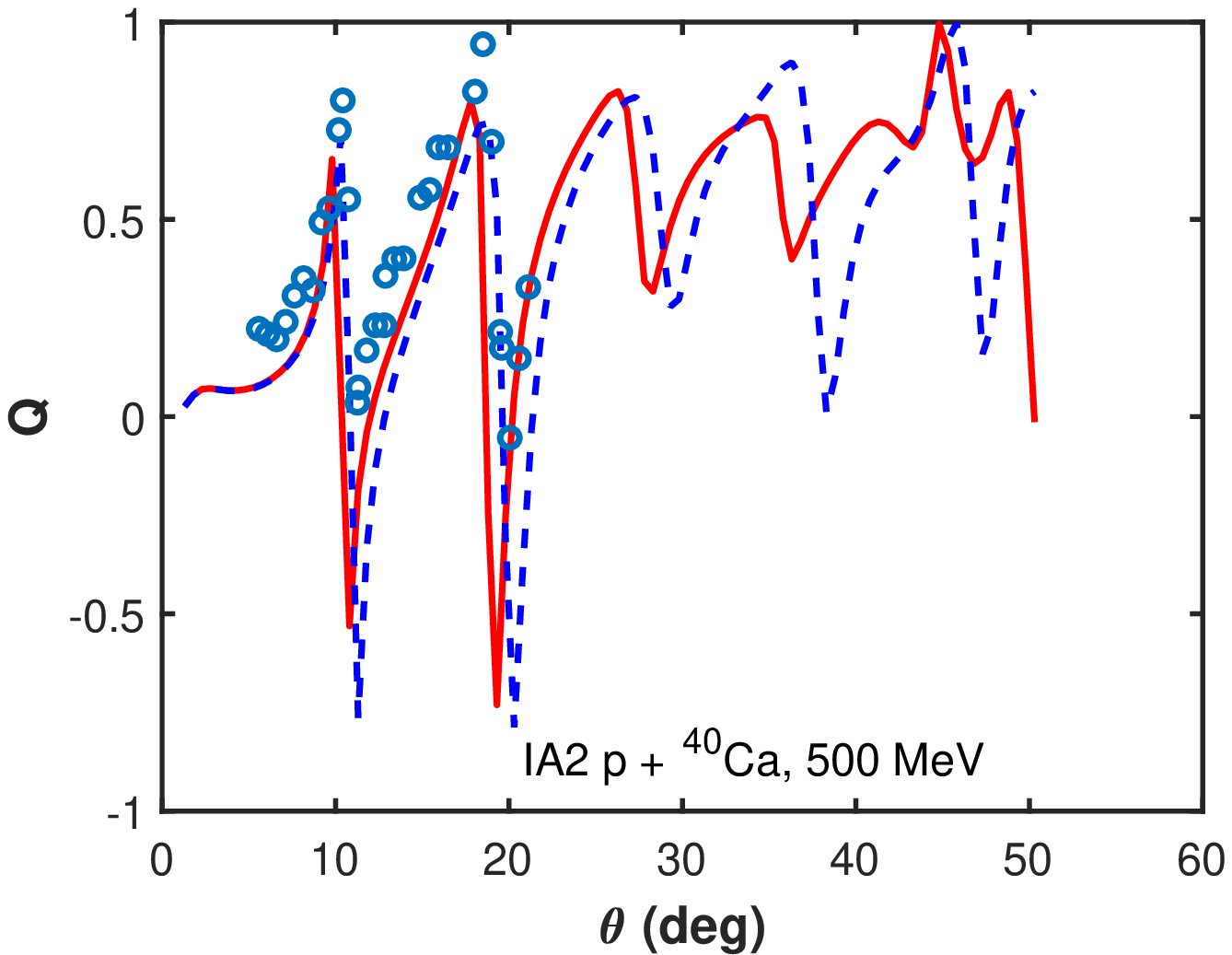}
	\caption{Same as in figure \ref{RMFCa40_200} except at $T_{\mathrm{lab}} = 500$MeV.}
	\label{RMFCa40_500}
\end{figure}

\begin{figure}
	\centering
	\includegraphics[width=0.49\linewidth]{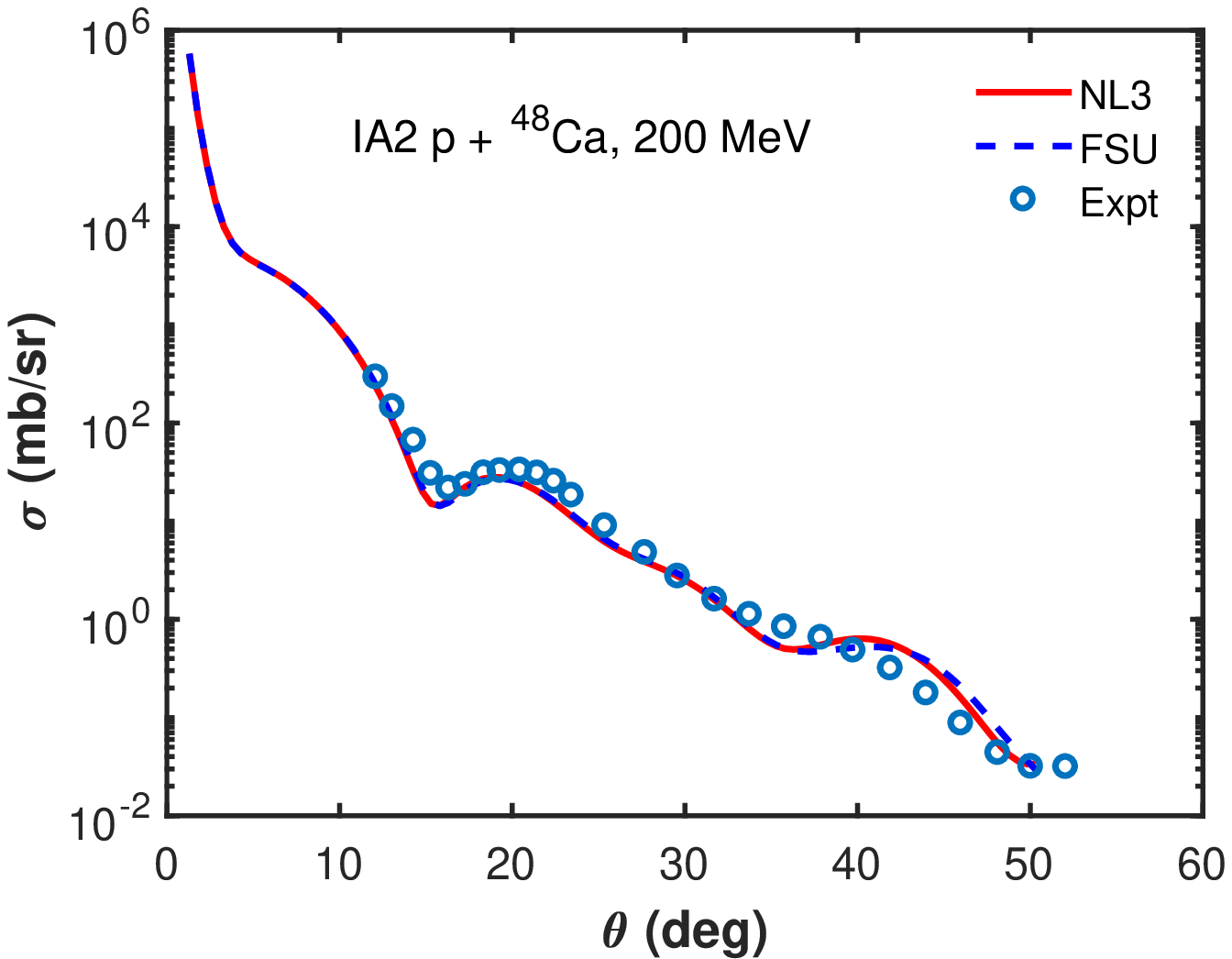}
	\includegraphics[width=0.49\linewidth]{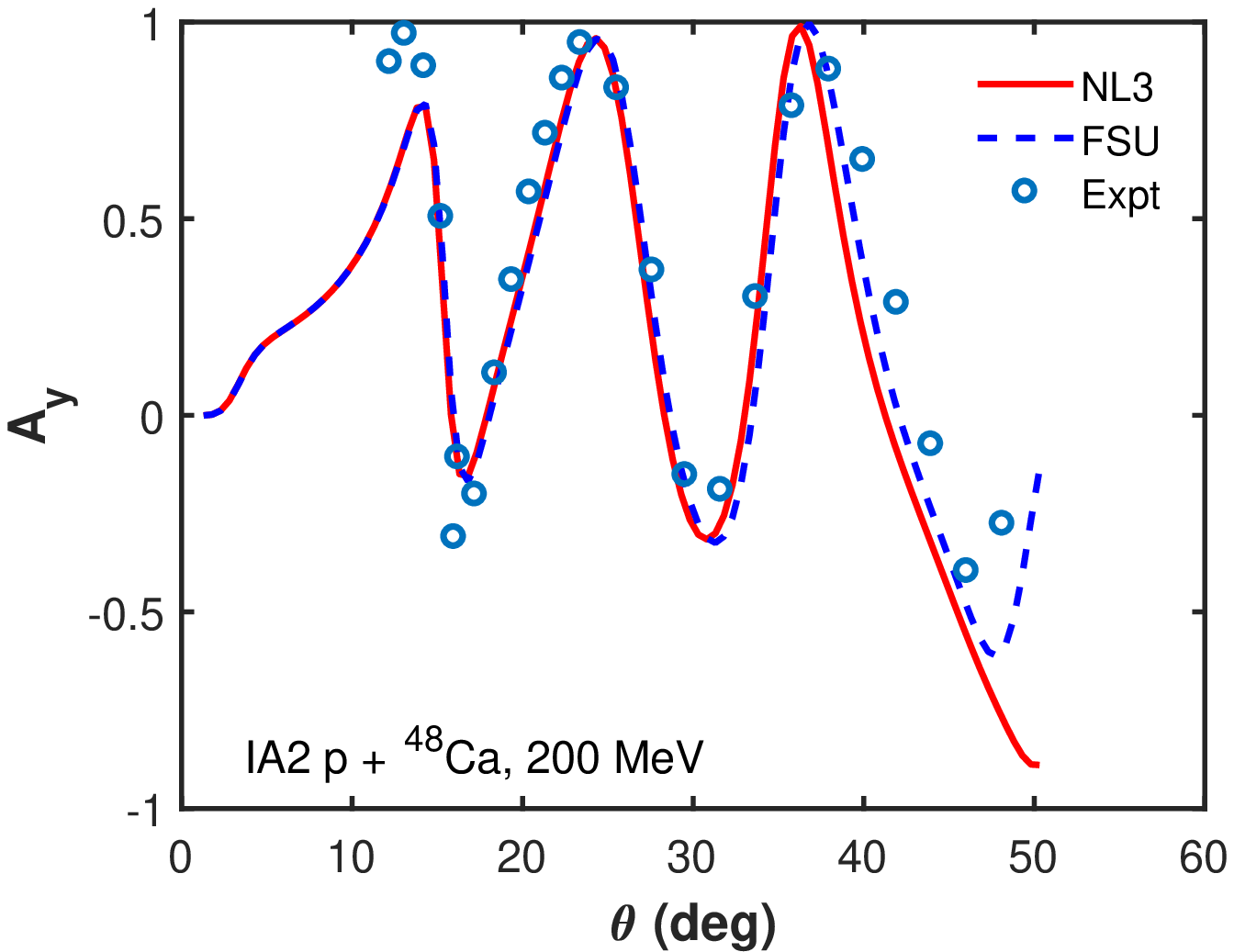}
	\includegraphics[width=0.49\linewidth]{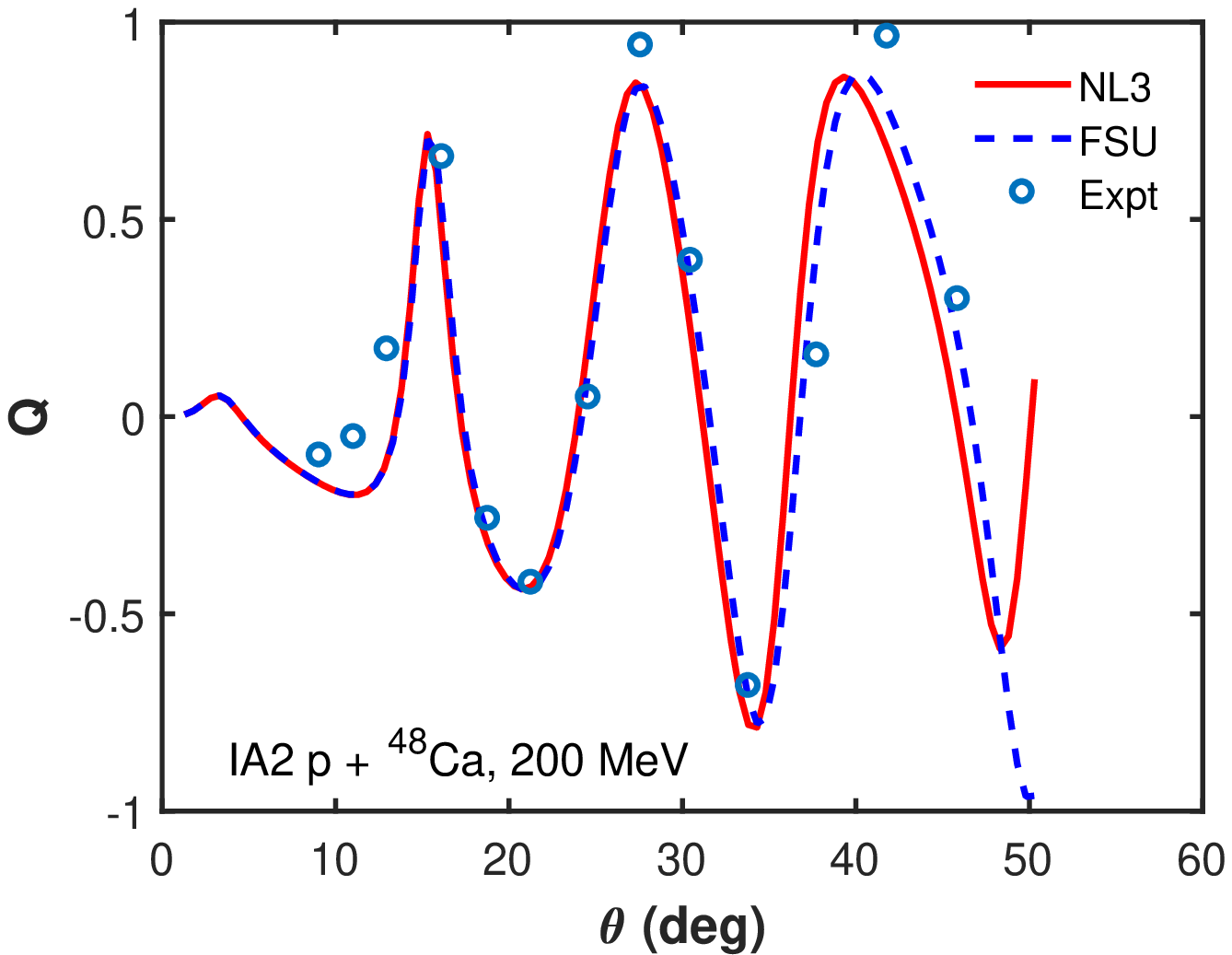}
	\caption{Same as in figure \ref{RMFCa40_200} except for $^{48}$Ca at $T_{\mathrm{lab}} = 200$MeV.}
	\label{RMFCa48_200}
\end{figure}

\begin{figure}
	\centering
	\includegraphics[width=0.49\linewidth]{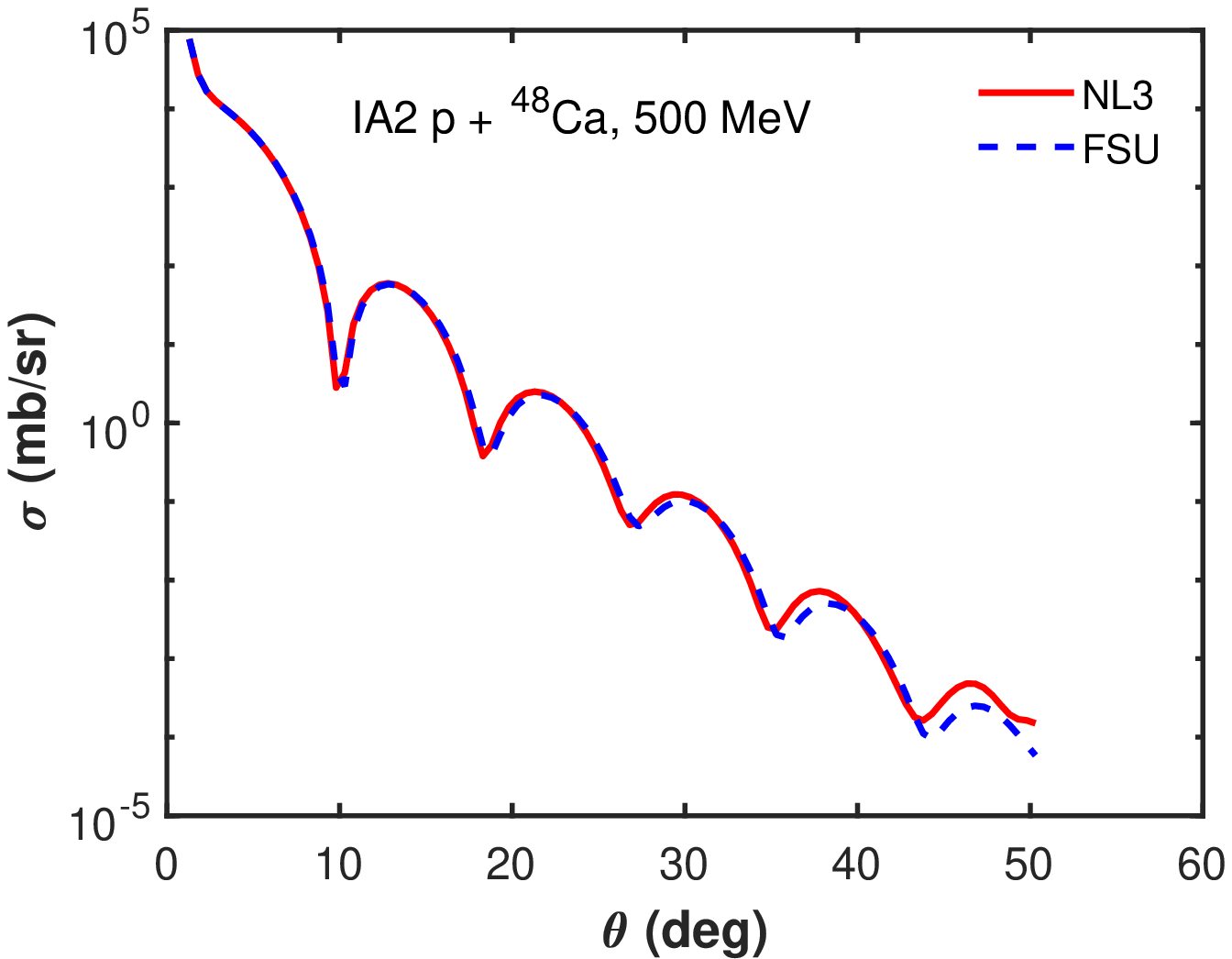}
	\includegraphics[width=0.49\linewidth]{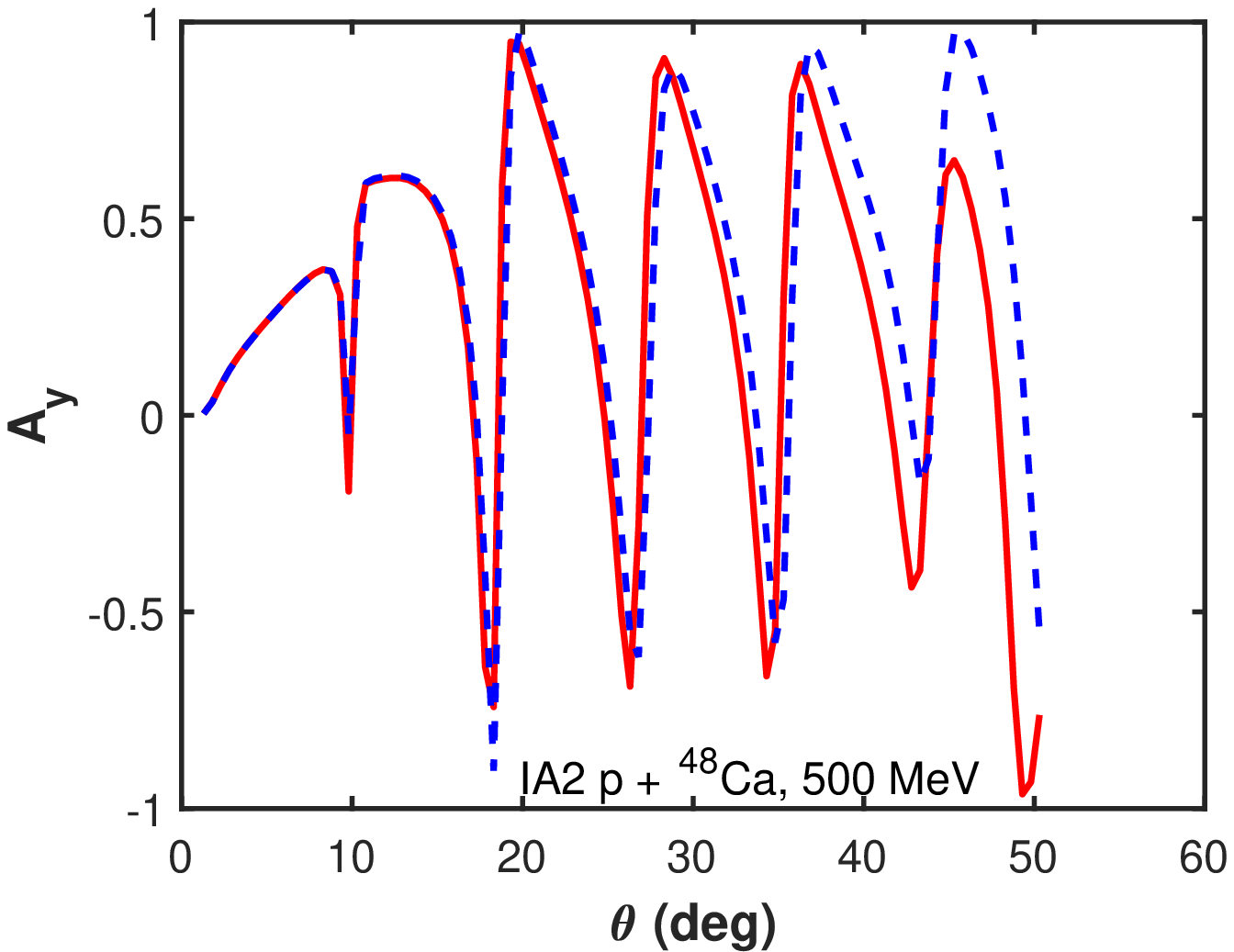}
	\includegraphics[width=0.49\linewidth]{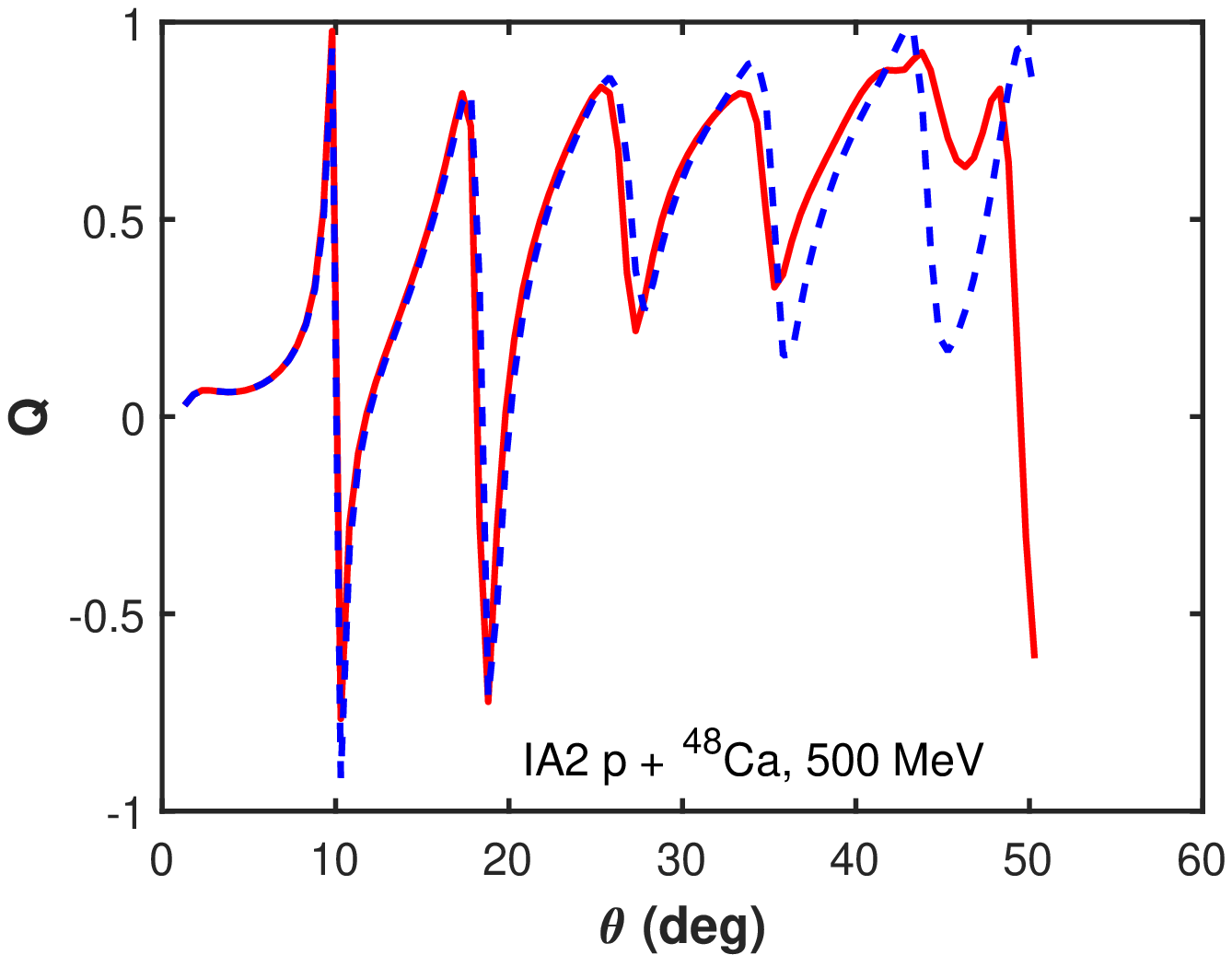}
	\caption{Same as in figure \ref{RMFCa40_200} except for $^{48}$Ca at $T_{\mathrm{lab}} = 500$MeV.}
	\label{RMFCa48_500}
\end{figure}

\begin{figure}
	\centering
	\includegraphics[width=0.49\linewidth]{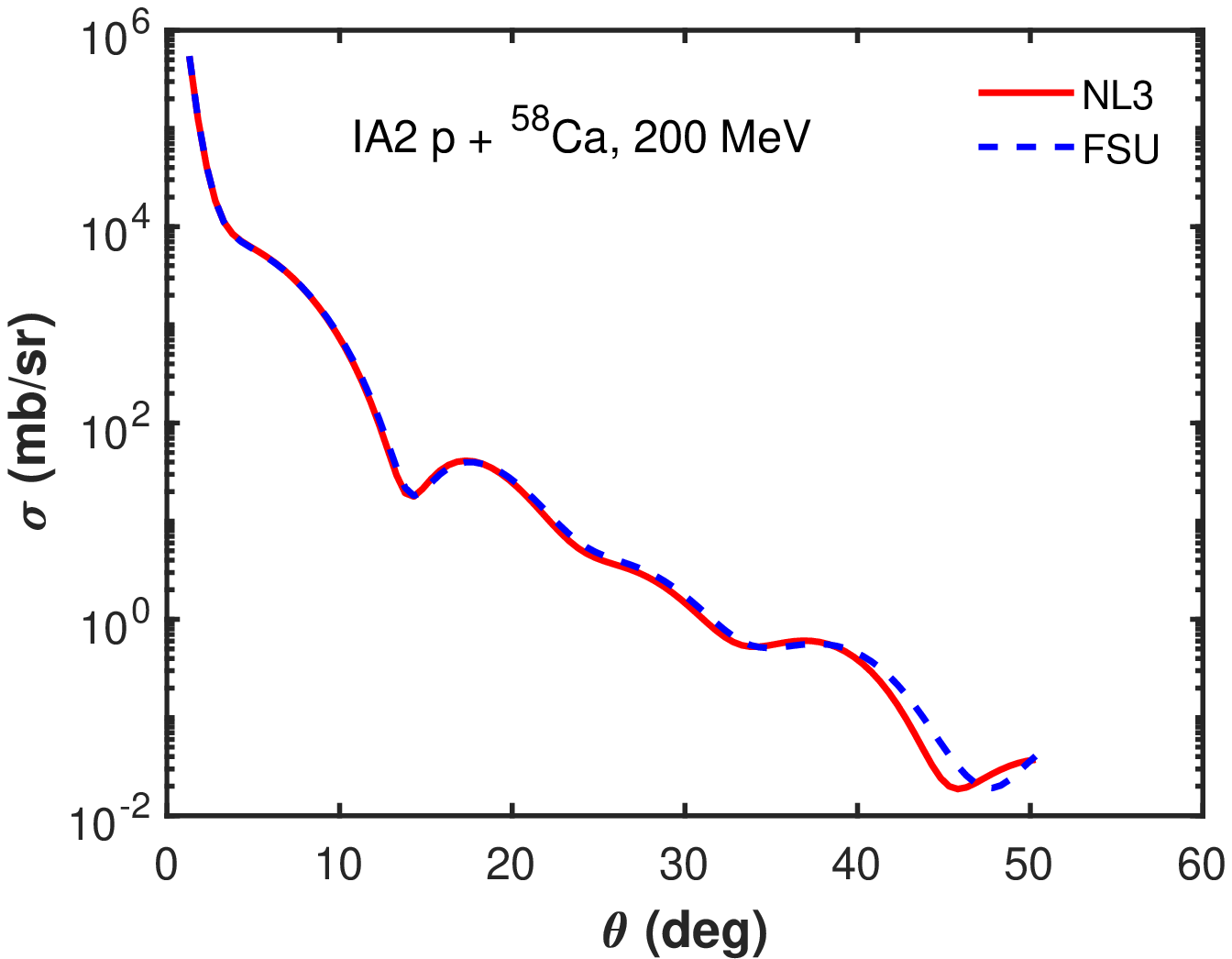}
	\includegraphics[width=0.49\linewidth]{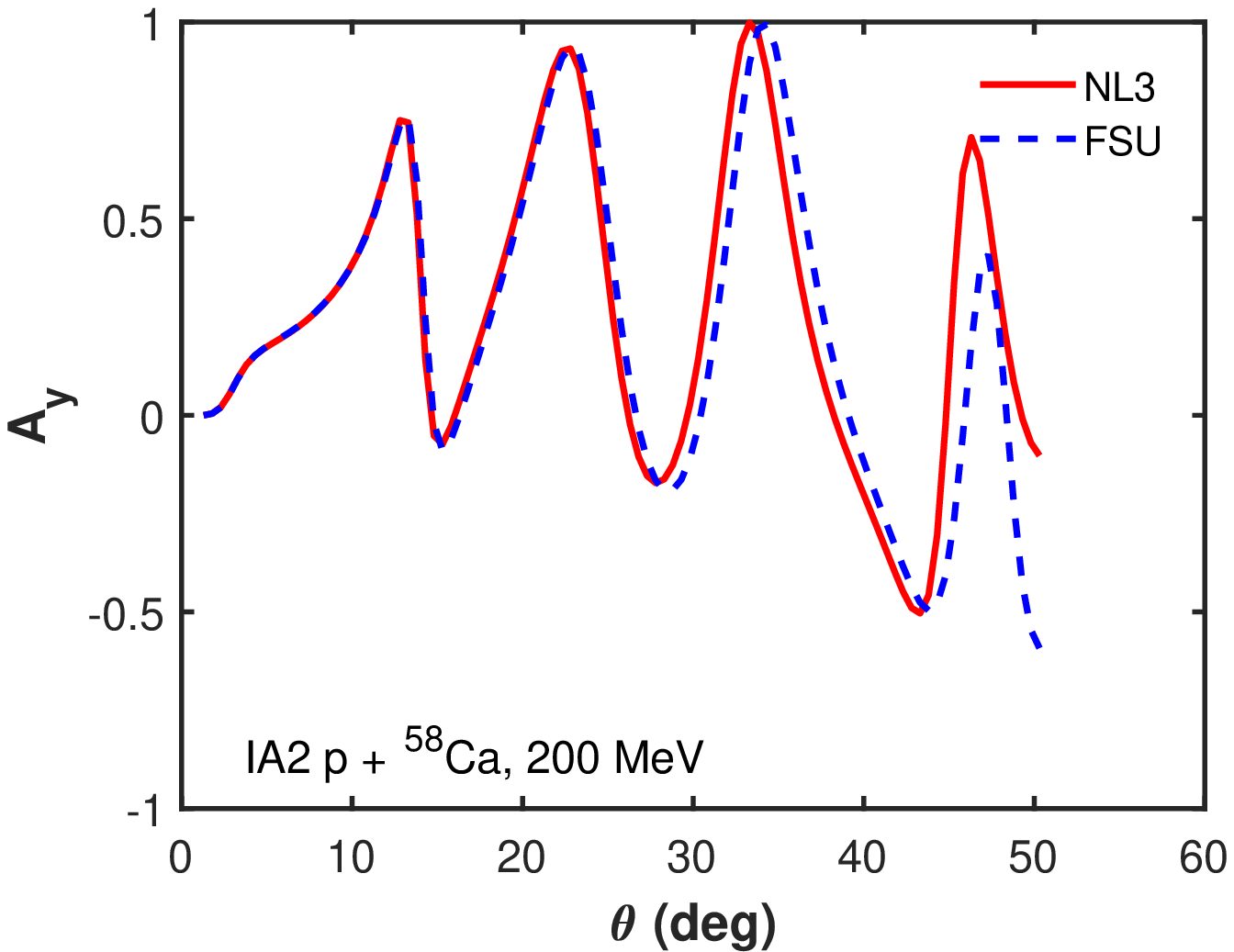}
	\includegraphics[width=0.49\linewidth]{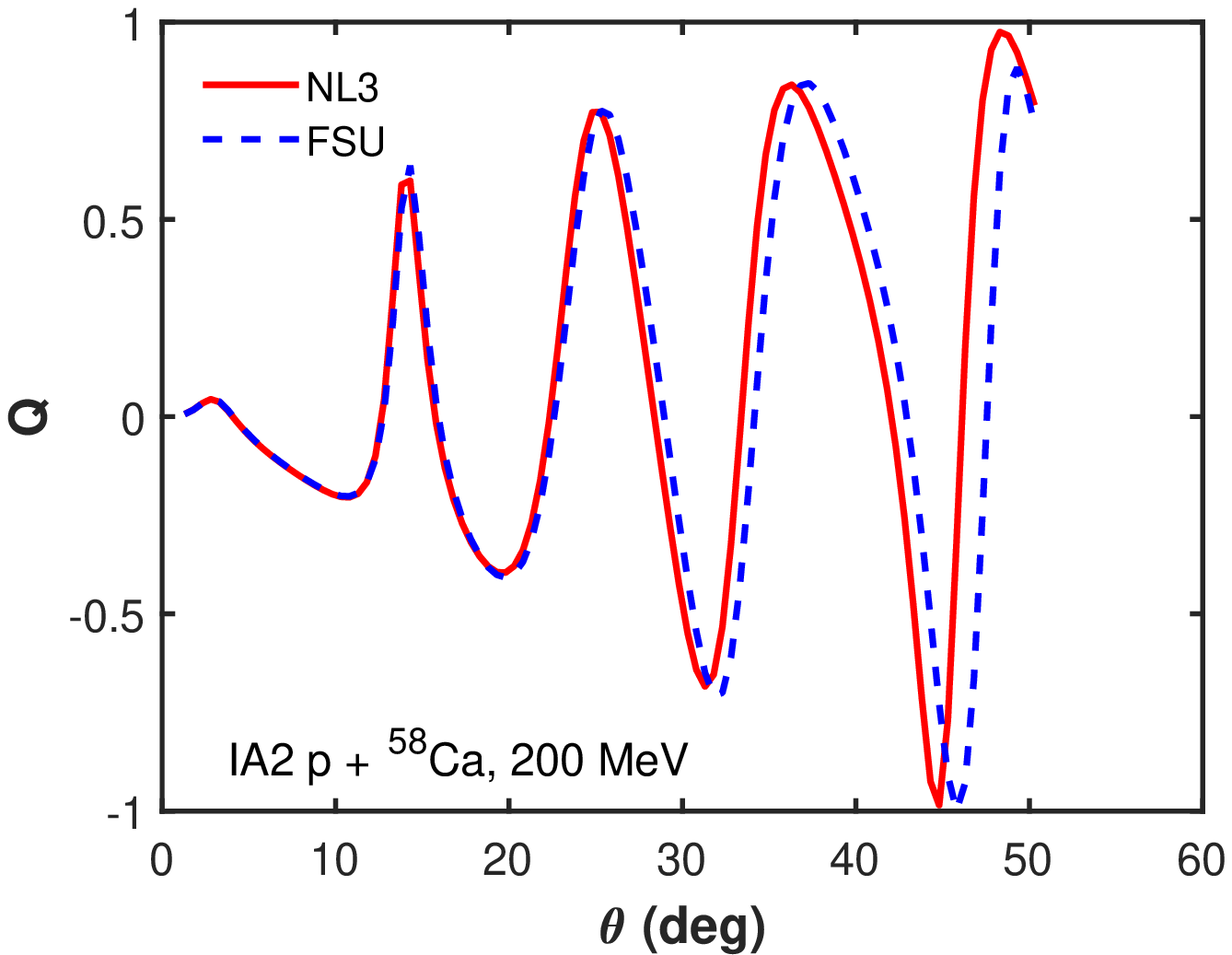}
	\caption{Same as in figure \ref{RMFCa40_200} except for $^{58}$Ca at $T_{\mathrm{lab}} = 200$MeV.}
	\label{RMFCa58_200}
\end{figure}

\begin{figure}
	\centering
	\includegraphics[width=0.49\linewidth]{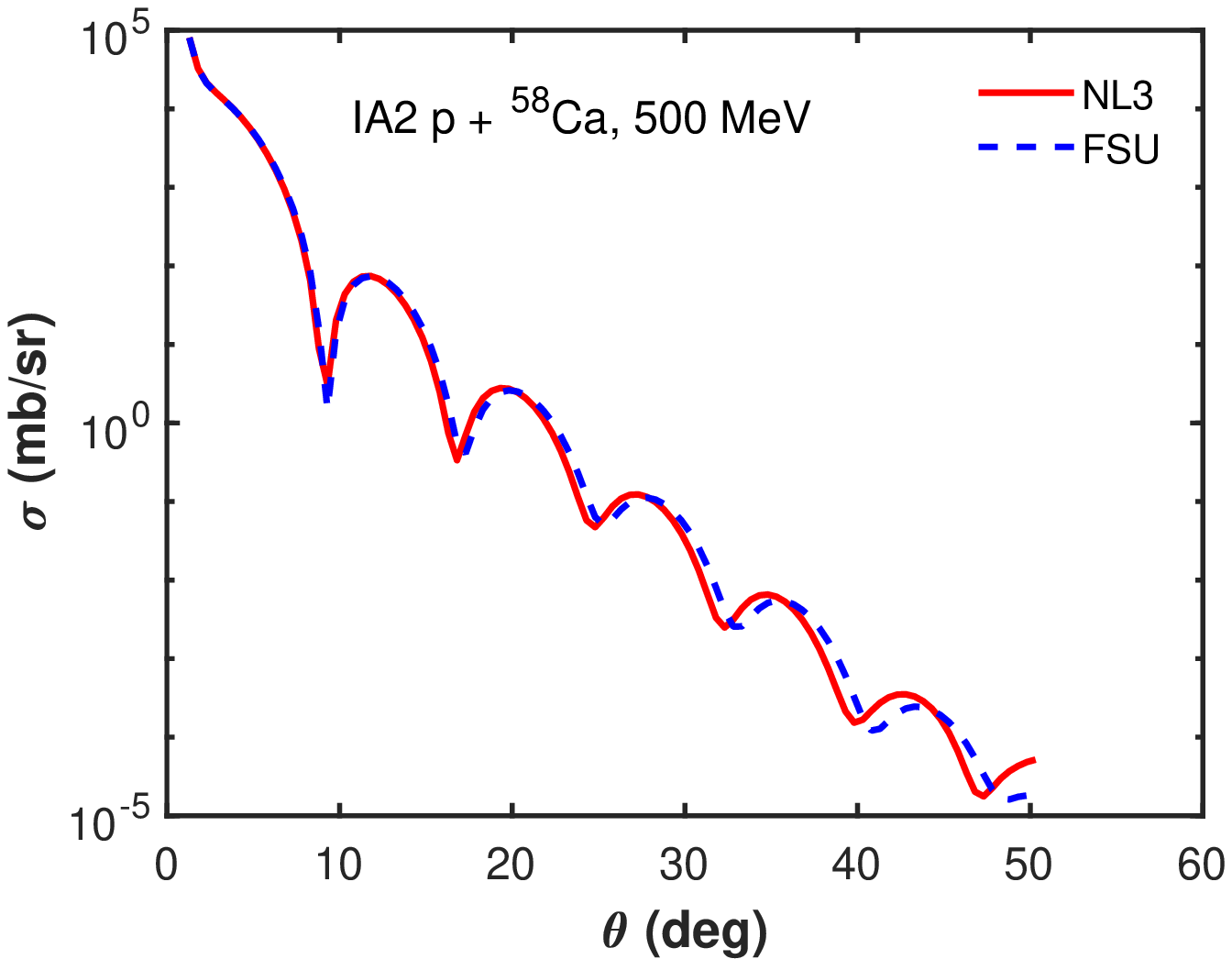}
	\includegraphics[width=0.49\linewidth]{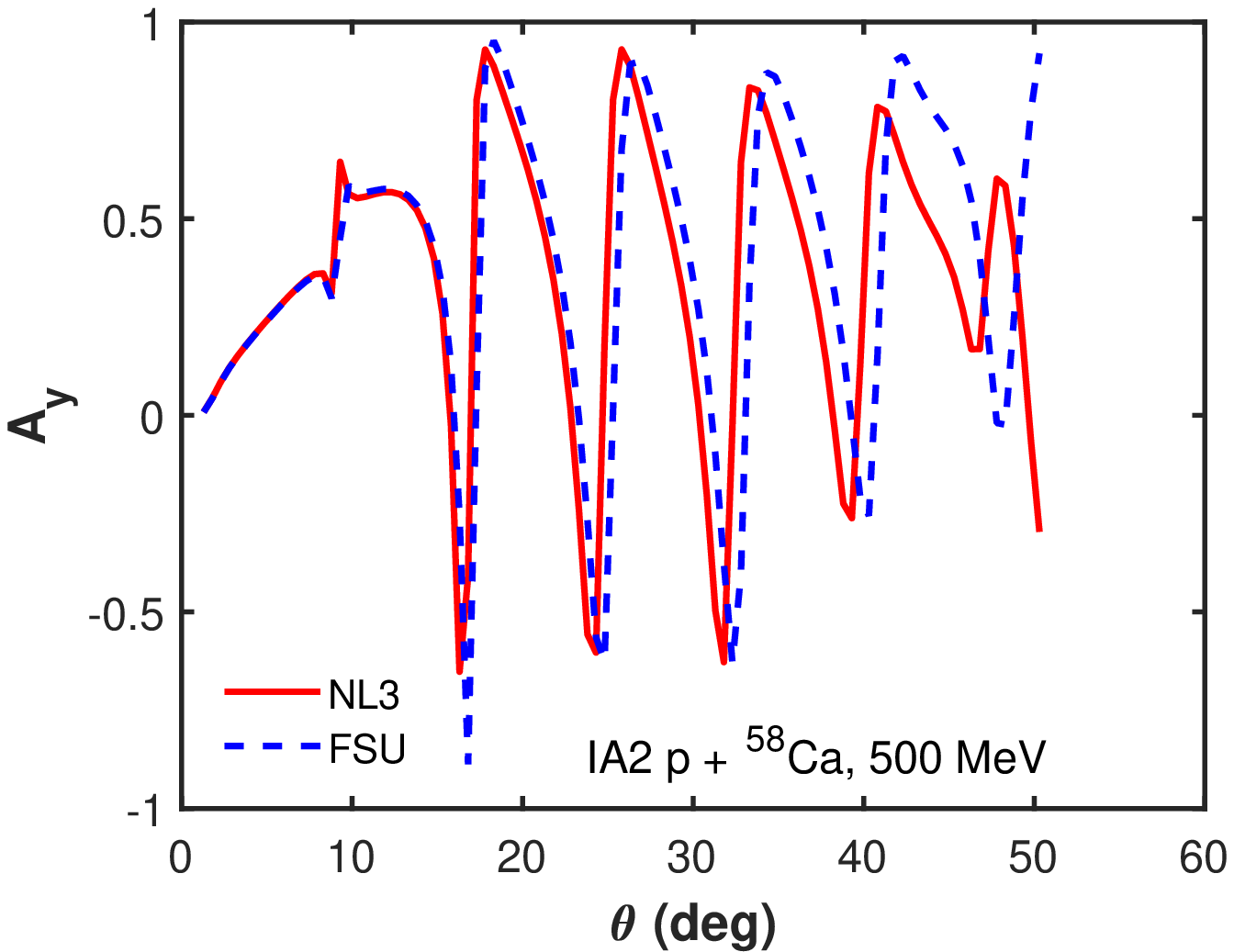}
	\includegraphics[width=0.49\linewidth]{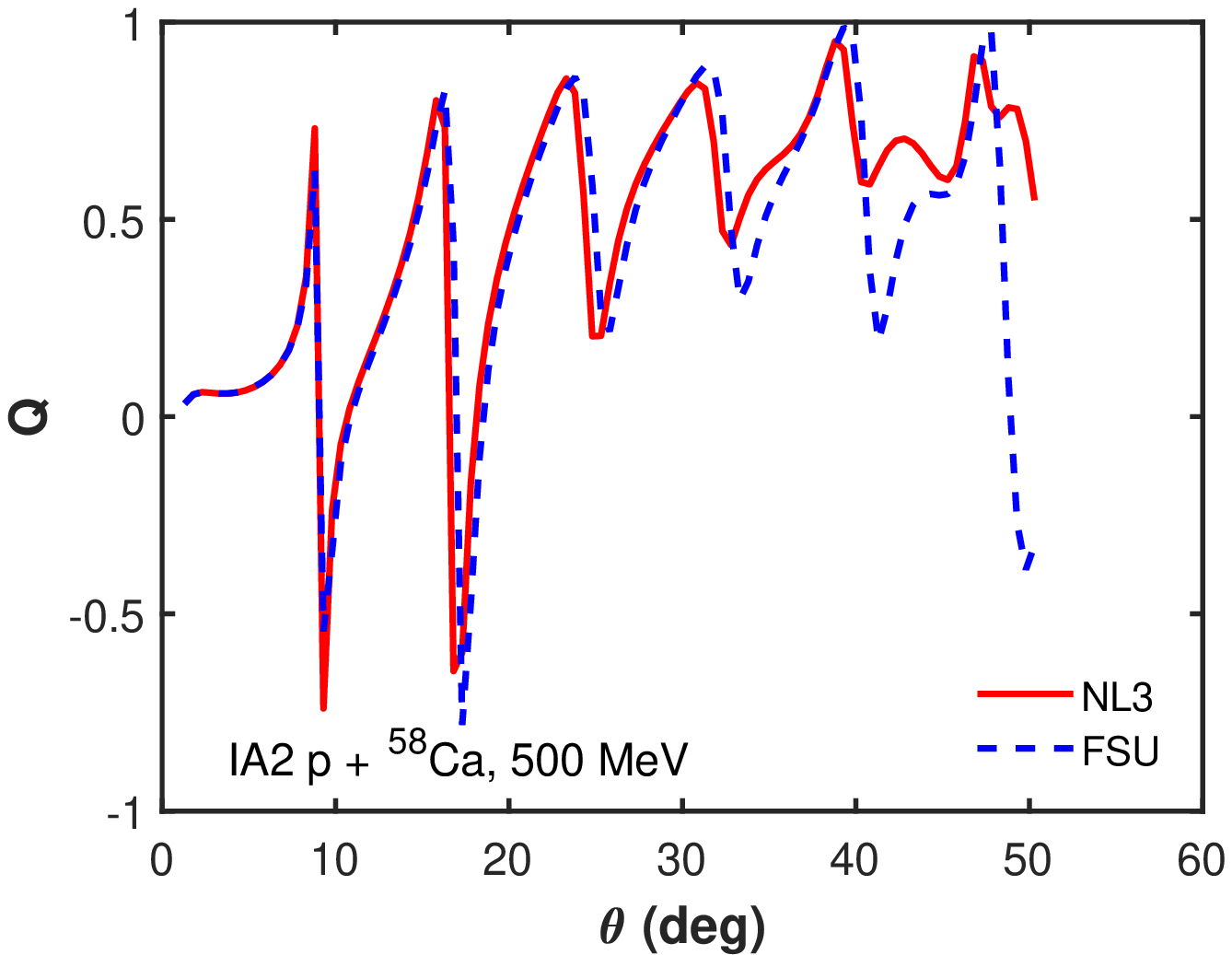}
	\caption{Same as in figure \ref{RMFCa40_200} except for $^{58}$Ca at $T_{\mathrm{lab}} = 500$MeV.}
	\label{RMFCa58_500}
\end{figure}

\begin{figure}
	\centering
	\includegraphics[width=0.49\linewidth]{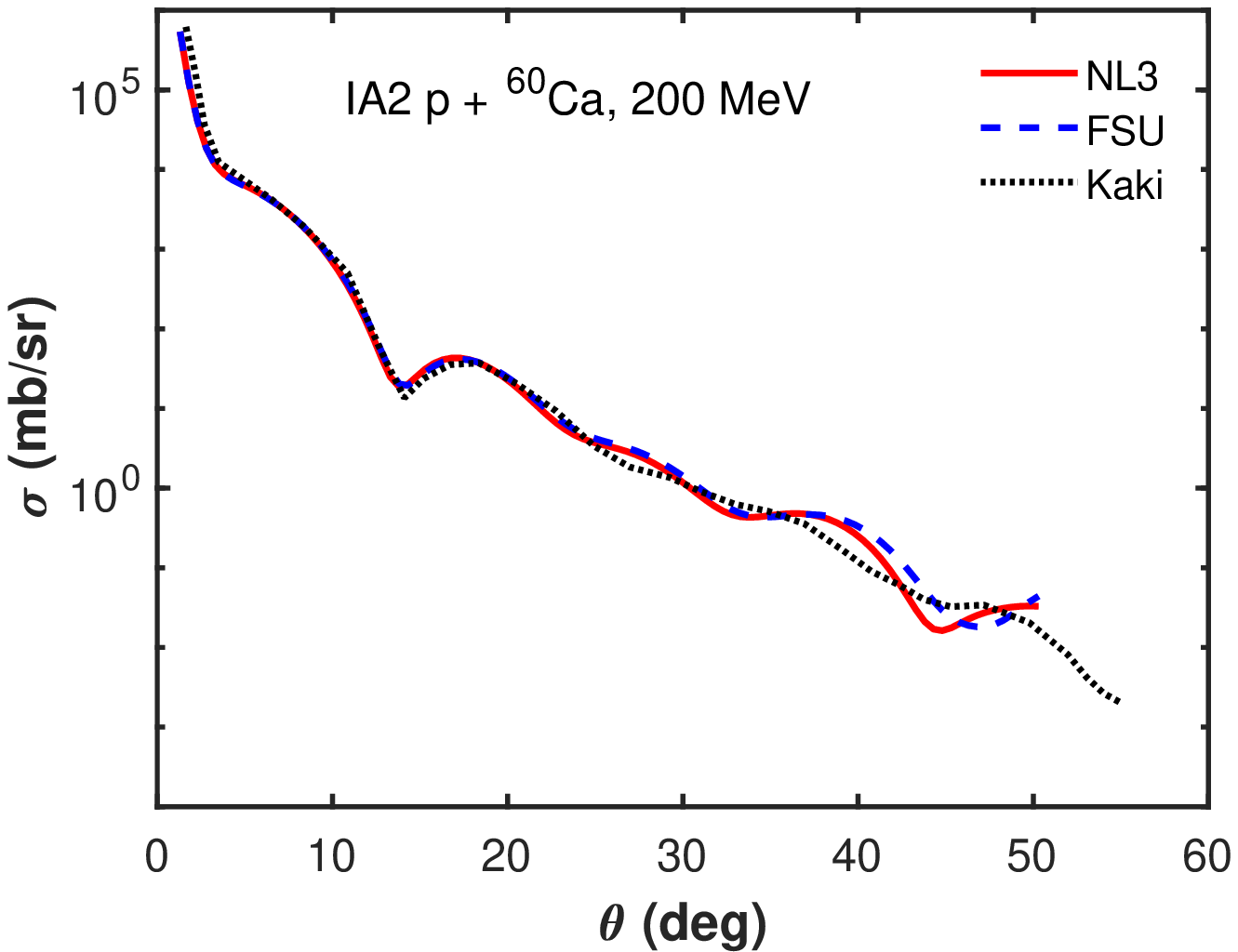}
	\includegraphics[width=0.49\linewidth]{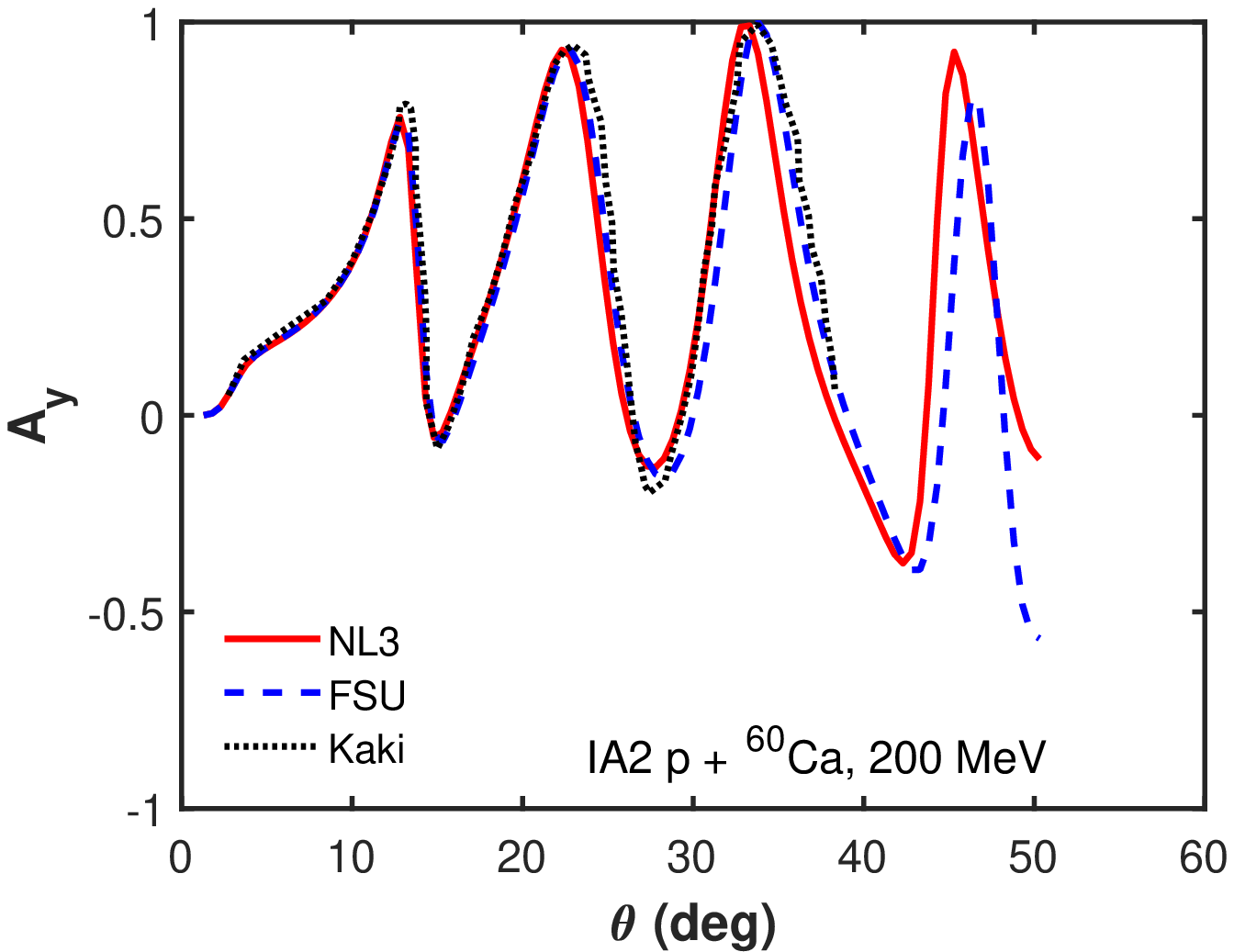}
	\includegraphics[width=0.49\linewidth]{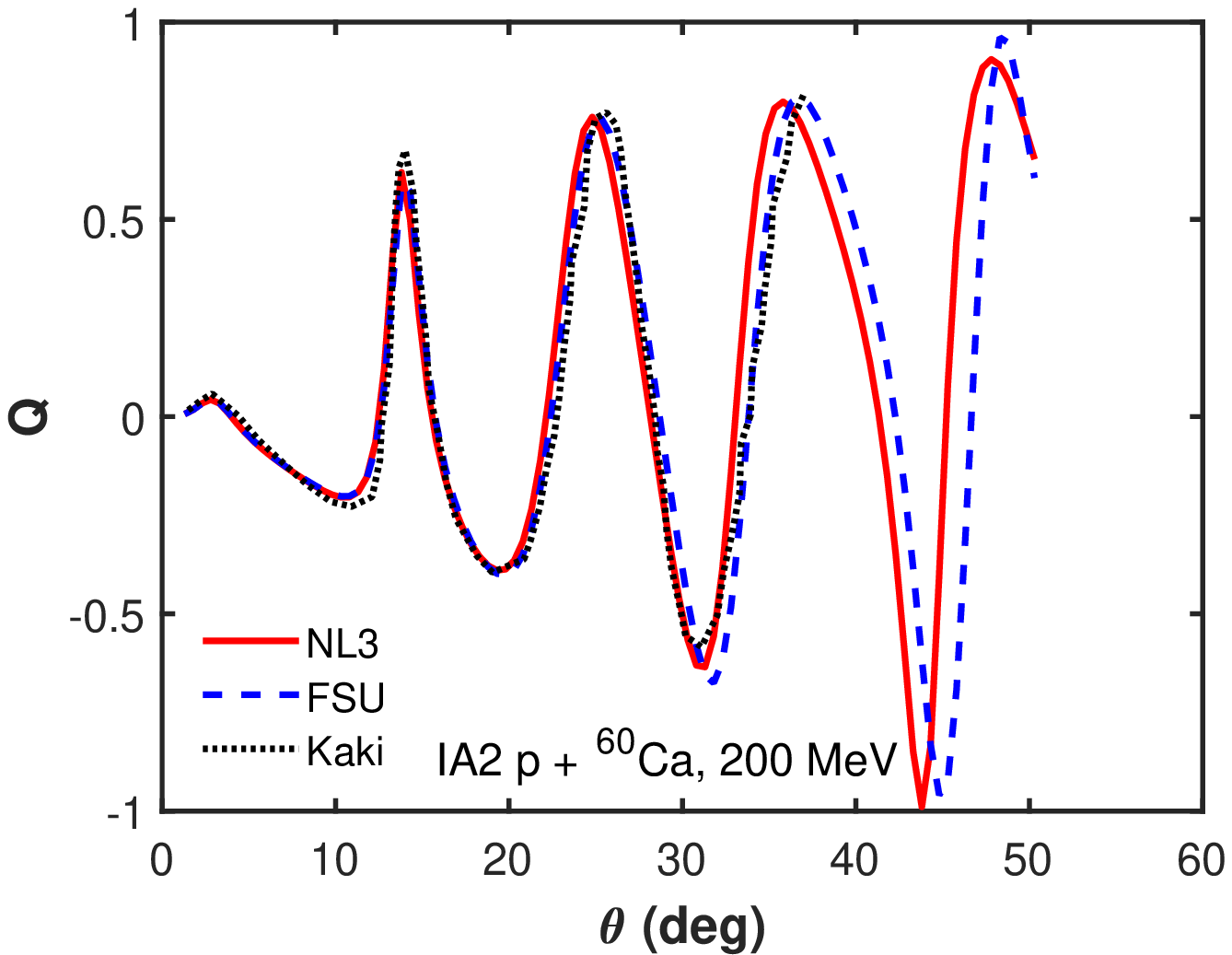}
	\caption{Same as in figure \ref{RMFCa40_200} except for $^{60}$Ca at $T_{\mathrm{lab}} = 200$MeV.}
	\label{RMFCa60_200}
\end{figure}

\begin{figure}
	\centering
	\includegraphics[width=0.49\linewidth]{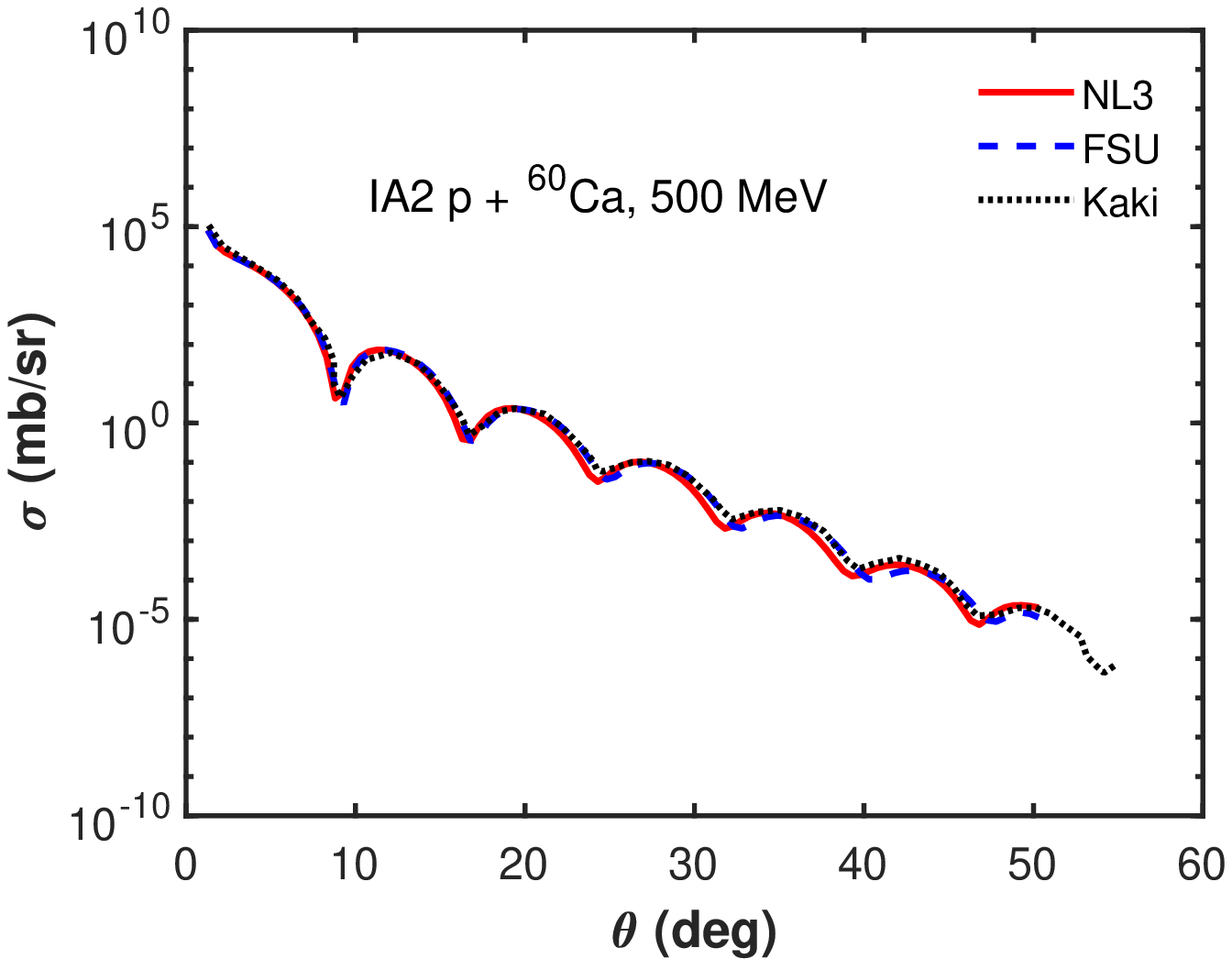}
	\includegraphics[width=0.49\linewidth]{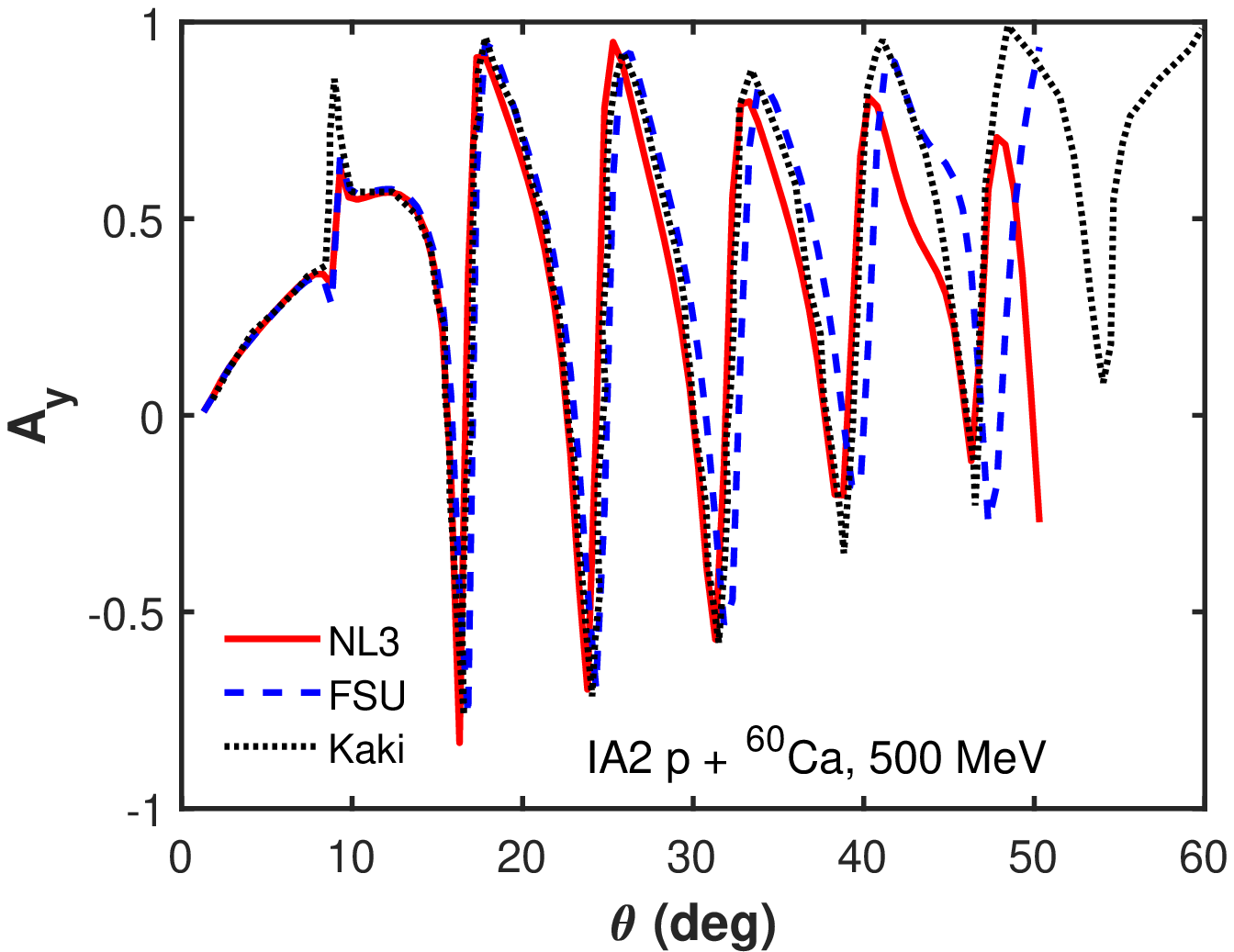}
	\includegraphics[width=0.49\linewidth]{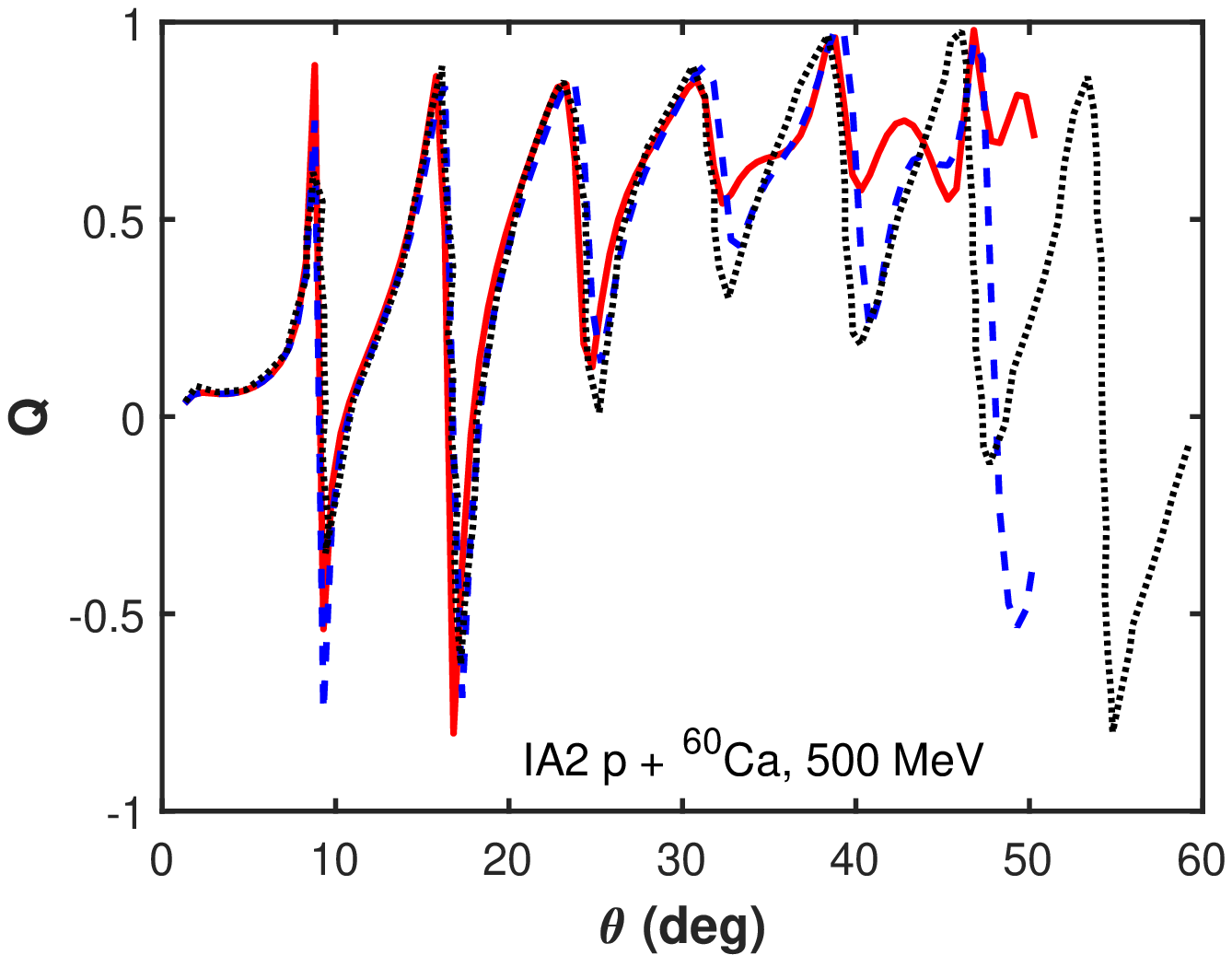}
	\caption{Same as in figure \ref{RMFCa40_200} except for $^{60}$Ca at $T_{\mathrm{lab}} = 500$MeV.}
	\label{RMFCa60_500}
\end{figure}

Figures \ref{RMFCa40_200} to \ref{RMFCa60_500} show the scattering observables for elastic proton scattering from $^{40-60}$Ca nucleus (at $T_{\mathrm{lab}} = 200$ and $500$ MeV) as functions of the centre of mass scattering angle $\theta$ calculated using the NL3, and FSUGold parameter sets with the IA2 relativistic optical potentials. The top left panel shows the scattering cross section results, the top right panel for analysing power, and the bottom plots show results for the spin rotation parameters. The results obtained with the NL3 parameter set are shown in solid lines, while the results obtained using the FSUGold parameter set are shown in dashed lines.

One can observe from figure \ref{RMFCa40_200} that the two RMF models (NL3 and FSUGold) give very good descriptions of the scattering observables for $p + ^{40}$Ca at $T_{\mathrm{lab}} = 200$MeV. However, the FSUGold parameter set gives better descriptions of the analysing power and spin rotation parameter at large scattering angles. The same conclusions can be drawn at incident projectile energy of $500$ MeV shown in figure \ref{RMFCa40_500}.

The two RMF models give very good descriptions of the scattering observables for $p + ^{48}$Ca at $T_{\mathrm{lab}} = 200$MeV as shown in figure \ref{RMFCa48_200}. Unlike the case of $^{40}$Ca where the FSUGold parameter set gave better descriptions of the spin observables, the two RMF models give comparable descriptions of the spin observables for $^{48}$Ca. The case for incident projectile energy of $500$ MeV is shown in figure \ref{RMFCa48_500}, where the difference between the two RMF models can be observed at large scattering angles $\theta \gtrapprox 35^{\circ}$.

For $p + ^{58}$Ca at $T_{\mathrm{lab}} = 200$ and $500$ MeV shown in figures \ref{RMFCa58_200} and \ref{RMFCa58_500}, it can be observed that the two models also give similar descriptions of the scattering observables. There is a slight difference however, observed at large scattering angles for incident projectile energy of $500$ MeV. 

Figures \ref{RMFCa60_200} and \ref{RMFCa60_500} show the scattering observables calculated for $p + ^{60}$Ca at incident proton energies of $200$ and $500$ MeV, respectively. There is no available experimental data now, but we compare with the theoretical calculations presented in Ref. \cite{kak09} by Kaki. There is a very good agreement between our results and that of Kaki, as the two results give identical values of scattering cross section at the first minimum. It should be noted that the IA2 formalism was also employed by Kaki, but with the use of different Lagrangian densities.

\begin{figure}
	\centering
	\includegraphics[width=0.49\linewidth]{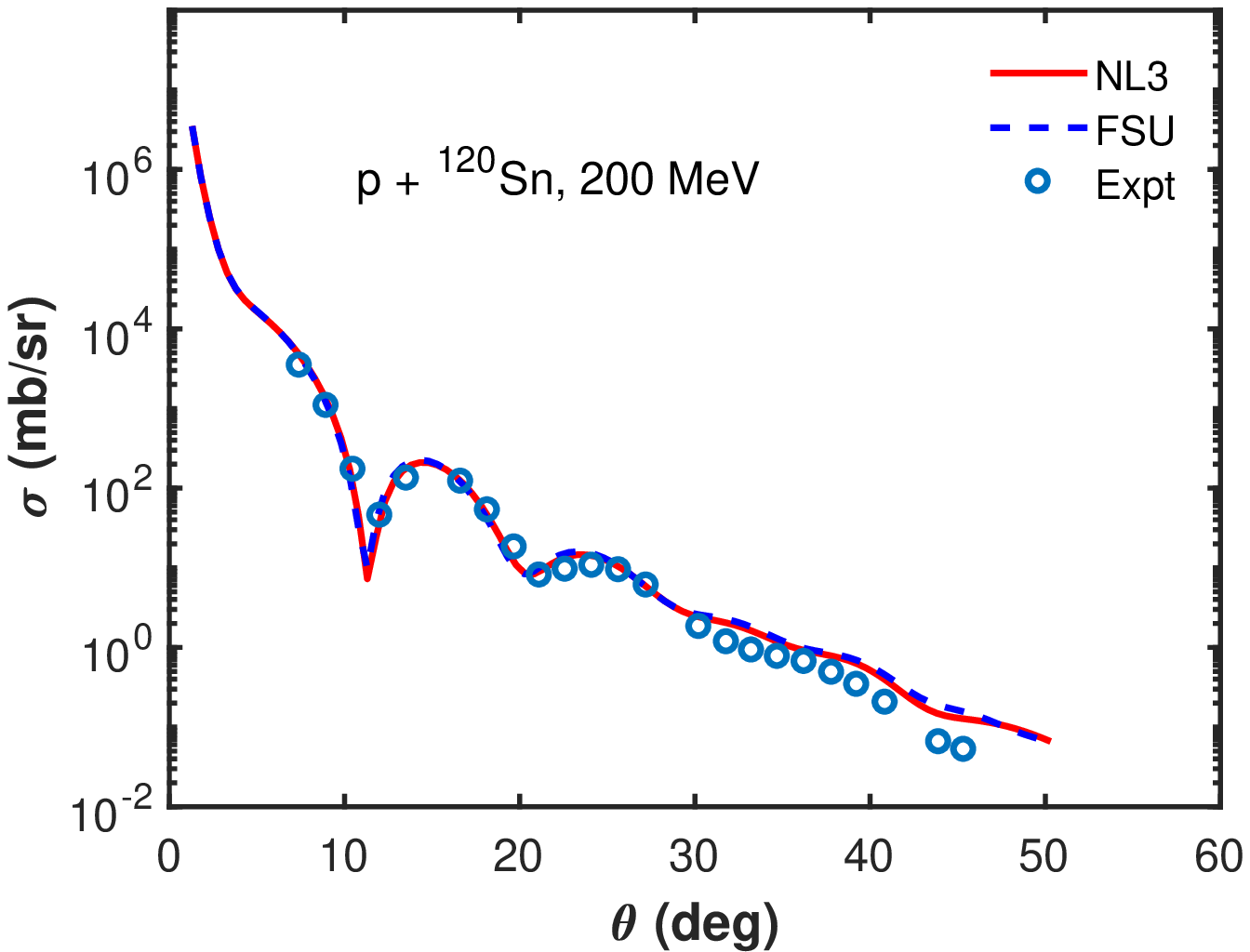}
	\includegraphics[width=0.49\linewidth]{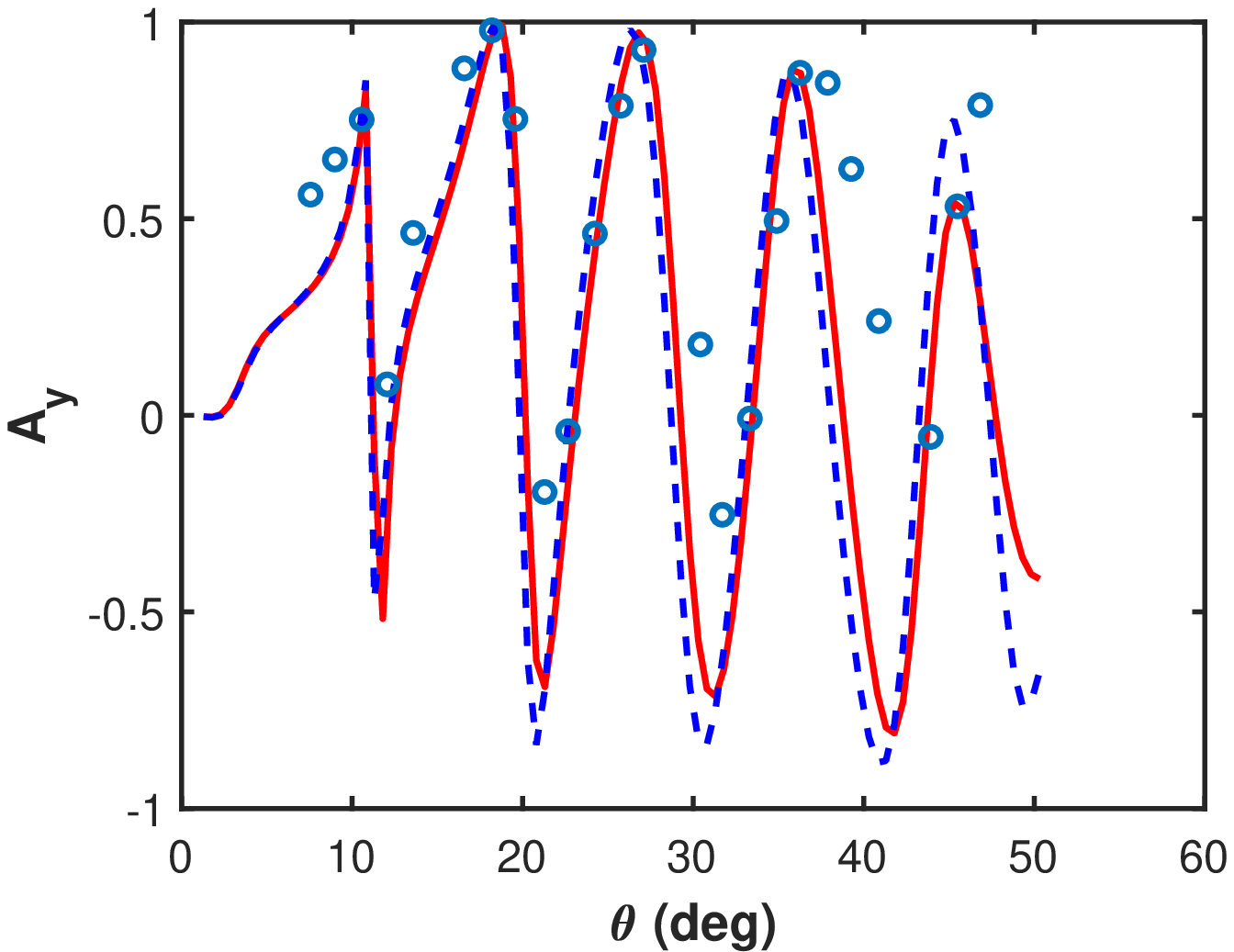}
	\includegraphics[width=0.49\linewidth]{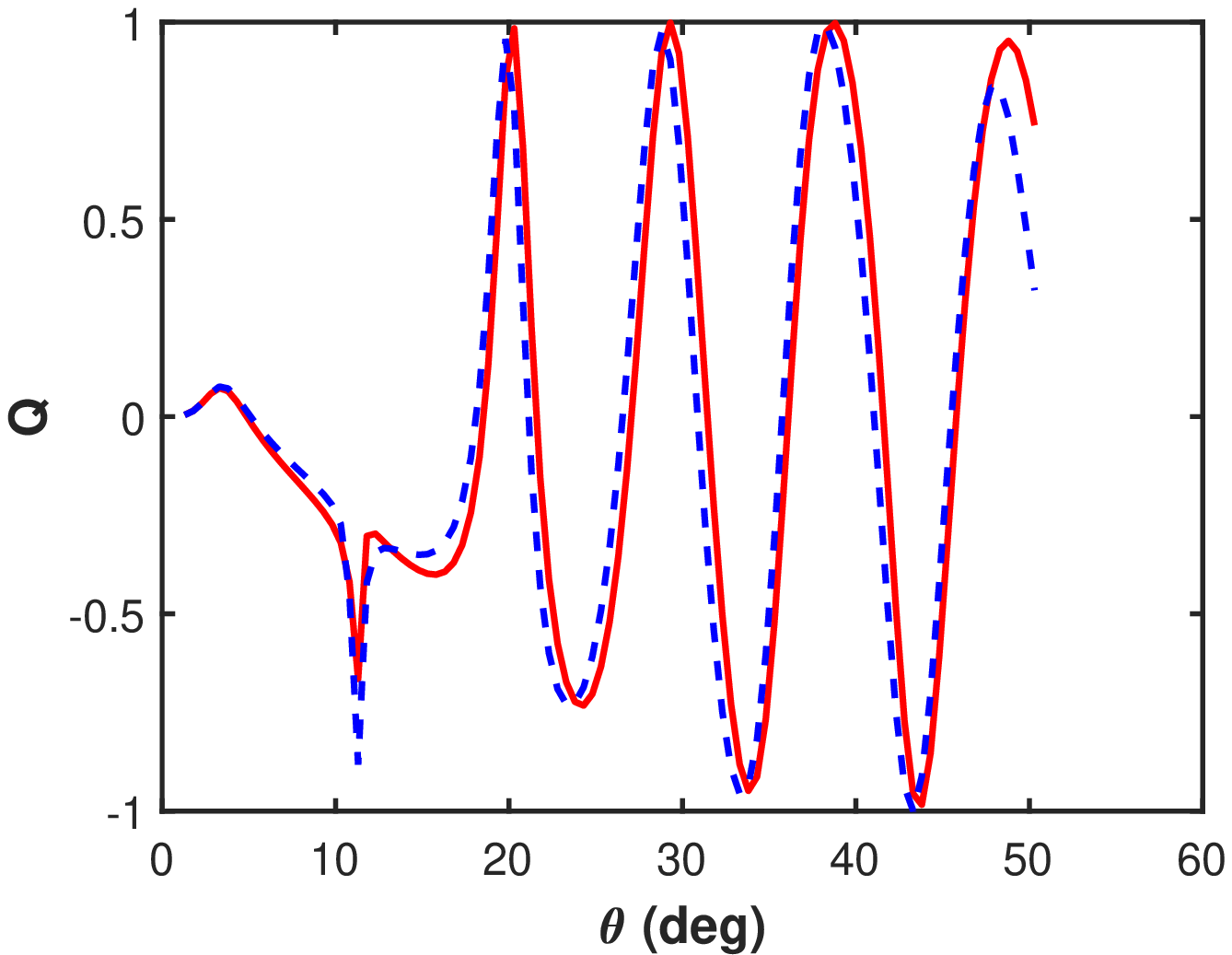}
	\caption{Same as in figure \ref{RMFCa40_200} except for $^{120}$Sn at $T_{\mathrm{lab}} = 200$MeV. Data taken from Ref. \cite{kak01}.}
	\label{RMFSn120_200}
\end{figure}

\begin{figure}
	\centering
	\includegraphics[width=0.49\linewidth]{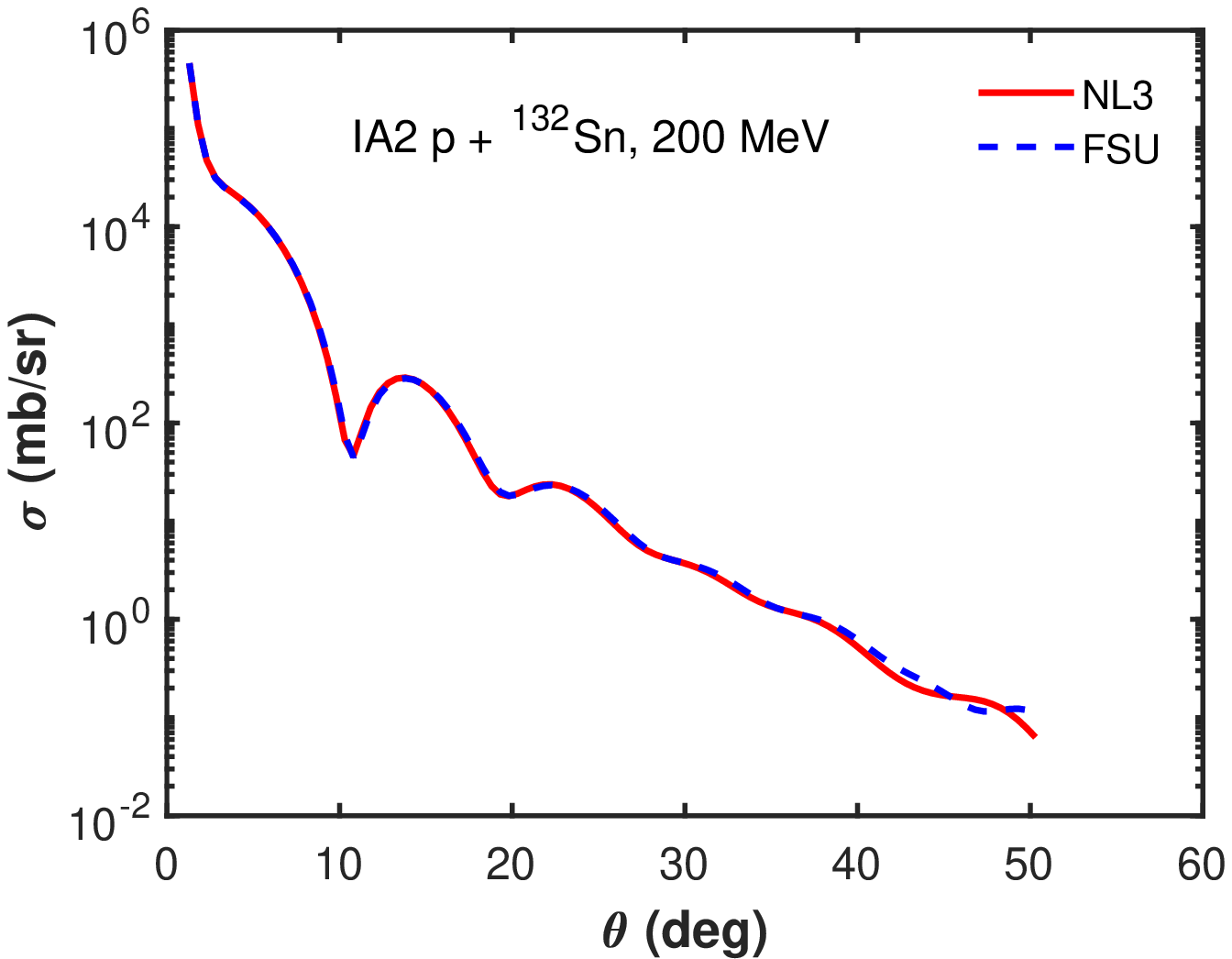}
	\includegraphics[width=0.49\linewidth]{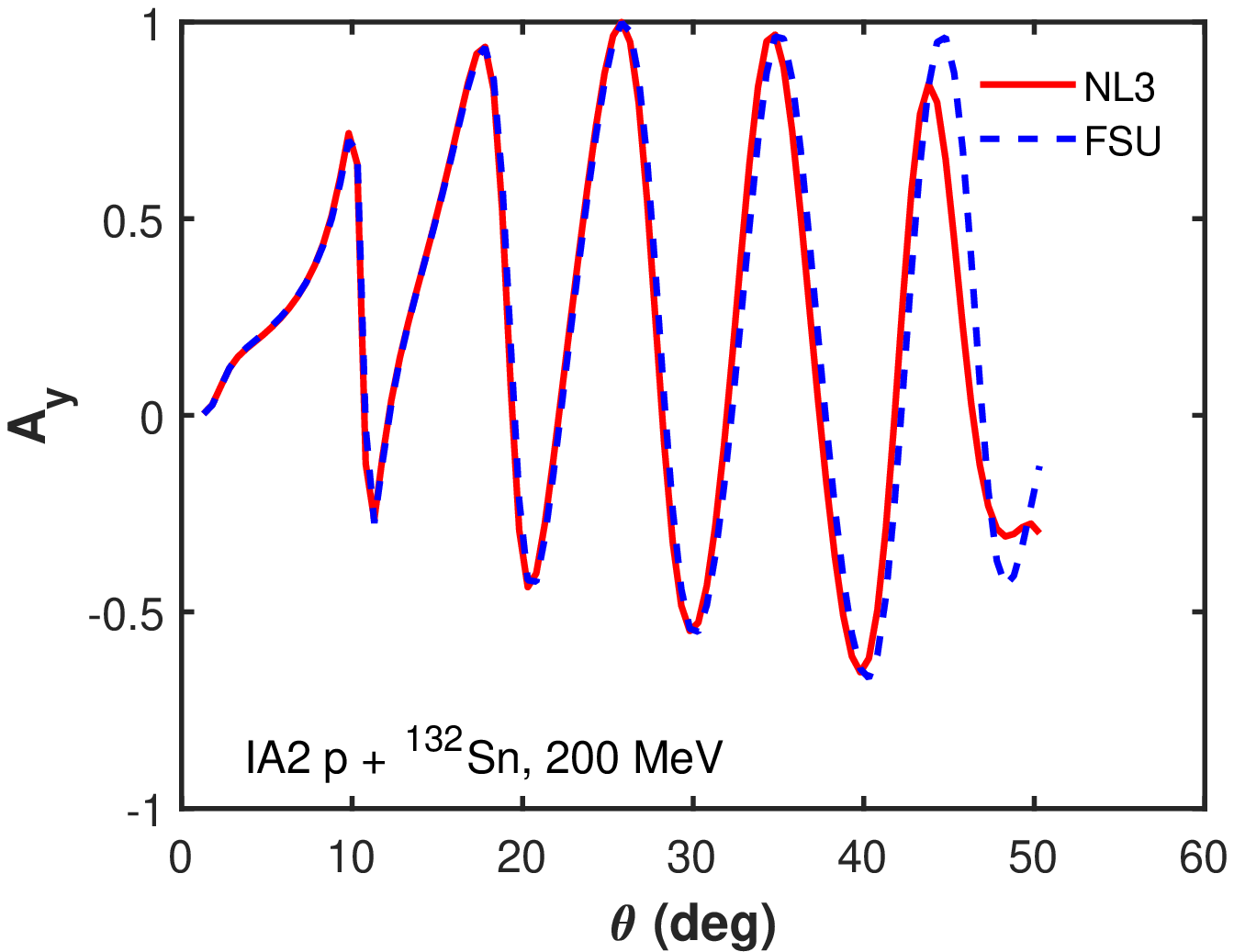}
	\includegraphics[width=0.49\linewidth]{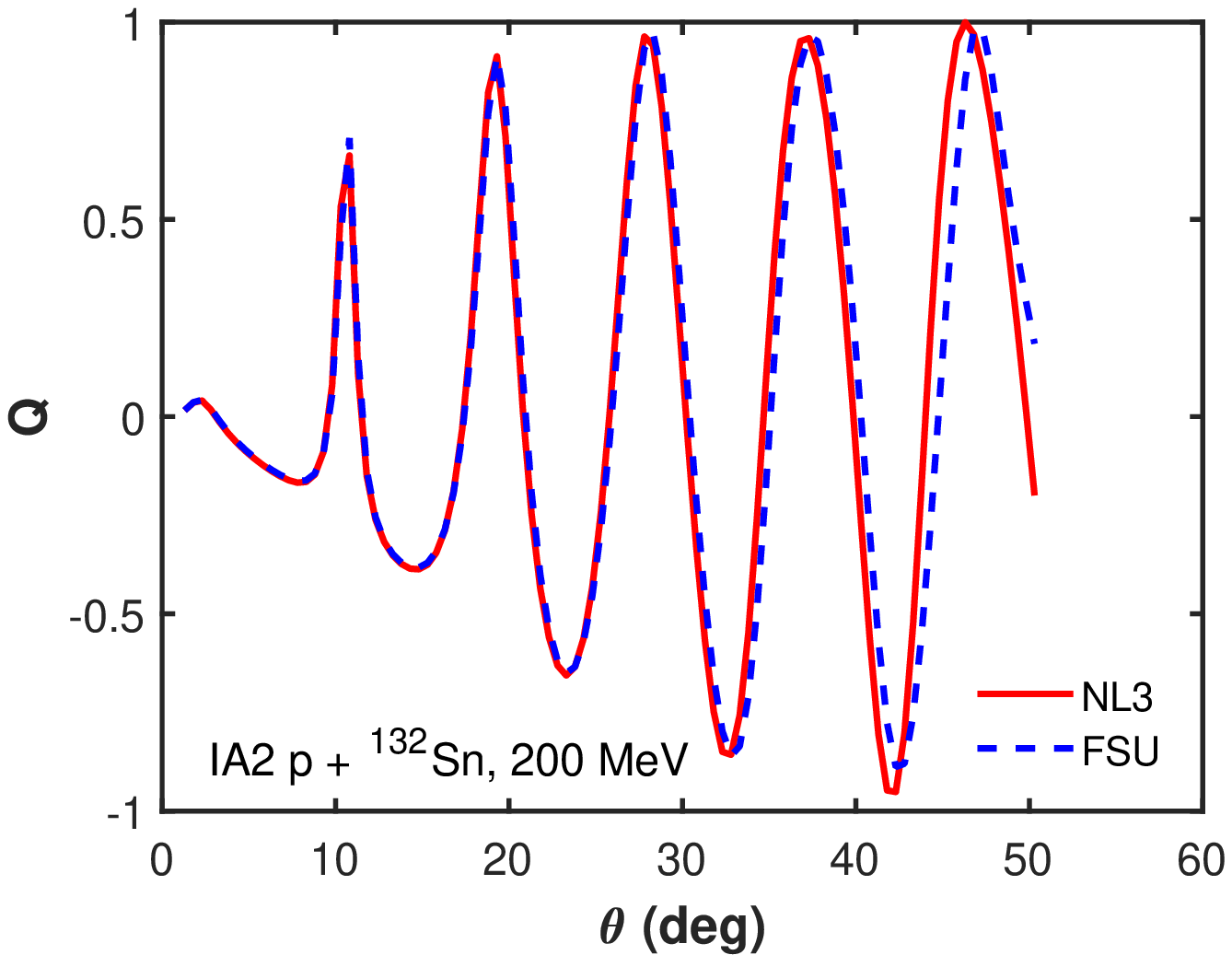}
	\caption{Same as in figure \ref{RMFCa40_200} except for $^{132}$Sn at $T_{\mathrm{lab}} = 200$MeV.}
	\label{RMFSn132_200}
\end{figure}

We also show in figures \ref{RMFSn120_200} and \ref{RMFSn132_200} the scattering observables for $p + ^{120}$Sn and $p + ^{132}$Sn, respectively at $T_{\mathrm{lab}} = 200$MeV calculated using both NL3 and FSUGold parameter sets. The two models give similar descriptions of the scattering observables. It can be observed from figure \ref{RMFSn120_200} that both NL3 and FSUGold models give very good descriptions of the differential cross section and analysing power data for proton scattering from $^{120}$Sn. Data are taken from Ref. \cite{kak01}

\subsection{IA1 and IA2}
In this sub-section, the results of the scattering cross section and spin observables obtained using the IA1 and IA2 formalisms are presented and compared. The calculations have been carried out using the NL3 parameter set.

\begin{figure}
	\centering
	\includegraphics[width=0.49\linewidth]{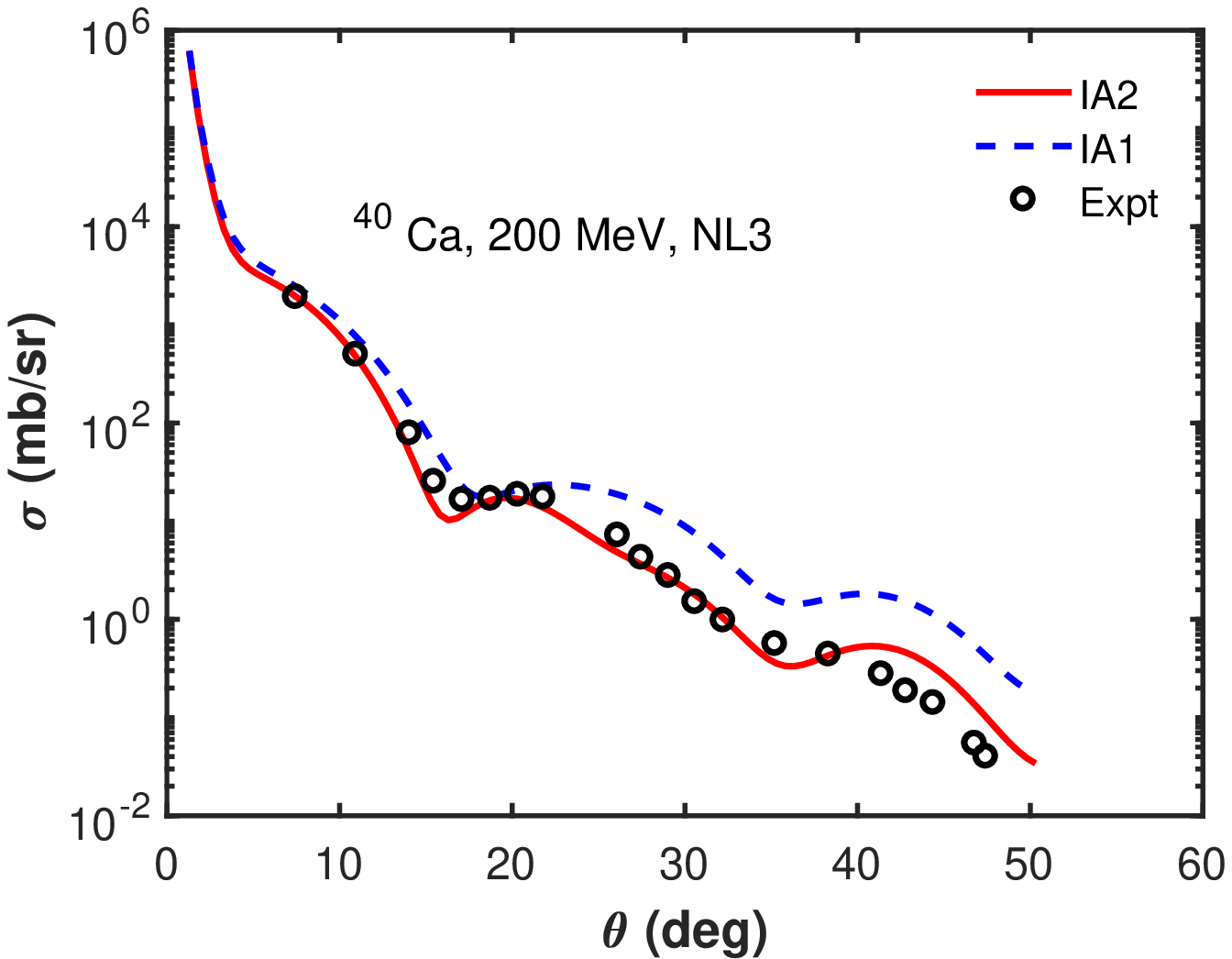}
	\includegraphics[width=0.49\linewidth]{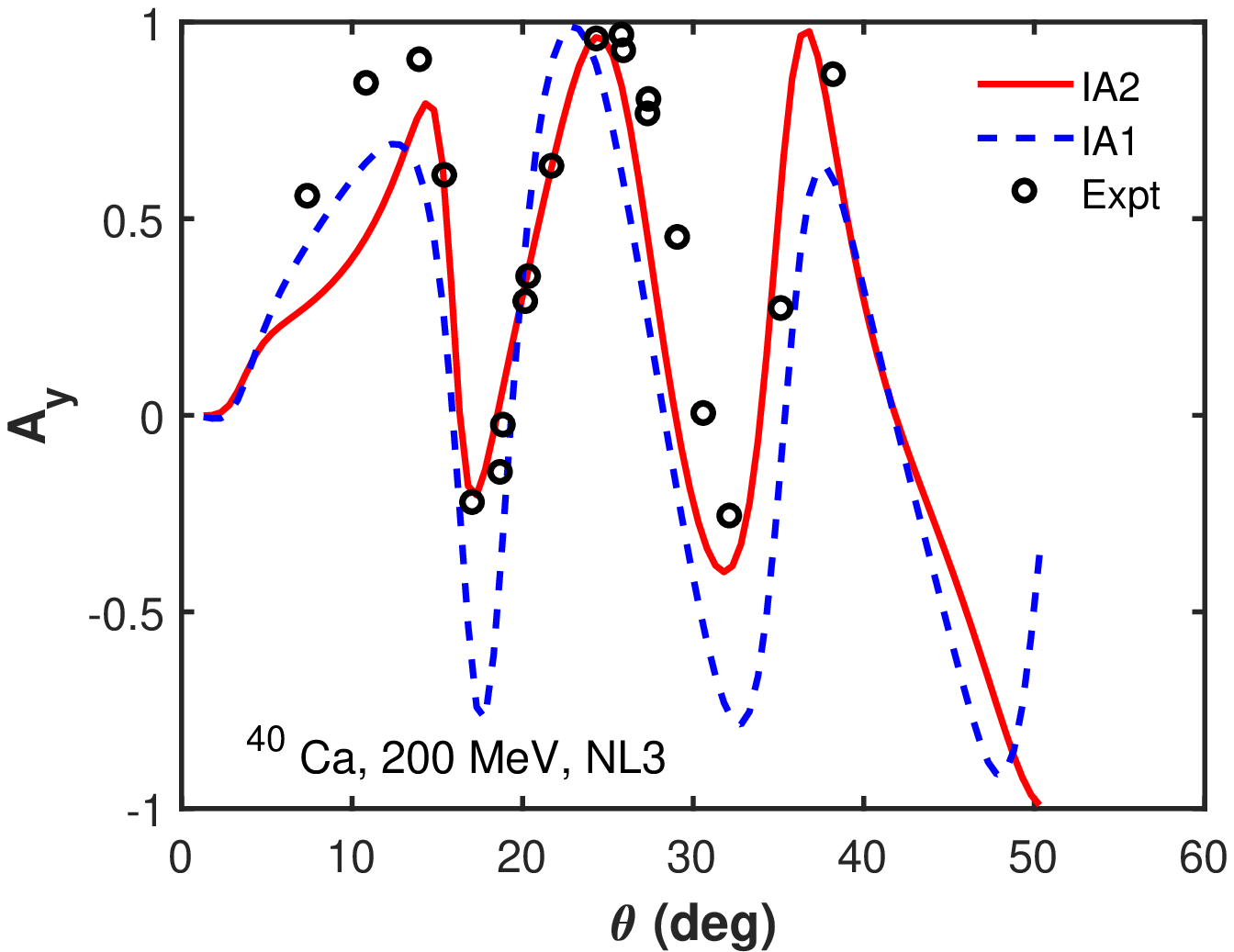}
	\includegraphics[width=0.49\linewidth]{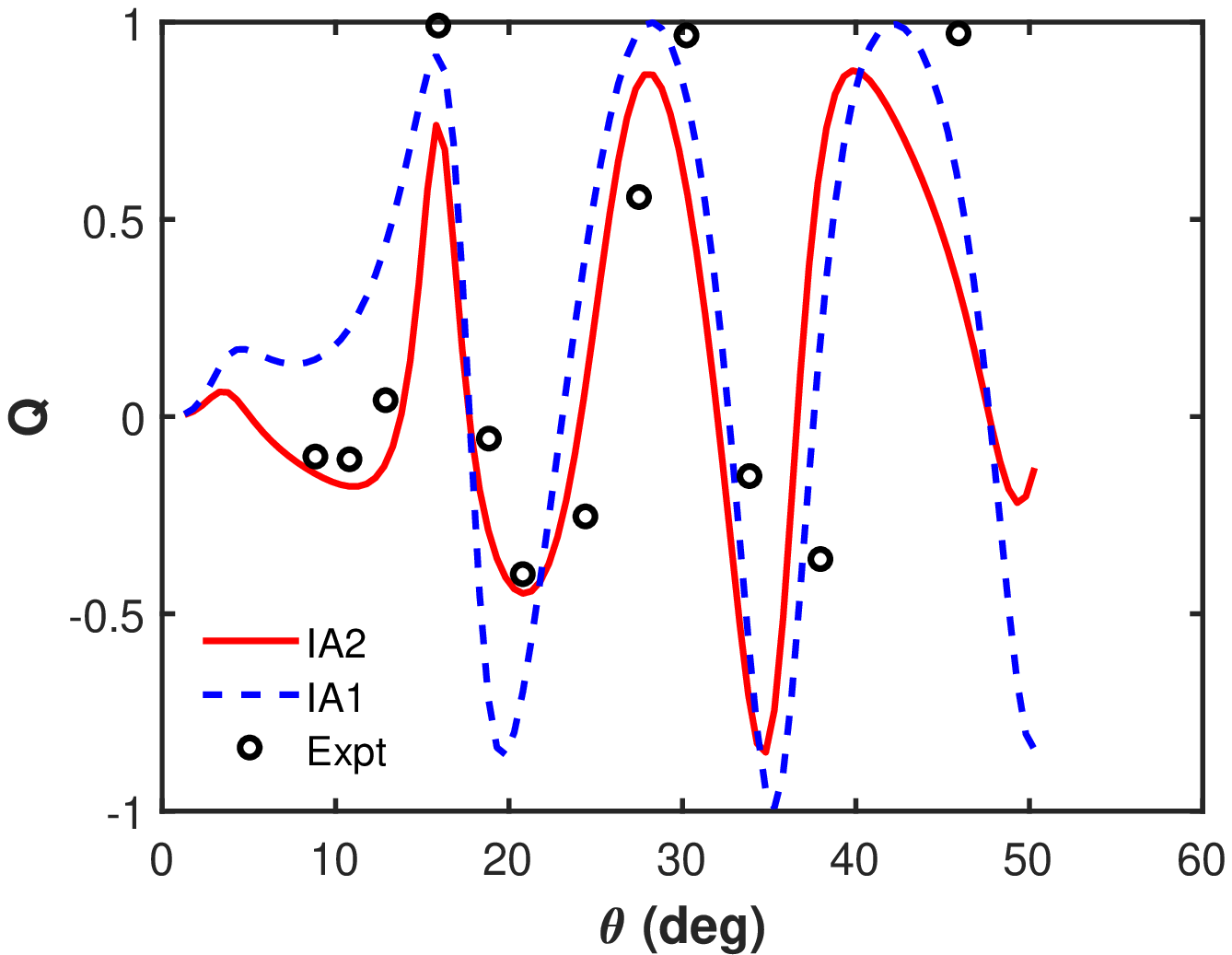}
	\caption{$^{40}$Ca scattering observables calculated with the NL3 parametrisation using IA1 and IA2 formalisms at $T_{\mathrm{lab}} = 200$MeV. The IA2 results are shown in solid lines, while the IA1 results are shown in dashed lines.}
	\label{obsIA1vsIA2Ca40_200}
\end{figure}

\begin{figure}
	\centering
	\includegraphics[width=0.49\linewidth]{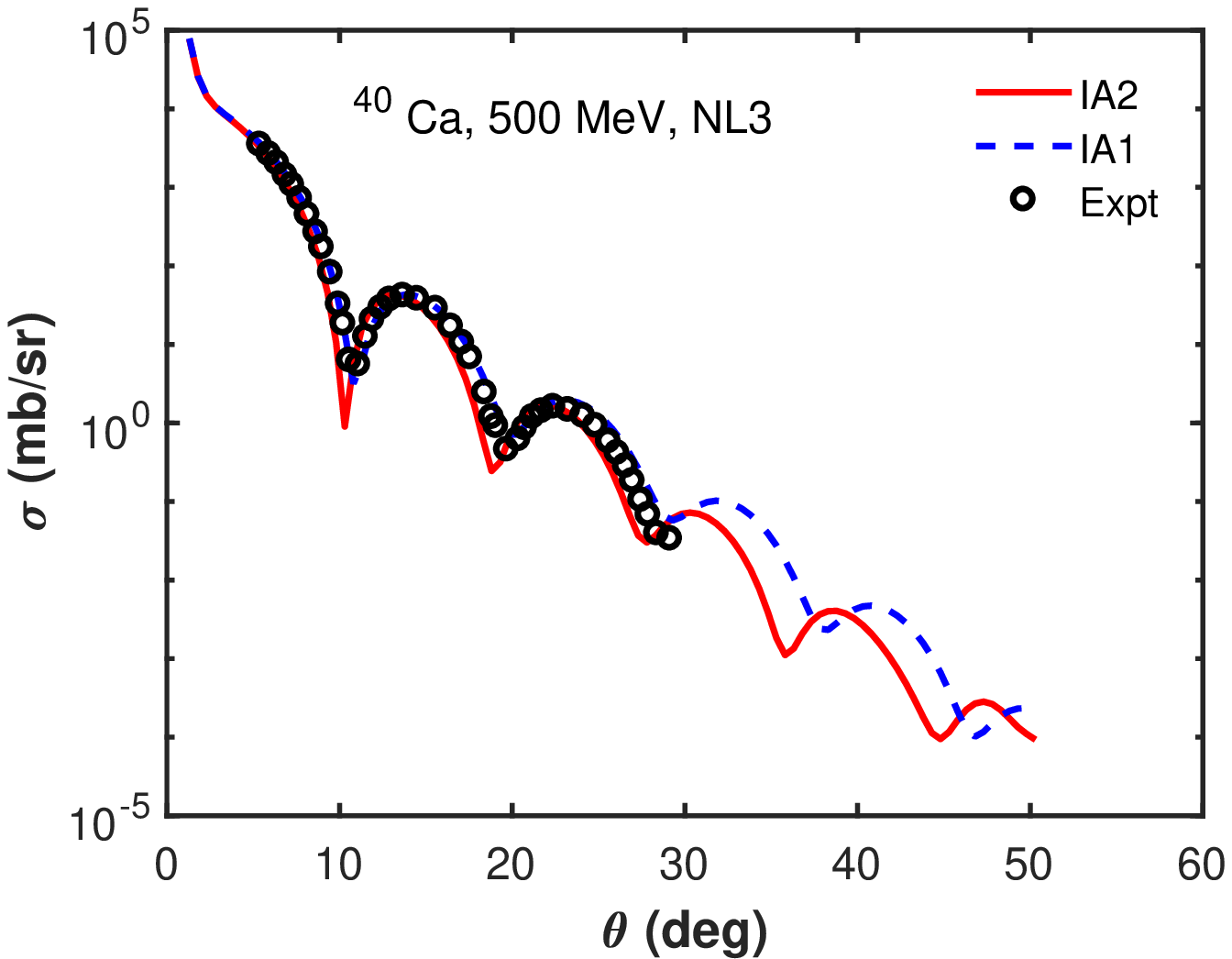}
	\includegraphics[width=0.49\linewidth]{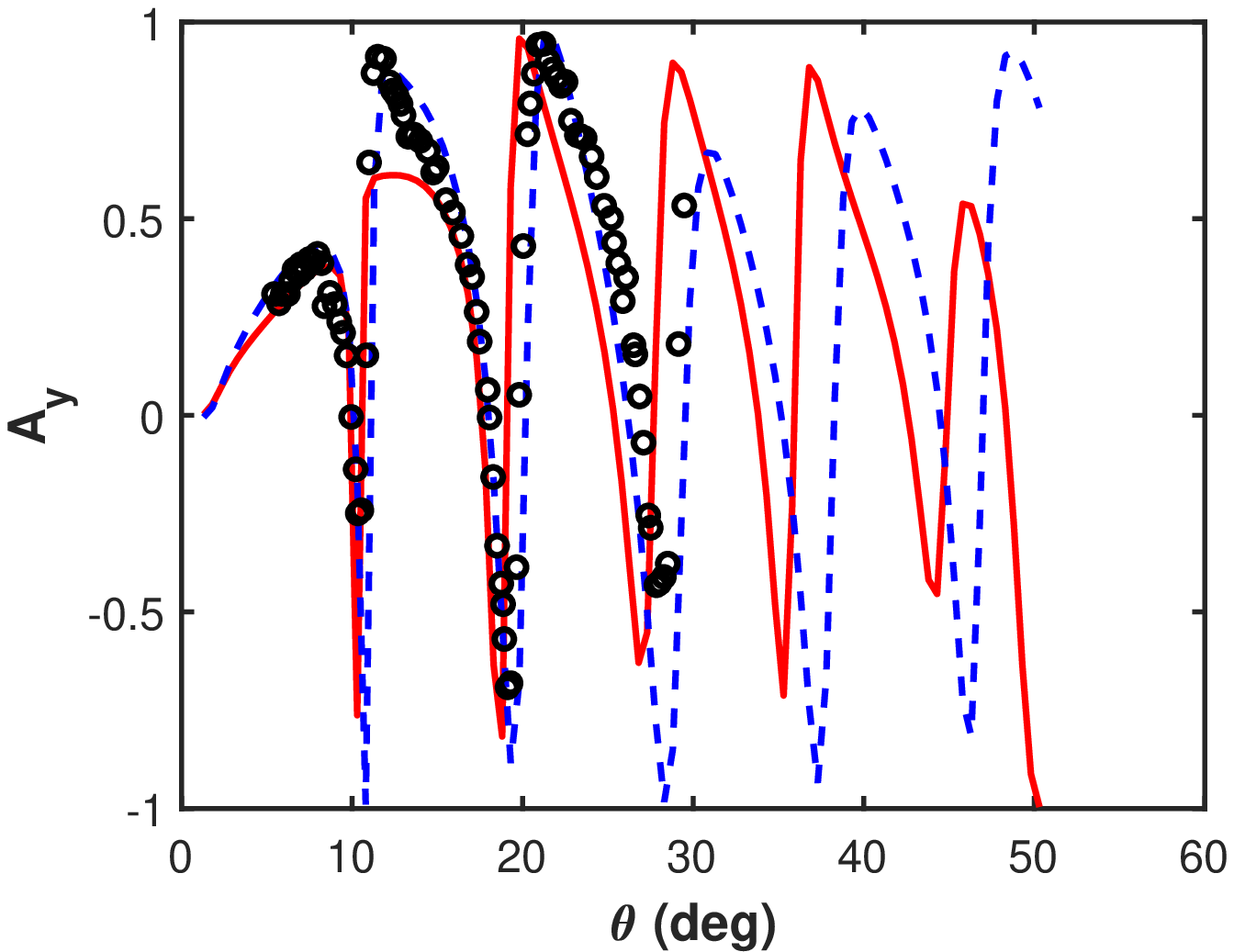}
	\includegraphics[width=0.49\linewidth]{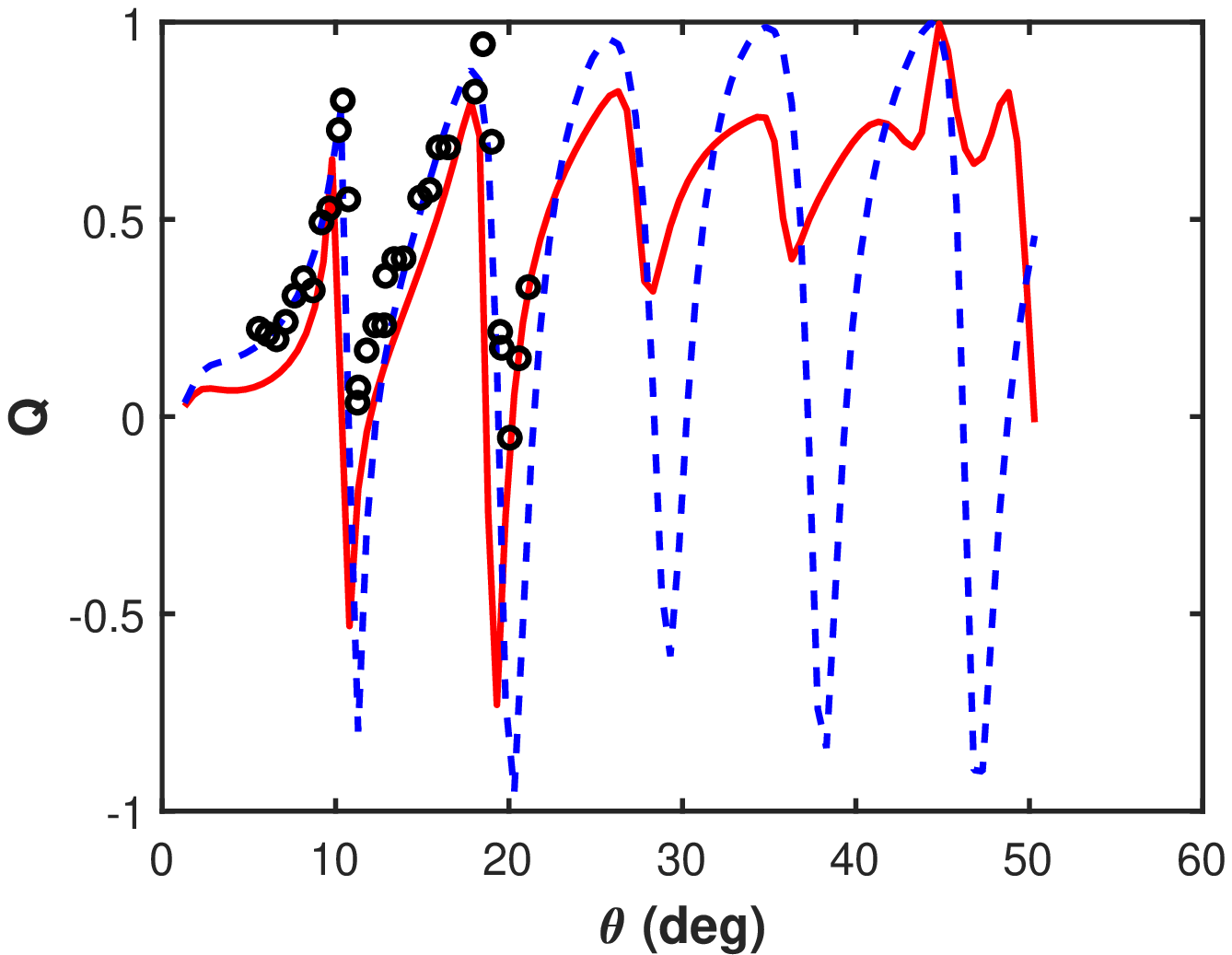}
	\caption{$^{40}$Ca scattering observables calculated with the NL3 parametrisation using IA1 and IA2 formalisms at $T_{\mathrm{lab}} = 500$MeV. The expressions of lines is the same as in figure \ref{obsIA1vsIA2Ca40_200}.}
	\label{obsIA1vsIA2Ca40_500}
\end{figure}

\begin{figure}
	\centering
	\includegraphics[width=0.49\linewidth]{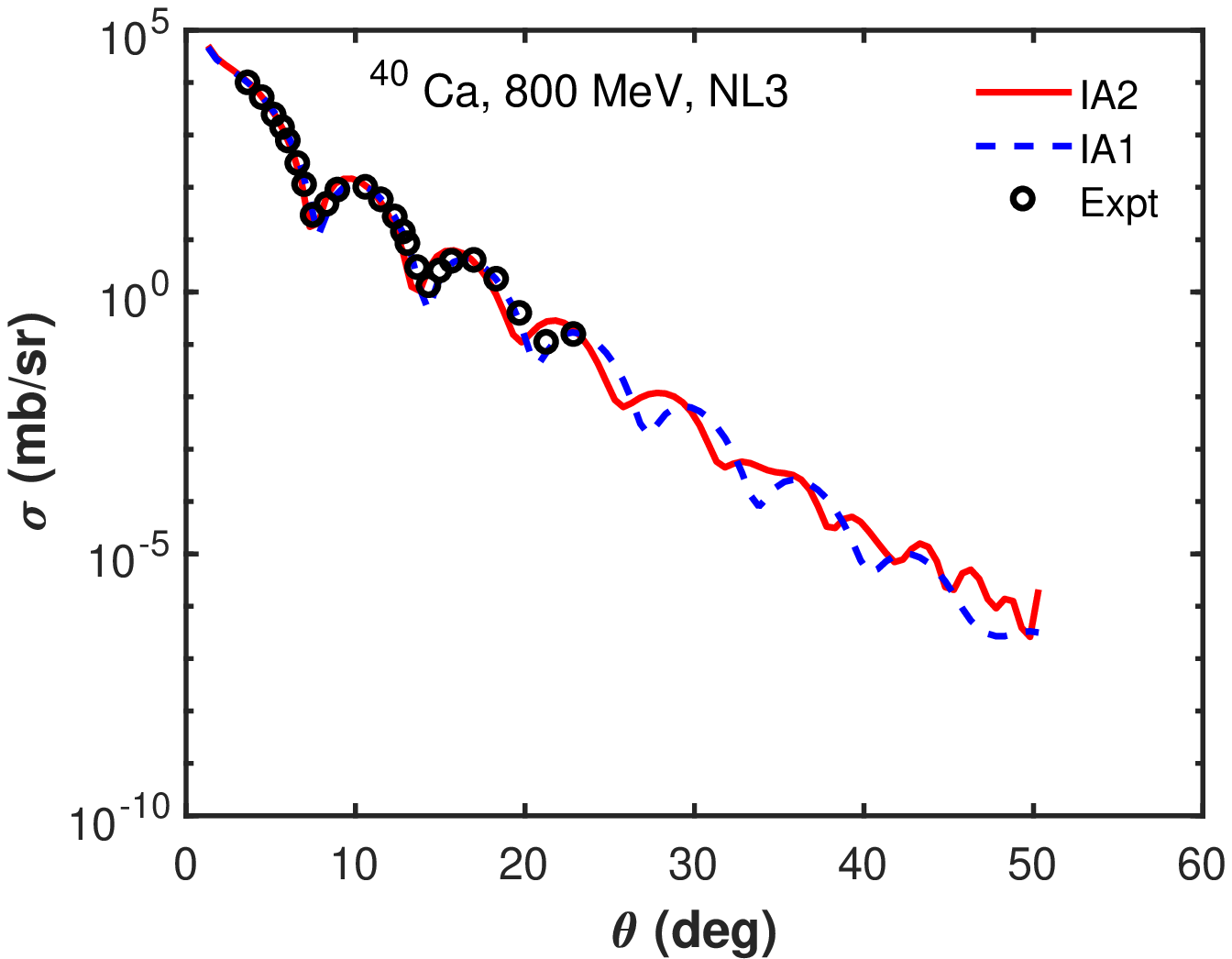}
	\includegraphics[width=0.49\linewidth]{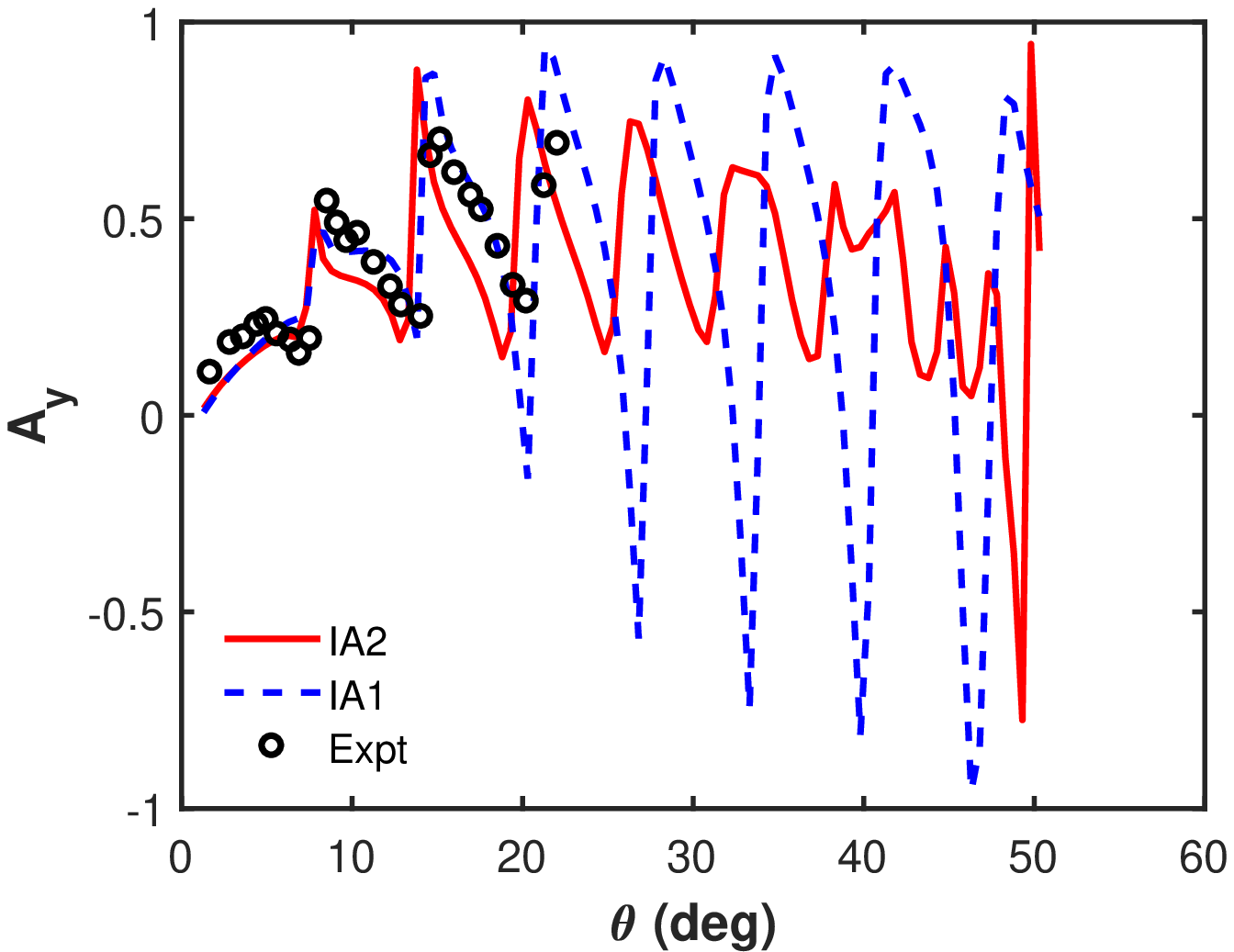}
	\includegraphics[width=0.49\linewidth]{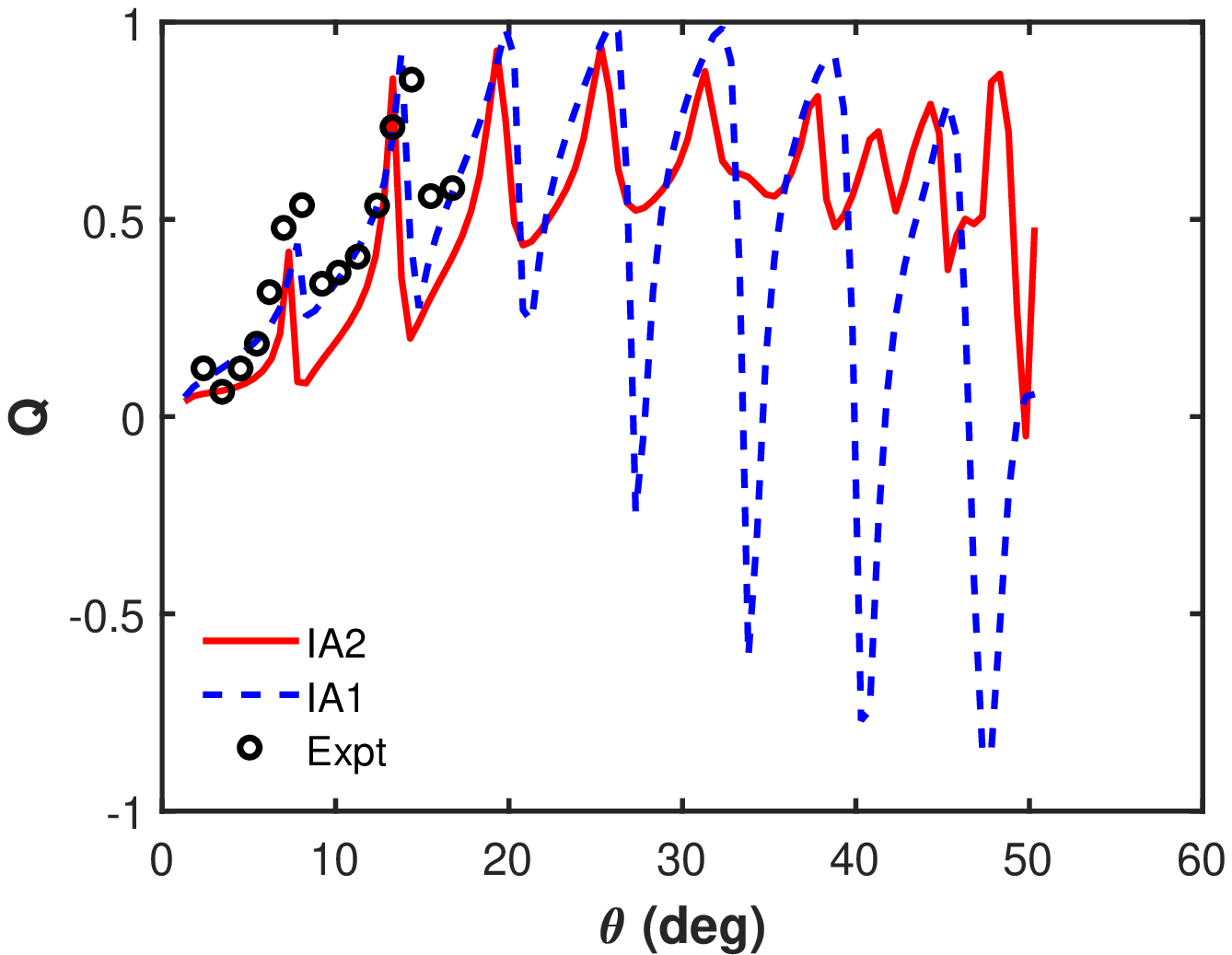}
	\caption{$^{40}$Ca scattering observables calculated with the NL3 parametrisation using IA1 and IA2 formalisms at $T_{\mathrm{lab}} = 800$MeV. The expressions of lines is the same as in figure \ref{obsIA1vsIA2Ca40_200}.}
	\label{obsIA1vsIA2Ca40_800}
\end{figure}

Figure \ref{obsIA1vsIA2Ca40_200} shows the plots of the scattering cross section, analysing power and spin rotation parameter against centre of mass scattering angle $\theta$ for elastic proton scattering from $^{40}$Ca at $T_{\mathrm{lab}} = 200$ MeV using the NL3 parameter set. The figure shows comparison of the IA1 and IA2 formalisms with experimental data. The top left panel shows the plots for the scattering cross section, the top right panel for analysing power, and the bottom panel for spin rotation function. The same scattering observables are shown in figure \ref{obsIA1vsIA2Ca40_500} for $^{40}$Ca at $T_{\mathrm{lab}} = 500$ MeV. One can observe that at $T_{\mathrm{lab}} = 500$ MeV there is competition between the two formalisms in describing the experimental data for the three scattering observables. The difference between the three formalisms is noticed at large scattering angles. At $T_{\mathrm{lab}} = 200$ MeV, the IA2 formalism gives a very good description of the scattering observables especially scattering cross section and spin rotation parameter. The IA1 formalism overestimates the scattering cross section, and failed to give correct descriptions of the minima and maxima in the case of the analysing power and spin rotation parameter. This follows from the overly large scalar and vector optical potentials given by the IA1 formalism at this incident projectile laboratory energy. One should note that the IA2 formalism also did not properly describe the analysing power at low scattering angle $\theta \lessapprox 13^{\circ}$, but give proper description at $ \theta \gtrapprox 13^{\circ}$. Figure \ref{obsIA1vsIA2Ca40_800} shows the plots of the scattering cross section, analysing power and spin rotation parameter against centre of mass scattering angle $\theta$ for elastic proton scattering from $^{40}$Ca at $T_{\mathrm{lab}} = 800$ MeV using the NL3 parameter set.

\begin{figure}
	\centering
	\includegraphics[width=0.49\linewidth]{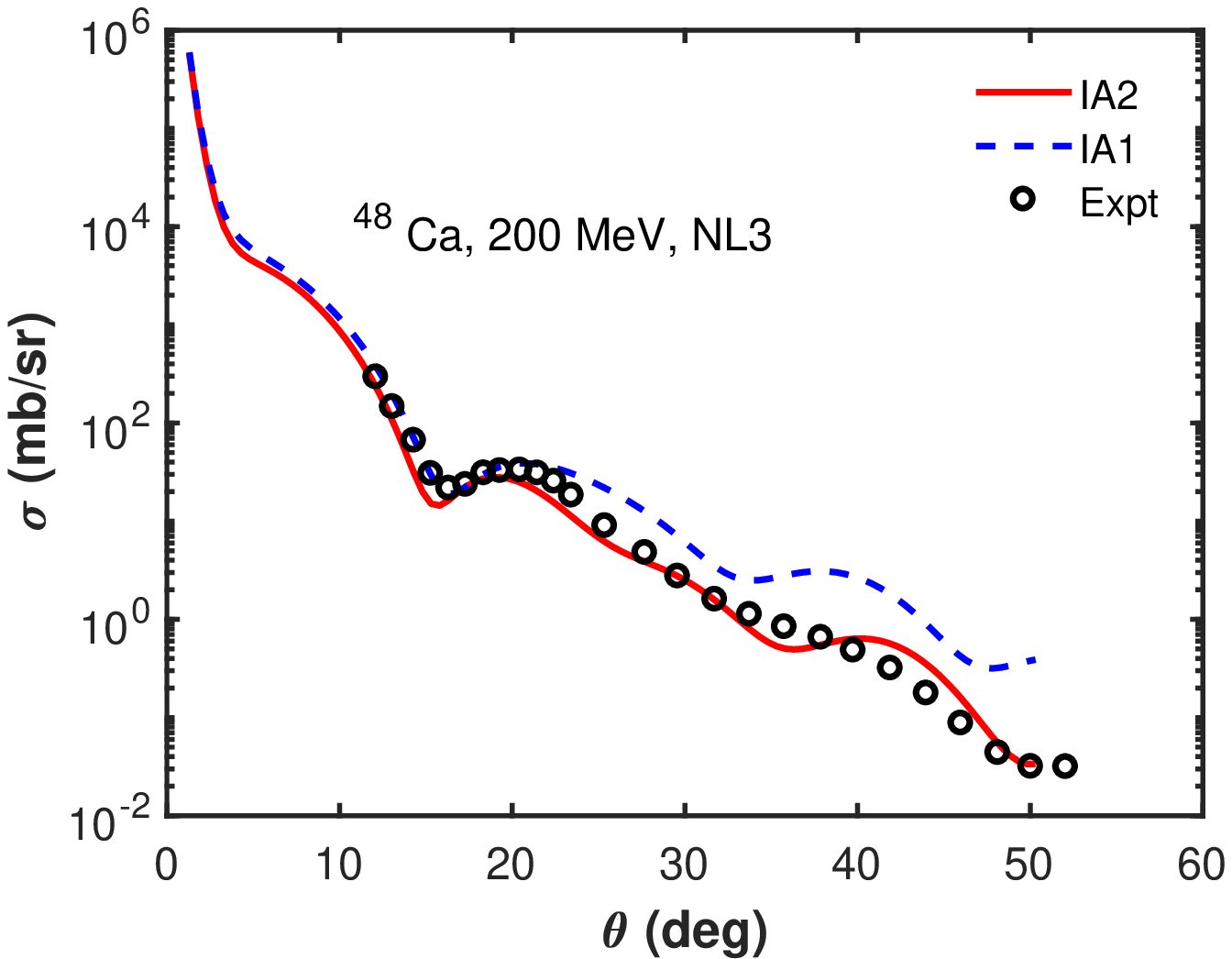}
	\includegraphics[width=0.49\linewidth]{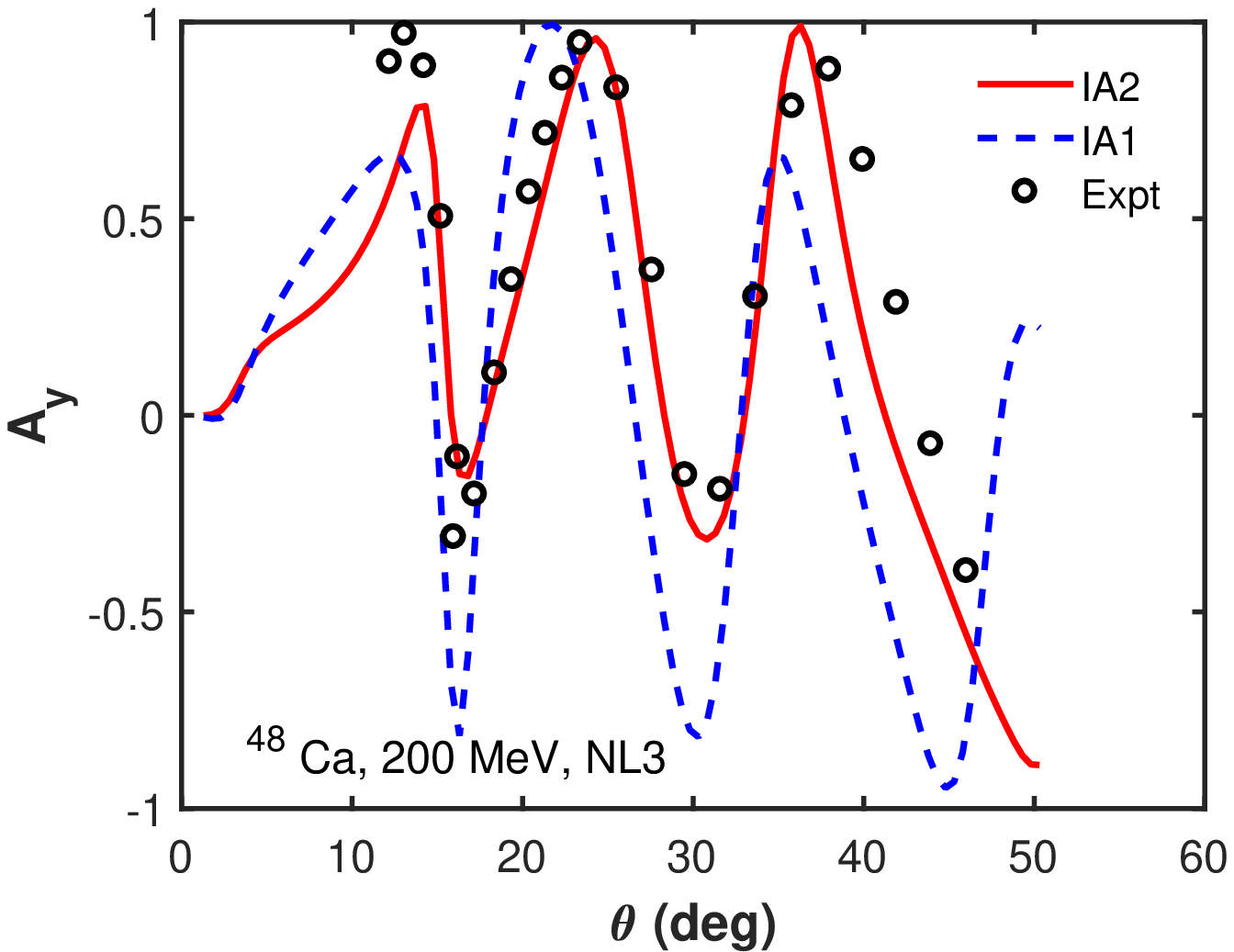}
	\includegraphics[width=0.49\linewidth]{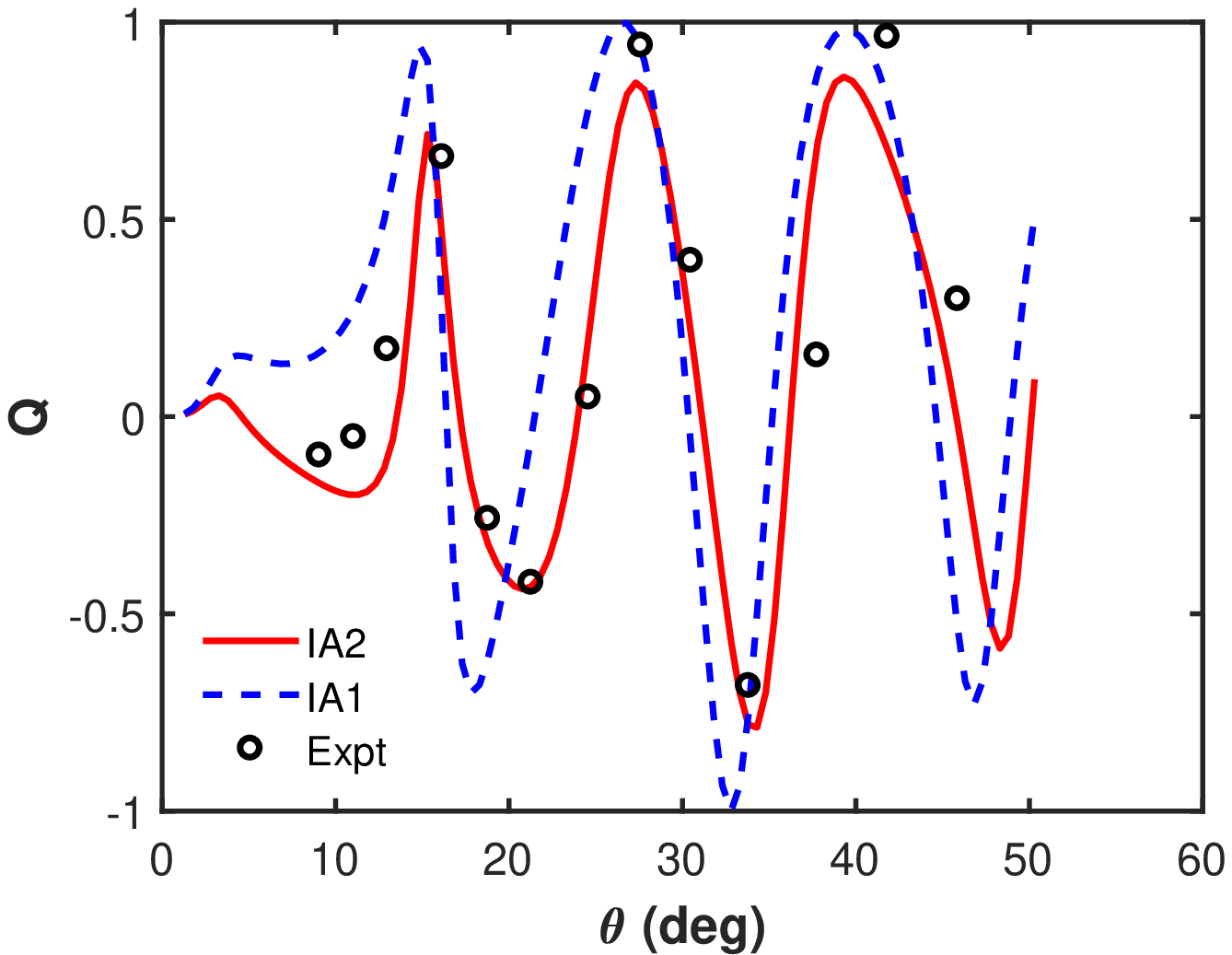}
	\caption{$^{48}$Ca scattering observables calculated with the NL3 parametrisation using IA1 and IA2 formalisms at $T_{\mathrm{lab}} = 200$MeV. The expressions of lines is the same as in figure \ref{obsIA1vsIA2Ca40_200}.}
	\label{obsIA1vsIA2Ca48_200}
\end{figure}

\begin{figure}
	\centering
	\includegraphics[width=0.49\linewidth]{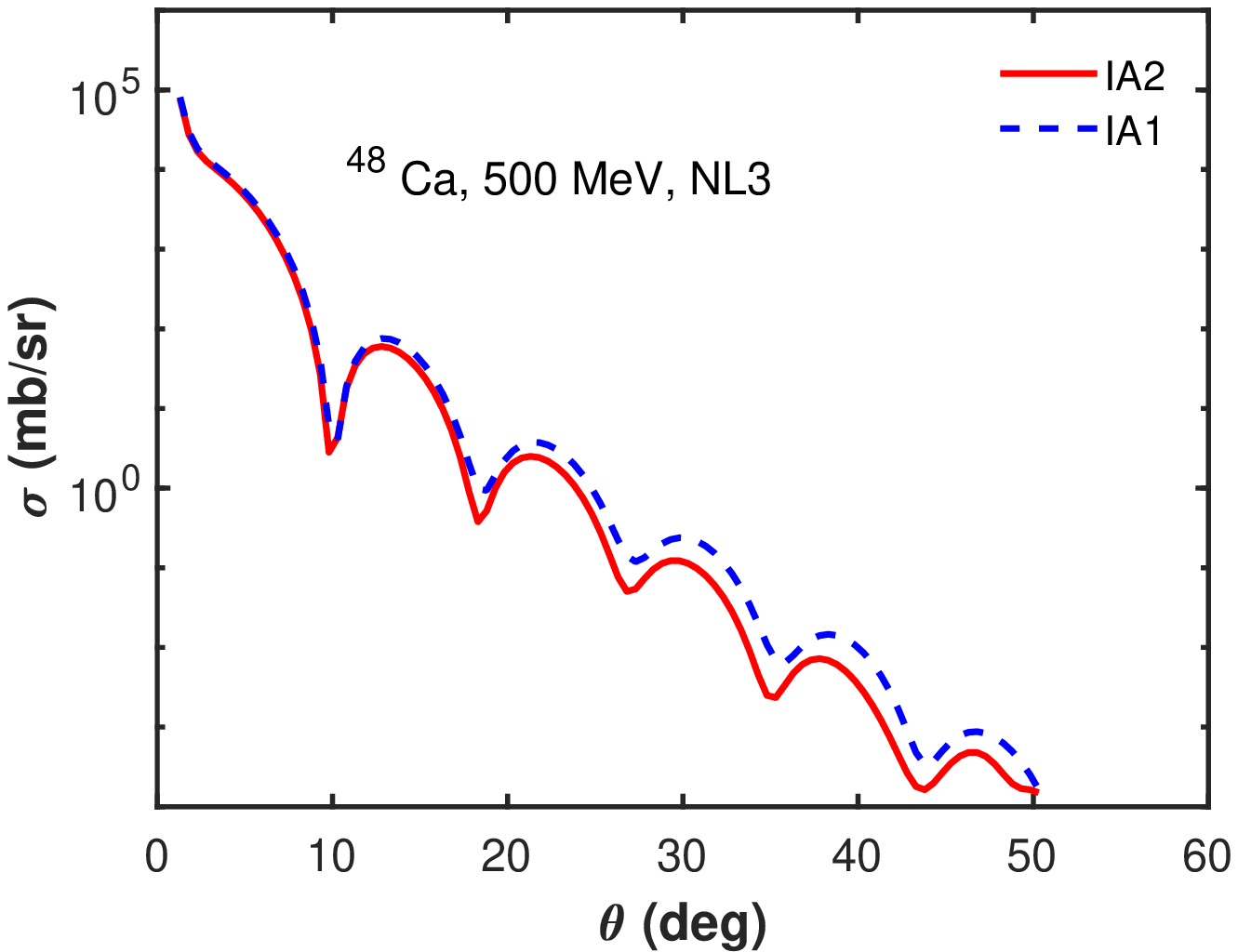}
	\includegraphics[width=0.49\linewidth]{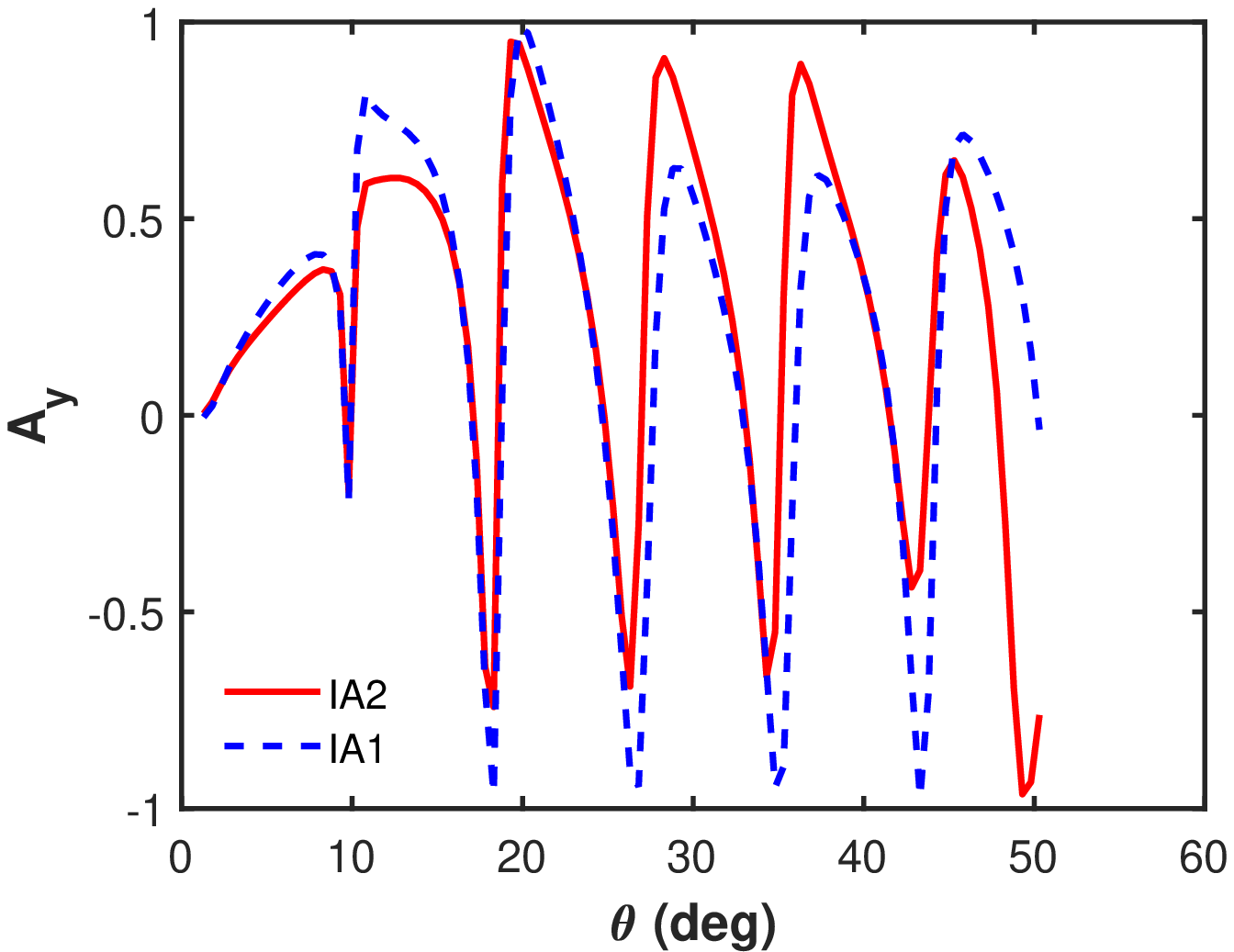}
	\includegraphics[width=0.49\linewidth]{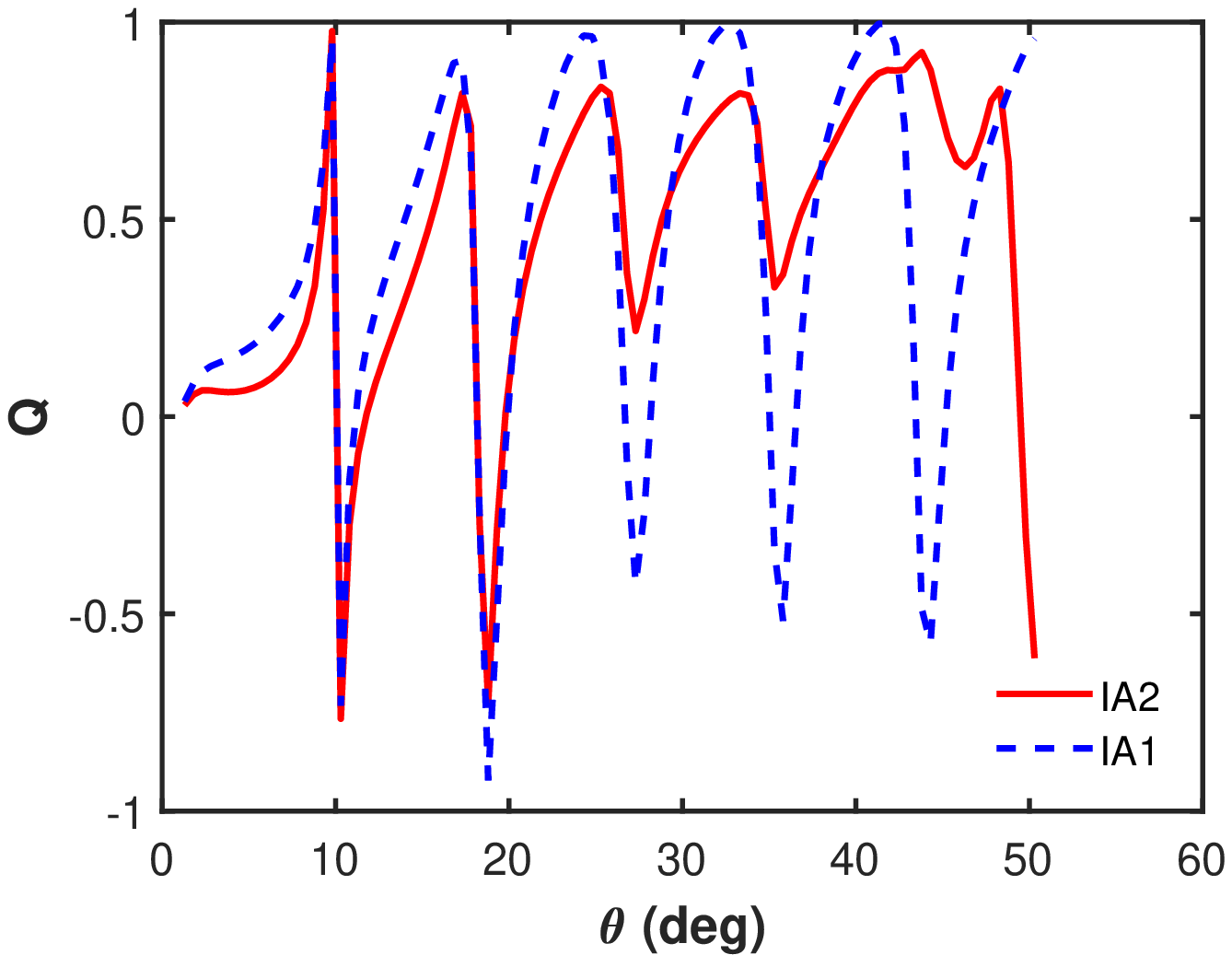}
	\caption{$^{48}$Ca scattering observables calculated with the NL3 parametrisation using IA1 and IA2 formalisms at $T_{\mathrm{lab}} = 500$MeV. The expressions of lines is the same as in figure \ref{obsIA1vsIA2Ca40_200}.}
	\label{obsIA1vsIA2Ca48_500}
\end{figure}

\begin{figure}
	\centering
	\includegraphics[width=0.49\linewidth]{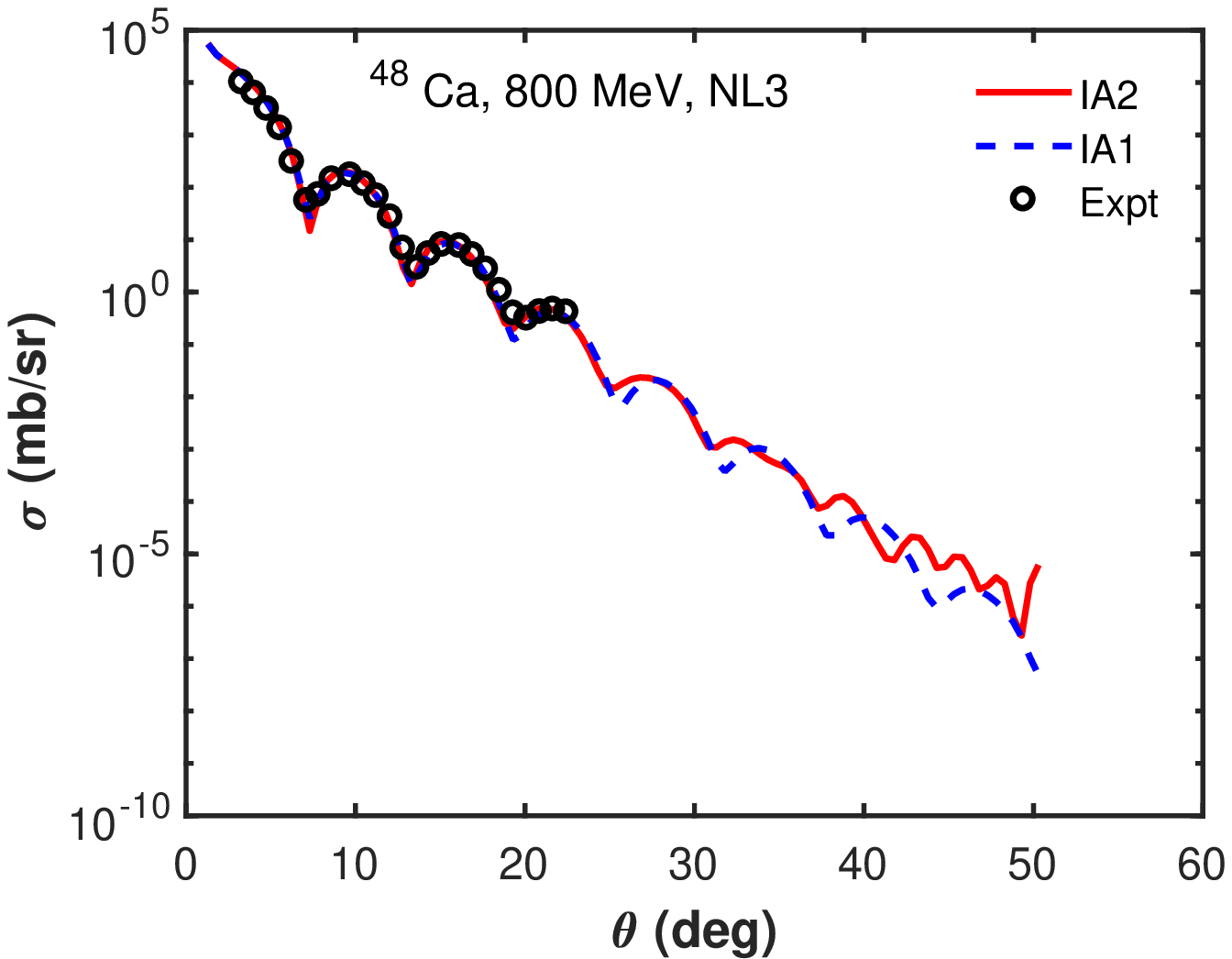}
	\includegraphics[width=0.49\linewidth]{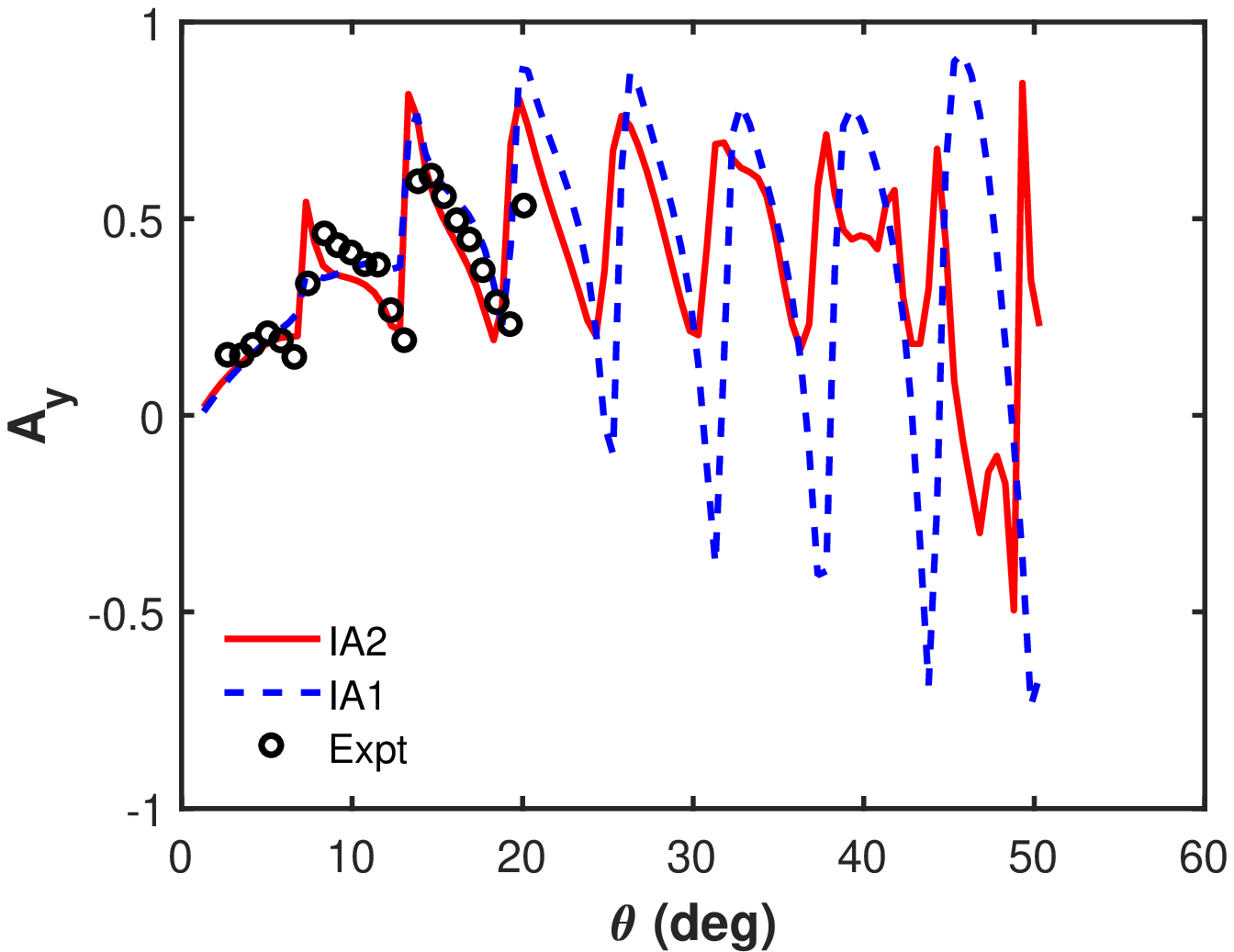}
	\includegraphics[width=0.49\linewidth]{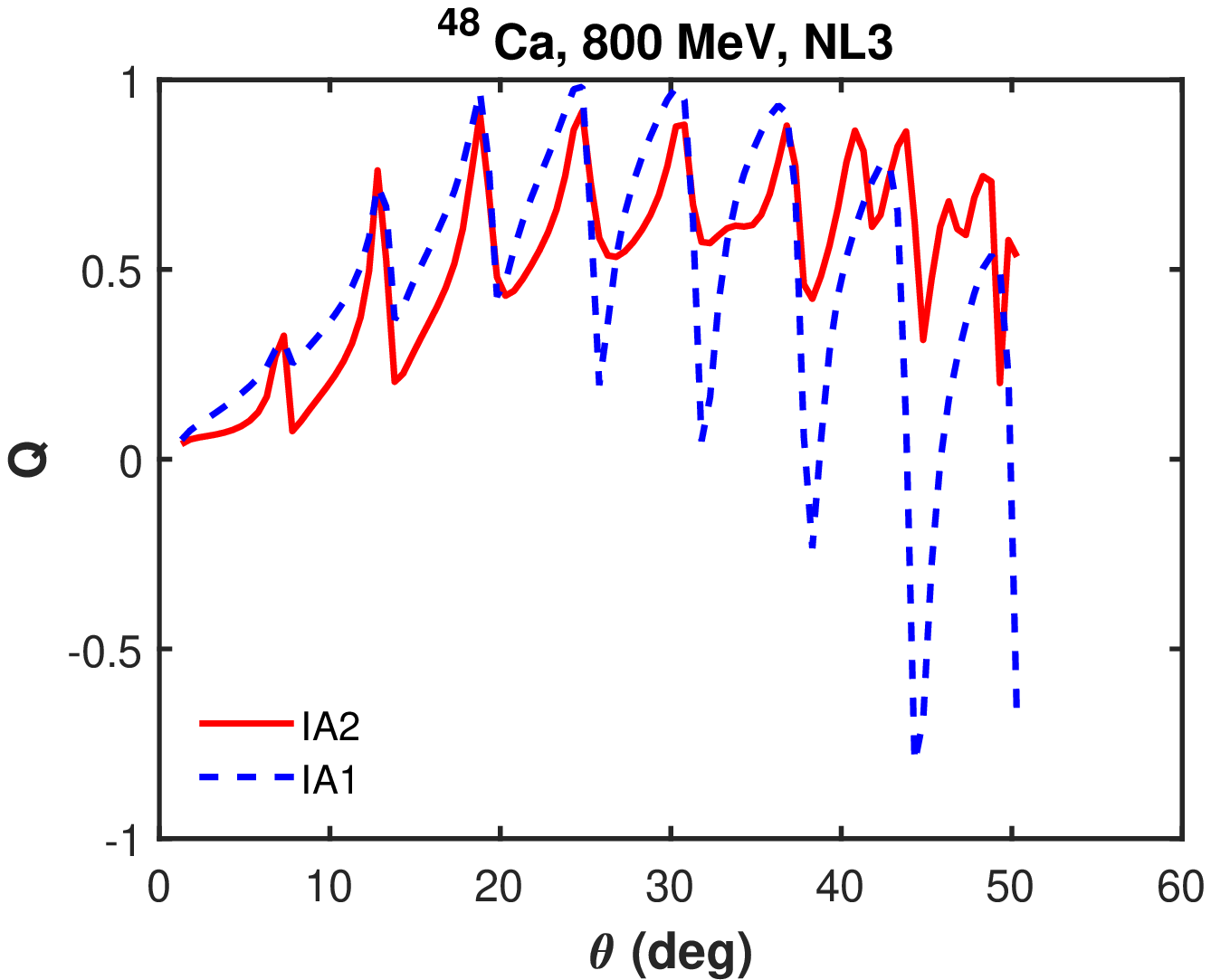}
	\caption{$^{48}$Ca scattering observables calculated with the NL3 parametrisation using IA1 and IA2 formalisms at $T_{\mathrm{lab}} = 800$MeV. The expressions of lines is the same as in figure \ref{obsIA1vsIA2Ca40_200}.}
	\label{obsIA1vsIA2Ca48_800}
\end{figure}

In Figures \ref{obsIA1vsIA2Ca48_200}, \ref{obsIA1vsIA2Ca48_500}, and \ref{obsIA1vsIA2Ca48_800} the plots of the elastic scattering observables are plotted against centre of mass scattering angle for $p+^{48}$Ca at $T_{\mathrm{lab}} = 200$ MeV, $500$ MeV and 800 MeV, respectively. It can be observed that at $T_{\mathrm{lab}} = 200$ MeV, the IA2 formalism gives very good descriptions of the three scattering observables. The IA1 formalism, apart from giving the correct first minimum, overestimates the scattering cross section data, and did not accurately predict the minima and maxima in analysing power and spin rotation parameter data. The two formalisms give similar descriptions of the scattering observables at $T_{\mathrm{lab}} = 500$ MeV, but at large scattering angles, one begins to notice the difference between them.

\begin{figure}
	\centering
	\includegraphics[width=0.49\linewidth]{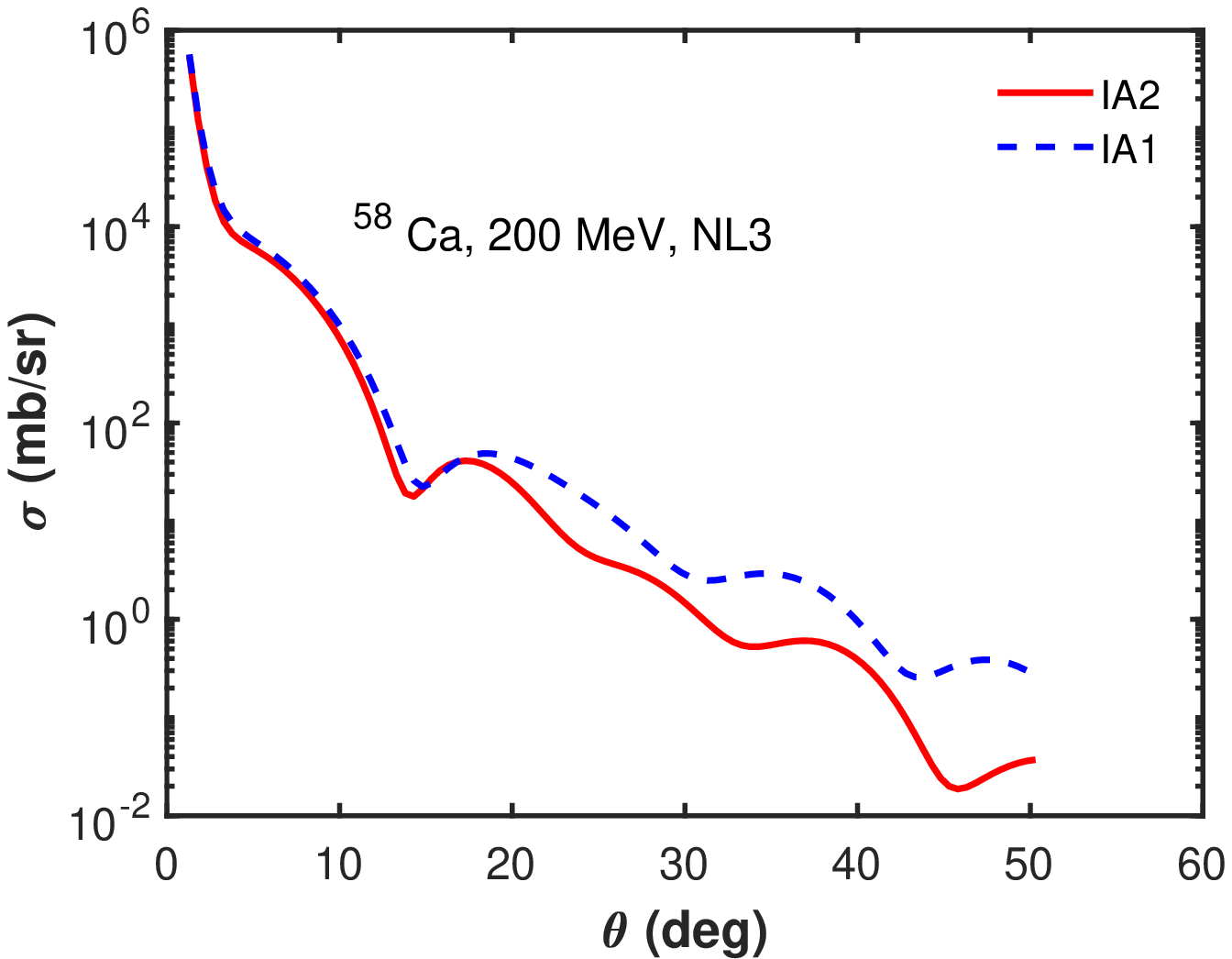}
	\includegraphics[width=0.49\linewidth]{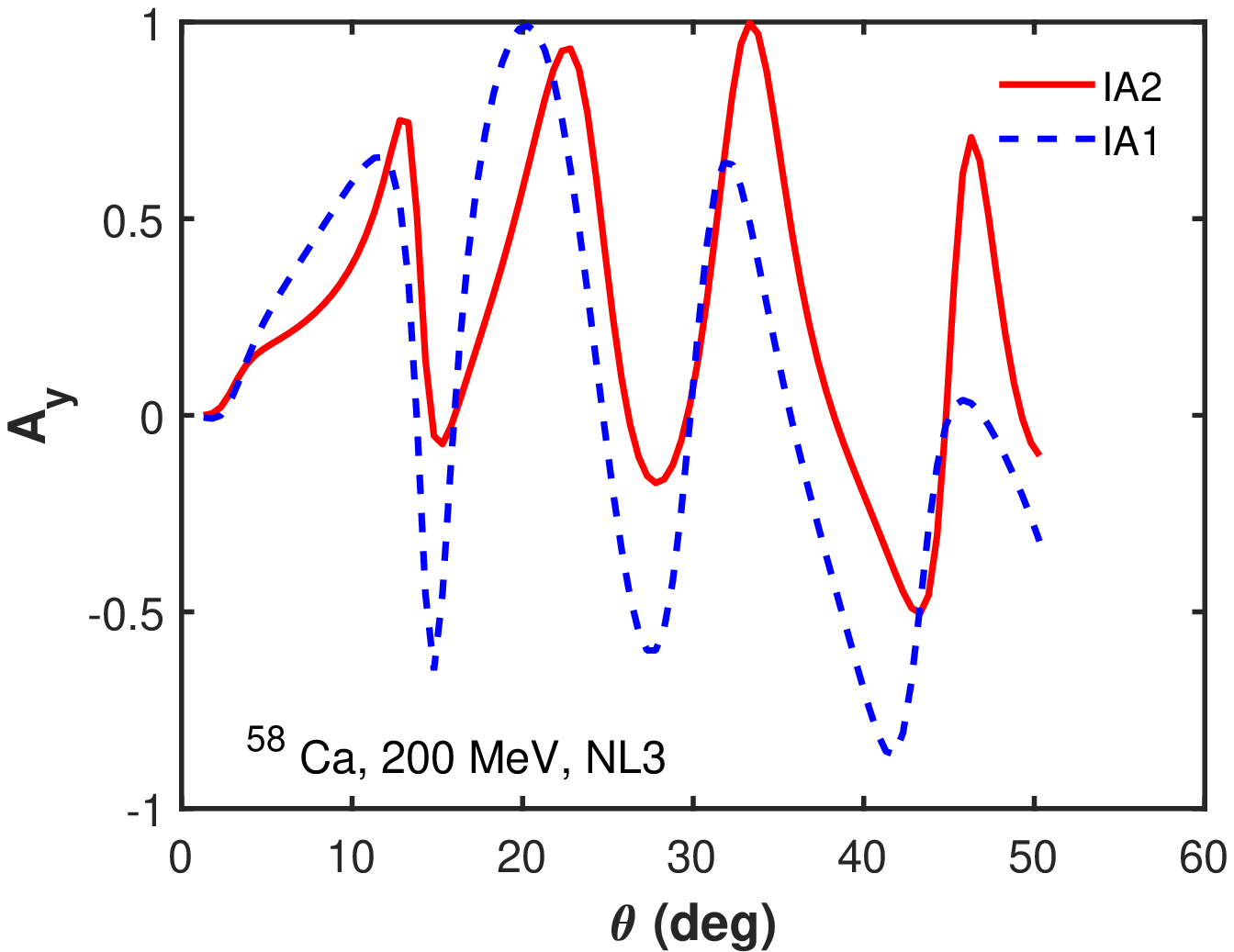}
	\includegraphics[width=0.49\linewidth]{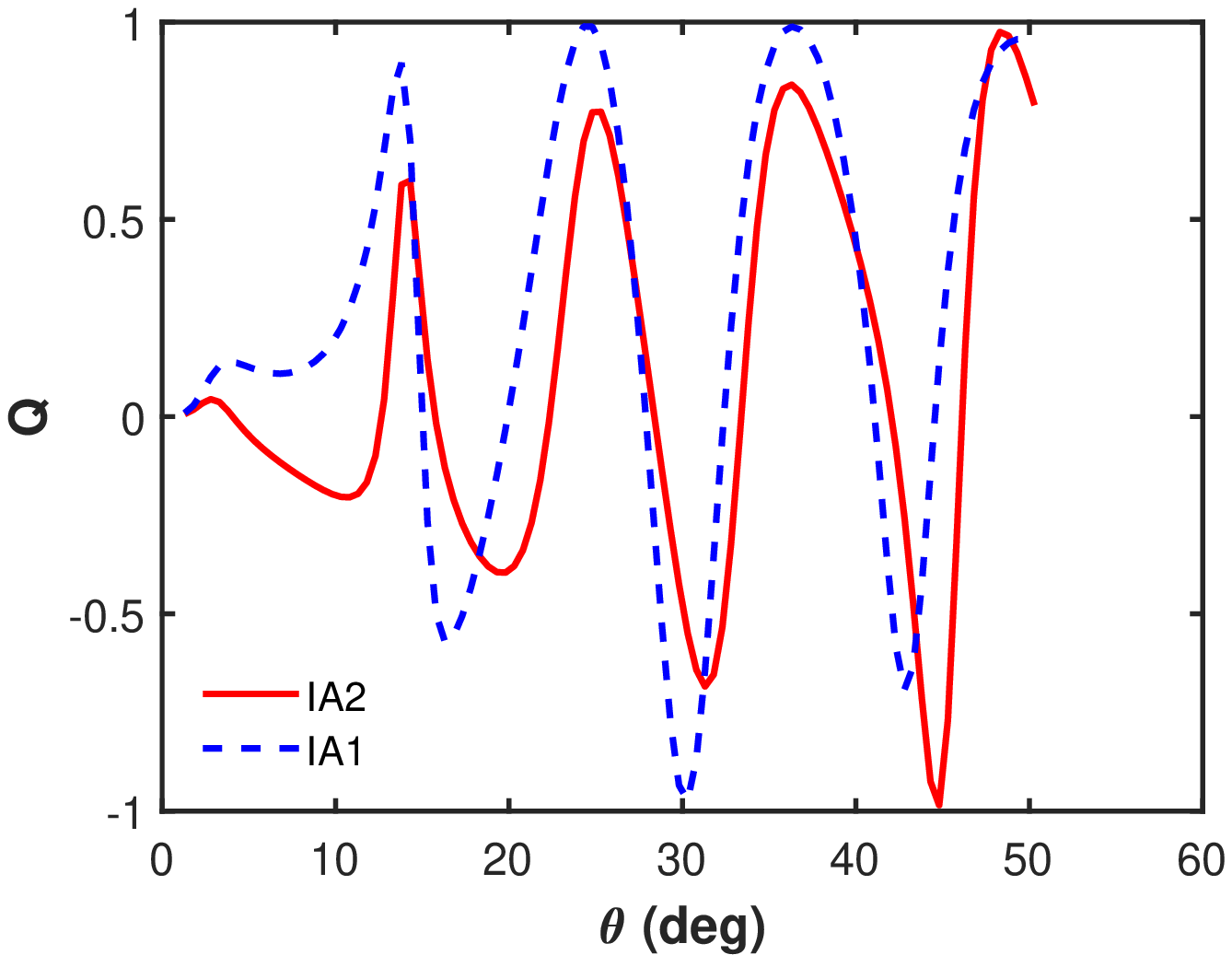}
	\caption{$^{58}$Ca scattering observables calculated with the NL3 parametrisation using IA1 and IA2 formalisms at $T_{\mathrm{lab}} = 200$MeV. The expressions of lines is the same as in figure \ref{obsIA1vsIA2Ca40_200}.}
	\label{obsIA1vsIA2Ca58_200}
\end{figure}

\begin{figure}
	\centering
	\includegraphics[width=0.49\linewidth]{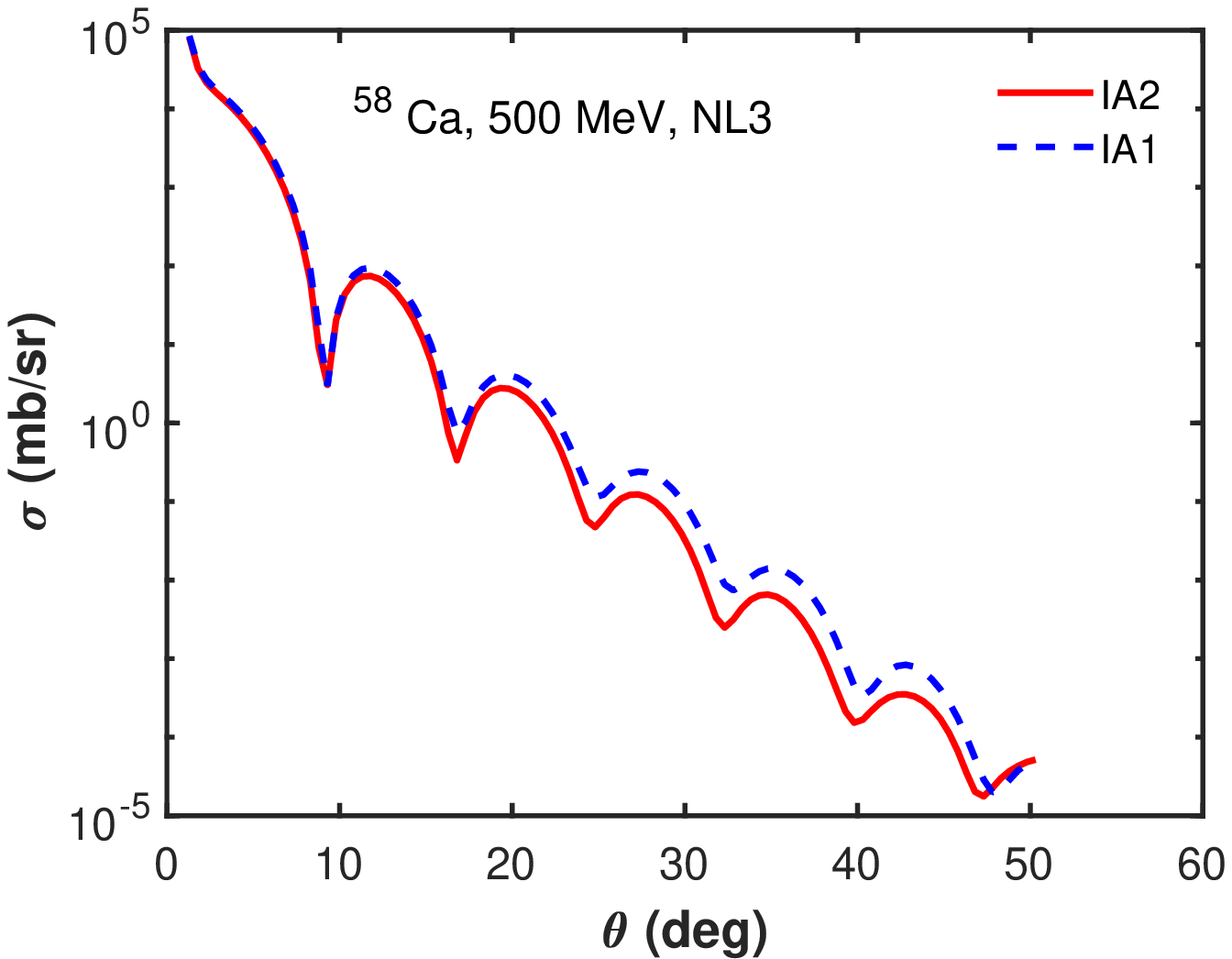}
	\includegraphics[width=0.49\linewidth]{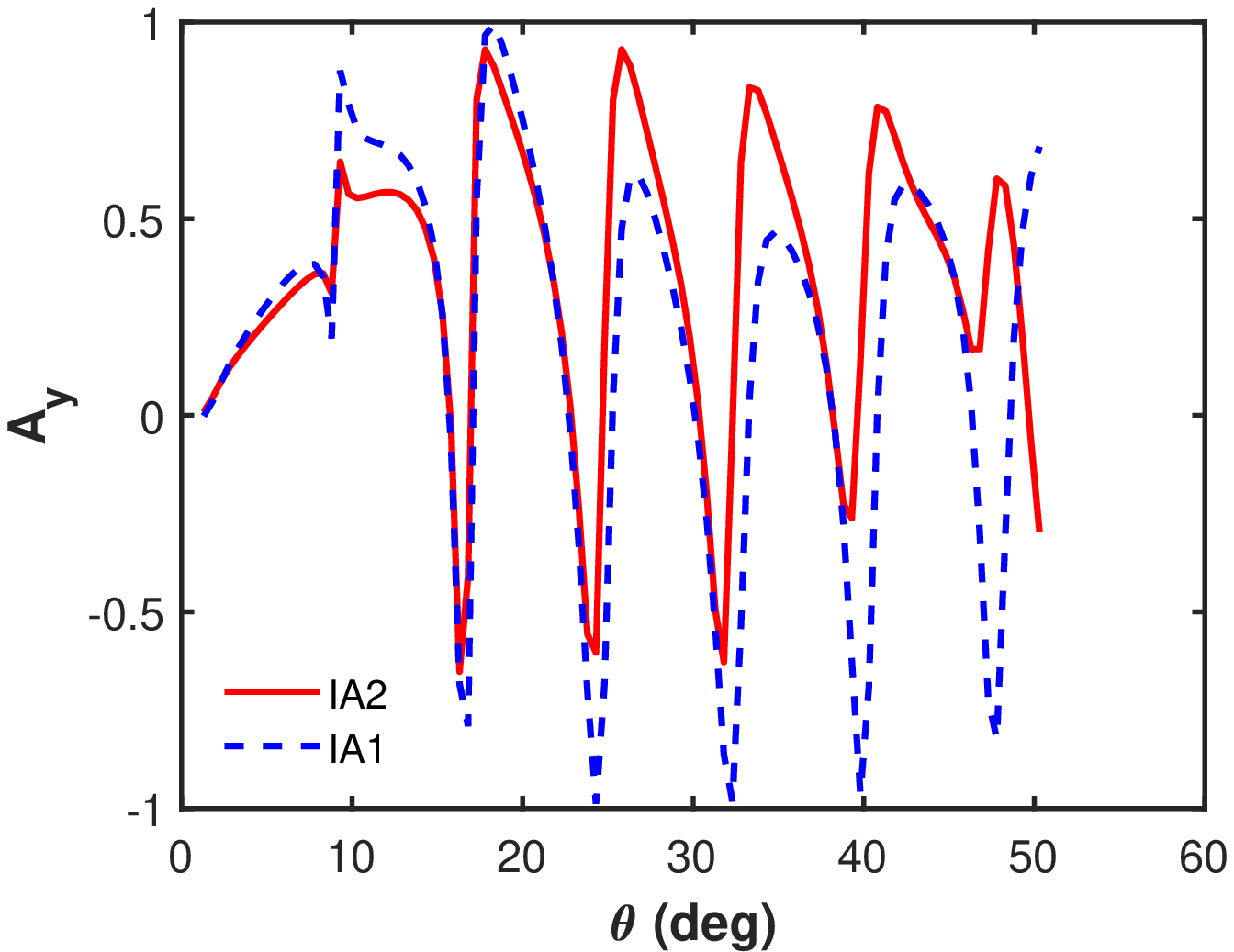}
	\includegraphics[width=0.49\linewidth]{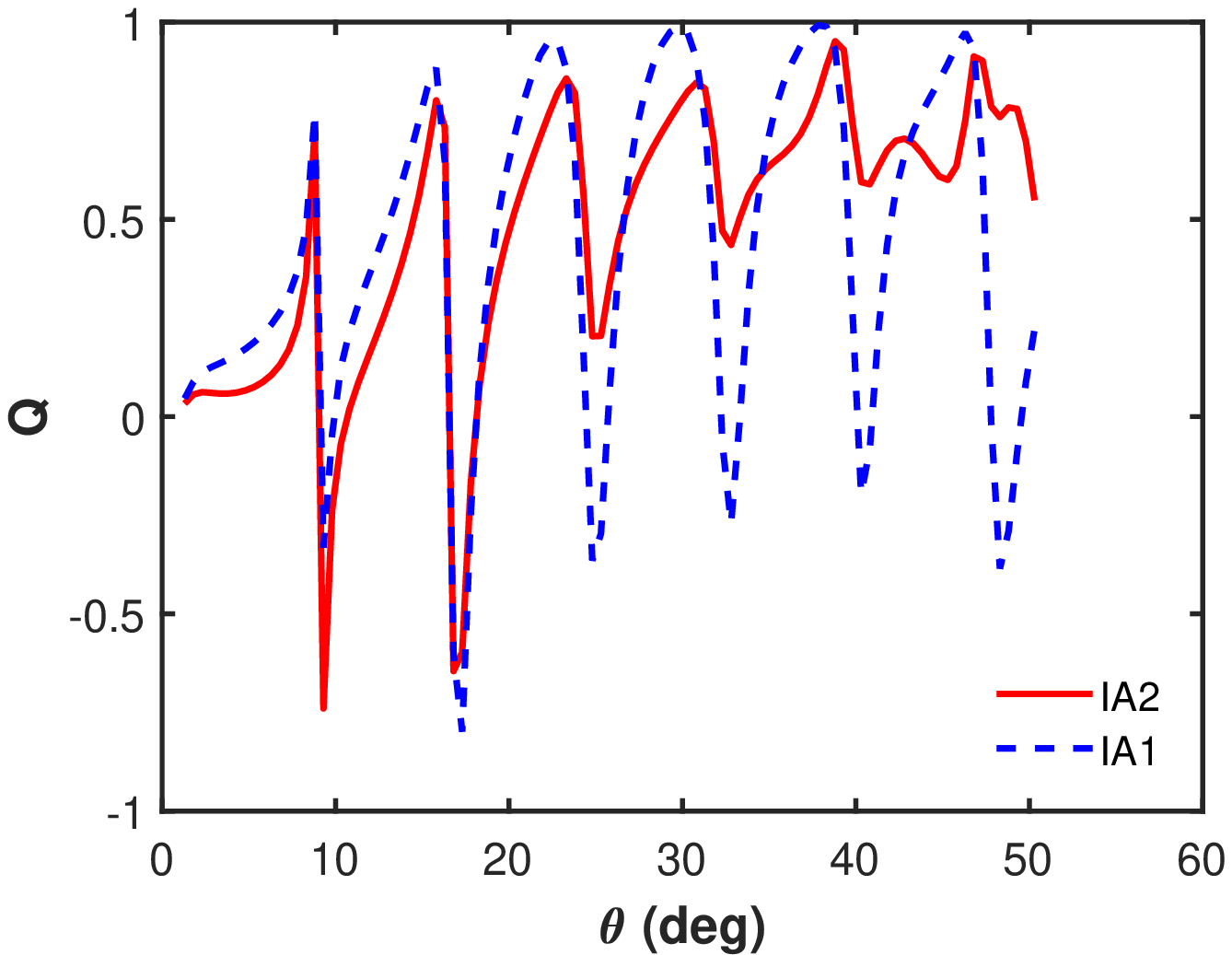}
	\caption{$^{58}$Ca scattering observables calculated with the NL3 parametrisation using IA1 and IA2 formalisms at $T_{\mathrm{lab}} = 500$MeV. The expressions of lines is the same as in figure \ref{obsIA1vsIA2Ca40_200}.}
	\label{obsIA1vsIA2Ca58_500}
\end{figure}

Figures \ref{obsIA1vsIA2Ca58_200} and \ref{obsIA1vsIA2Ca58_500} show the plots of the elastic proton scattering observables against centre of mass scattering angle for the unstable $^{58}$Ca target at $T_{\mathrm{lab}} = 200$ MeV and 500 MeV, respectively. As expected, the IA1 calculation gives larger values after the first dip, compared with the IA2 calculations. Unlike the case of stable nuclei, there is no good agreement between the IA1 and IA2 descriptions of the scattering observables at $T_{\mathrm{lab}} = 500$ MeV. Apart from the minimum at $\theta \lessapprox 10^{\circ}$, the IA1 gives deeper minima of analysing power and spin rotation and larger scattering cross section compared to the IA2 formalism.

\subsection{Effect of full folding versus optimally factorised optical potential on scattering observables}

\begin{figure}
	\centering
	\includegraphics[width=0.49\linewidth]{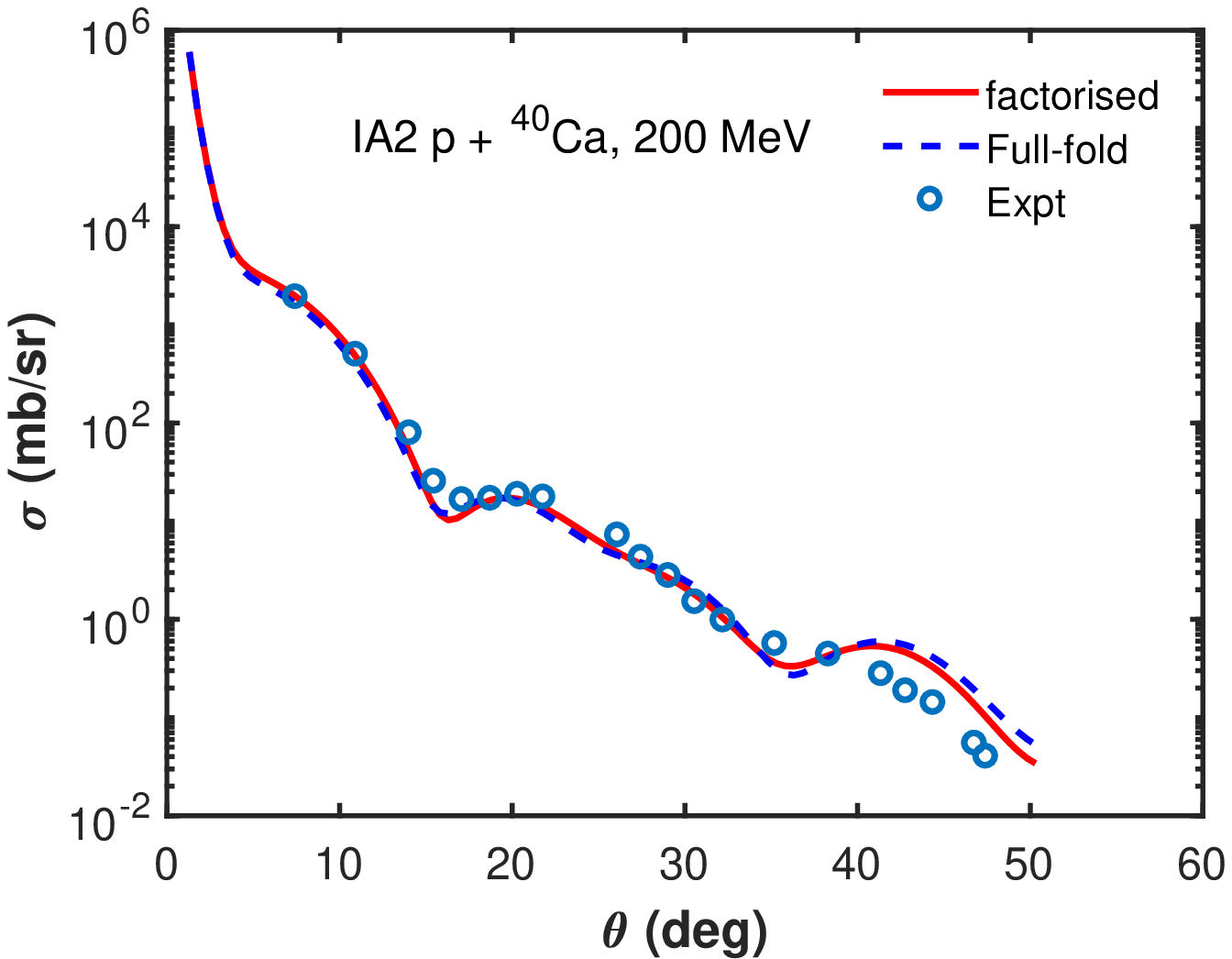}
	\includegraphics[width=0.49\linewidth]{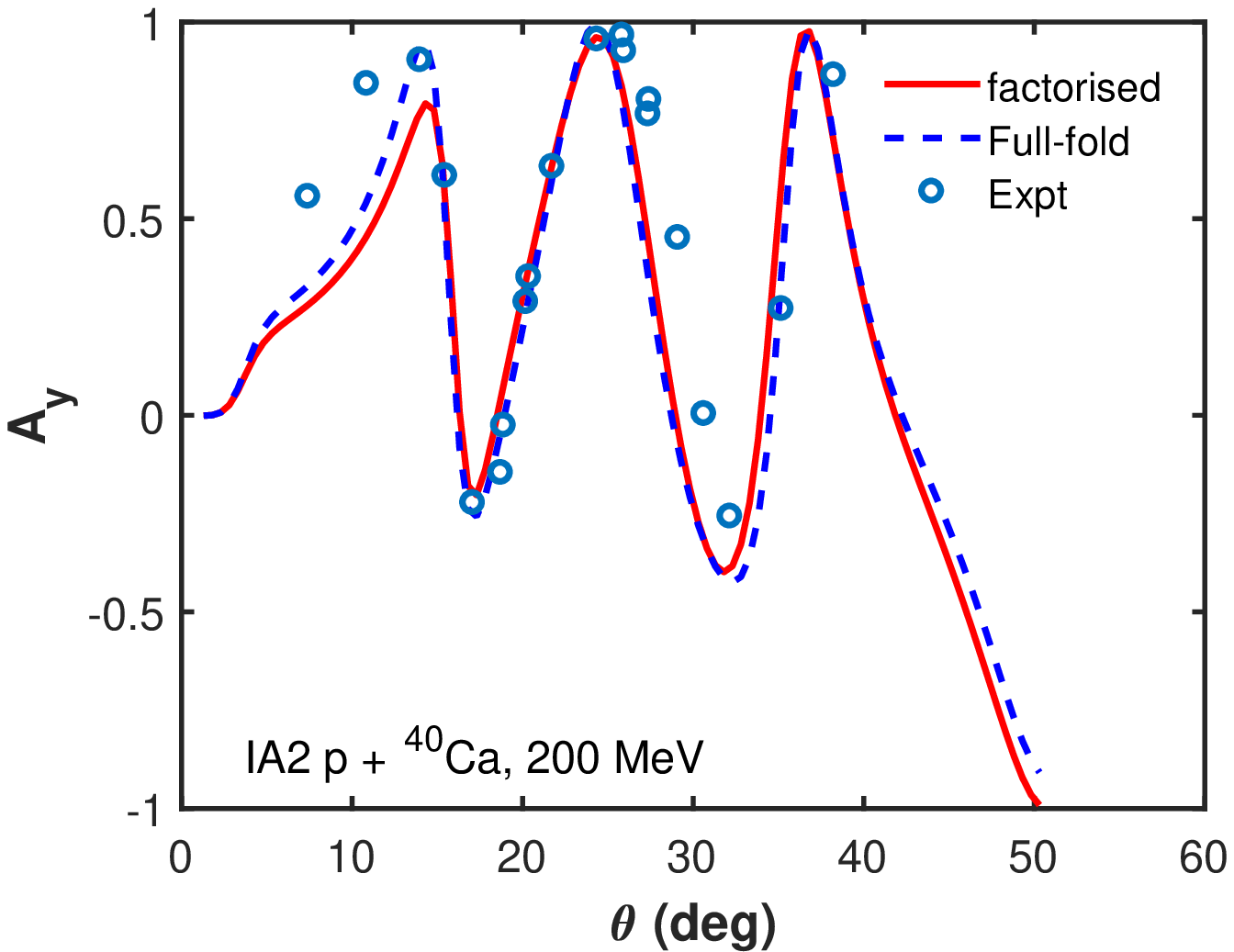}
	\includegraphics[width=0.49\linewidth]{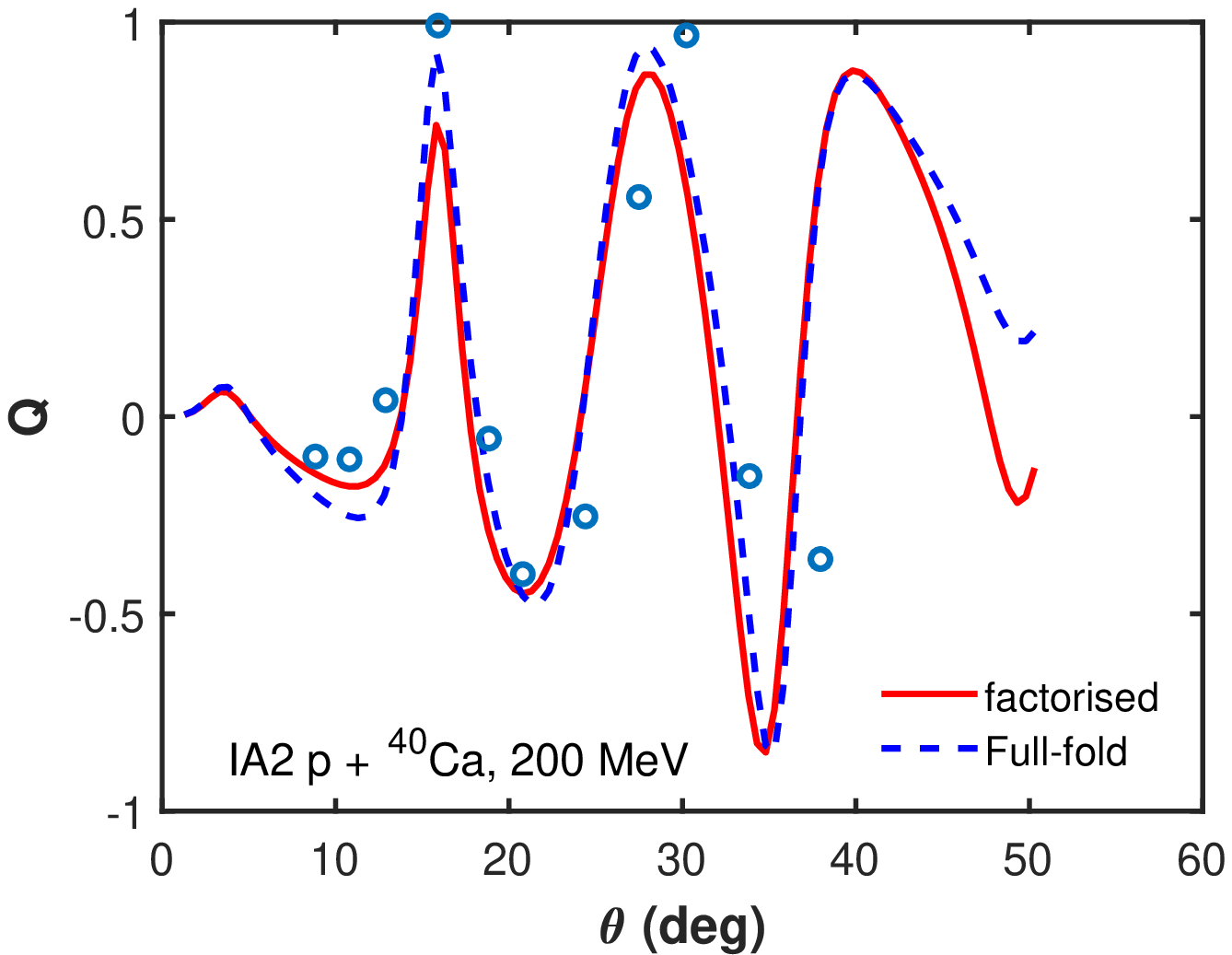}
	\caption{$^{40}$Ca scattering observables calculated with the NL3 parametrisation using optimally factorised and full-folding IA2 optical potentials at $T_{\mathrm{lab}} = 200$MeV. The results obtained using factorised optical potentials are shown in solid lines while dashed lines indicate results calculated by including medium effects.}
	\label{fullfold_Ca40_200}
\end{figure}

\begin{figure}
	\centering
	\includegraphics[width=0.49\linewidth]{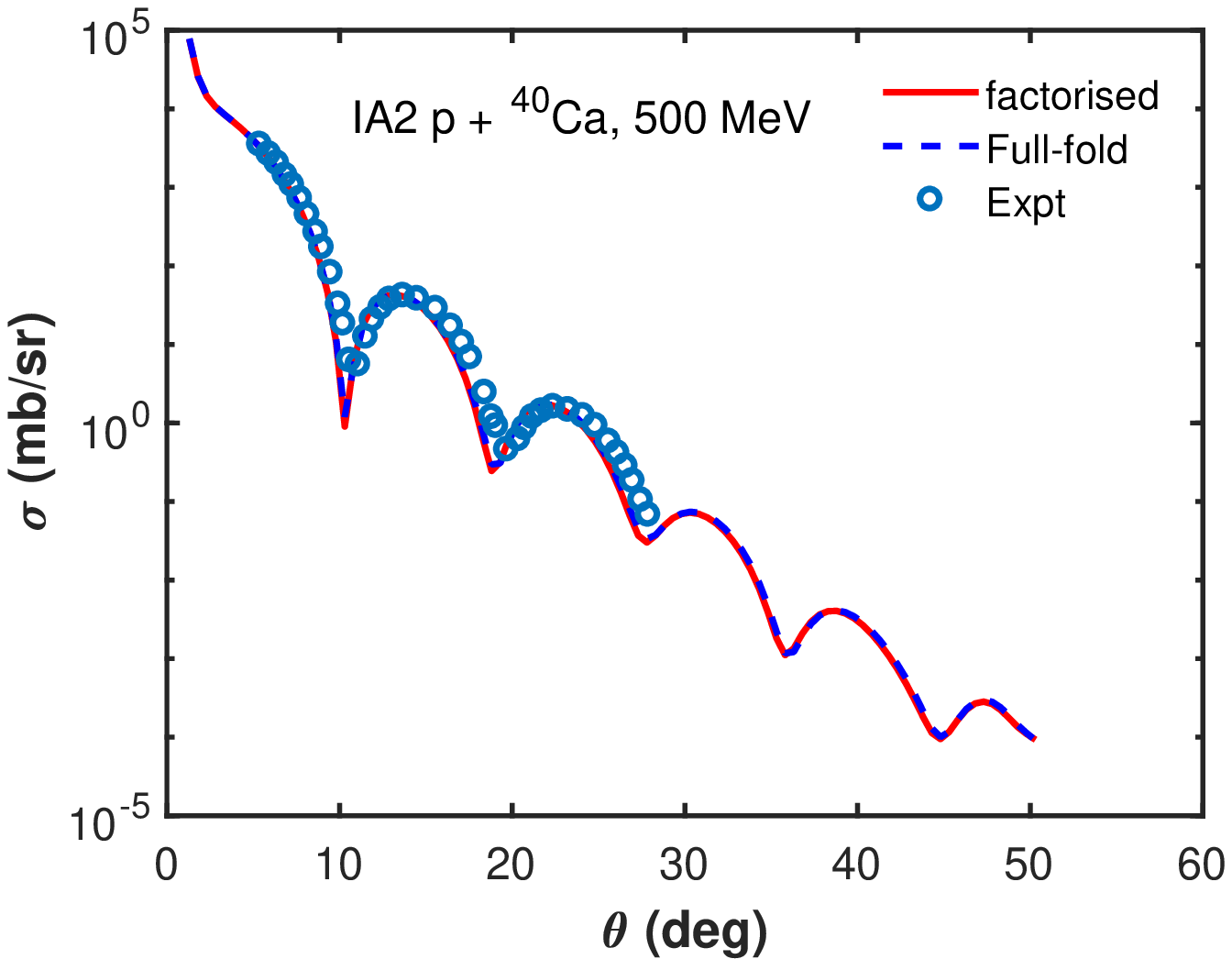}
	\includegraphics[width=0.49\linewidth]{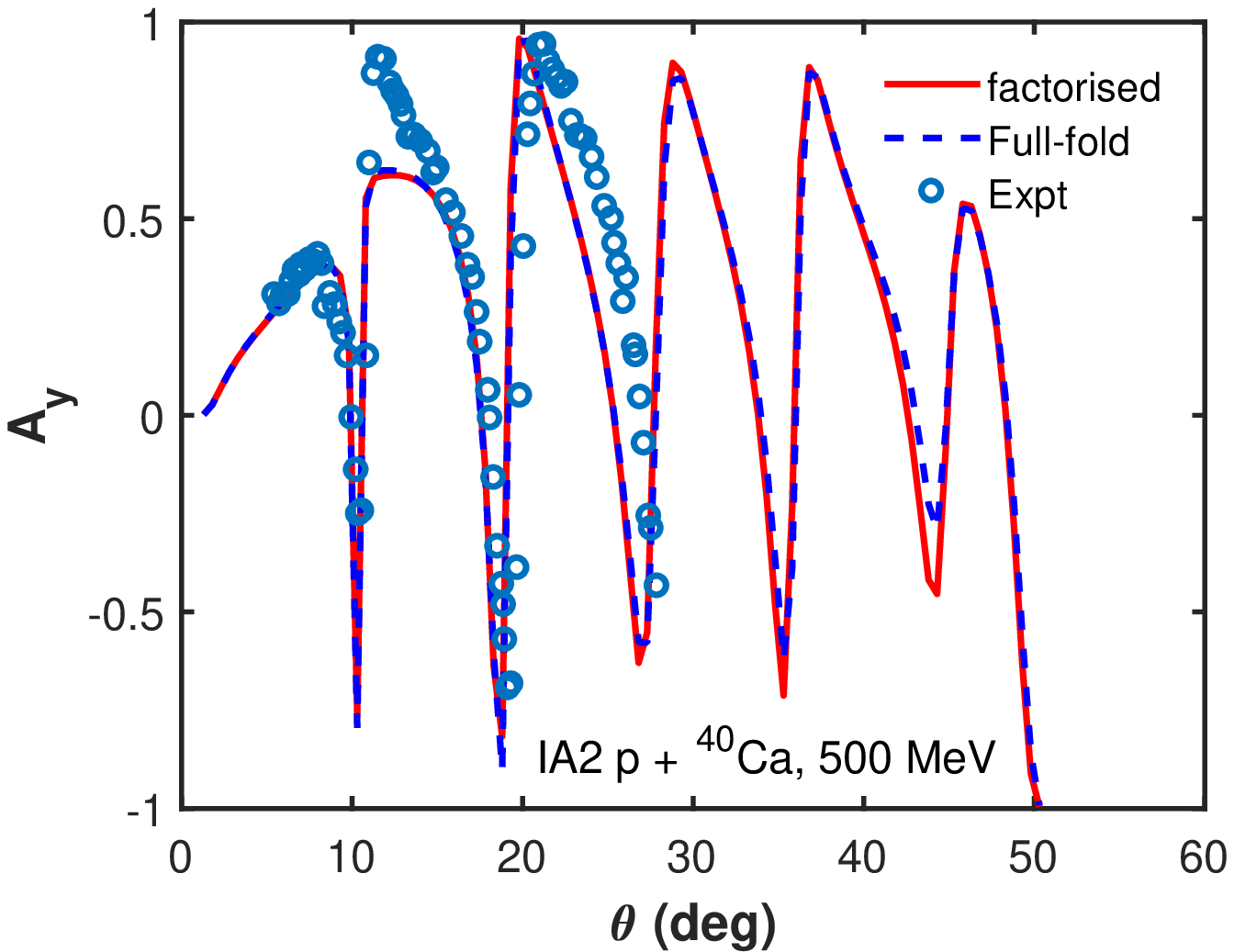}
	\includegraphics[width=0.49\linewidth]{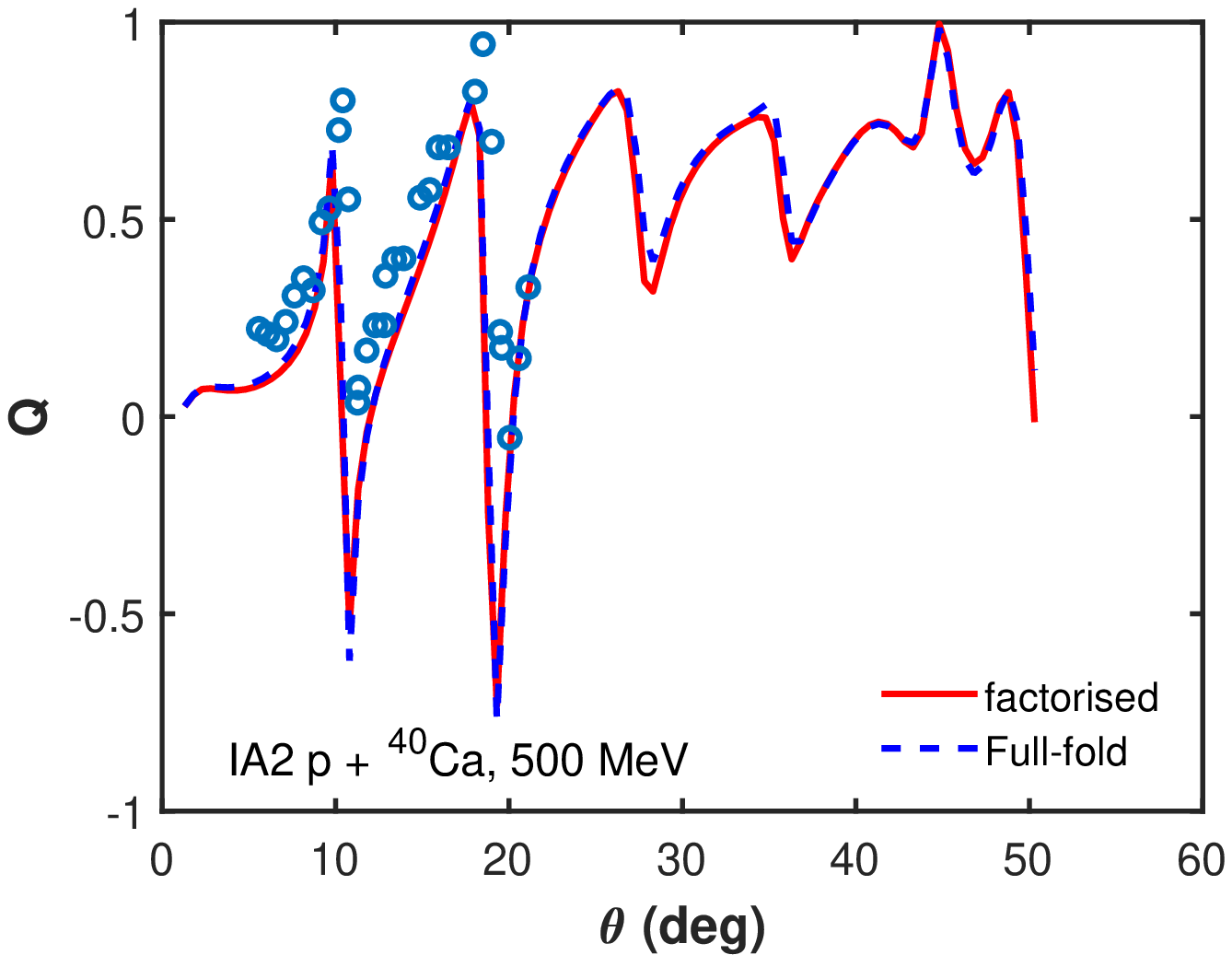}
	\caption{$^{40}$Ca scattering observables calculated with the NL3 parametrisation using optimally factorised and full-folding IA2 optical potentials at $T_{\mathrm{lab}} = 500$MeV. The expressions of lines is the same as in figure \ref{fullfold_Ca40_200}.}
	\label{fullfold_Ca40_500}
\end{figure}

\begin{figure}
	\centering
	\includegraphics[width=0.49\linewidth]{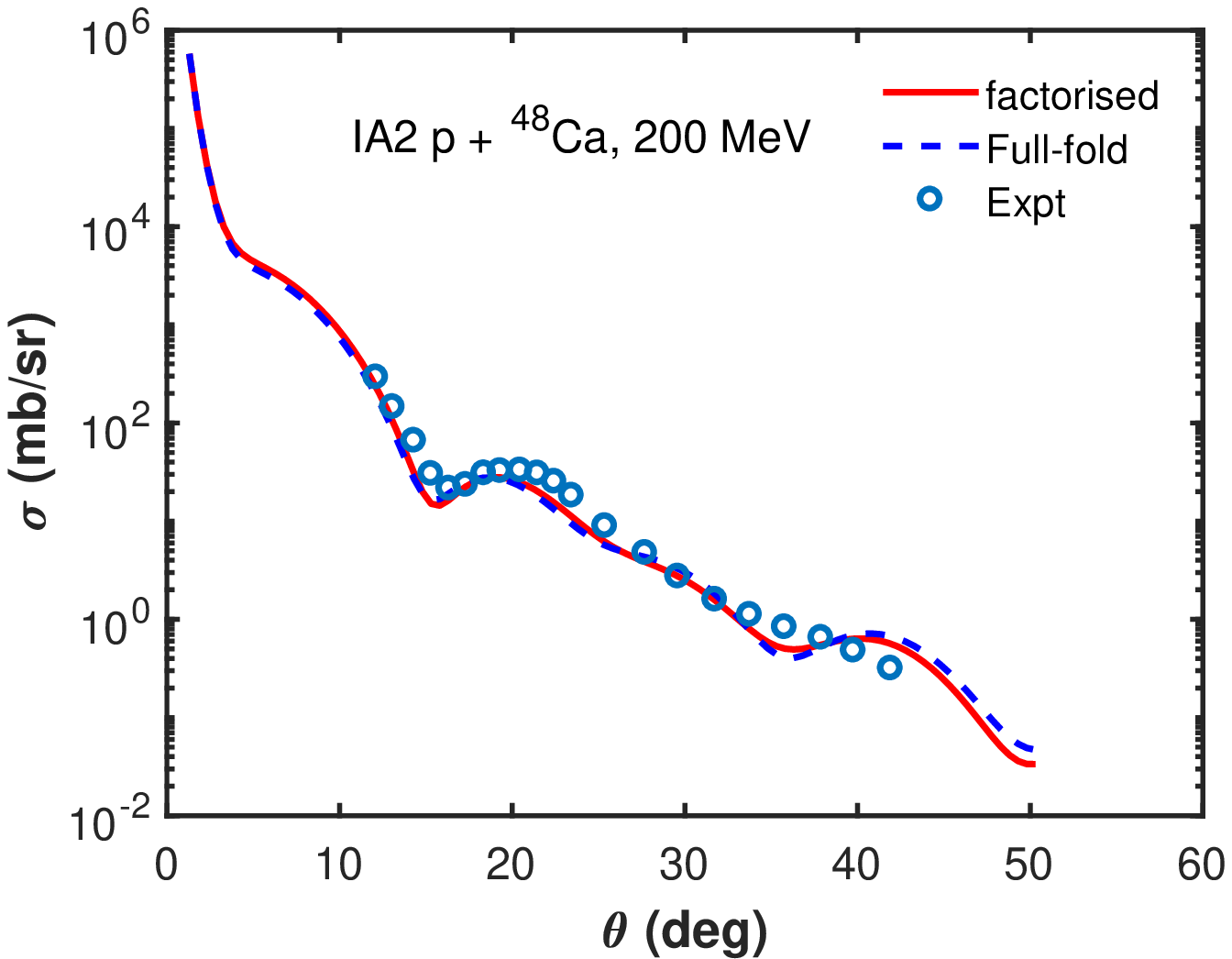}
	\includegraphics[width=0.49\linewidth]{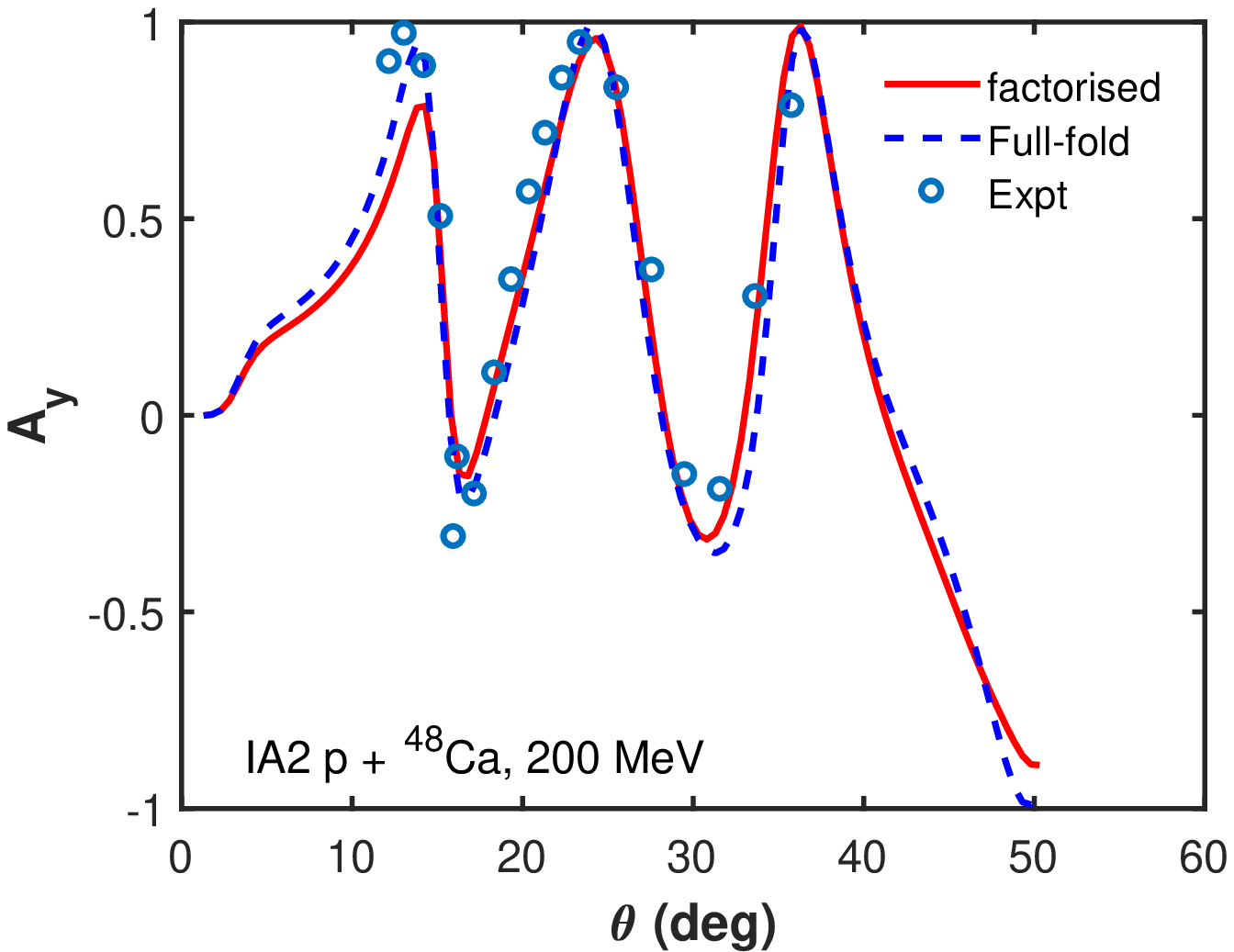}
	\includegraphics[width=0.49\linewidth]{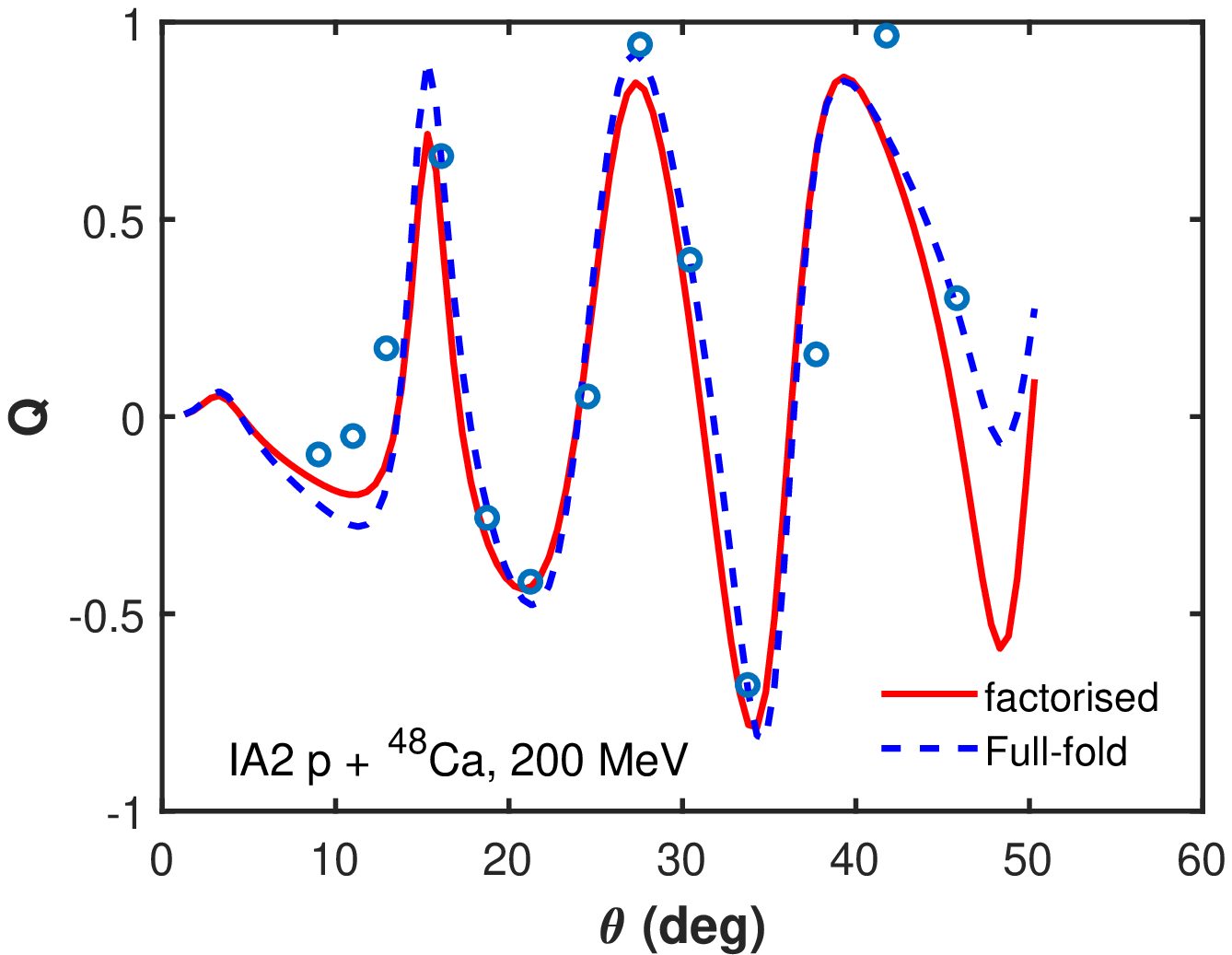}
	\caption{$^{48}$Ca scattering observables calculated with the NL3 parametrisation using optimally factorised and full-folding IA2 optical potentials at $T_{\mathrm{lab}} = 200$MeV. The expressions of lines is the same as in figure \ref{fullfold_Ca40_200}.}
	\label{fullfold_Ca48_200}
\end{figure}

\begin{figure}
	\centering
	\includegraphics[width=0.49\linewidth]{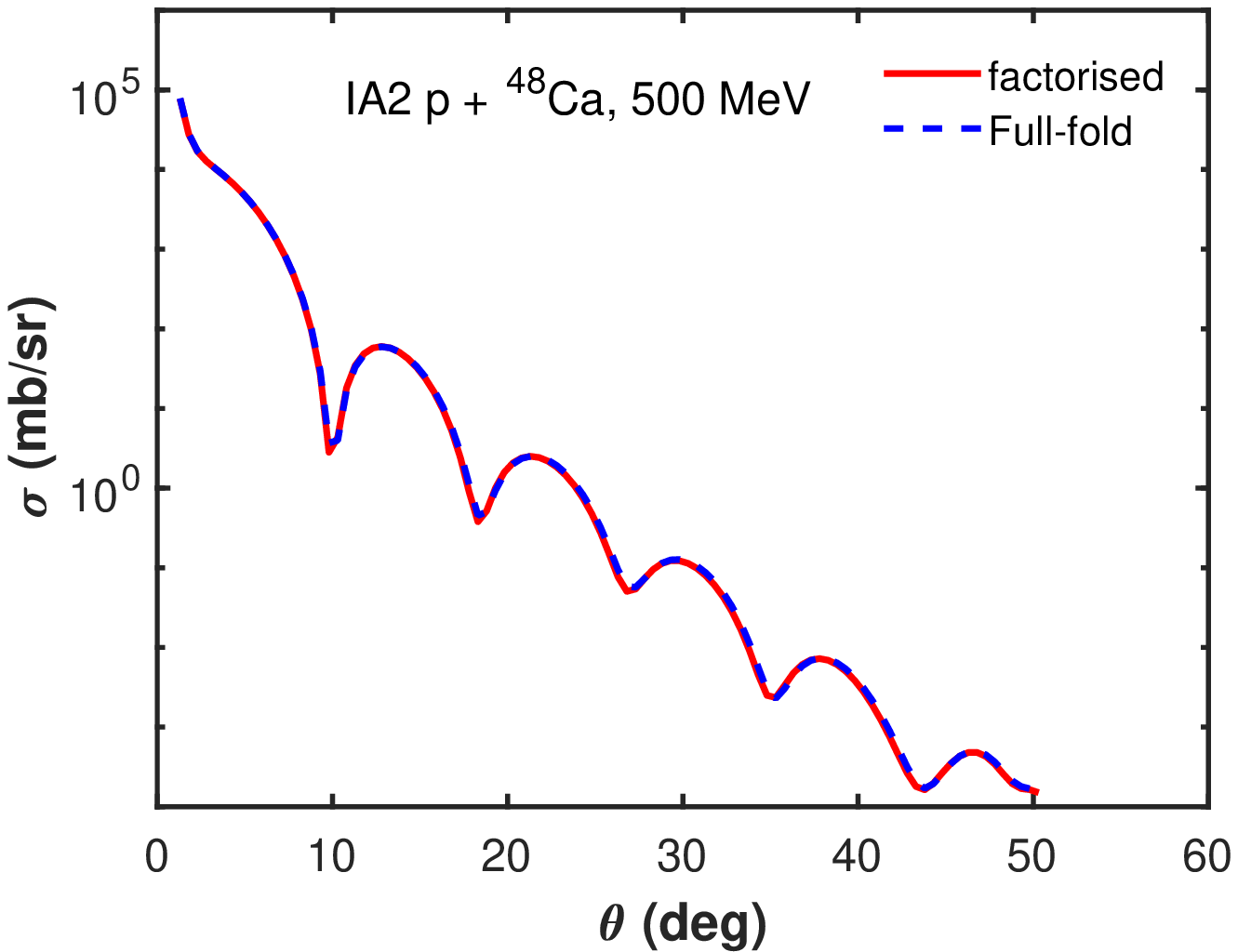}
	\includegraphics[width=0.49\linewidth]{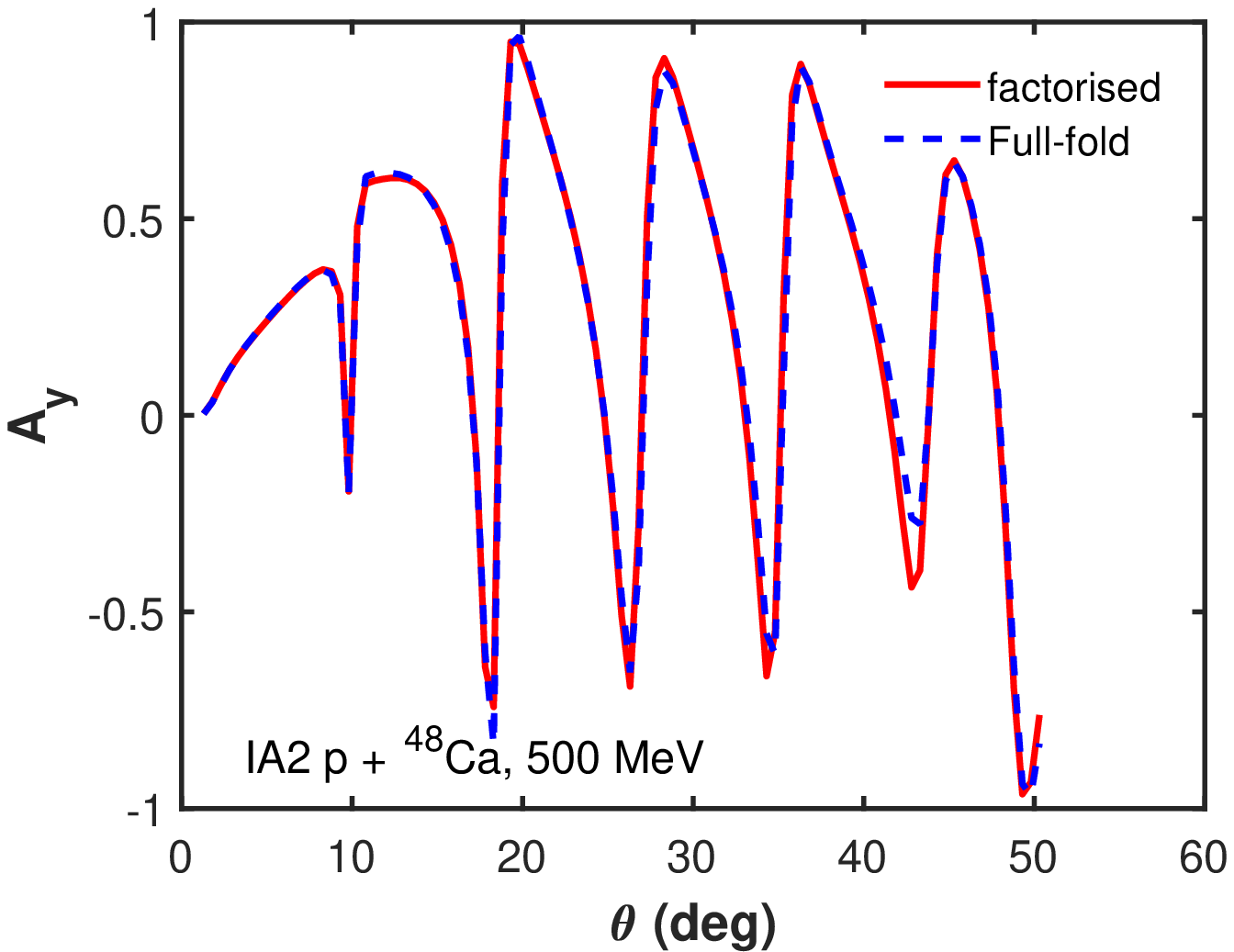}
	\includegraphics[width=0.49\linewidth]{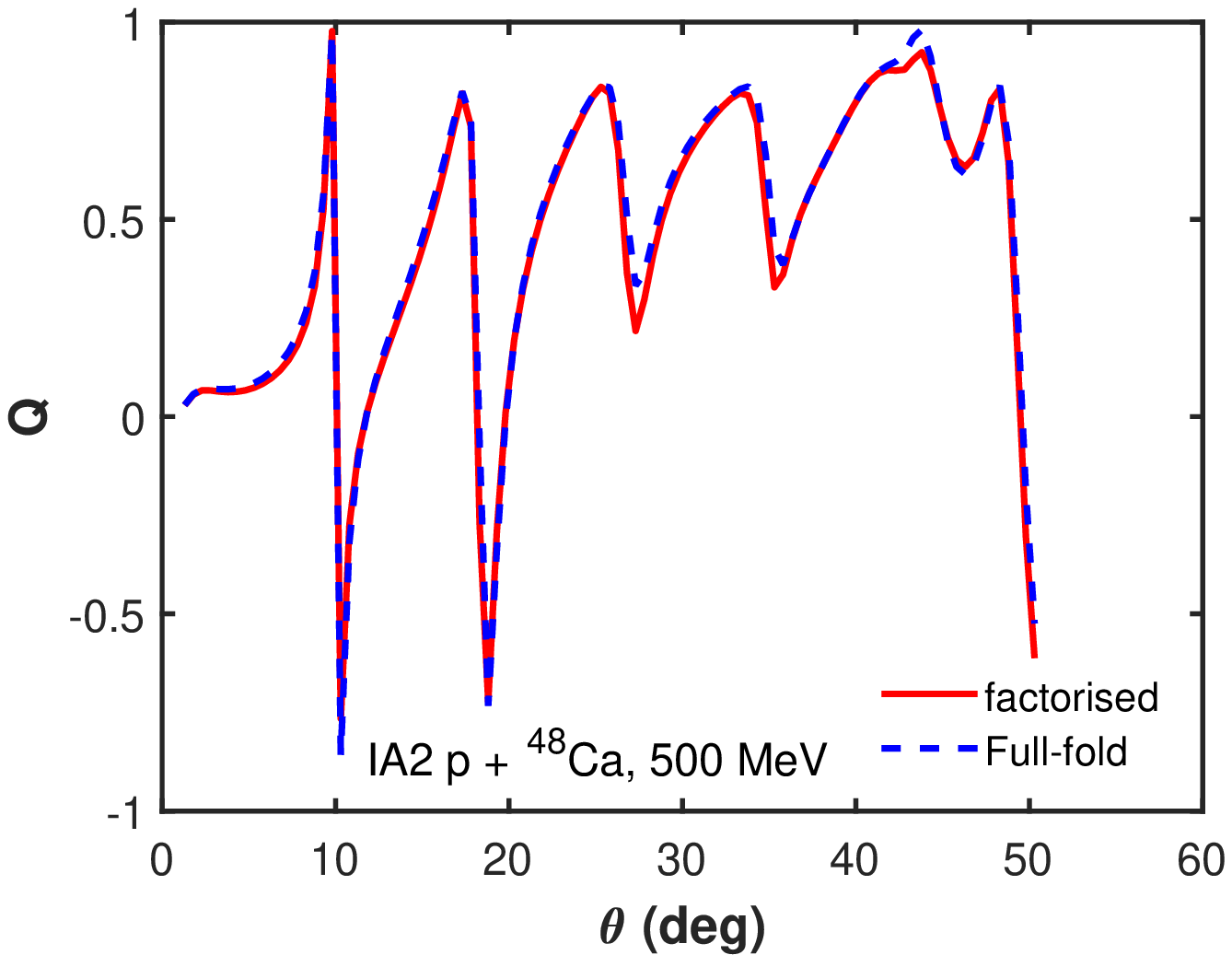}
	\caption{$^{48}$Ca scattering observables calculated with the NL3 parametrisation using optimally factorised and full-folding IA2 optical potentials at $T_{\mathrm{lab}} = 500$MeV. The expressions of lines is the same as in figure \ref{fullfold_Ca40_200}.}
	\label{fullfold_Ca48_500}
\end{figure}

\begin{figure}
	\centering
	\includegraphics[width=0.49\linewidth]{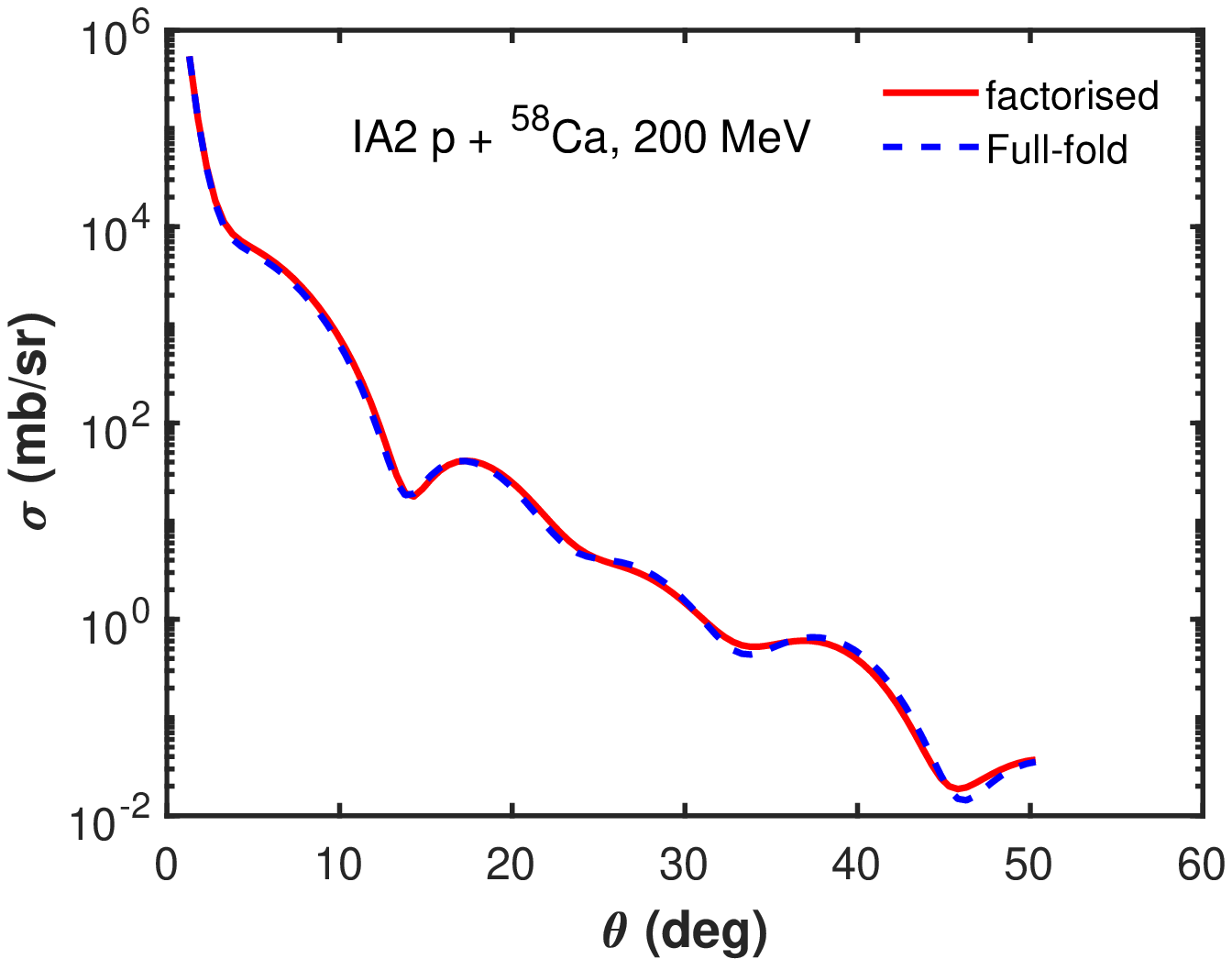}
	\includegraphics[width=0.49\linewidth]{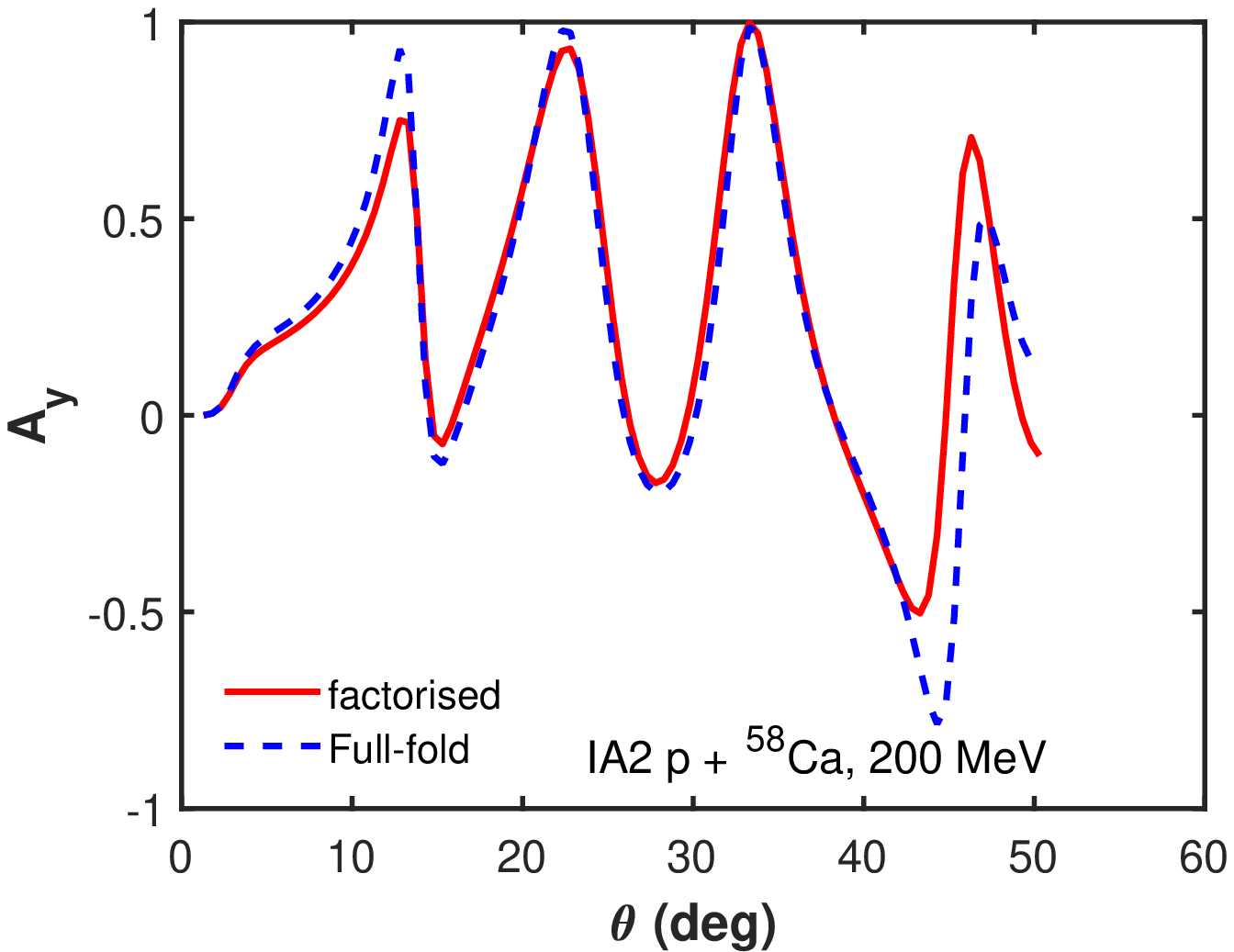}
	\includegraphics[width=0.49\linewidth]{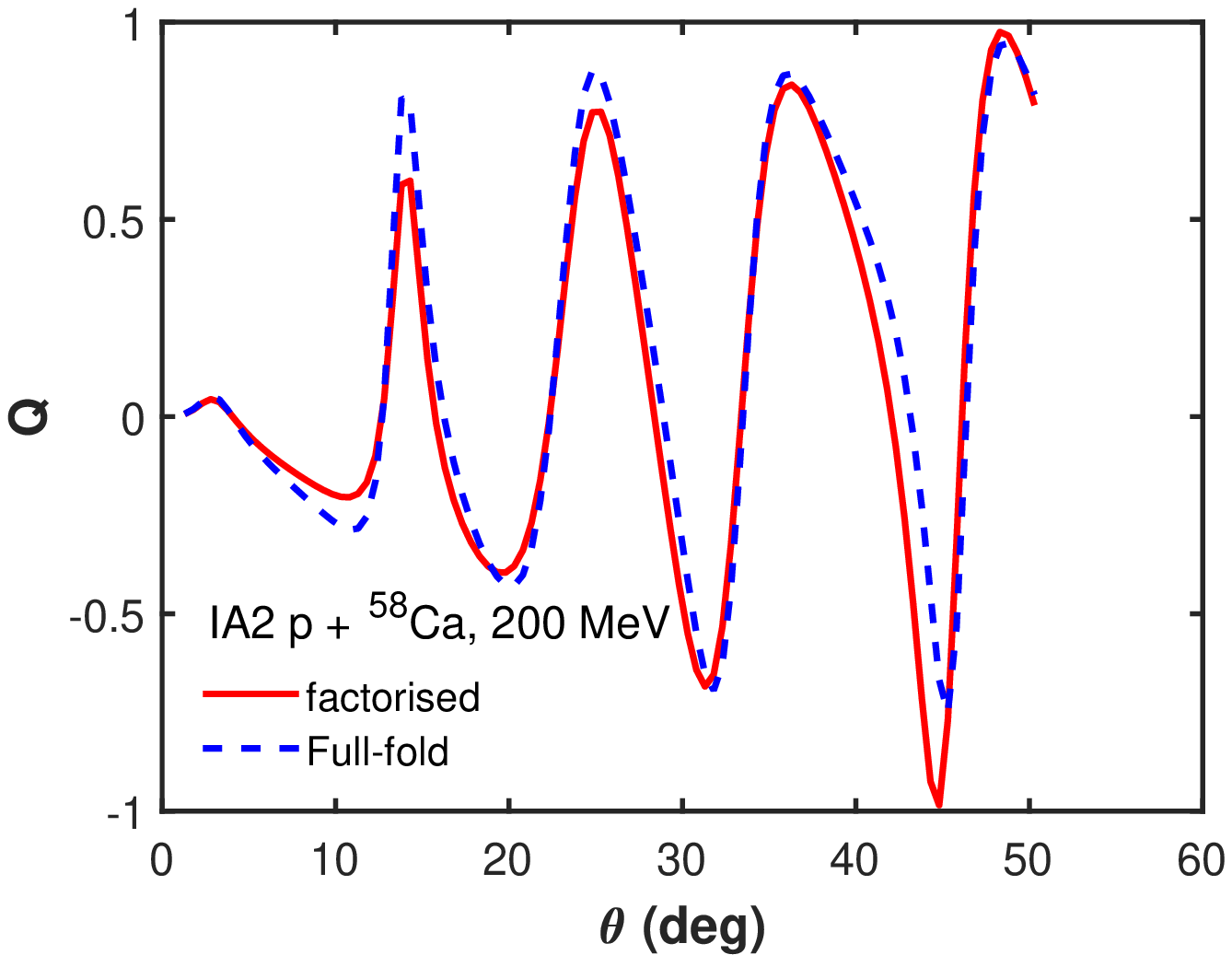}
	\caption{$^{58}$Ca scattering observables calculated with the NL3 parametrisation using optimally factorised and full-folding IA2 optical potentials at $T_{\mathrm{lab}} = 200$MeV. The expressions of lines is the same as in figure \ref{fullfold_Ca40_200}.}
	\label{fullfold_Ca58_200}
\end{figure}

\begin{figure}
	\centering
	\includegraphics[width=0.49\linewidth]{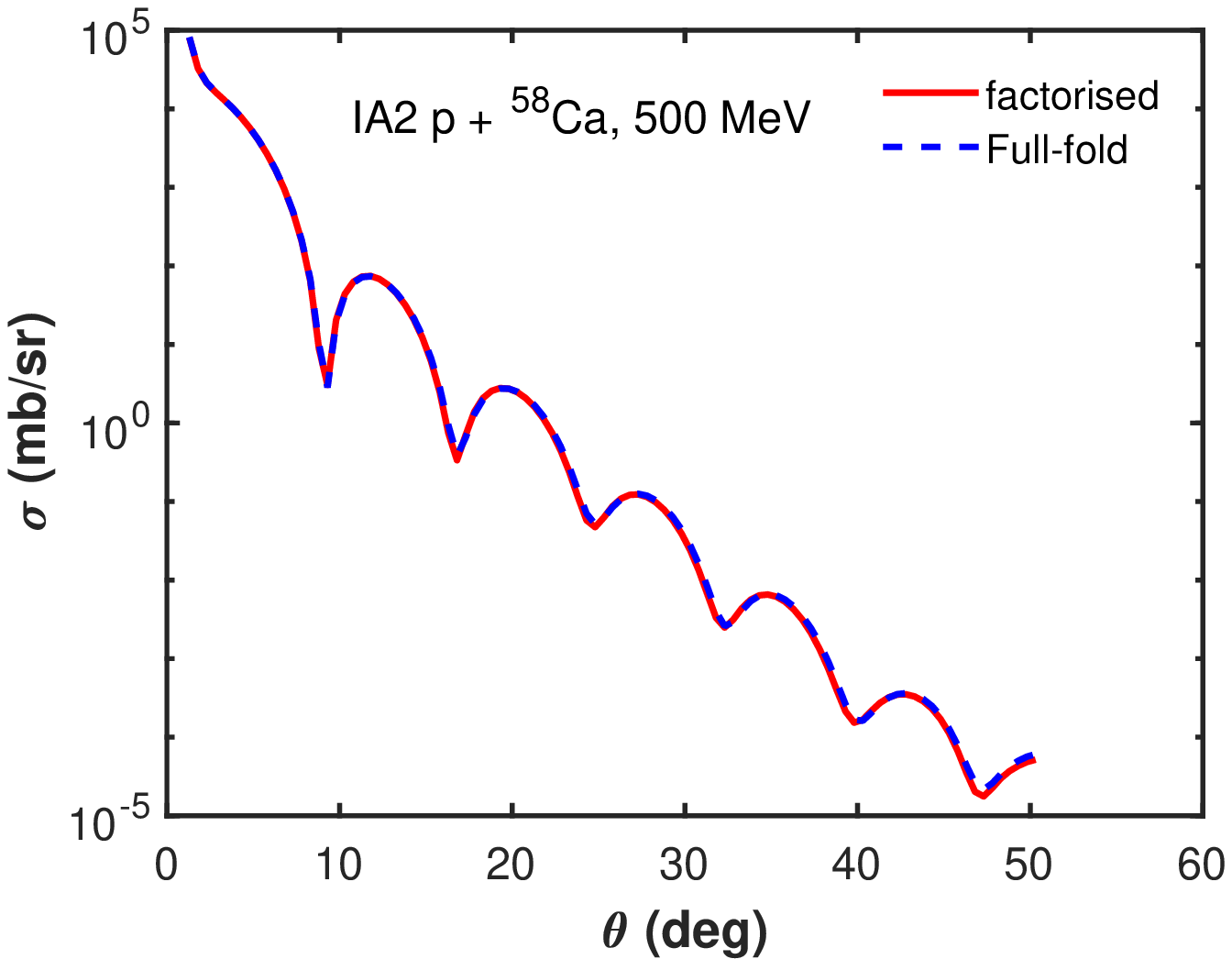}
	\includegraphics[width=0.49\linewidth]{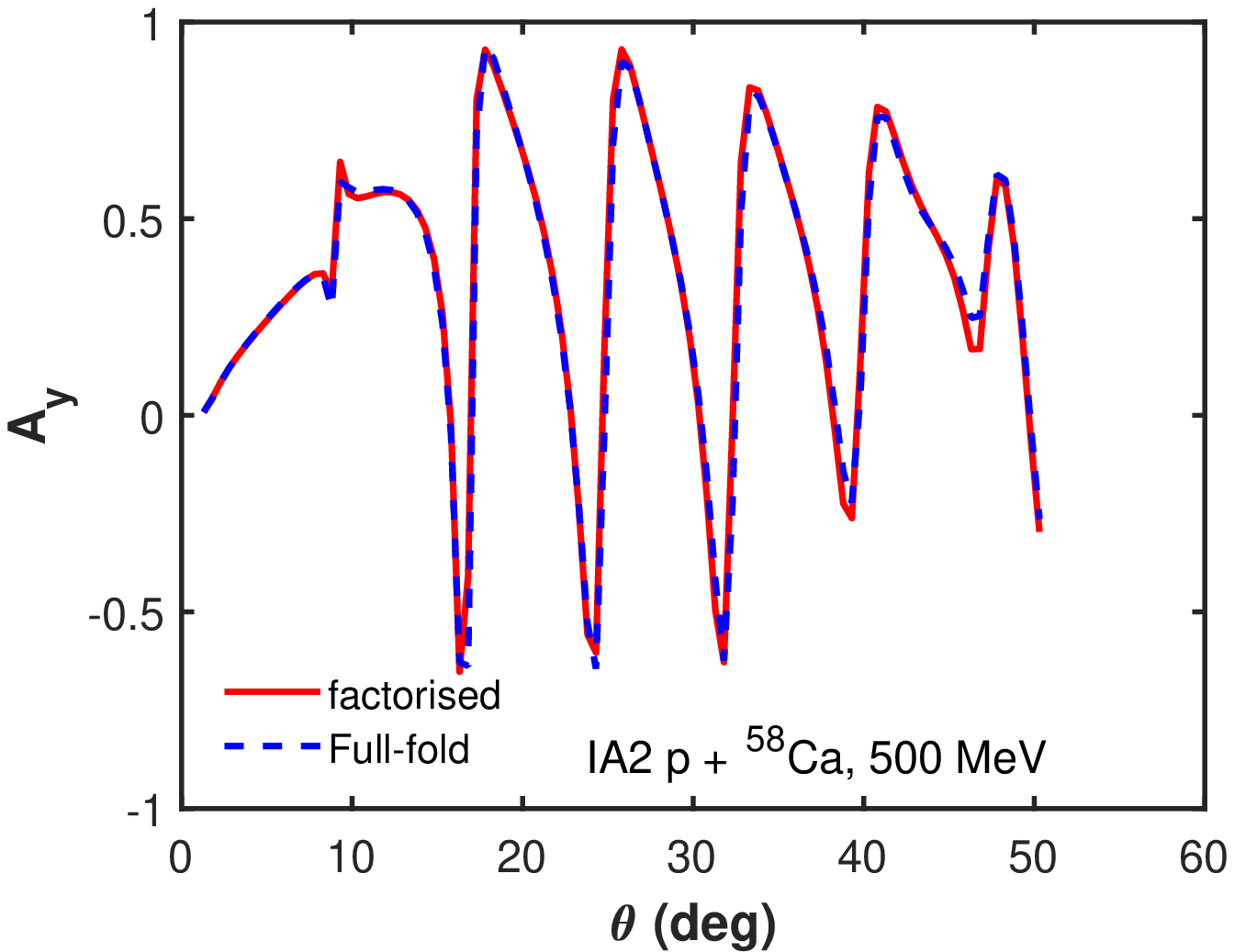}
	\includegraphics[width=0.49\linewidth]{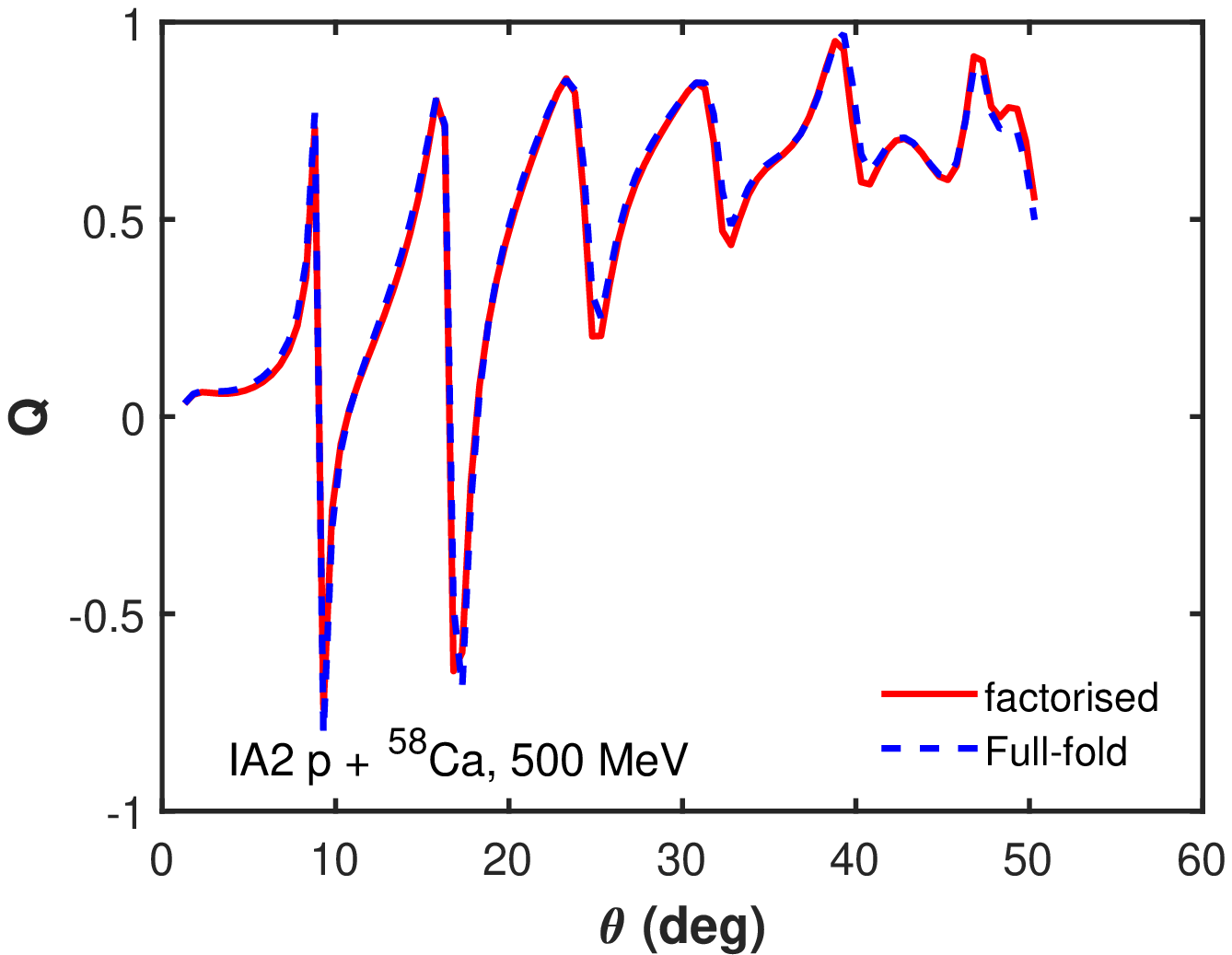}
	\caption{$^{58}$Ca scattering observables calculated with the NL3 parametrisation using optimally factorised and full-folding IA2 optical potentials at $T_{\mathrm{lab}} = 500$MeV. The expressions of lines is the same as in figure \ref{fullfold_Ca40_200}.}
	\label{fullfold_Ca58_500}
\end{figure}

The results of the scattering observables calculated using optimally factorised optical potentials and full folding optical potentials will be presented and compared here; this will show the effect of medium contributions on the scattering observables. Figures \ref{fullfold_Ca40_200} -- \ref{fullfold_Ca58_500} show the scattering cross sections, analysing powers and spin rotation functions for elastic proton scattering from $^{48,54,58}$Ca targets at $T_{\mathrm{lab}} = 200$ MeV and $500$ MeV. The calculations obtained using optimally factorised optical potentials (denoted as 'factorised' and shown in solid lines) are compared with the calculations that incorporate medium effects (denoted as 'full-fold' and shown in dashed lines). In figures \ref{fullfold_Ca40_200} and \ref{fullfold_Ca48_200}, the scattering observables are shown against scattering angles for $^{40,48}$Ca nuclei at $T_{\mathrm{lab}} = 200$ MeV. There is not much effect of including medium modifications on the cross sections at this incident energy. There is a conspicuous effect however, on the analysing power; the analysing power data at the first maximum are better reproduced. The use of optimally factorised optical potential could not correctly reproduce the first maximum of the analysing power data, as it underestimates it. Medium effects are also seen on the spin rotation function at large scattering angles and first minimum. Figures \ref{fullfold_Ca58_200} show the case for $^{58}$Ca at $T_{\mathrm{lab}} = 200$ MeV. One can observe that there is no contribution of medium effect to the scattering cross sections for this target. The contributions are seen in the analysing power and spin rotation function. Medium modifications increase the value of the first analysing power maximum and increase the depth of the third minimum. For the spin rotation function, medium effects increase the value of the first minimum and maximum and reduce the depth of the third minimum.

\begin{figure}
	\centering
	\includegraphics[width=0.49\linewidth]{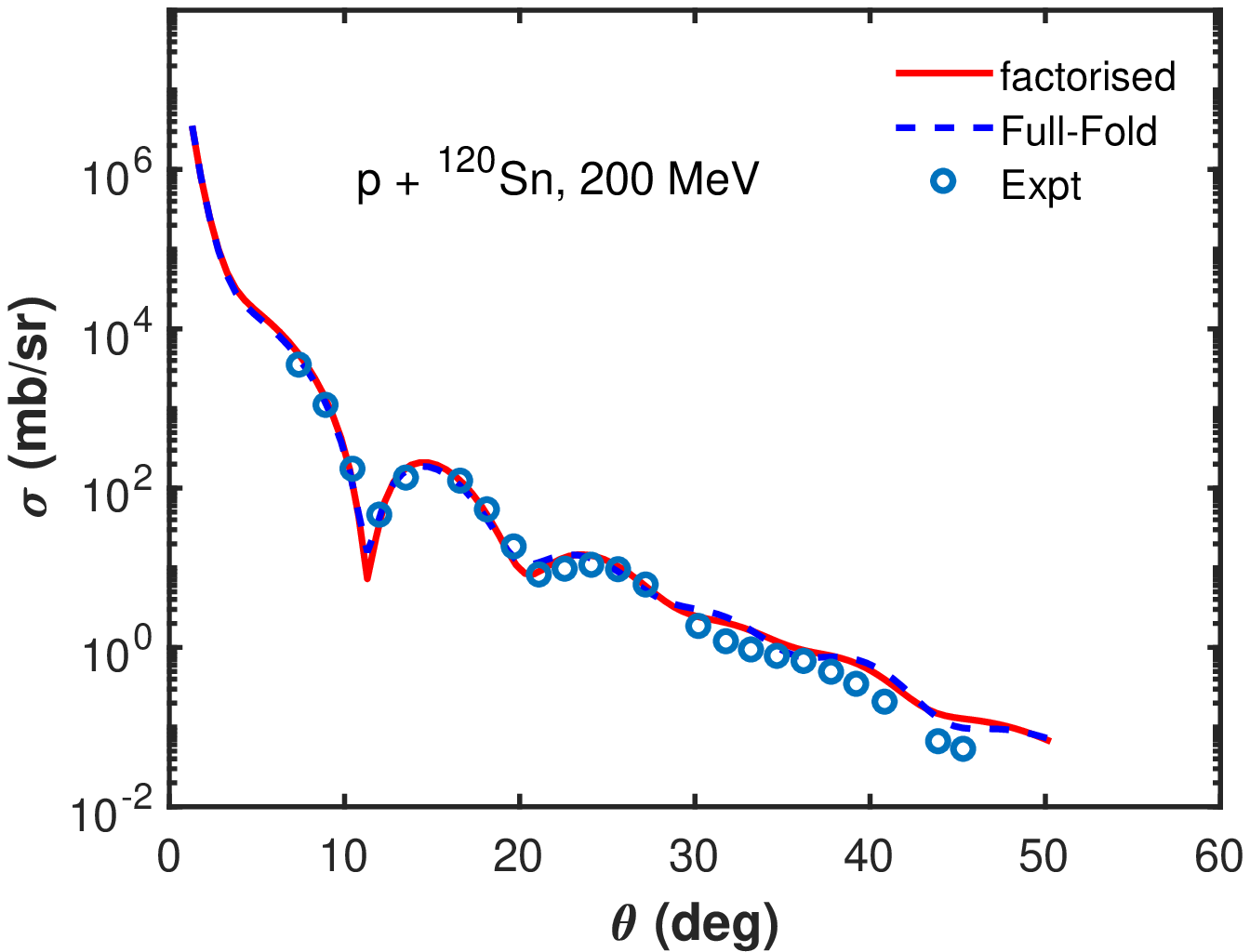}
	\includegraphics[width=0.49\linewidth]{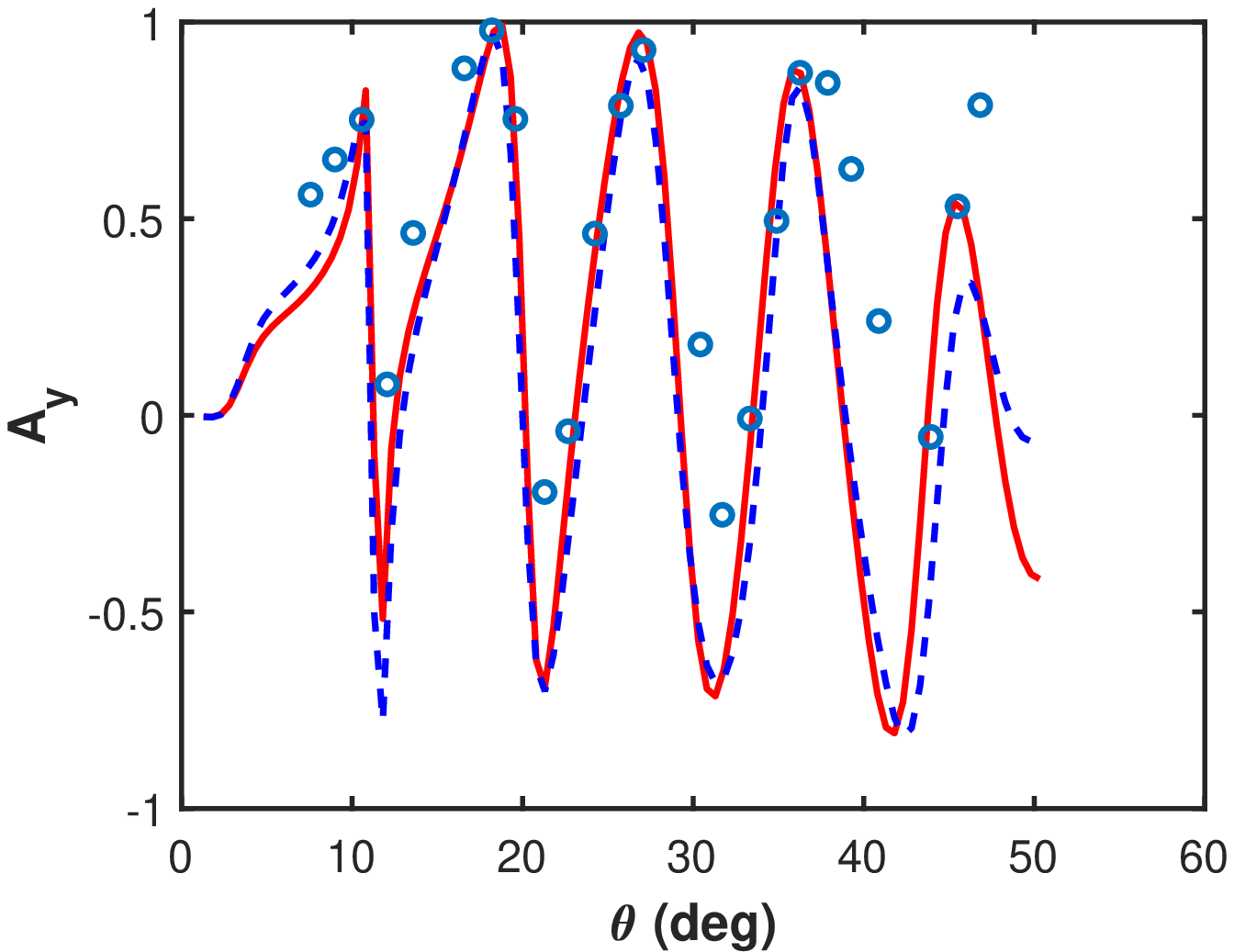}
	\includegraphics[width=0.49\linewidth]{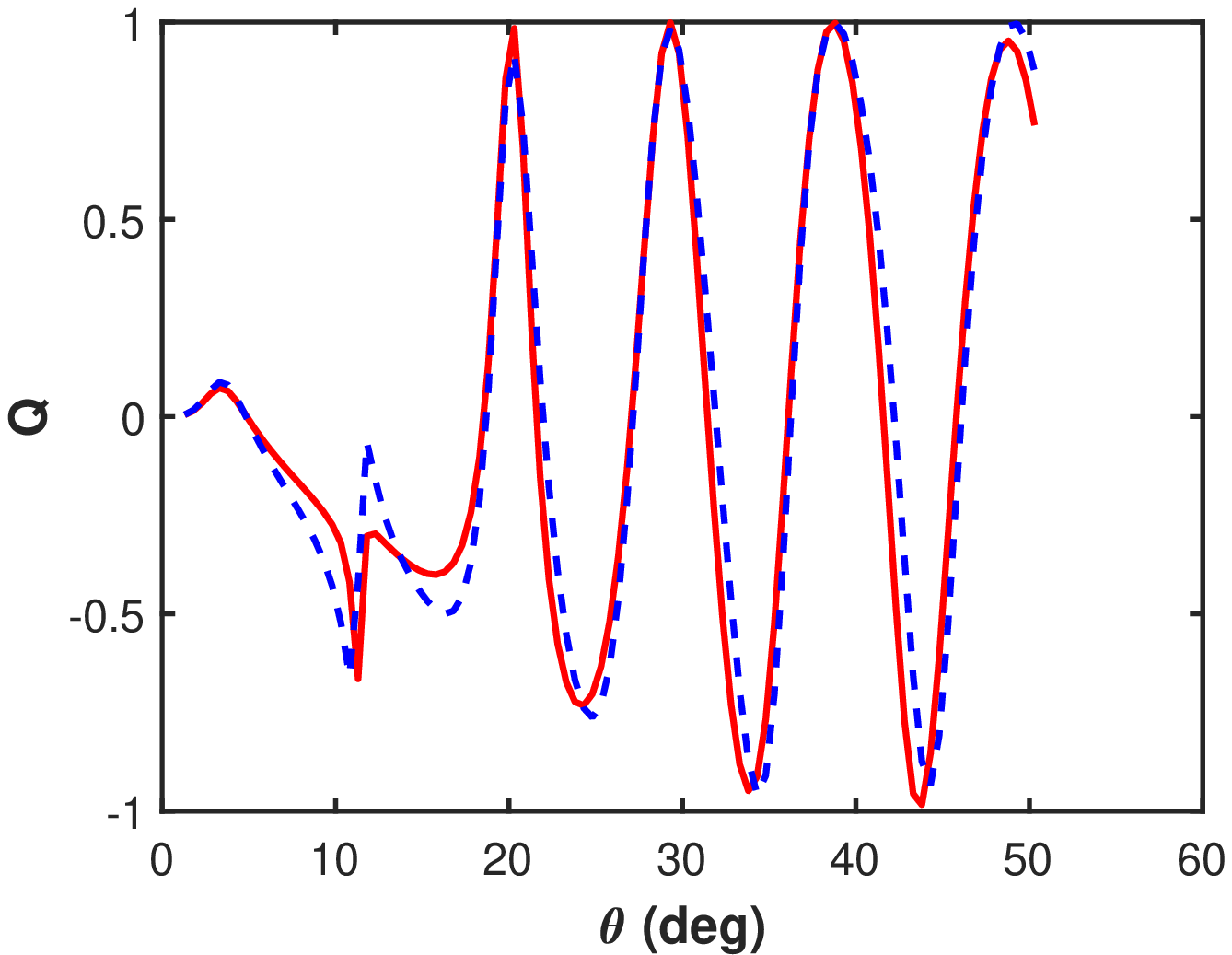}
	\caption{$^{120}$Sn scattering observables calculated with the NL3 parametrisation using optimally factorised and full-folding IA2 optical potentials at $T_{\mathrm{lab}} = 200$MeV. The expressions of lines is the same as in figure \ref{fullfold_Ca40_200}.}
	\label{fullfold_Sn120_200}
\end{figure}

In figure \ref{fullfold_Sn120_200}, the scattering observables for elastic proton scattering from $^{120}$Sn at $T_{\mathrm{lab}} = 200$ MeV calculated using both optimally factorised and full folding optical potentials are shown. In this case there is no obvious difference for the differential cross section and analysing power data. A small difference can however, be seen at small scattering angles for spin rotation parameter $Q$.

In all the three $^{40,48,58}$Ca targets, there are no noticeable contributions of medium effects on the scattering observables at $T_{\mathrm{lab}} = 500$ MeV. This is observed from Figures \ref{fullfold_Ca40_500}, \ref{fullfold_Ca48_500}, and \ref{fullfold_Ca58_500}.  In summary, medium effects have contributions at $T_{\mathrm{lab}} = 200$ MeV and not at $T_{\mathrm{lab}} = 500$ MeV for the calcium isotopes considered here. In the case of $p + ^{120}$Sn (a heavier nucleus), there is not much difference in the scattering observables even at $T_{\mathrm{lab}} = 200$ MeV.

\subsection{Results of scattering observables calculated using local and non-local optical potentials}
Results of the differential scattering cross section, analysing power and spin rotation function calculated using the non-local optical potentials in the coupled Lippmann-Schwinger-like equations are presented in this section.

\begin{figure}
	\centering
	\includegraphics[width=0.49\linewidth]{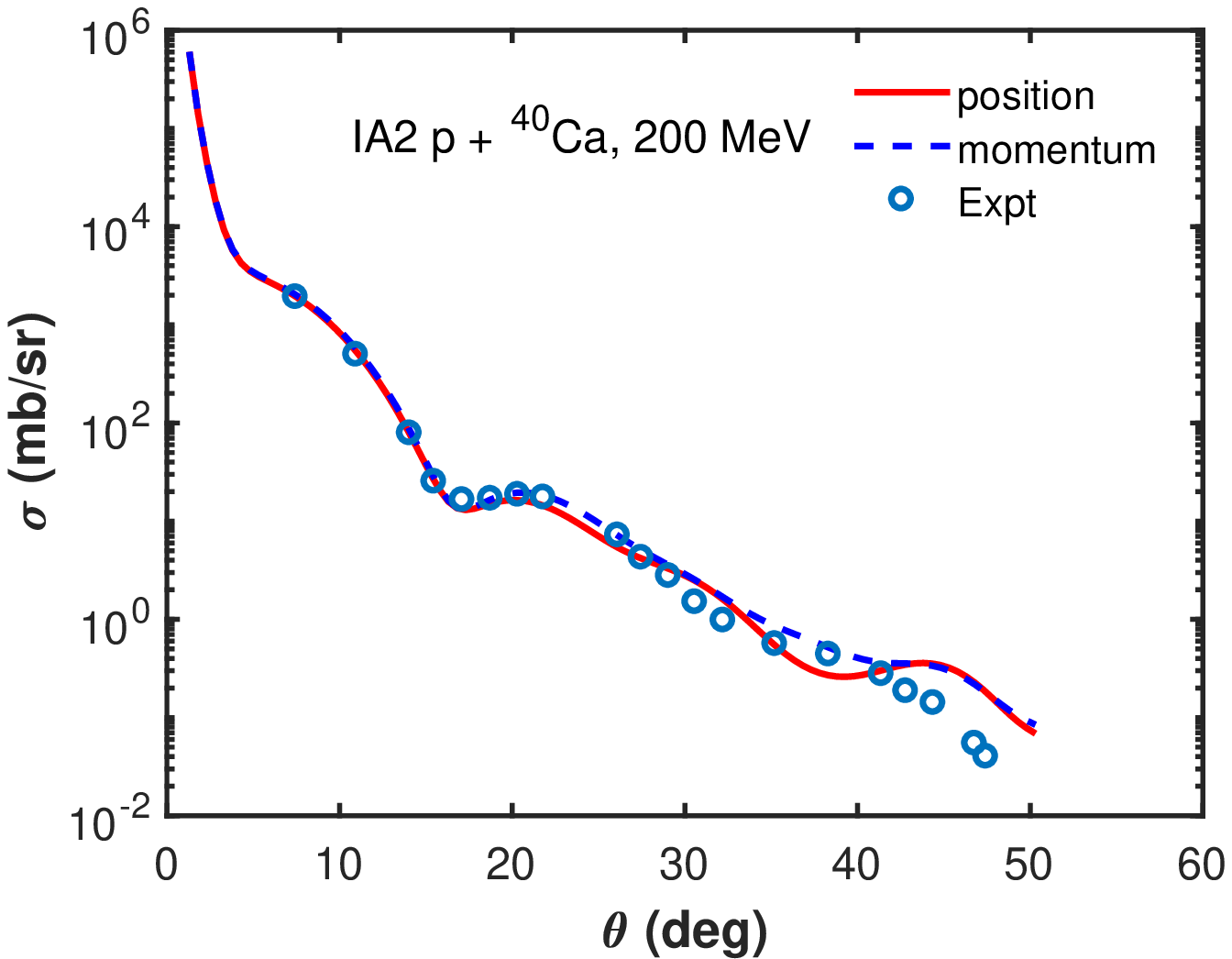}
	\includegraphics[width=0.49\linewidth]{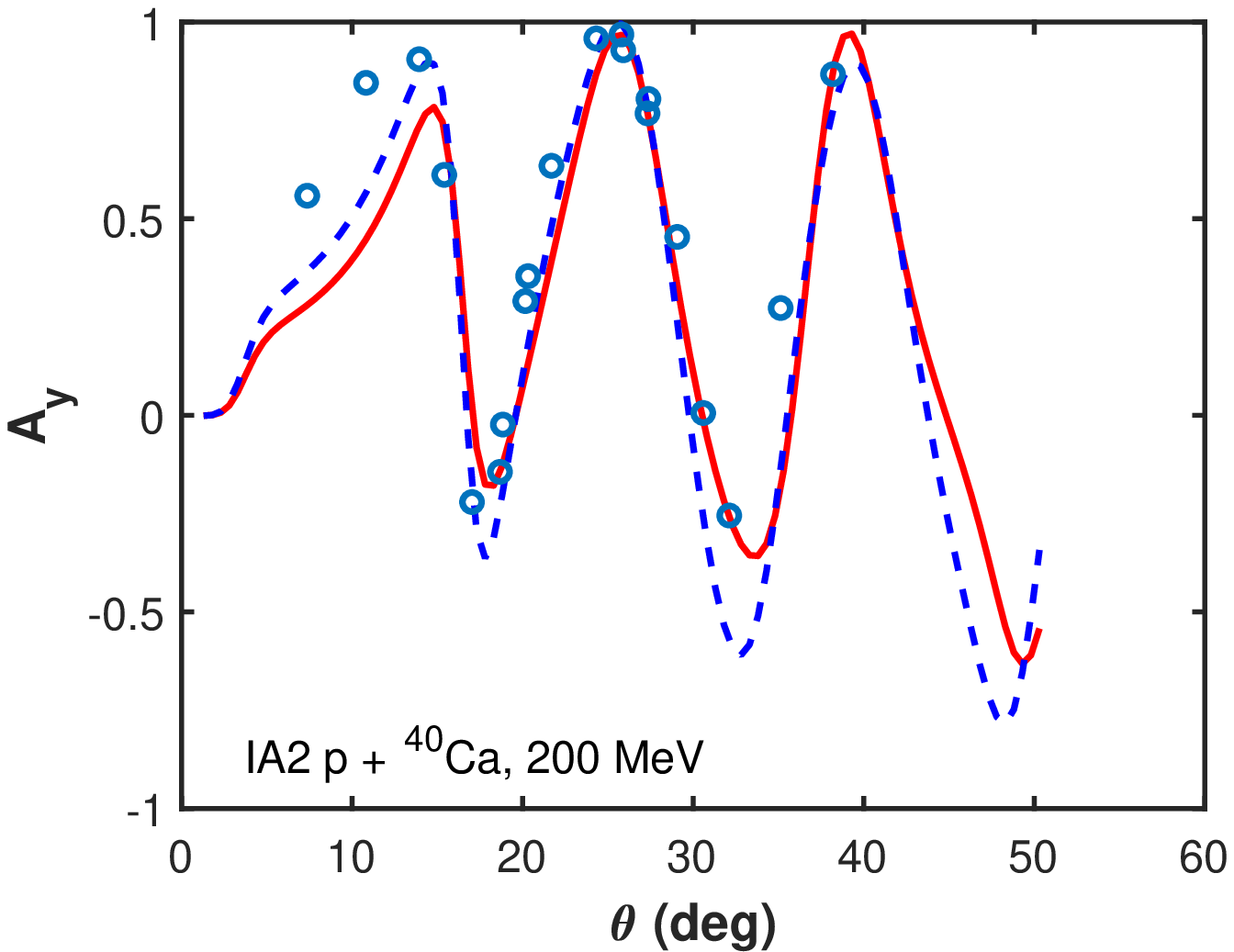}
	\includegraphics[width=0.49\linewidth]{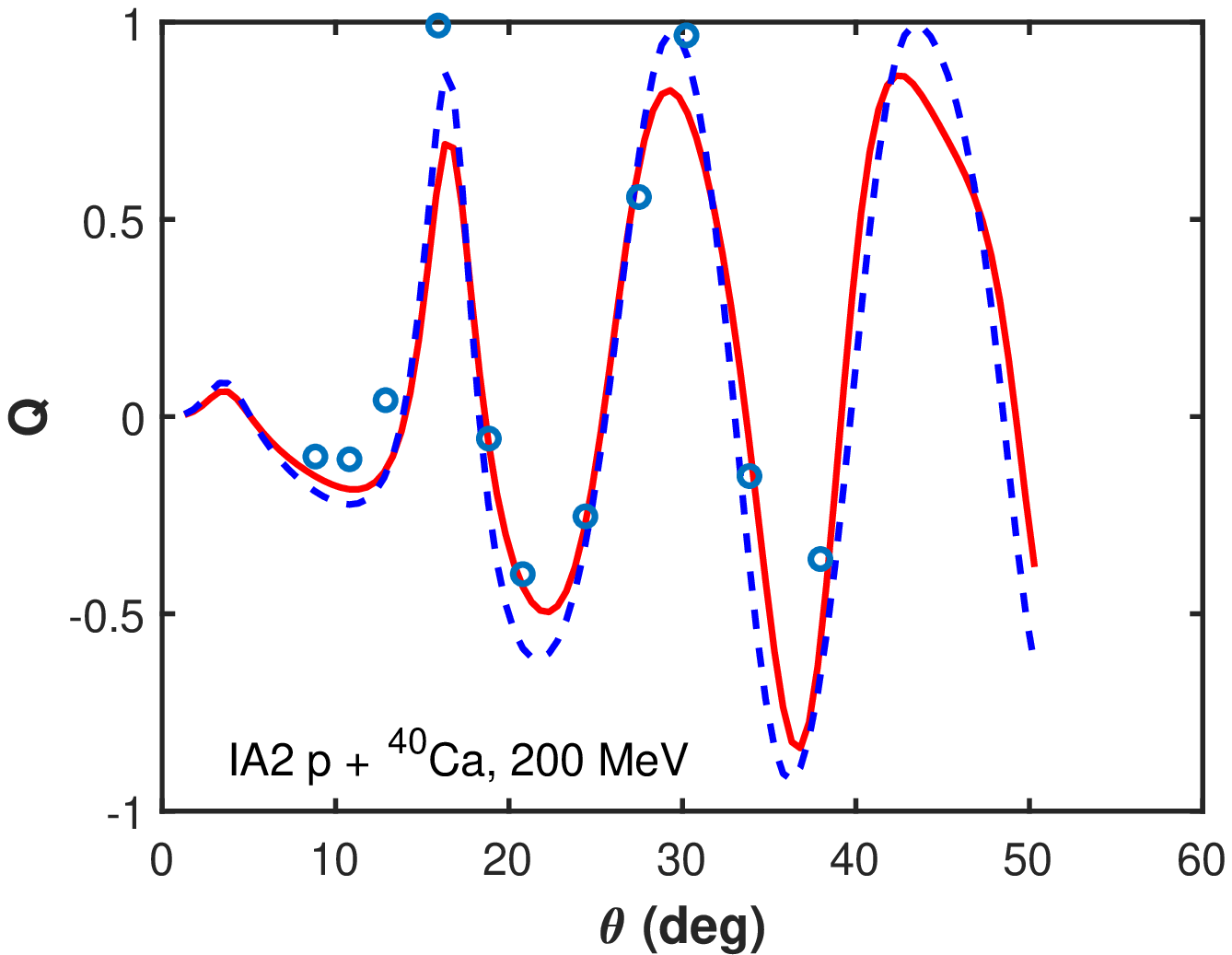}
	\caption{Scattering observables calculated in position space and momentum space for $p + ^{40}$Ca at $T_{\mathrm{lab}} = 200$ MeV using FSUGold parametrisation. solid lines indicate momentum space calculations using non-local potentials while dashed lines indicate position space calculations using localised potentials. Experimental data are shown in circles.}
	\label{sigma_posvsmomCa40_200}
\end{figure}

\begin{figure}
	\centering
	\includegraphics[width=0.49\linewidth]{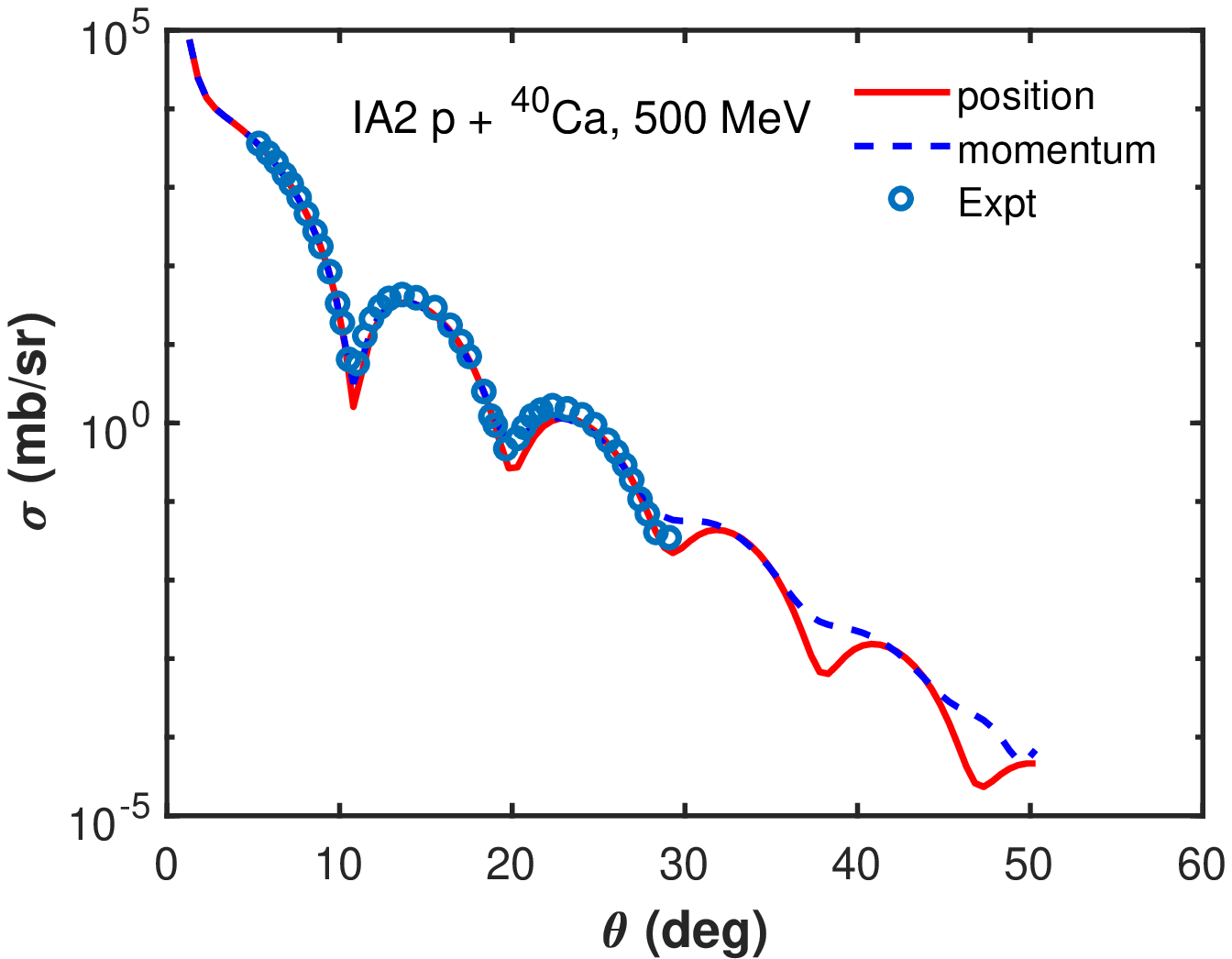}
	\includegraphics[width=0.49\linewidth]{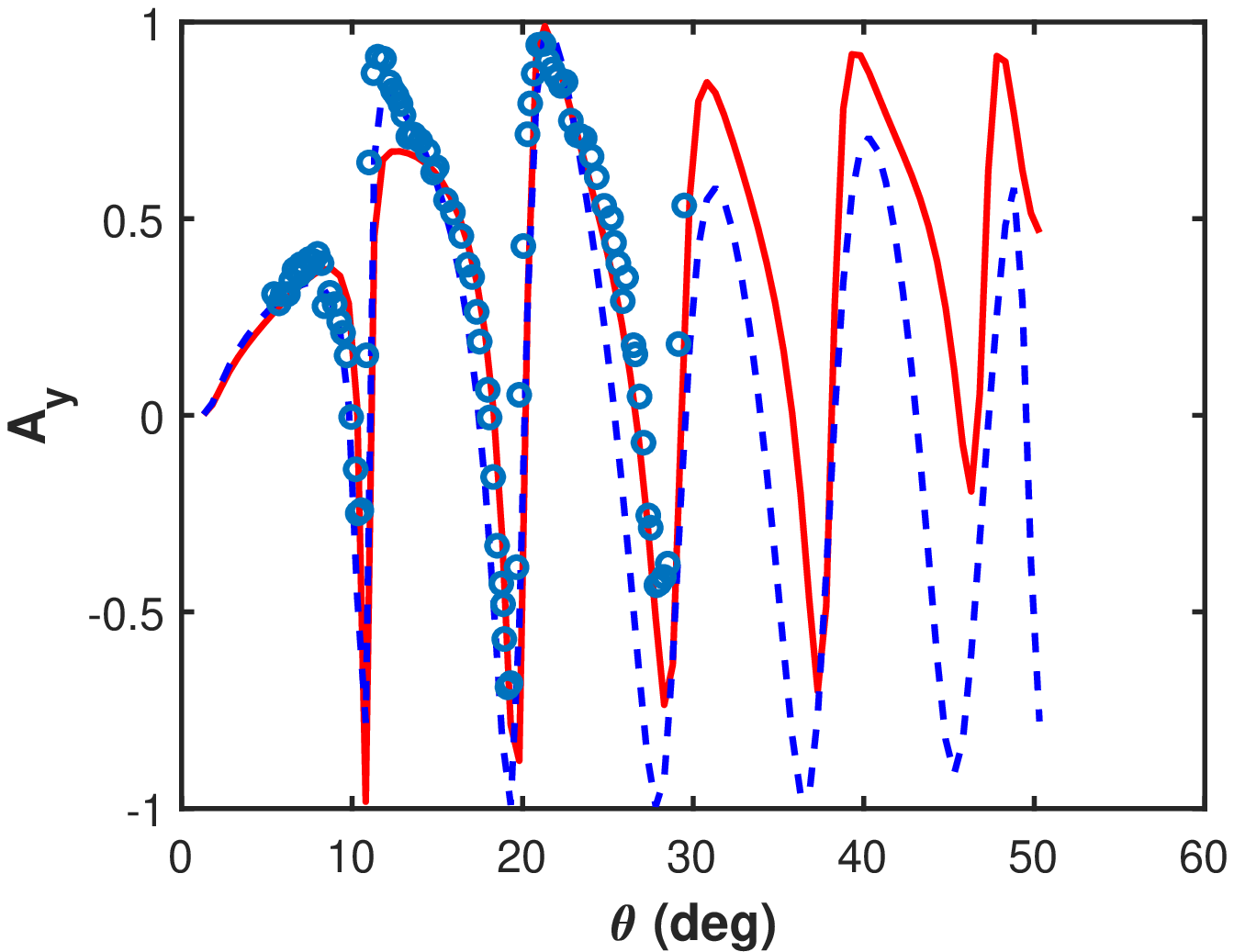}
	\includegraphics[width=0.49\linewidth]{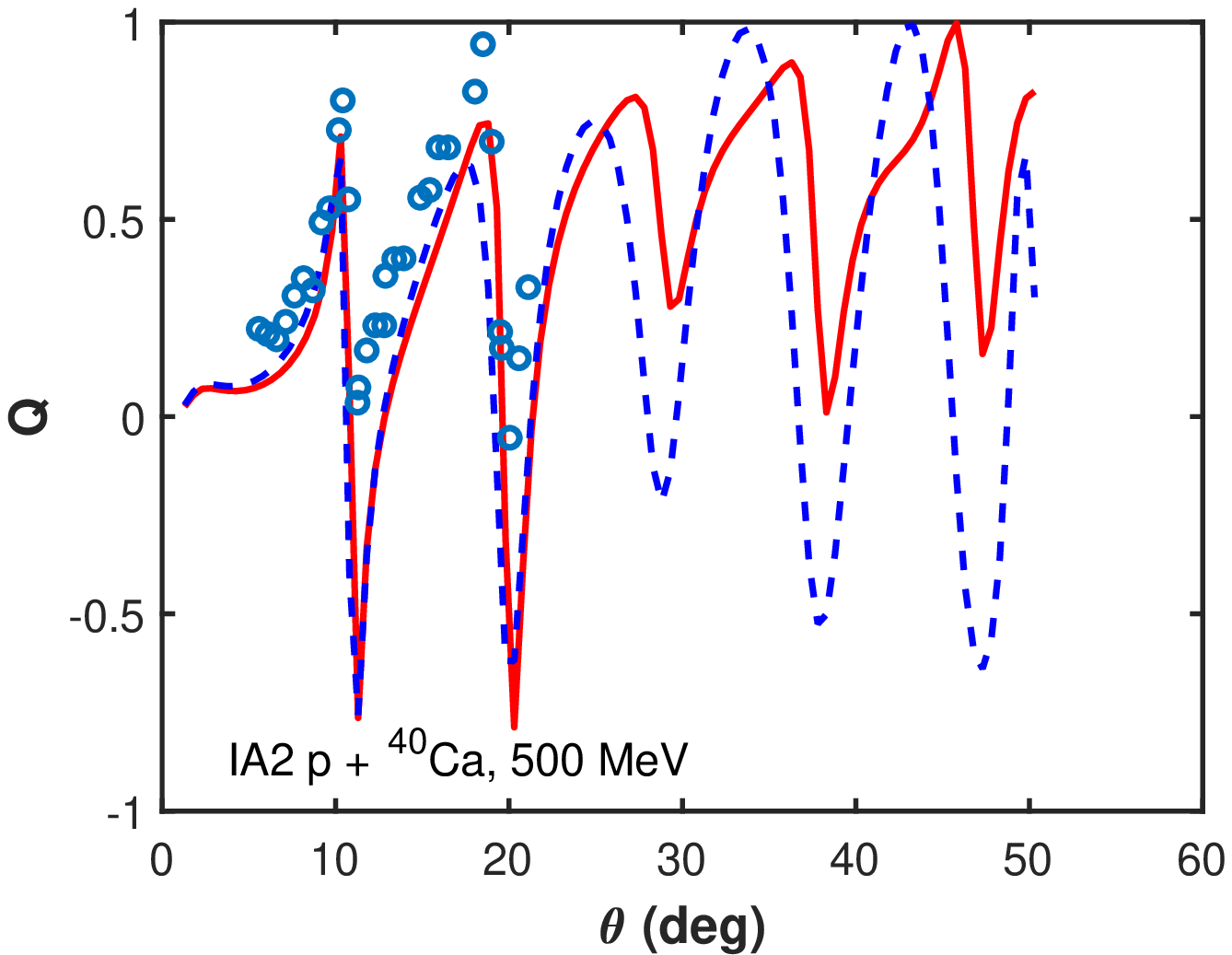}
	\caption{Same as in figure \ref{sigma_posvsmomCa40_200} except for $T_{\mathrm{lab}} = 500$ MeV.}
	\label{sigma_posvsmomCa40_500}
\end{figure}

\begin{figure}
	\centering
	\includegraphics[width=0.49\linewidth]{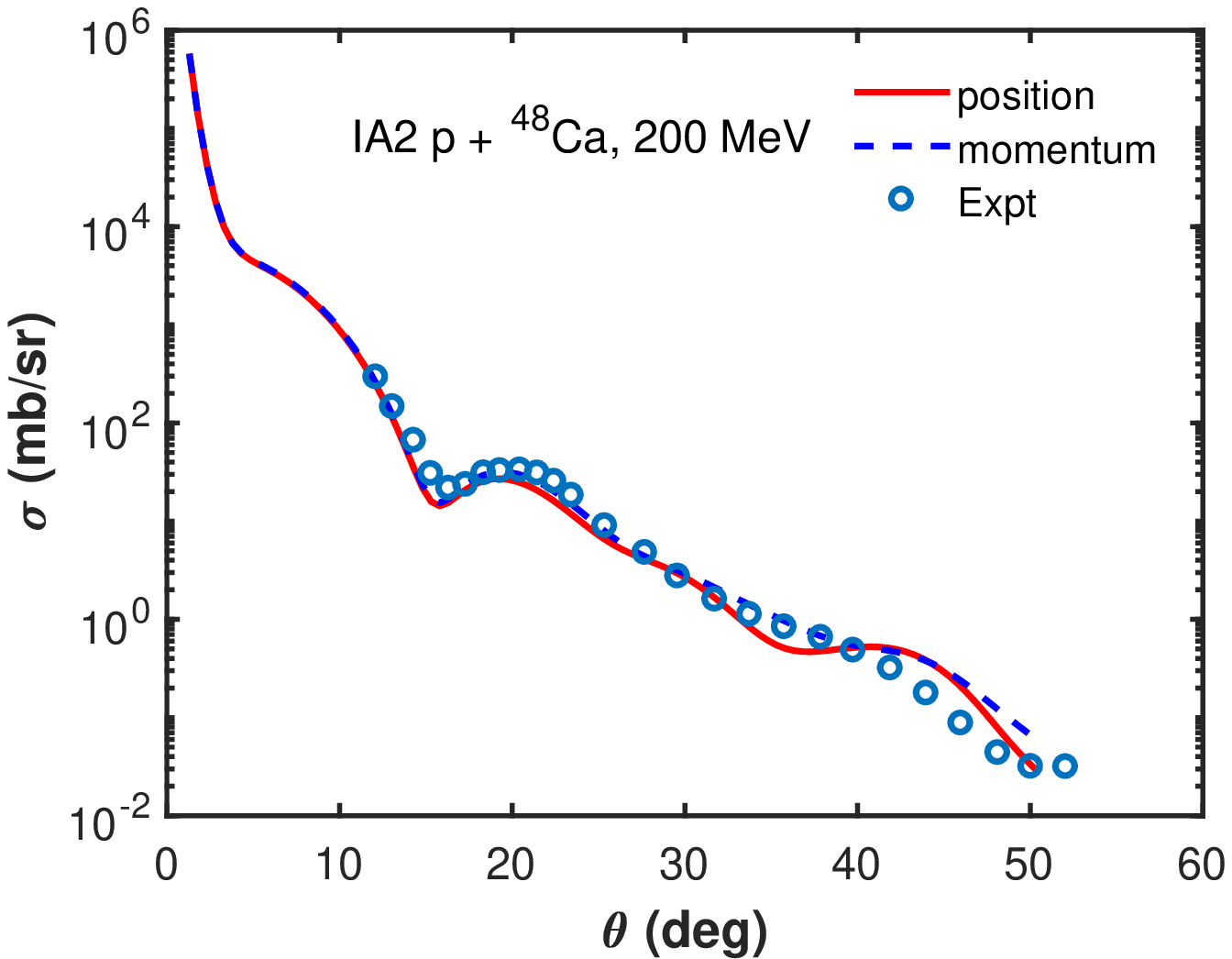}
	\includegraphics[width=0.49\linewidth]{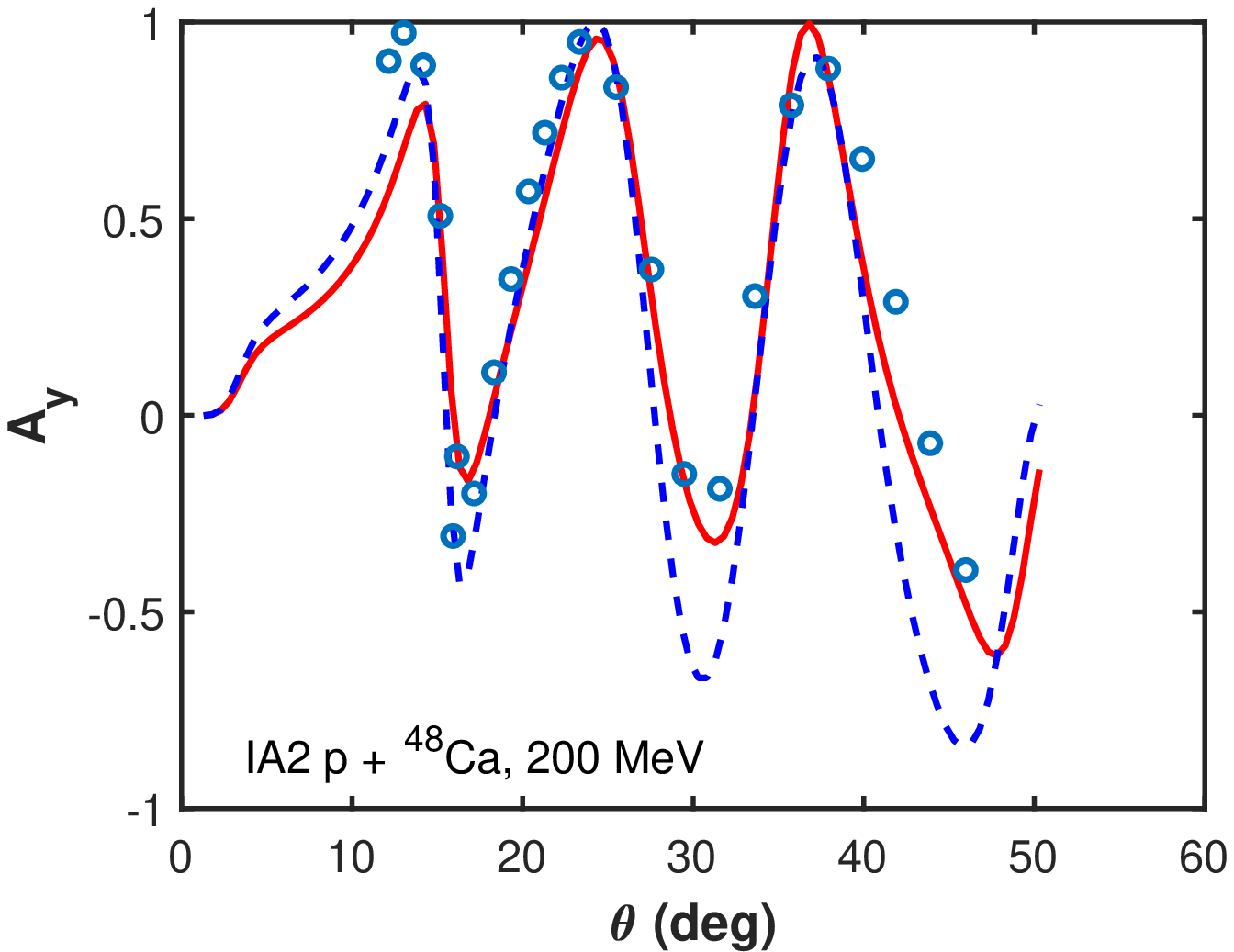}
	\includegraphics[width=0.49\linewidth]{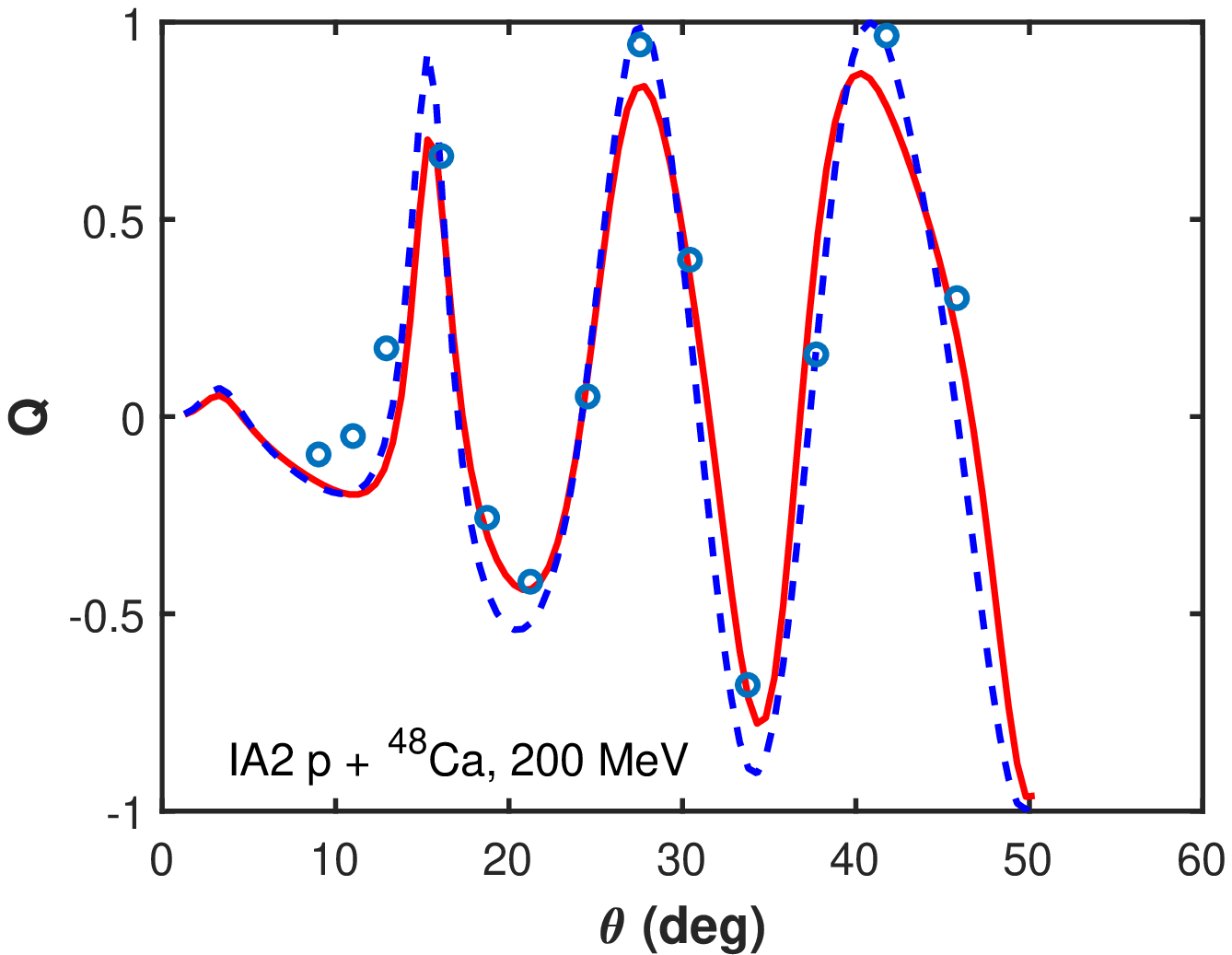}
	\caption{Same as in figure \ref{sigma_posvsmomCa40_200} except for $p + ^{48}$Ca at $T_{\mathrm{lab}} = 200$ MeV.}
	\label{sigma_posvsmomCa48_200}
\end{figure}

\begin{figure}
	\centering
	\includegraphics[width=0.49\linewidth]{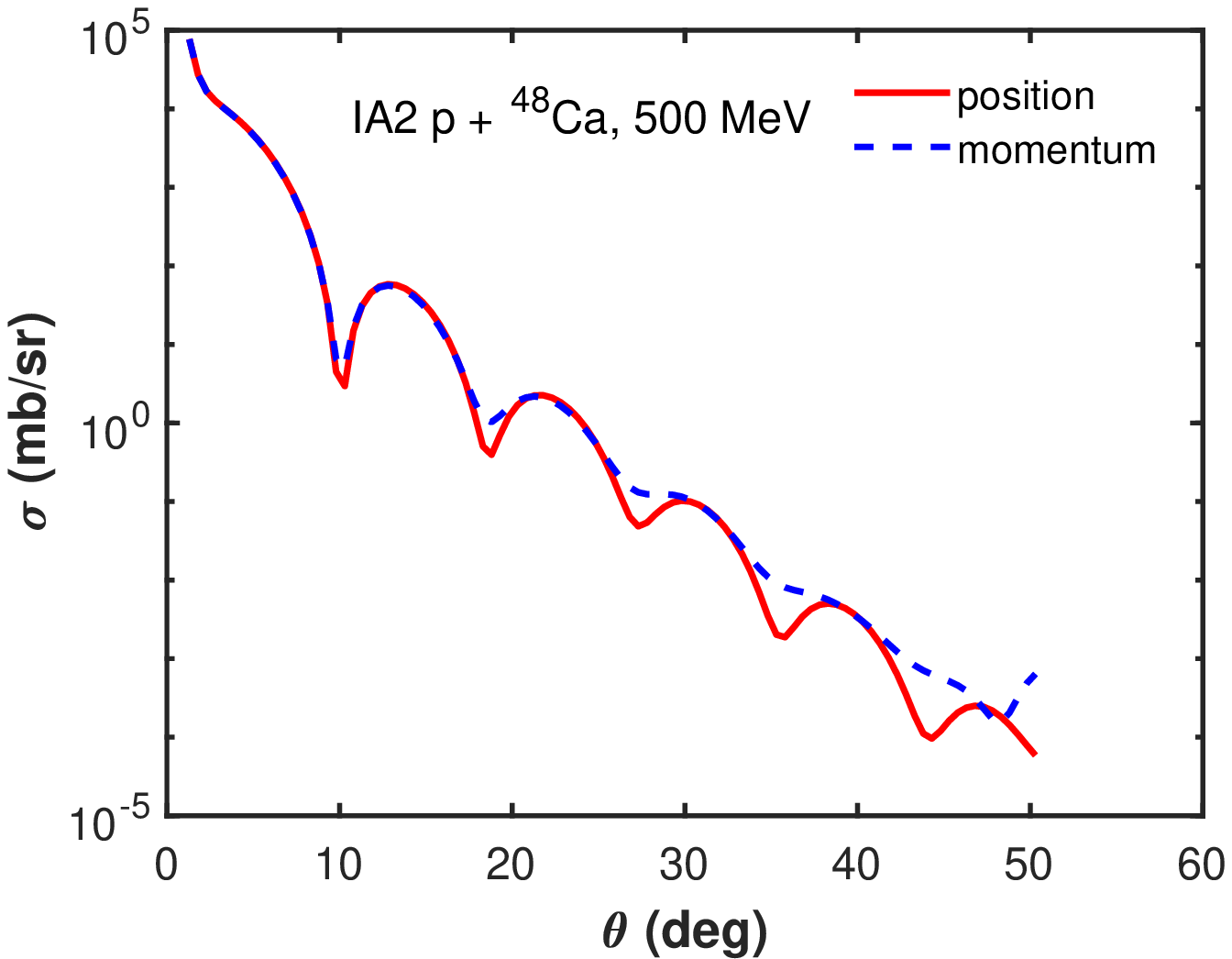}
	\includegraphics[width=0.49\linewidth]{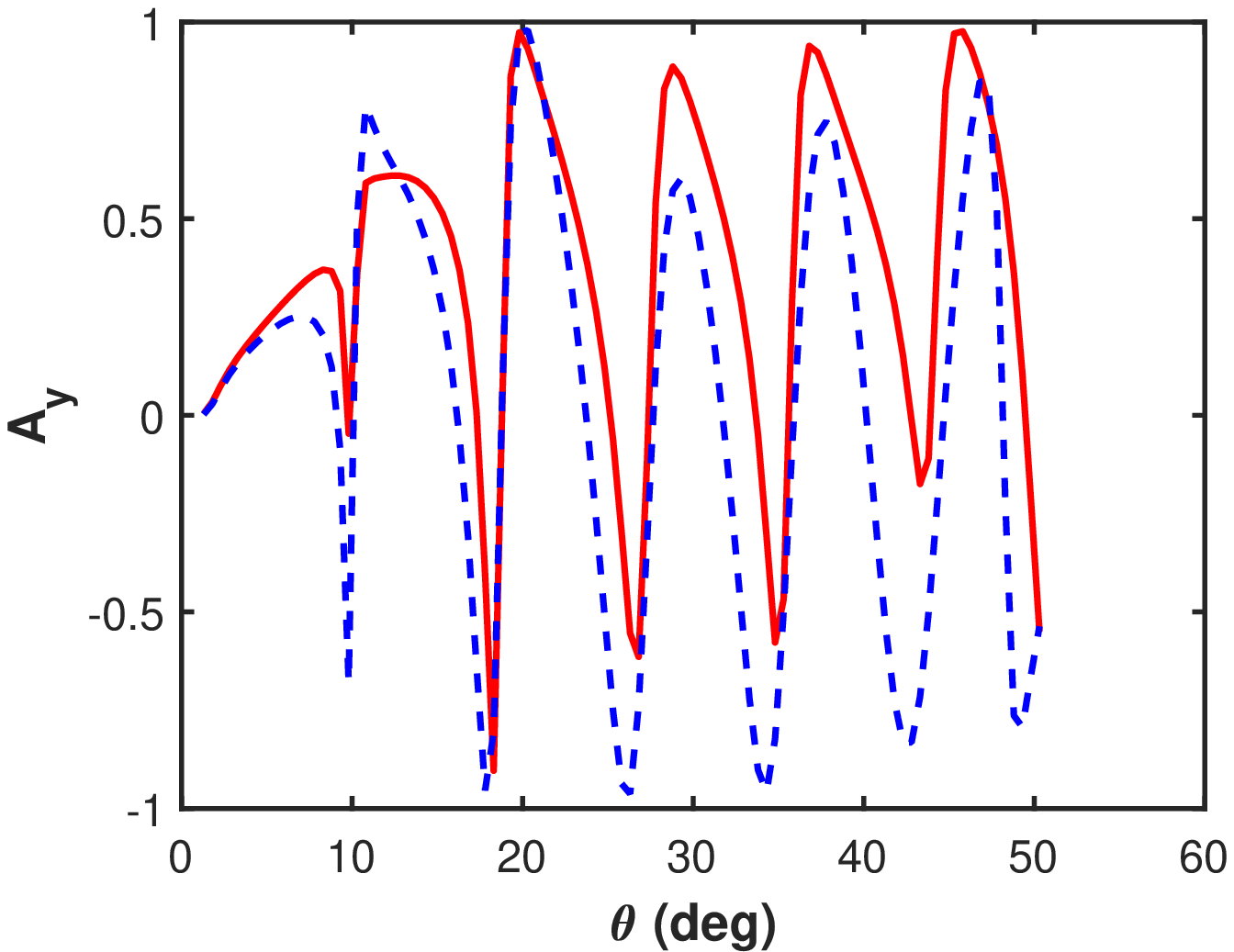}
	\includegraphics[width=0.49\linewidth]{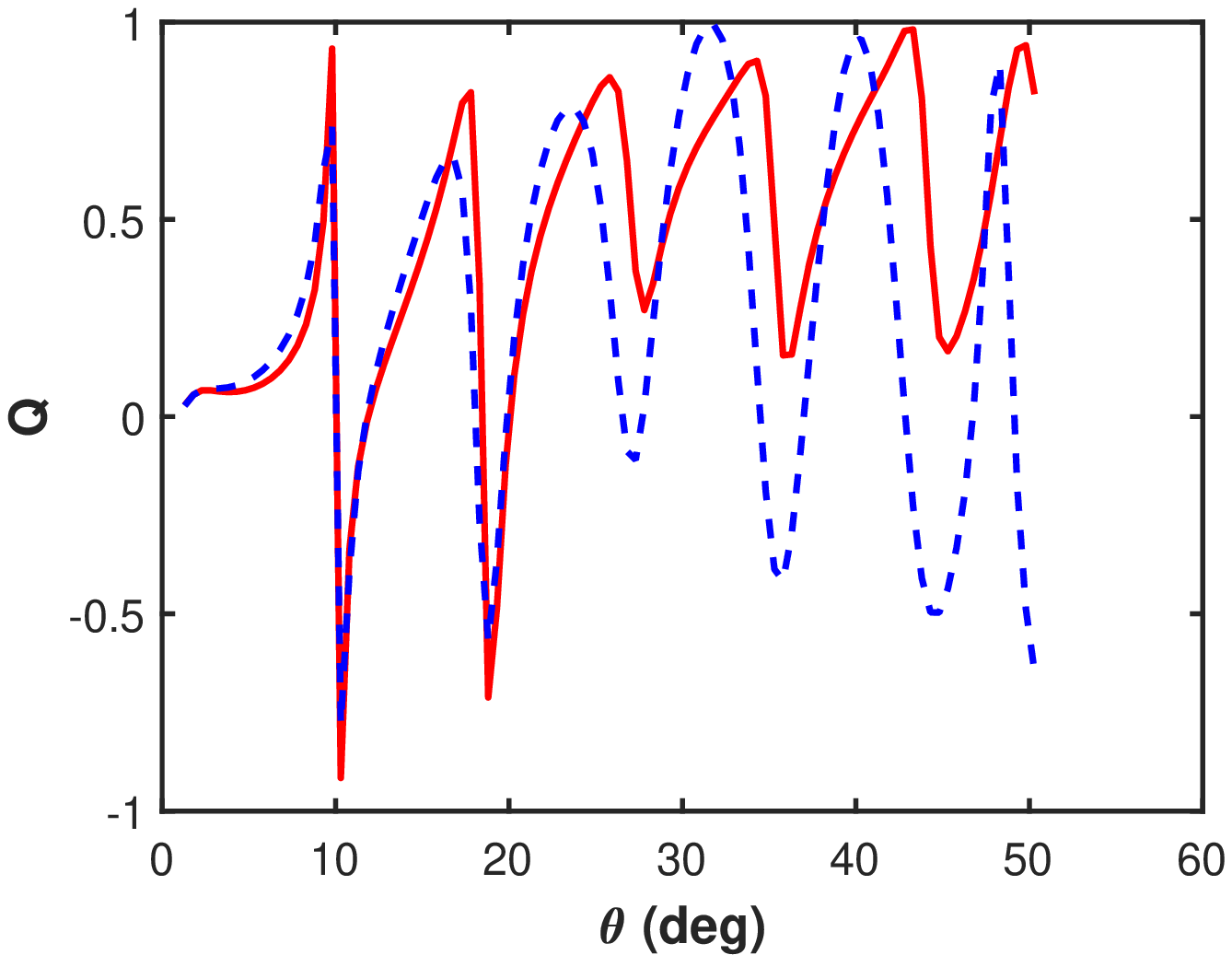}
	\caption{Same as in figure \ref{sigma_posvsmomCa40_200} except for $p + ^{48}$Ca at $T_{\mathrm{lab}} = 500$ MeV.}
	\label{sigma_posvsmomCa48_500}
\end{figure}

\begin{figure}
	\centering
	\includegraphics[width=0.49\linewidth]{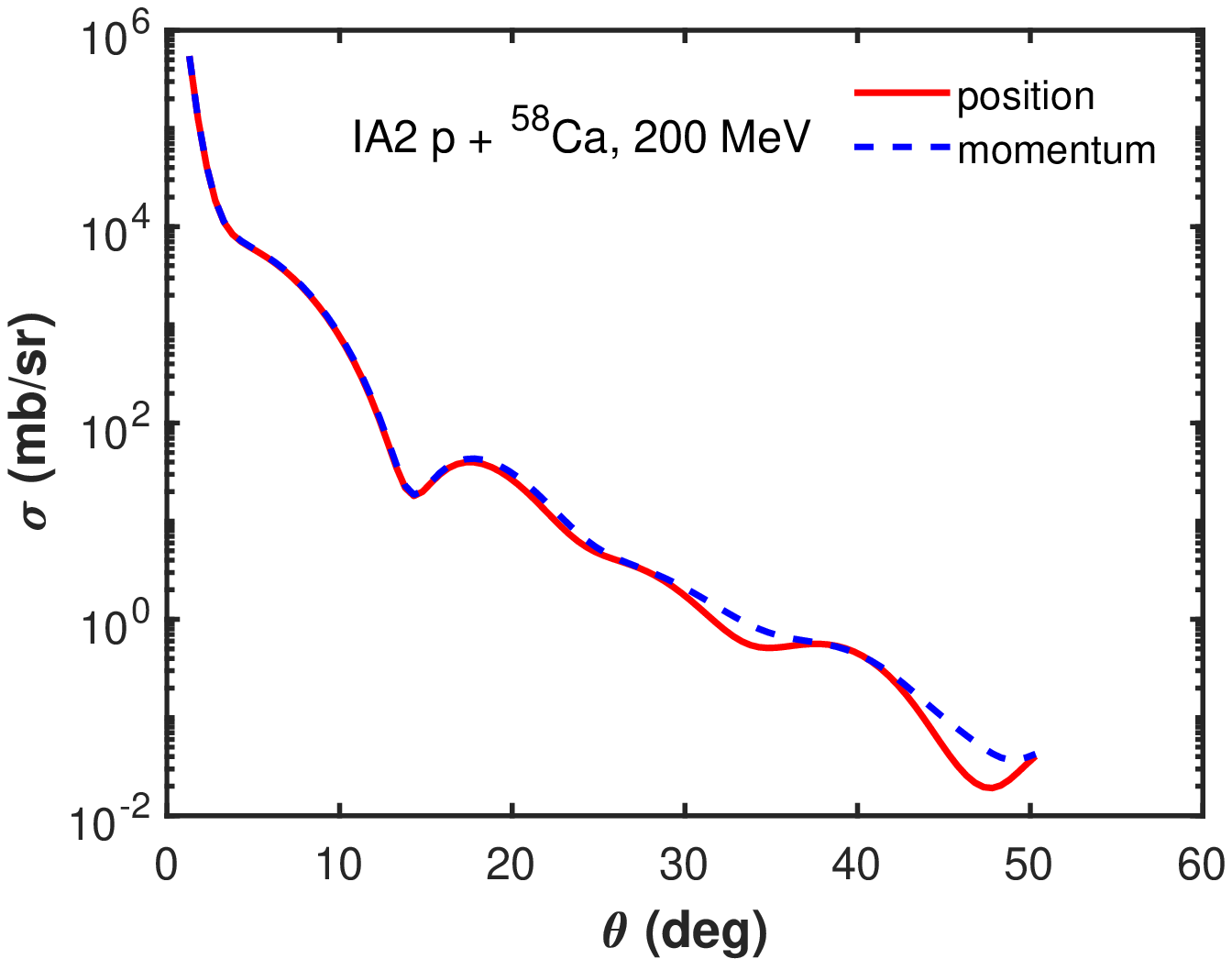}
	\includegraphics[width=0.49\linewidth]{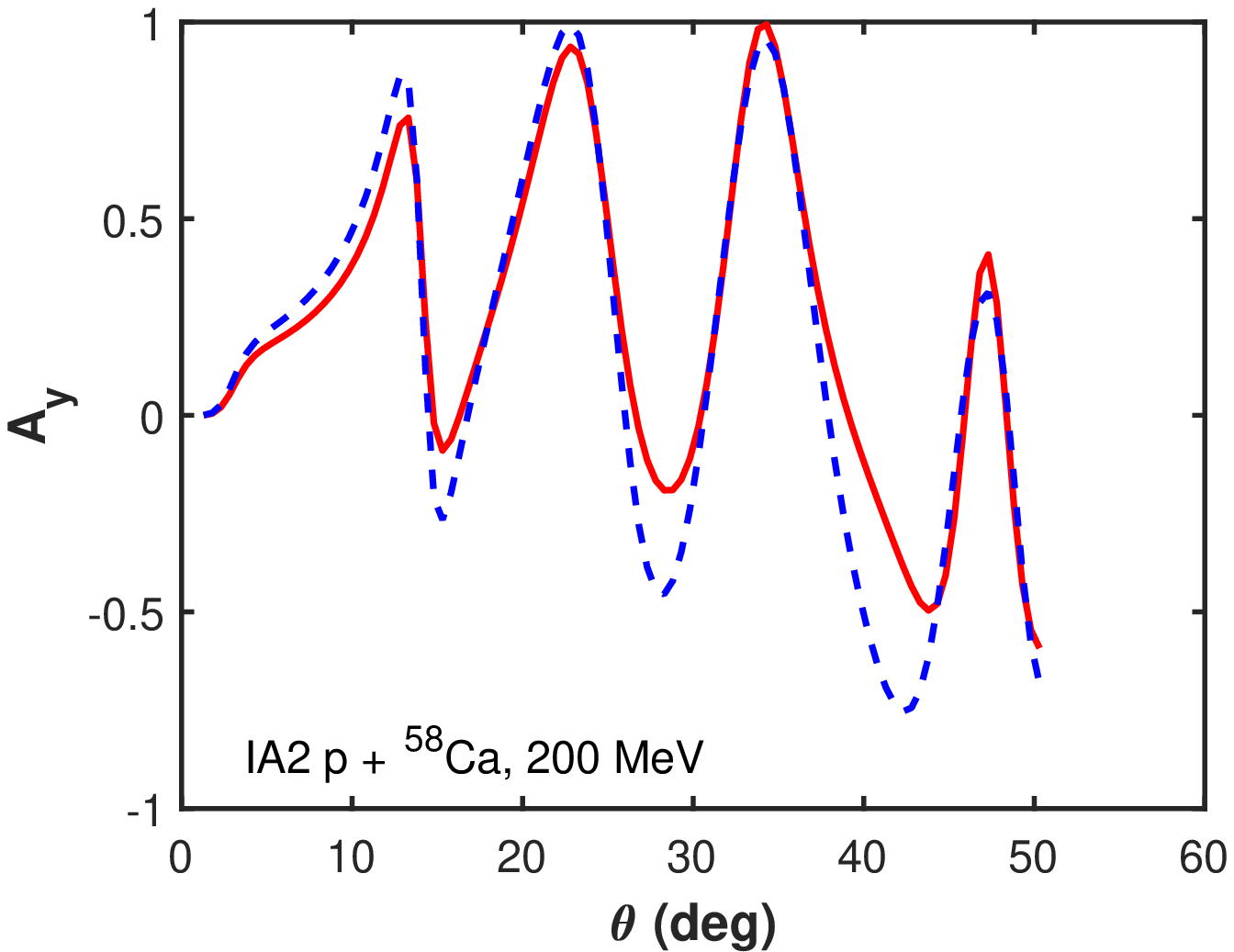}
	\includegraphics[width=0.49\linewidth]{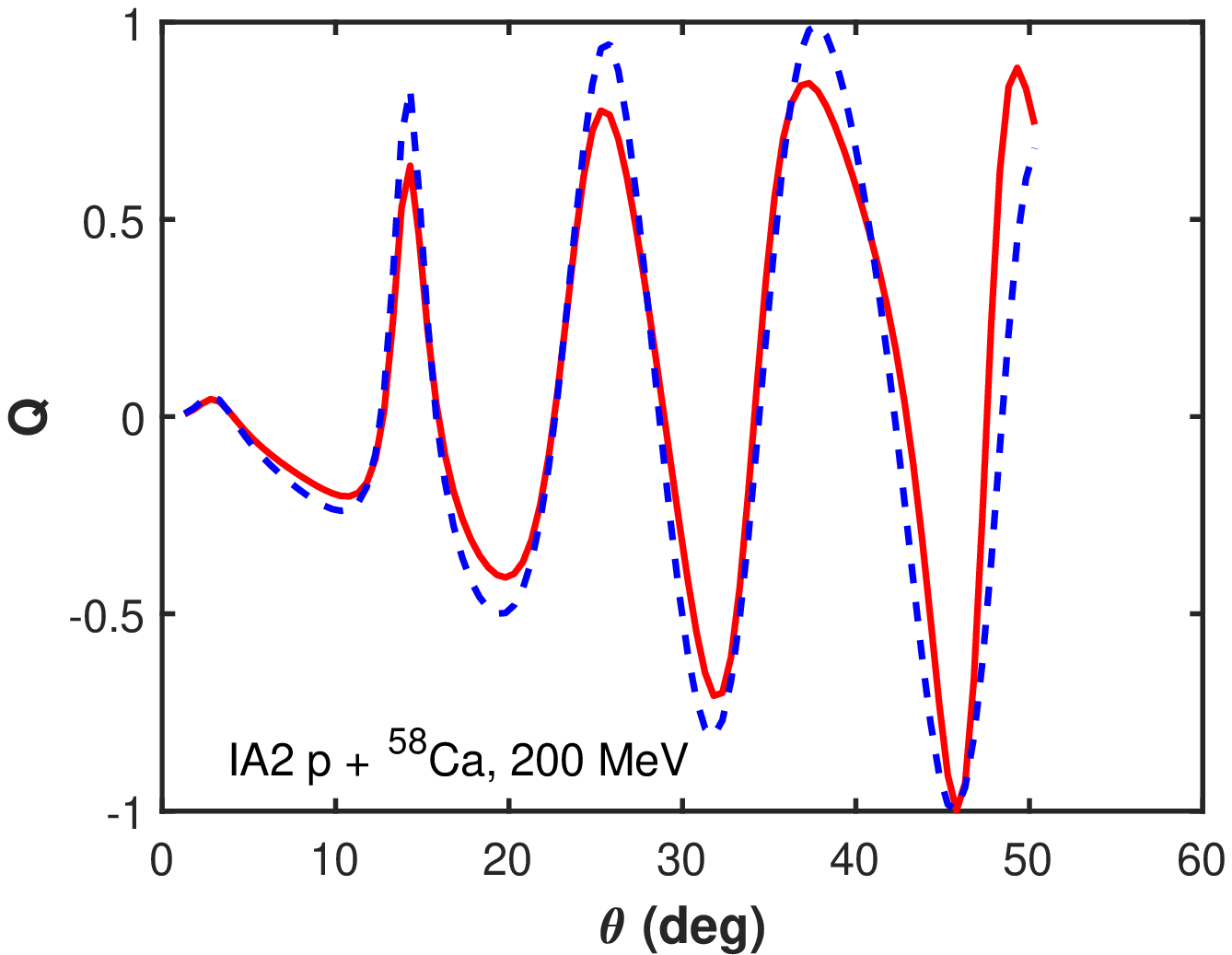}
	\caption{Same as in figure \ref{sigma_posvsmomCa40_200} except for $p + ^{58}$Ca at $T_{\mathrm{lab}} = 200$ MeV.}
	\label{sigma_posvsmomCa58_200}
\end{figure}

\begin{figure}
	\centering
	\includegraphics[width=0.49\linewidth]{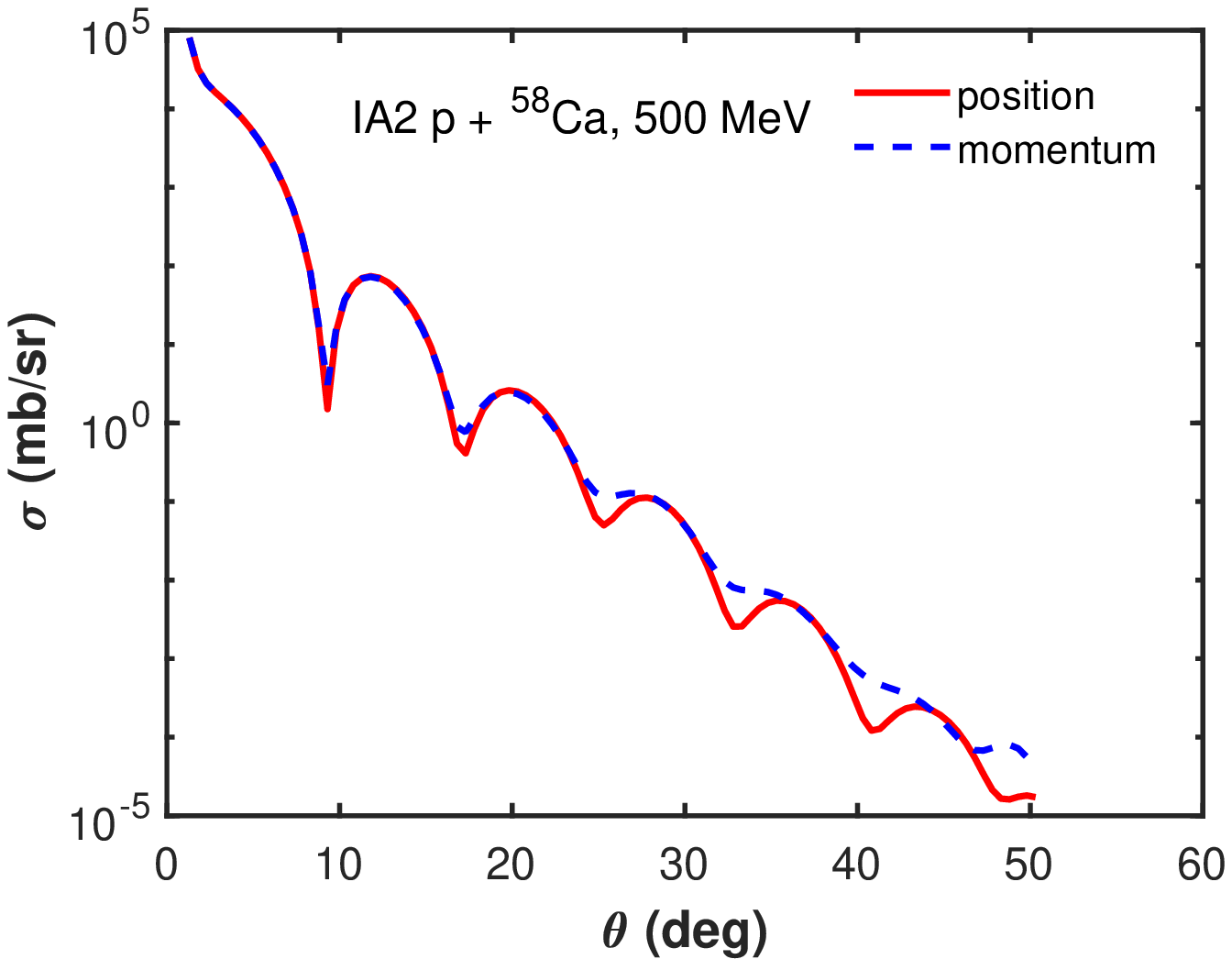}
	\includegraphics[width=0.49\linewidth]{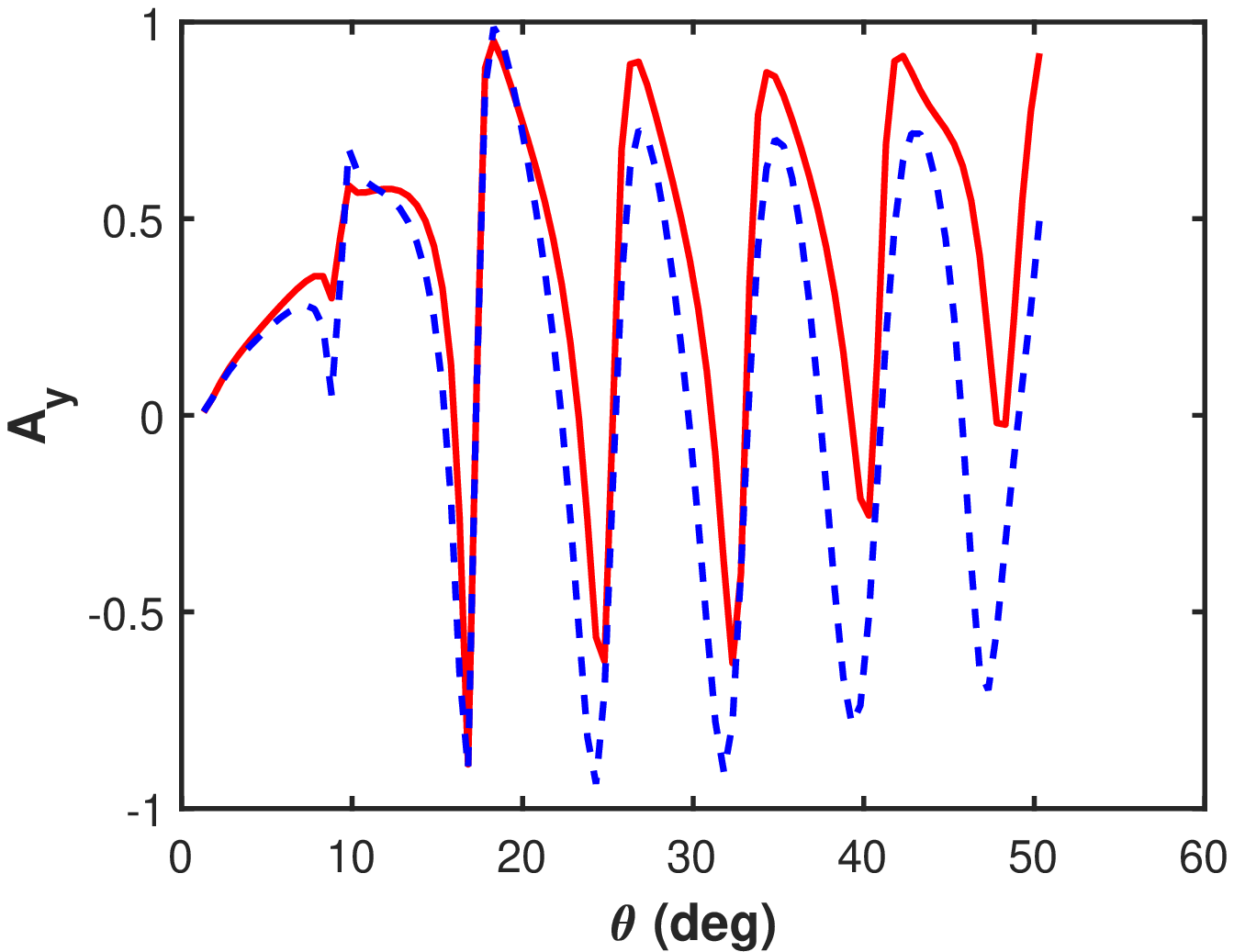}
	\includegraphics[width=0.49\linewidth]{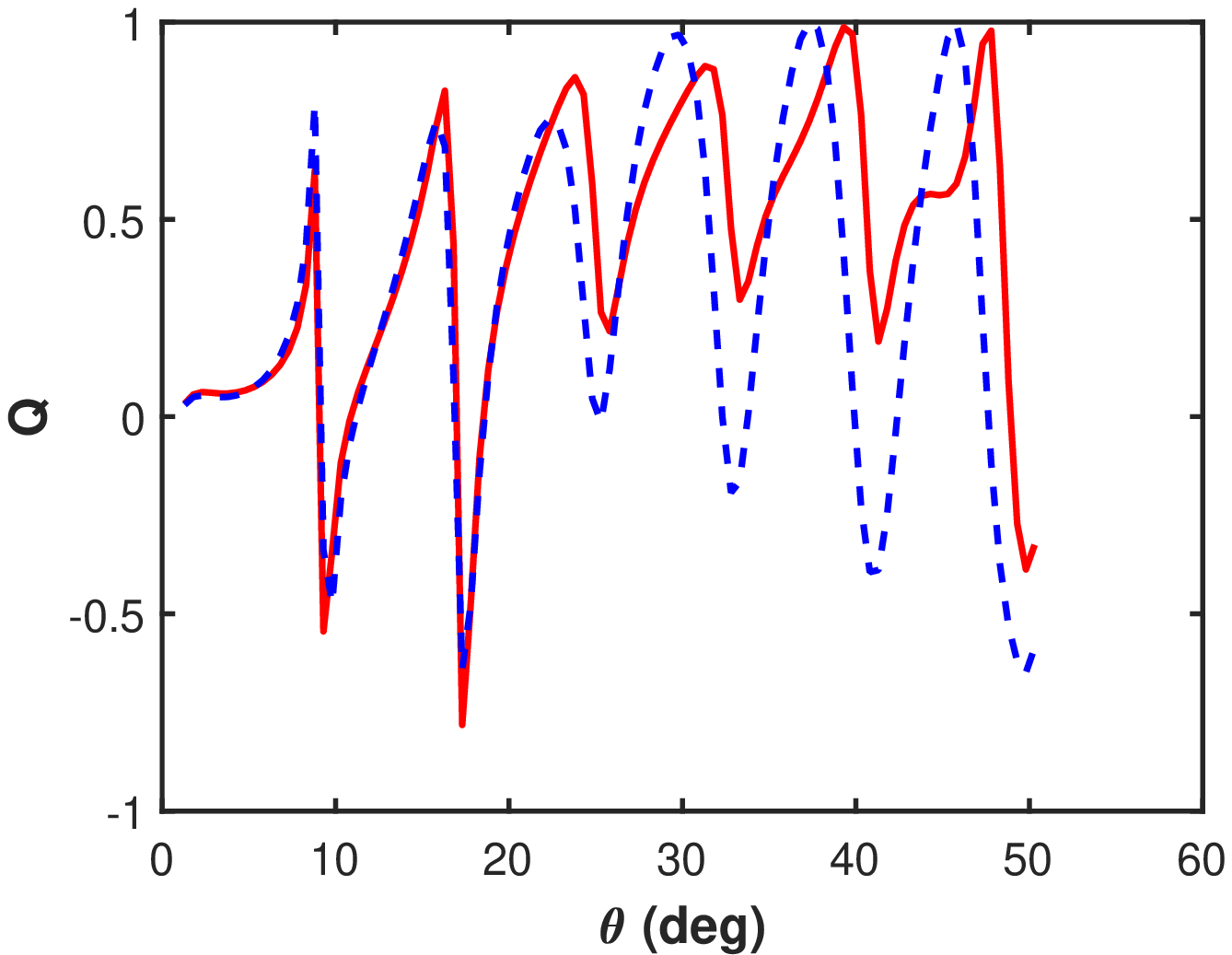}
	\caption{Same as in figure \ref{sigma_posvsmomCa40_200} except for $p + ^{58}$Ca at $T_{\mathrm{lab}} = 500$ MeV.}
	\label{sigma_posvsmomCa58_500}
\end{figure}

\begin{figure}
	\centering
	\includegraphics[width=0.49\linewidth]{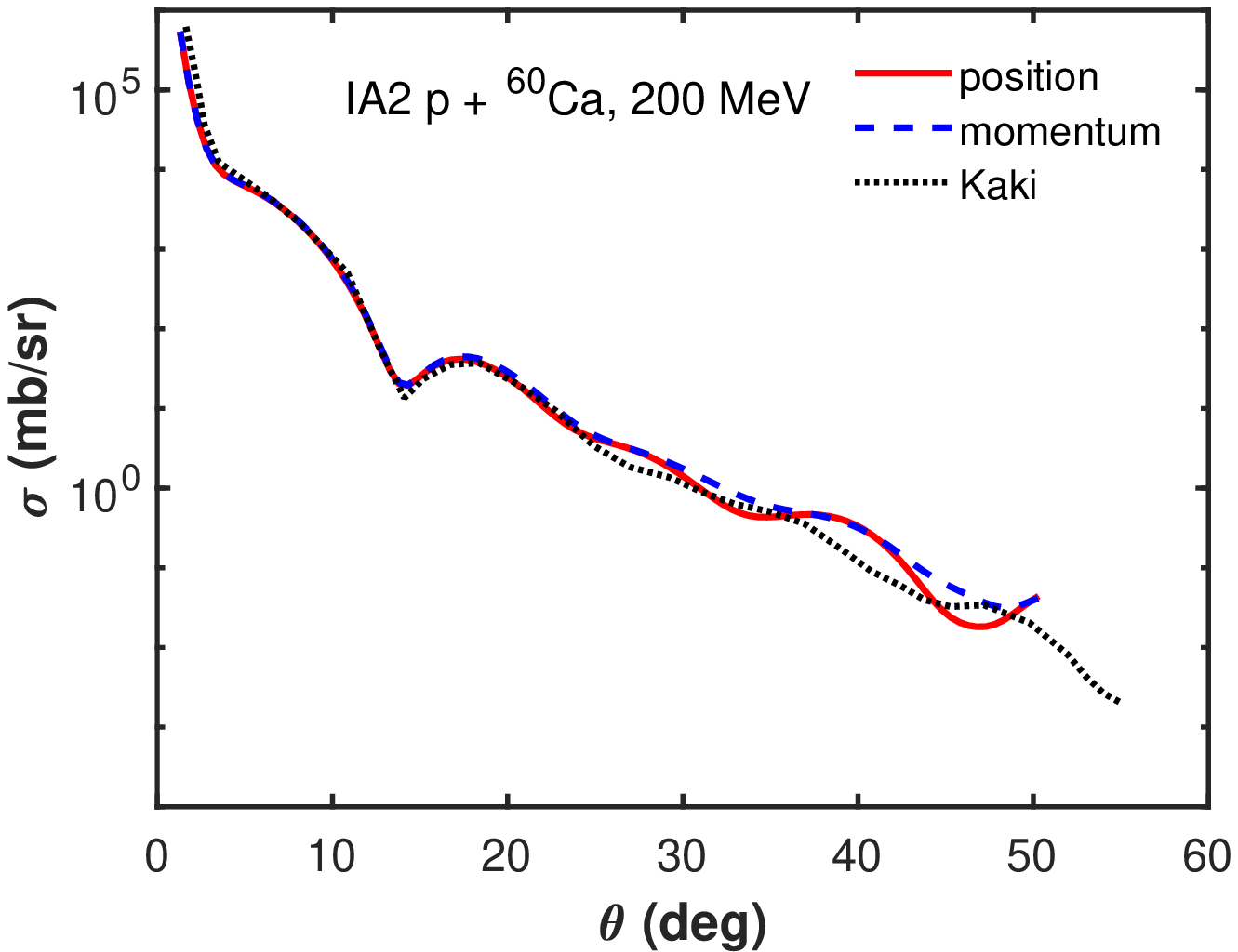}
	\includegraphics[width=0.49\linewidth]{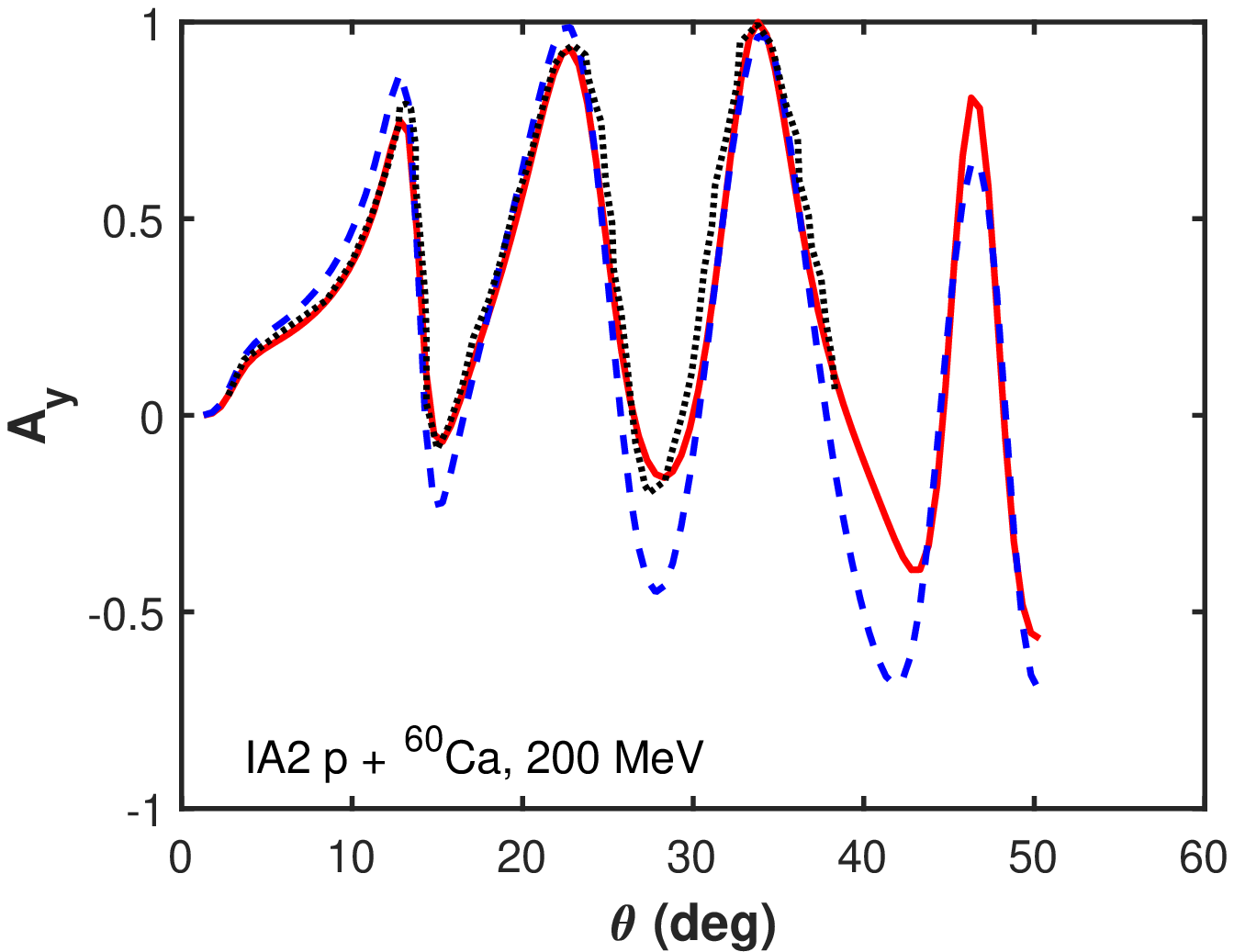}
	\includegraphics[width=0.49\linewidth]{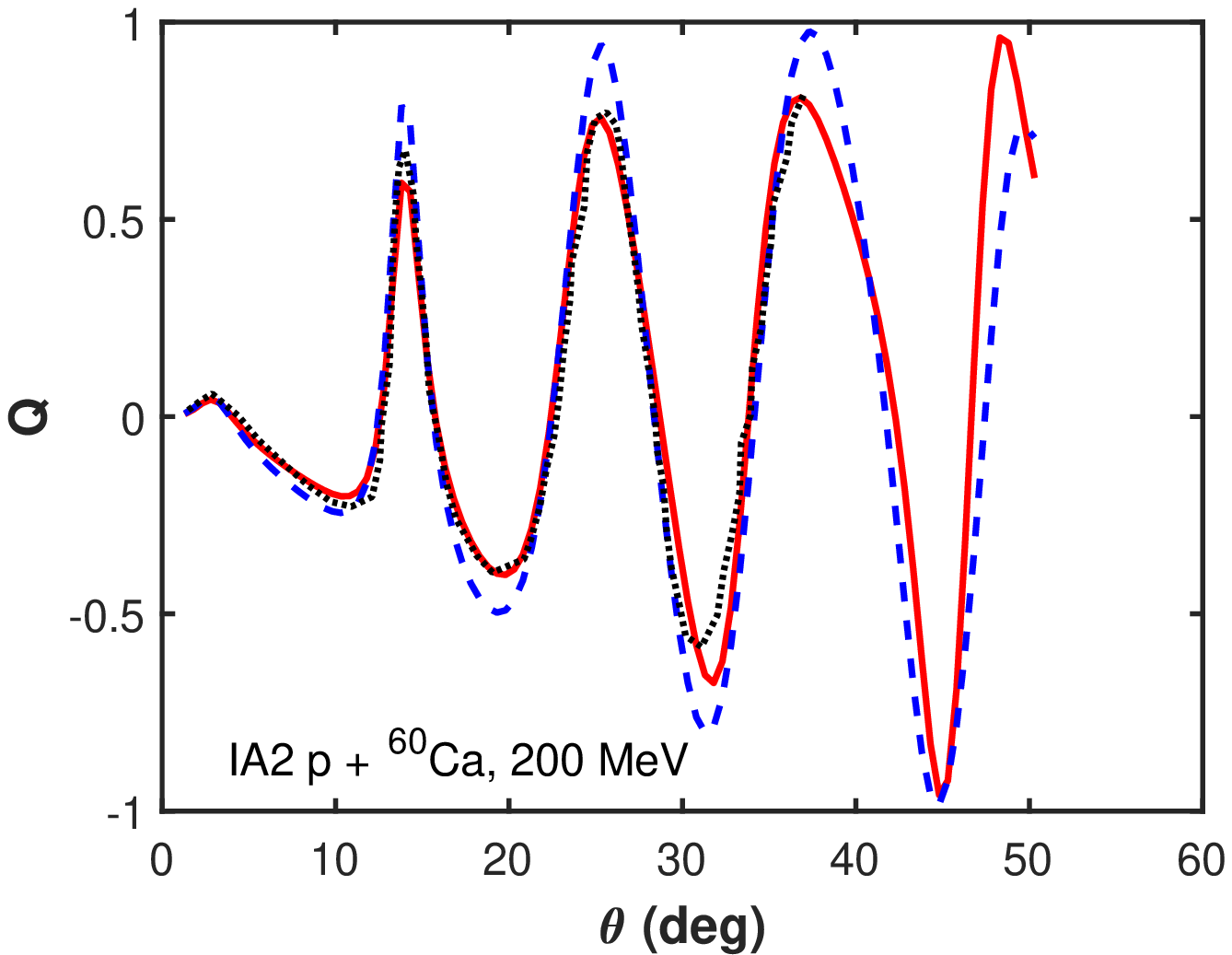}
	\caption{Same as in figure \ref{sigma_posvsmomCa40_200} except for $p + ^{60}$Ca at $T_{\mathrm{lab}} = 200$ MeV.}
	\label{sigma_posvsmomCa60_200}
\end{figure}

\begin{figure}
	\centering
	\includegraphics[width=0.49\linewidth]{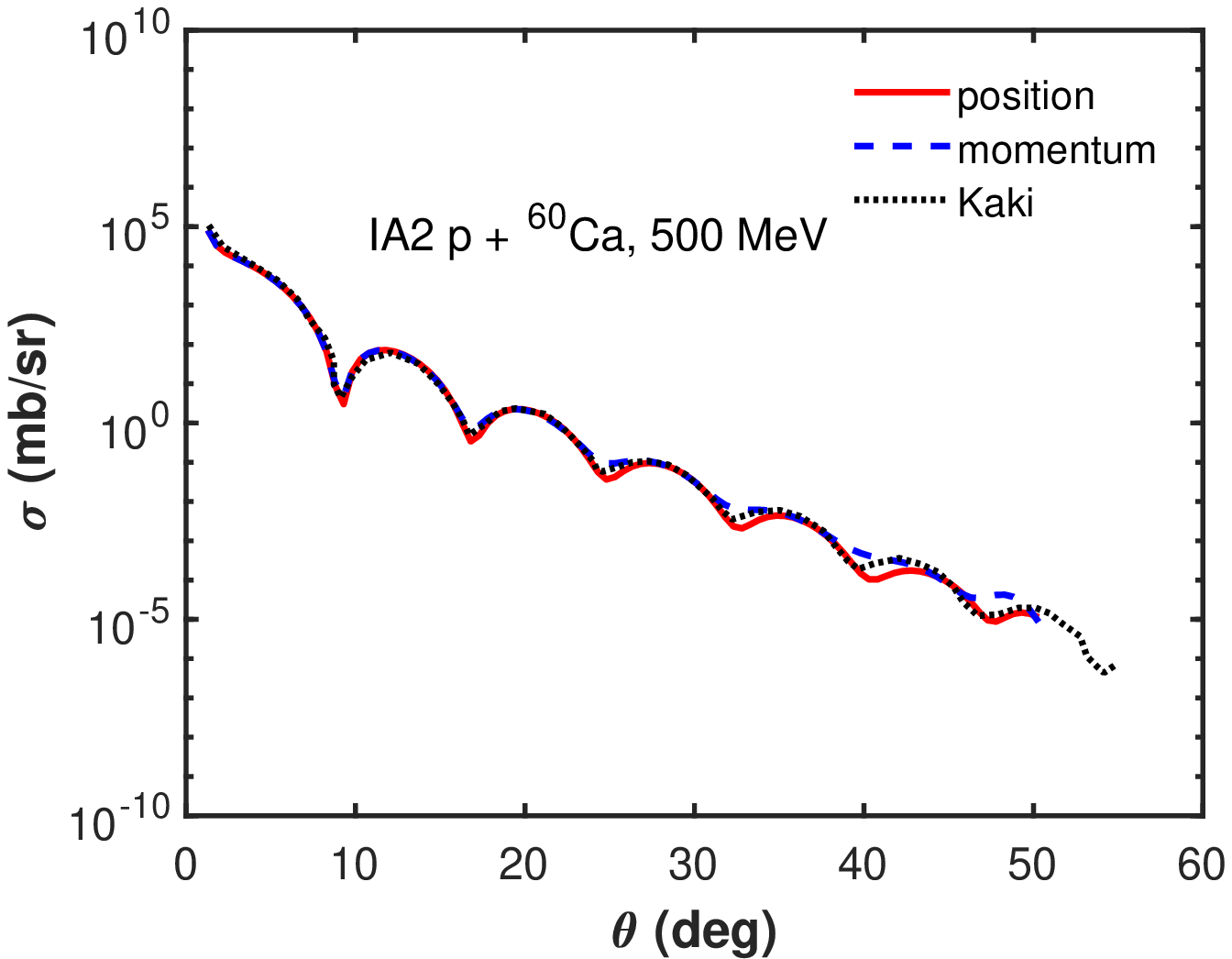}
	\includegraphics[width=0.49\linewidth]{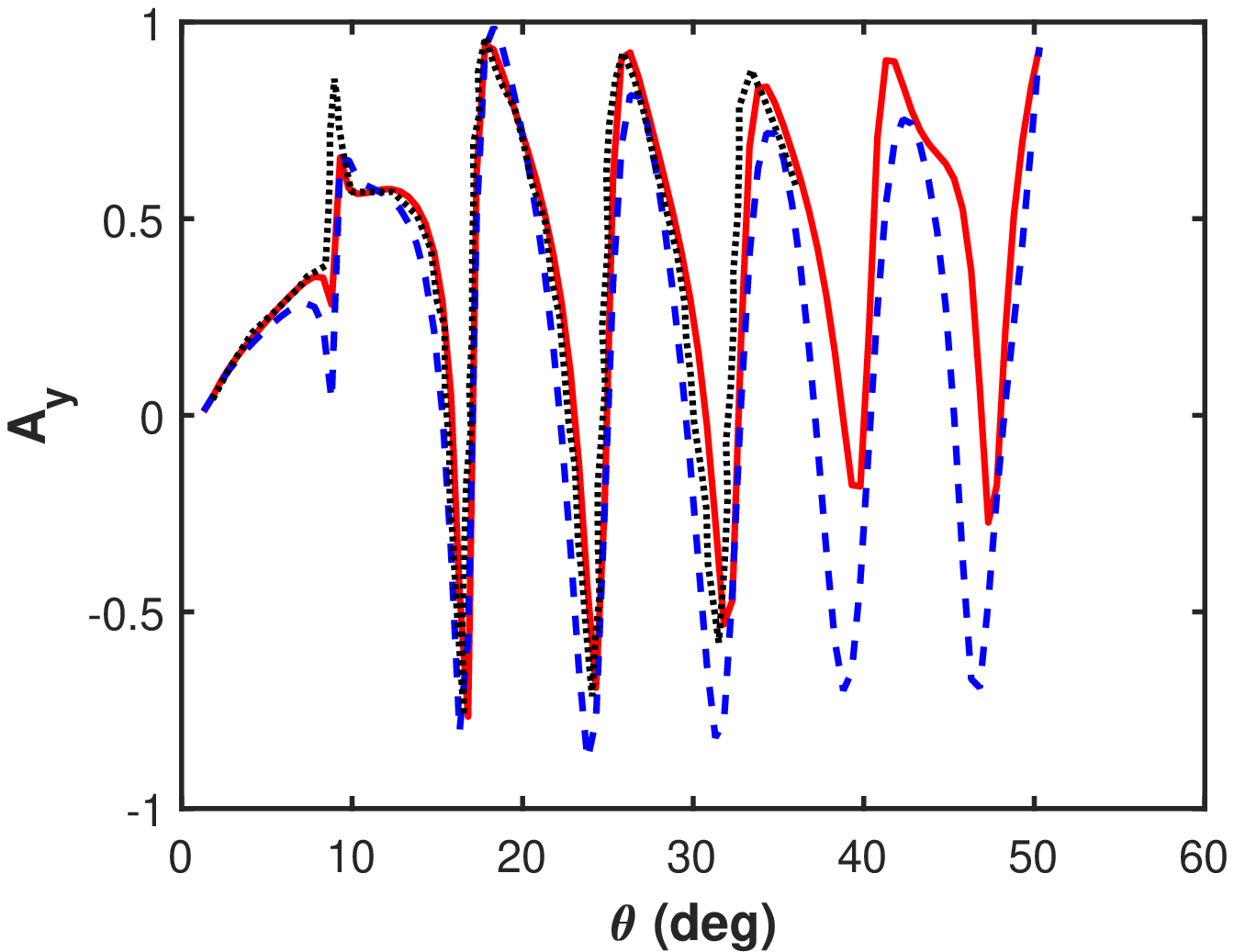}
	\includegraphics[width=0.49\linewidth]{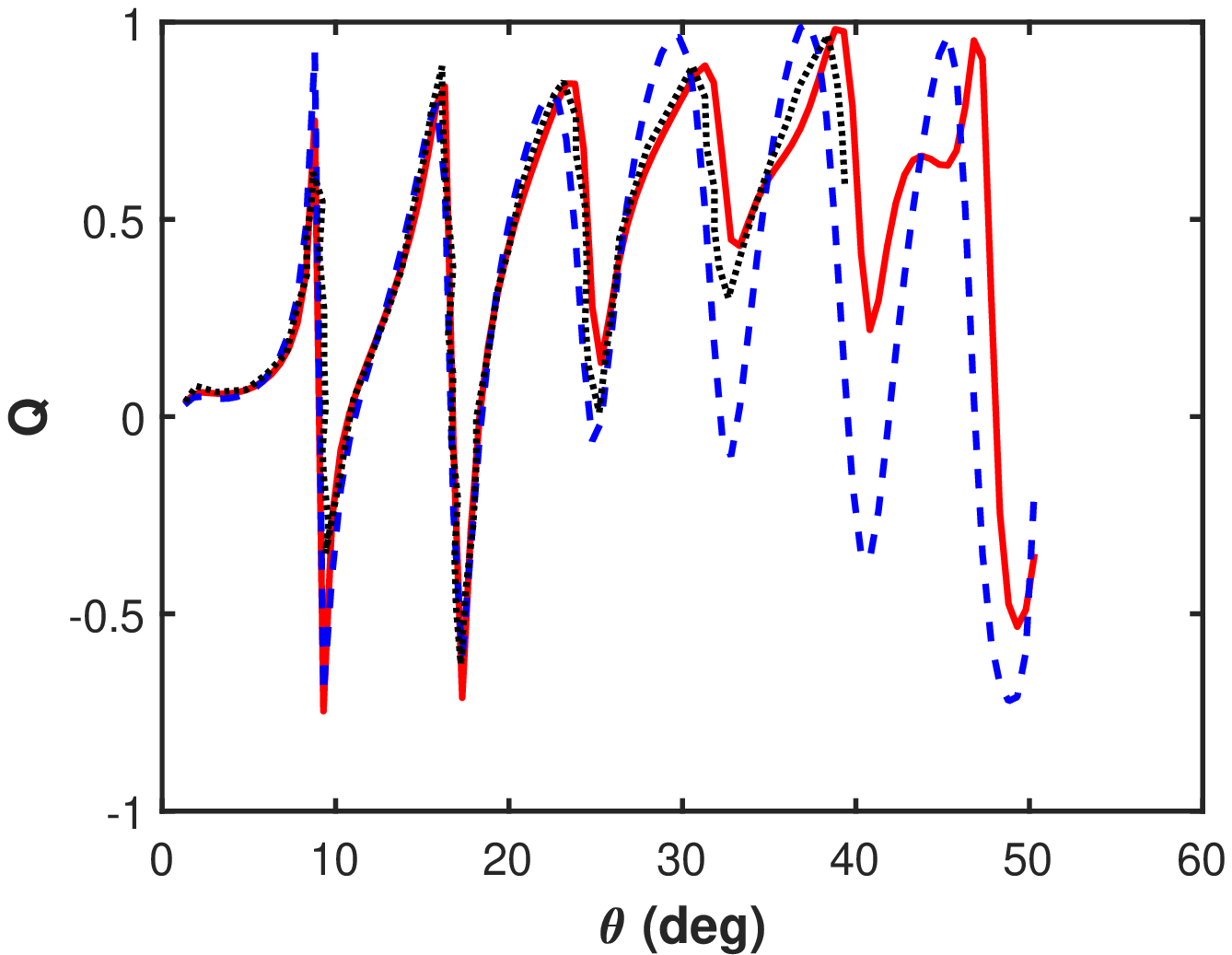}
	\caption{Same as in figure \ref{sigma_posvsmomCa40_200} except for $p + ^{60}$Ca at $T_{\mathrm{lab}} = 500$ MeV.}
	\label{sigma_posvsmomCa60_500}
\end{figure}

Figures \ref{sigma_posvsmomCa40_200} -- \ref{sigma_posvsmomCa60_500} show results of the proton elastic scattering observables calculated in position space (using localised IA2 optical potentials) and in momentum space (using non-local IA2 optical potentials) for $^{40,48,58,60}$Ca targets at incident projectile energies of 200 and 500 MeV. The FSUGold parametrisation was used in the calculations of the relativistic densities. Solid lines indicate position space calculations using localised optical potentials while dashed lines indicate momentum space calculations using non-local optical potentials. Experimental data are shown in black circles. The top left plots in each figure show the scattering cross section results, the top right show the analysing power results, while the bottom plots show the spin rotation parameters. 

One observes from figures \ref{sigma_posvsmomCa40_200} and \ref{sigma_posvsmomCa48_200} that for $^{40,48}$Ca targets, both local and non-local optical potentials give very good descriptions of the differential cross section data, however there are competitive descriptions of the spin observables. The use of non-local optical potentials give better descriptions of the analysing powers at first maxima and minima but the local potentials give better descriptions afterwards. At incident projectile energy of $500$ MeV shown in figures  \ref{sigma_posvsmomCa40_200} and \ref{sigma_posvsmomCa48_200}, both approaches give similar descriptions of the scattering observables data. The obvious difference between both formalisms in describing the scattering observables is noticed at large scattering angles where there is no available experimental data. The same conclusions can be drawn for $p + ^{58,60}$Ca at incident projectile energies of $200$ MeV and $500$ MeV shown in figures \ref{sigma_posvsmomCa58_200}, \ref{sigma_posvsmomCa58_500}, \ref{sigma_posvsmomCa60_200} and \ref{sigma_posvsmomCa60_500}. There are similar descriptions of the three scattering observables at incident projectile energy of $200$ MeV. There is an obvious difference between the two methods in describing the spin observables at large scattering angles when the incident proton energy is $500$ MeV.

\section{Conclusion}
We have presented a microscopic study of proton elastic scattering from unstable nuclei using a relativistic formalism. The densities are calculated using bound state wave functions obtained from relativistic mean field theory, employing the NL3 and FSUGold parameter sets. The different RMF parameter sets give different descriptions of the neutron densities at and close to the interior of the unstable nuclei studied. Up to one decimal place, the RMF models give similar values of the charge densities for calcium isotopes considered here.

The microscopic relativistic optical potentials are calculated using both the IA1 and IA2 formalisms. A comparison of the IA1, IA2, and Dirac phenomenology optical potentials shows that the IA2 formalism gives the lowest scalar and vector potential strengths at incident projectile energies of 200 MeV and 500 MeV. At 200 MeV, the IA1 formalism gives potential strengths stronger than Dirac phenomenology for both stable and unstable nuclei. The overly strong scalar and vector optical potentials produced by the IA1 formalism at this low energy (200 MeV) has been attributed to the implicit incorporation of pseudoscalar pion coupling. The optical potentials calculated using optimal factorization are also compared with those obtained with full-folding optical potentials. The effect of using full-folding optical potentials is found at an incident projectile energy of 200 MeV, while there is no noticeable difference at 500 MeV and above.

The calculated optical potentials are used as inputs in the Dirac equation. The non-local optical potentials are used in the momentum space Dirac equation while the localised optical potentials are substituted into the coordinate-space Dirac equation. We have decided to use the two approaches to investigate the effect of using non-local optical potentials on the elastic scattering observables for unstable nuclei. After solving the position space Dirac equation, elastic scattering observables were calculated for $^{40,48,58,60}$Ca targets. In order to check the sensitivity of elastic scattering observables to different RMF densities, we showed plots of the scattering observables with two RMF densities. Except at large scattering angles, the two model densities give similar descriptions of the elastic scattering observables for both stable and unstable nuclei considered in this work.

The results of elastic scattering observables computed using IA1 and IA2 formalisms are compared. At incident projectile energy of 500 MeV, both formalisms give similar descriptions of the elastic scattering observables for both stable and unstable nuclei at low scattering angles, but at large scattering angles, the difference between both formalisms becomes obvious. At incident projectile energy of 200 MeV however, the IA2 formalism gives a better description of the scattering observables for both stable and unstable nuclei. The inability of IA1 formalism to give proper descriptions of the scattering observables at incident projectile energies $\lessapprox$ 200 MeV is due to the large scalar and vector optical potentials it gives at low energies.

We also discussed effect of full-folding optical potentials on the scattering observables compared with the calculations using optimally factorised optical potentials. We found that the use of full-folding optical potentials improve the spin observables (analysing power and spin rotation function) at incident projectile energy of 200 MeV for the calcium isotopes, while there is no discernible difference at 500 MeV. However for $^{120}$Sn, there is not much difference in the scattering observables calculated using optimally factorised and full-folding optical potentials.

Finally, we studied elastic scattering observables calculated using non-local optical potentials. To achieve this we substituted the non-local optical potential into momentum space Dirac equation, which is then transformed to two coupled integral equations. The transformation is necessary because the scattering observables are connected to the T-matrix. The treatment of solutions at high angular momentum states is done using high-order global adaptive quadratures to solve the oscillatory integrals encountered at high angular momenta. This approach is sufficient for the nuclei studied in this work and at incident projectile energies up to $\lessapprox 500$ MeV. Matrix inversion technique was used to solve the coupled integral equations, from which the elastic scattering observables are computed. We observed that results of momentum space calculations using non-local optical potentials give better descriptions of the spin observables at incident projectile energy of 200 MeV. There is a competitive description of the scattering observables data at incident projectile energy of 500 MeV between the two approaches.


\begin{acknowledgments}
	W. A. Yahya acknowledges iThemba LABS partial funding, and appreciates Kwara State University for granting permission to carry out research at Stellenbosch University. 
\end{acknowledgments}

\bibliography{yahya}

\end{document}